# The Evolution of Security Prices Is Not Stochastic but Governed by a Physicomathematical Law

Wally Tzara[1]

Draft date: July 2019

## ABSTRACT

Since Louis Bachelier's thesis in 1900 (in which he lays the foundation of the stochastic process, or Brownian motion, as a model of stock price changes, and thus making him the father of mathematical finance), attempts at understanding the nature of stock market prices and at predicting them have failed. Statistical methods have only found minor regularities and anomalies, and other mathematical and physical approaches do not work. This leads researchers to consider that the evolution of security prices is basically random, and, thus, inherently not predictable. We show for the first time that the evolution of security prices is not at all a stochastic process but is largely deterministic and governed by a physical law. The law takes the form of a physicomathematical theory centered around a purely mathematical function. This function is unrelated to models and to statistical methods. It can be described as an "isodense" network of "moving" regression curves of an order greater than or equal to 1. The salient aspect of the function is that, when inputting a time series of any security into the function, new mathematical objects emerge spontaneously, and these objects exhibit the unique property of attracting and repelling the quantity. The graphical representation of the function is called a "topological network" because of the preeminence of shapes over metrics, and the emergent objects are called "characteristic figures" (the main ones named "cords", due to their appearance). The attraction and repulsion of the price by the cords results in the price bouncing from cord to cord. Therefore, the price has to be considered as driven by the cords in a semi-deterministic manner (leaning towards deterministic). Having a function that describes the evolution of the price, we now understand the reason behind each price movement and can also predict stock prices both qualitatively and quantitatively. The function is universal, does not rely on any fitting, and, due to its extreme sensitivity, reveals the hidden order present in financial time series data that existing research never uncovered. The extreme sensitivity of the function can be put to use to understand and predict the evolution of other physical quantities.

In order to limit the file size of this document, figures in the appendices are of reduced image quality. For a version with better quality images (30 MB file), please contact the author.

[1] Independent researcher. Email: wally@tzaranet.com

# Table of Contents

# The Evolution of Security Prices Is Not Stochastic but Governed by a Physicomathematical Law

## I.    Introduction

Since Bachelier's thesis in 1900 [1], and up until today, the field of mathematical finance has relied on one hypothesis: security price fluctuations are stochastic (known as the "random walk hypothesis") [2-4]. This hypothesis was reinforced by the formulation in 1970 of the Efficient Market Hypothesis (EMH), which has become the current paradigm. It states that security prices fully reflect all available information, which is considered by some as close to the random walk hypothesis. In more recent years, physicists have become more interested in the problem [5-10]. Normally, physicists are not looking for models, but for empirical laws and theories; however, the econophysicists too are working under the assumption that the evolution of financial time series follows a random walk [11] (so much so that Voit [7] calls it "the standard model of finance"), using statistical models, despite evidence that these models do not account for what is observed in reality [12]. This could be due to the fact that, when considering any given financial time series, the research cannot show that its evolution is not random. Needless to say that under such a hypothesis, the evolution of any financial time series cannot be predicted.

It is clear that as long as one has not found a physical law that governs the evolution of security prices, and considering the results of the statistical tests, the temptation to conclude that it is random is great. However, the evolution of security prices being random is an assertion by default. But this conclusion is logically incorrect (the absence of proof of non-randomness is not proof of the absence of non-randomness, that is, proof of randomness). The conclusion should rather read: for now, nothing leads us to believe that the evolution of security prices is not non-random.

Some researchers reject the hypothesis of a random evolution of security prices, not because they proved that the evolution is non-random, but because they showed that it is slightly not random for certain securities at certain times. For example, certain statistical tests reveal a certain degree of dependence (positive autocorrelations) in specific stock market returns, but they correspond only to weak trends over short timeframes [13]. Moreover, as soon as these tests are applied to individual stocks, the autocorrelations become statistically insignificant [13]. As for models, in-sample and out-of-sample testing suggests that they do not work at predicting at all [14] or that they "work", but to such a minute and limited extent [15] that it does not make much of a difference. Independently, and from a practical viewpoint, the professional investment managers do not outperform the market (in other words, they cannot predict), which also does not suggest that the market is not non-random [16]. In any case, the rejection of the hypothesis is weakly supported and everyone agrees that one cannot predict the evolution of prices. To summarize the situation, to this day, the nature of the stock market and, more specifically, the nature of security prices are still unknown, there is no known way to predict security prices and other financial time series, and, more generally, the random walk hypothesis is regarded as true. Nonetheless, even though the autocorrelation function (mentioned above) decreases exponentially, there is still a small positive autocorrelation found in some time series [17-18], which needs to be accounted for. But, even more in contradiction to the random walk hypothesis, are stock market crash fluctuations, such as the one of Black Monday of October 1987, which represents a fluctuation of 34 standard deviations, that is, a probability of $10^{-267/2}$ of occurring [12], not to say zero. This is totally incompatible with the random walk model. Over a 35-year period, the S&P 500 index had 30-40 fluctuations greater than 5 standard deviations, which has essentially zero chance of happening in a random walk model [12]. All this should have made researchers doubt that the evolution of the quantity conforms at all to a random walk model, even a biased one. The function presented in this paper, as we will see, proves that the decrease of the actual dependence is all but exponential since even the 10,000$^{th}$ previous data point still has an influence on the price at instant $t$.

This article presents a new function that describes the evolution of security prices, and other physical quantities, and shows that the evolution of prices is governed by a physical law. This function is accompanied by a theory. The two are unrelated to statistical physics and have no relationship with existing work (because, in existing work, as we will see, there is neither emergence of mathematical objects nor the price being driven in a very deterministic way), and are not included in or linked to other existing theories. The theory and the function encompass, directly or indirectly, a large part of the economy, that is, the part of the economy for which there are quantitative data in the form of time series. We show here that not only is the evolution of security prices not random but it is largely deterministic, which proves, incidentally, that the random walk hypothesis is false. Moreover, security prices are so well governed by this physical law that it applies to any stock prices, whether they be indices or individual stocks (recall that current research could not even find positive autocorrelations in individual stocks [13]), whatever the timeframe considered.

Because we claim the existence of a physical law, there is no point discussing the random walk hypothesis further. To put it trivially, if a new function can predict the trajectory of a "drunken walker", trying to figure out if he is walking randomly, calculating the probability that he reaches a certain perimeter, or trying to mimic his walking with a mathematical model, no matter how elaborate (see [19] for an overview), becomes much less interesting. This is the reason why we should normally only discuss literature that deals with the existence of a physical law driving the market, and we are not aware of any. In addition, since the function and theory have not been described before, their description alone makes for a long article already.

Considering that we are presenting a new theory, largely unrelated to existing work, the standard structure (Introduction, Methods, Results, Discussion) is not suitable for this article. A description of the structure used in this article follows. Section II briefly presents the new function and theory. The mathematics of the function and its result, the "topological network", is presented in Section III. Section IV describes the laws that constitute the theory. Section V reminds one that, besides the lack of practical interest of a theory that cannot predict, a theory is only scientifically valid if it can allow predicting. In light of this, the weaknesses of existing approaches are discussed, and a basic description of how to predict using the function and the theory is provided. In Section VI, a final independent demonstration is provided. This section also points out the fundamental problem with the expectations of investors in light of what the function and theory teach. Section VII discusses the contributions of the function and the theory and, then, concludes.

## II. Introduction to the Tool and the Theory

### *A. The Tool and the Theory*

Because the function and the theory cannot be apprehended without the representation of the function (the "topological network"), we need to refer to both the function and its representation at the same time. Thus, we will refer to the function and its representation as the "tool". The tool consists of a network of curves (mathematically described in Section III.A) in which new mathematical objects, called "characteristic figures" emerge spontaneously (that is, they are not mathematically constructed). The network can be represented graphically and can be interpreted relatively easily. These characteristic figures possess the unique property of driving the price. Graphically speaking, the price tends to jump from one characteristic figure to another one, following rules, which constitute the empirical theory. The tool and theory lead us to claim the existence of a physical law that governs security prices. It should be noted that the tool and theory are not based on a model and are unrelated to statistical methods. The tool does not rely on "fitting" or calibration. The tool is universal (refer to App. Q and S) and applies, unchanged, to any time series (stocks, commodities, foreign exchange rates, bonds, futures, etc.), provided there is enough data.

### *B. What the Tool Allows*

The mathematical tool allows not only predicting the evolution of physical quantities with a high level of precision and confidence, but also better understanding the phenomenon of the stock market and other economic phenomena. Thanks to this tool, the very nature of the stock market is much better understood. The tool also sheds new light on a certain number of subjects that are commonly debated.

Looking at the chart in Fig. 1, which is a graphical representation of the tool, one can clearly observe that the price bounces off thick "lines" (called "cords"), in particular, at the points labeled 1 to 4. <u>The mere fact that no other tool or method to this day is capable of knowing why and how the price turned around at the 4 points proves that the tool is important</u>. Indeed, the tool allows knowing where (retrospectively or in advance) the turning points are or will be, no matter the time granularity, and with great precision. Refer to App. A for other examples. <u>It should be strongly emphasized that the examples have been randomly chosen; the bouncing of the price off cords is systematic and not anecdotal</u> (refer to App. Q, R, and S). We will examine what these strange "attractors" are further on.

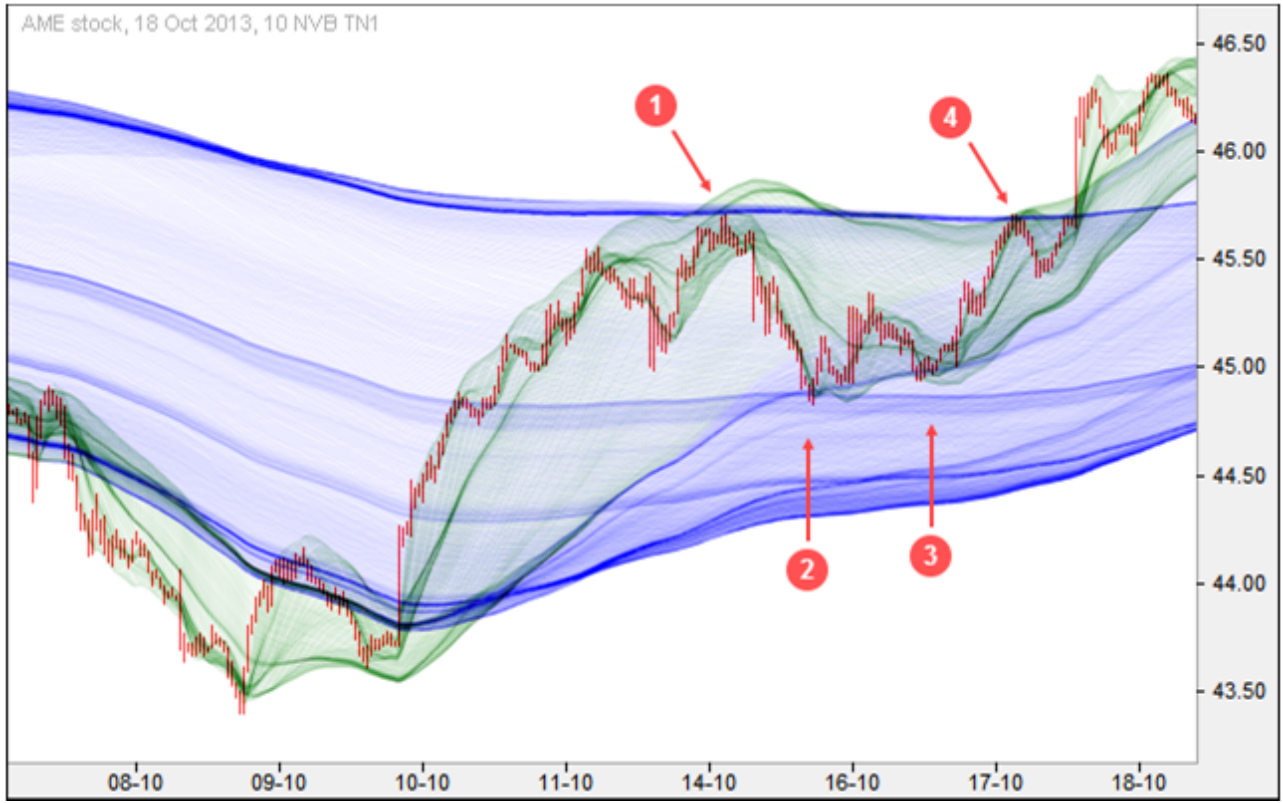

**FIG. 1. Chart representing the topological network.** A topological network from the data of a stock. The numerical data used to build the network have a granularity equivalent to 10 minutes (see Section III.J, "Data Concerned"). The network presented in this figure is based on first order regressions, denoted by TN1. The arrows point out 4 particularly striking local extrema having interacted with the characteristic figures of the network.

### C. A Predictive and Explicative Empirical Theory

To this day, nothing satisfactory has been proposed in terms of prediction of quantities, whether financial or economic, and we believe that the tool presented here remedies the situation. Let us keep in mind that nothing attests to the quality of a theory or a mathematical tool more compellingly than its capacity to predict. It is even the principal criterion to which a scientific theory needs to comply. That said, we are not here talking about a mathematical theory in the constructivist sense of the term, but about an "empirical theory" (expression that seems the most suitable), validated empirically also, but in a manner difficult to contest, as we will try to demonstrate in what follows.

It is important to realize that the price movement (in Fig. 1) that took place between arrows 1 and 2 over a period of 6 hours is "determined" (as the notion will be defined later) by the "cords" (objects specific to the tool itself, which will be discussed further on) produced by

W. Tzara - The Evolution of Security Prices Is Not Stochastic but Governed by a Physicomathematical Law

the data going as far back as 1.5 months (38.5 trading days). Who would have imagined that a price movement lasting 2 months was determined by cords going as far back as 5 years and 4 months (referring to the movement between the two cords in the Fig. 13 example)? Or that <u>a price movement lasting 1 month was determined by cords going as far back as 12 years and 5 months</u> (referring to the last movement between the two cords in Fig. C5 of App. C)? Or that a price movement lasting 45 minutes was determined by cords going as far back as 4 days (referring to the movement between points 3 and 4 in Fig. A2 of App. A)?

### D. Introduction to the Mathematical Tool

The tool, which we will refer to as TN, can be conveniently described as a dense network of "moving" regression curves of an order (degree) greater than or equal to 1 (zero order is neither interesting nor results in "characteristic figures", which are described further on). For convenience sake, we will use TN to refer to the tool as well as the result produced by the tool (that is, the topological network) since the tool is apprehended through its graphical representation, unless there is a reason not to confuse the two. When displayed on a computer monitor, the topological network built from numerical data from a quantity, such as the price of a stock or a commodity, clearly shows ultra-dense formations (appearing spontaneously), which act like true magnets on the quantity. Even if the proof to this day remains empirical, it is hard to deny since it is observed day after day, hour after hour, minute after minute (which allusively introduces the notions of granularity (or "resolution") and self-similarity). Refer to App. Q and S.

To dispel possible assumptions, misunderstandings, or misinterpretations that might arise early on, let us immediately make (even though it violates a bit the chronology of the presentation) several remarks:

- When new data points are added, the past network (that is, the network before the new data points are added), of course, does not change. In other words, new data points simply prolong the existing network (see Section III.A, "Construction of the Topological Network", to understand the mathematics).
- Characteristic figures are not mathematical objects that "join" or "connect" local extrema. A basic examination of the mathematics of TN allows clearly understanding that the past local extrema have nothing to do with the construction of the network. App. P addresses this misconception.

- Characteristic figures are not mathematically drawn; they emerge spontaneously.
- The cords, and other characteristic figures, are not simple agglomerations of curves. They are objects resulting from complex crossings and intersections of many curves.
- The curve parameters ($n_k$) are fixed (see Section III.A, "Construction of the Topological Network"), and have nothing to do with the data that is input. In other words, all the charts generated are based on the same set of parameters. Note that there is nothing magical about the chosen set of parameters; it is just properly chosen to ensure a uniform (see algebraic formula in the aforementioned section) and exploitable network. Another similar set (for example, with a different value for *N*) also works. See also App. O, "The Non-Pertinence of In- and Out-of-Sample Testing".

### E. How Can One Predict?

The possibility to predict with TN, and predict with a high degree of confidence, relies on the analysis of the action of dense formations on the quantity. This exercise leaves actually little room for subjective interpretations provided the "analyst" is methodical and experienced.

### F. To Reassure the Reader that the Examples Are Not Well-Chosen Examples

The charts which are presented in what follows are not chance ad hoc examples. They are examples among an infinite number of others (see App. I). Similar examples are encountered all the time. Evidence that they are not ad hoc is that even during extraordinary events, such as the Brexit referendum (see Fig. B1 in App. B), the price still bounces off cords. Refer to App. Q, R, and S for a more formal proof.

### G. Terminology

Numerous new terms have been introduced for the needs of the TN empirical theory (characteristic figures, cords, topological network, etc.). These terms are not flowery jargon, but constitute the semantic framework of the empirical theory.

### H. About the Discovery of TN

Before going on, let us say a word on how something so simple could work as well as it is claimed. First, the simplicity is only in appearance. Second, the usefulness of a scientific discovery is not measured by its mathematical sophistication. It is true that all new important scientific discoveries today are very advanced, but there is no reason to assume that laws or equations that do not rely on very abstract mathematical formalism could not describe some phenomena. We do not know how many other "simple" scientific discoveries have not yet been found, but they could be missed, and all the easier today, given that research is focused on very complicated things.

## I. *Why the Initial Apprehension of Some Researchers?*

The few researchers, economists and econophysicists, to whom TN was briefly presented immediately (wrongly) put TN in the group of "technical indicators" for the simple reason that this tool happens to be in the form of charts containing curves. Consequently, TN's fate was sealed in the eyes of these researchers. Technical analysis is based on a simple and gratuitous idea: curves and other graphical elements made from data of a stock would allow one to know if the price will rise or fall according to their positions or values. The very principle is obviously simplistic, and the indicators are often even fanciful, just as they look. What the few researchers failed to observe is that the only thing TN has in common with "technical indicators" is the simple presence of charts in both cases. But TN evidently does not rely on the idea that the position or value of a curve is associated with a rise or drop in the price. <u>On the contrary, TN relies on the principle (empirically justified) that the price is directly attracted or repelled, not by the curves, but by elements derived from the curves (which form spontaneously).</u> This fundamental aspect is obviously absent in "technical indicators"; it is completely new and completely counterintuitive. Possible instinctive intellectual apprehension will need to be overcome in order to consider and accept the TN theory.

## III. The Mathematical Tool

### A. *Construction of the Topological Network*

Without describing in great detail (namely, with mathematical functions) how these networks of dense curves are constructed (which is of no particular interest once the principle is understood), let us say that each curve of the dense network based on regressions of order $D$ is obtained by calculating, for each point, the regression curve of order $D$ over the last $n_k$ points of a times series. These curves can be called "moving regressions of order $D$". Suppose one wants to construct a first curve of the network of order $D$; for each point, only the last point of the regression curve of order $D$ is conserved, and one starts all over again for the next point of the series, over the last $n_k$ points. A curve of the network is thus obtained. This operation is repeated for the ensemble of the $N$ curves of the network (which we will name of order $D$, by extension), by varying $n_k$ in such a way as to obtain a network of "uniform density", which can be satisfactorily obtained thanks to the following algebraic formula:

$$n_k = \text{round}\left[n_1 + (k-1)a + \frac{k(k-1)}{N(N-1)}\left(n_N - n_1 - (N-1)a\right)\right], \quad (1)$$

where $k$, $N$, and $a$ represent the rank of the curve of the network, the number of curves of the network, and a real parameter to be determined, respectively. Variables $n_1$, $n_k$, and $n_N$ represent respectively the numbers of points of the regression curves of order $D$ for the first curve of the network, for the $k^{\text{th}}$ curve of the network, and for the last curve of the network. The algebraic formula is valid when $n_N - n_1 > a(N-1)$. This ensures an ad hoc distribution of $n_k$. For example, for a dense network based on regressions of order 3, the following distribution of 600 curves can be used: $n_k$ = (7, 9, 11, 13, 15, 17, 19, 22, 24, 26, …, 4448, 4461, 4474, 4487, 4500).

### B. Illustration of the Construction of the Network

Fig. 2(a) below illustrates how the first four curves (C1 to C4) of such a network are constructed iteratively using the data (curve C) from a physical, economic, or other phenomenon, where the time is represented on the x-axis and the value of the quantity on the y-axis. The abscissas 1 to 4 correspond to the last points conserved of the regression curve for each of the first points of the four first curves of the network. Abscissa 5 corresponds to the first point of the time series. The four shifts observed are due to the fact the number of points $n_k$ used for the regressions of the curves C$k$ grows with index $k$.

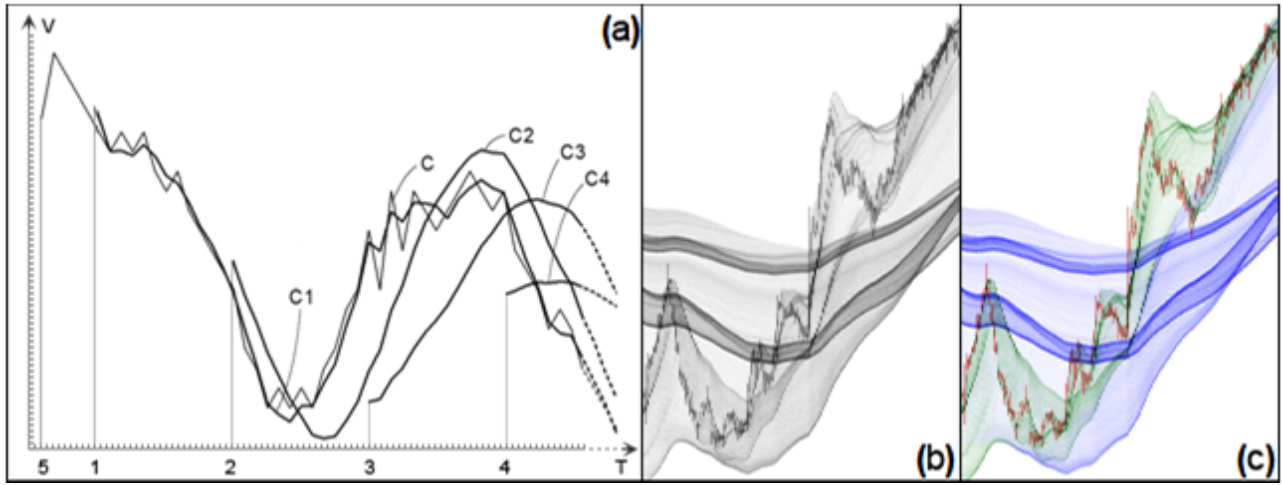

**FIG. 2. Construction of a topological network and computerized rendering of a chart.** (a) Construction of the first four curves (C1 to C4) of a topological network of order $D$. (b) Network in which the curves have been digitally processed in order to improve the rendering and therefore overall

readability. (c) Same network, with improved visual rendering thanks to the use of appropriate colors, allowing one, in particular, to distinguish the "short" curves from the "long" ones.

### C. Graphical Representation of TN

In Fig. 2(c), as well as in all the charts that are presented, the ensemble of curves, made more readable thanks to ad hoc graphical treatment, produces a dense network, called the "topological network". Each vertical bar represents a grouping of consecutive numerical data points from the considered time series, which can be seen as the average value and the dispersion ($M \pm d$) within each grouping. It is easy to see that dense formations are present inside and at the periphery of the network. The first chart that is presented in Fig. 1 has for the quantity the price of a stock over time, but TN works with many physical quantities evolving over time, and even with other types of quantities. Numerous examples relating to financial instruments are presented in this article.

The use of colors allows improved visual appearance, an essential characteristic, as we will see, allowing the most to be made of the tool. The chart in Fig. 2(c) presents the same topological network as in Fig. 2(b), making use of two very distinct colors for the drawing of the curves. Thus, the structure of each dense formation stands out better.

### D. Visual Observation of a Topological Network

Close examination of the first chart reveals something intriguing: the bars representing the quantity (quantity whose data are also used to construct the network) seem to "interact" in a unique manner with the dense formations mentioned before. It is this property that sets TN completely apart, notably from what are known as "technical analysis" tools, as will be discussed later.

### E. Characteristic Figures and Their Classification

The dense formations can be classified into three categories: cords, envelopes, and boltropes (see Fig. 4).
- A cord is a pronounced condensation, produced by curves, which stands out from a less dense background of curves.
- An envelope outlines the boundary of a group of curves of the network.
- A boltrope is both a cord and an envelope.

Cords, envelopes, and boltropes are called "characteristic figures". We will see that these characteristic figures "seem to exert" (to say "exert" would be semantically and logically incorrect, but, that said, we will ignore this precaution and take this linguistic liberty for the sake of brevity) a very strong action on the quantity.

### F. Curve Subtypes

TN refers to the topological network under all regression orders. In practice, we will limit ourselves to orders 1 to 5, since the zero order does not exhibit characteristic figures and since the fifth order is already exceedingly complicated. We will call a "subtype" the topological network under a given order, and we will denote a network under order $D$ by TN$D$.

No matter the subtype, the topological network has the same characteristics and exhibits the same behavior in relation to the quantity. Characteristic figures are common to all subtypes.

In general, the higher the subtype, the more numerous the characteristic figures. That said, more points are required to generate networks of higher subtypes that exhibit characteristic figures that look equivalent. This can be seen in the example in of Fig. 3 (refer to App. C to see a larger version of each subtype).

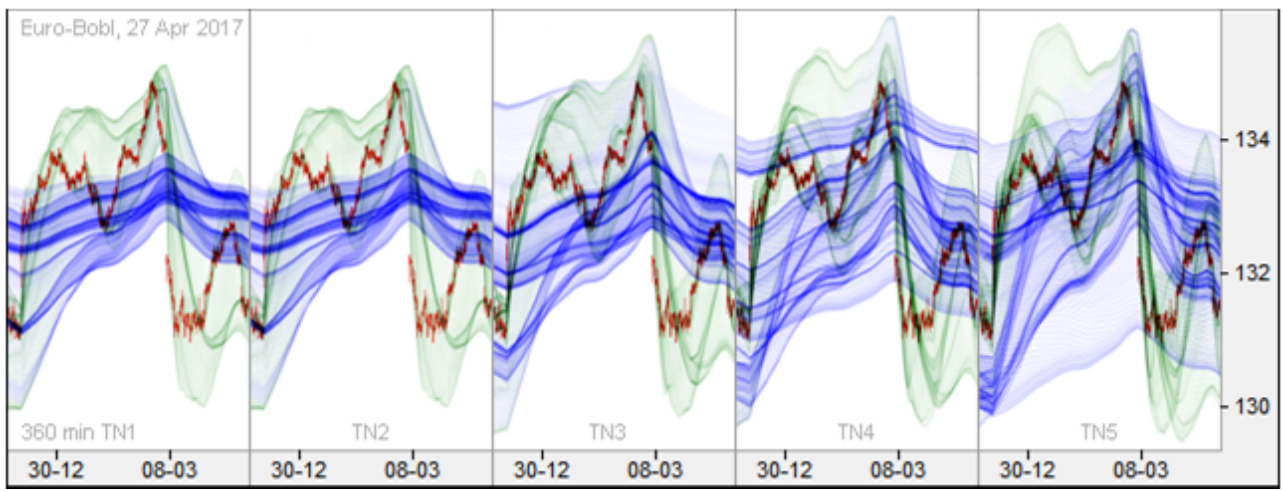

**FIG. 3. Five consecutive subtypes of the topological network.** Side-by-side comparison of charts of the same instrument under the subtypes, denoted TN1 to TN5, corresponding to first to fifth order regressions, respectively.

### G. The Action of Characteristic Figures

Basically, a characteristic figure attracts and repels (according to rules that will be discussed in this article) the quantity all the stronger as the characteristic figure is more pronounced. Remarkably, almost every characteristic figure, even the smallest, exerts an action on the price appearing as a more or less pronounced interaction (sticking and/or bouncing) between the two. To avoid all ambiguity, note that the physical reason behind why the characteristic figures exert an action on the quantity is unknown. This is not an argument for dismissing the tool as one does not know any better the reason why such and such a physical force is inversely proportional to the square of the distance, or, from a deeper standpoint, why the universe conforms to more general symmetry principles. This can be regarded as a law of nature. The TN theory, as any physical theory, teaches one about the how and not the why.

### H. An Interdependent System that Is Paradoxical in Appearance Only

Let us briefly look at the apparent paradox that TN seems to present and at the implications resulting from TN. We now know that characteristic figures drive (by attracting and repelling) the quantity. Yet, the quantity, through its numerical data, is at the origin of the curves, and thus the characteristic figures themselves. One is confronted with a strange pair where each element (the quantity on one hand and the characteristic figures on the other) determines the other. It is as if the characteristic figures determined the evolution of the quantity, which, in turn, determines the characteristic figures. This seems paradoxical at first, but it is not as it seems, as we will see. Regardless, it is the entire field of economy that needs to be apprehended in a whole new manner, since, clearly, if the price of a financial instrument is determined by characteristic figures (which are produced solely from past data), the role attributed to factors that are not intrinsically linked to the quantity, such as human psychology, the influence of information, etc., must be largely abandoned. Note that when viewing finance as a complex system, price changes, including crashes, have an endogenous cause, which is confirmed by TN. We will come back to the effect of extrinsic stimuli (such as news) on the topological network later.

### I. Analogies with Other Theories in Order to Better Understand

To understand the seemingly paradoxical role that the quantity and the characteristic figures seem play in relation to each other, it may be helpful to think about general relativity, not that TN has any relation to it, but because it offers a convenient conceptual perspective. General relativity can be understood thanks to an analogy consisting in imagining balls and a fabric stretched across a horizontal frame. The fabric represents space and the balls, planets or other massive celestial objects. After having been placed on the fabric, the balls start moving. So to speak, one can say that the balls move as a result of the deformation of the fabric, which plays the role of the gravitational field (or spacetime geometry). Conversely, one can say that the deformation of the fabric is the result of the action of the balls. Thus, the balls and the fabric form a whole that is inseparable. Returning to TN, the balls are the numerical data from the quantity and the fabric is the characteristic figures of the network. Thanks to this analogy, one can better understand the role that the characteristic figures "play" in relation to the quantity, and conversely. The situation can be summarized by saying that there is an interdependence between the cause (the data) and the effect (the result of the action of the characteristic figures on the data). But, one must understand that, here, the interdependence is temporal (in other words, it is not the data at instant $t$ that determine the cords which act on the same data at instant $t$). This is what makes things paradoxical in appearance only. The nature of this interdependence is specified further on.

It may also be worth dwelling on the term "to play". In quantum mechanics, in interpreting Young's double slit experiment, some go as far as to say that the photons passed through both slits at the same time in order to account for the observed effect. In fact, this is not the case, but our language can only reflect what we can conceive of. Characteristic figures "play" a role in relation to the quantity, just like the fabric makes the balls move, but, by linguistic liberty and for convenience sake again (nevertheless, without being misled), we will say that the characteristic figures "play" a role in relation to the quantity, and say even "exert an action" on the quantity, by attracting and repelling it.

The goal of these two perspectives is only to help one better understand the nature of TN, which is counterintuitive to the point of making many people uncomfortable.

### *J. Data Concerned*

TN works with any kind of data susceptible of producing a topological network, and, more precisely, characteristic figures. But this criterion is a posteriori. We can say that the types of data that can be used are numerous, ranging from financial data, to economic data, to data from physical quantities, and even to completely unexpected data. The tool is thus particularly versatile. The reason for this is presented in the conclusion of this article. The use of TN with data of different natures is beyond the scope of this article.

Note that the fact that TN works with many different types of data is in no way a valid argument for trying to refute TN; it is due precisely to its versatility. Even more importantly, it proves (in an unexpected way and not without consequences) that certain data (or quantities, by extension) possess "hidden order", either greater than previously thought or that was not known of before.

TN works also with any granularity and with different data representations of the phenomenon and, in particular, with "normalized volume bars" (NVB), which are commonly used in the examples found in this article.

## IV. The Empirical Theory

The empirical theory behind TN can be described well by nine laws. Some of them have already been evoked.

### A. First Law (Spontaneous Appearance of Characteristic Figures)

The first law can be expressed as: Dense formations (the characteristic figures, that is, cords, envelopes and boltropes) appear spontaneously within a network of curves.

See Section III.E, "Characteristic Figures and Their Classification", for a description of the characteristic figures.

### B. Second Law (Action of the Characteristic Figures)

The second law can be expressed as: Characteristic figures interact with the quantity, by attracting and repelling it.

We will now examine in detail the action of each of the three characteristic figures.

#### B.1. Cords

Cords, as we have seen, are condensations of curves, varying in density, markedness, and "length", and are the principle characteristic figures. Unlike the two other types, they act like dipole magnets, to use a striking and convenient image to represent their action. The quantity either 1) literally bounces off the cord (possibly after sticking to the cord) and takes off in the opposite direction, or 2) crosses the cord, after interacting to a greater or lesser degree with it (possibly sticking to the cord), and is then propelled in the same direction (often with an equal force). It is the crossing of the cord and the propelling in the same direction that corresponds to the second pole of the magnet. In Fig. 4(a), the main interactions between the cords and the price are indicated by arrows. The previously described notion of action at a local scale, as opposed to action at a larger scale, is clearly visible. The larger scale actions are indicated by the green arrows and the local scale actions by the pink arrows.

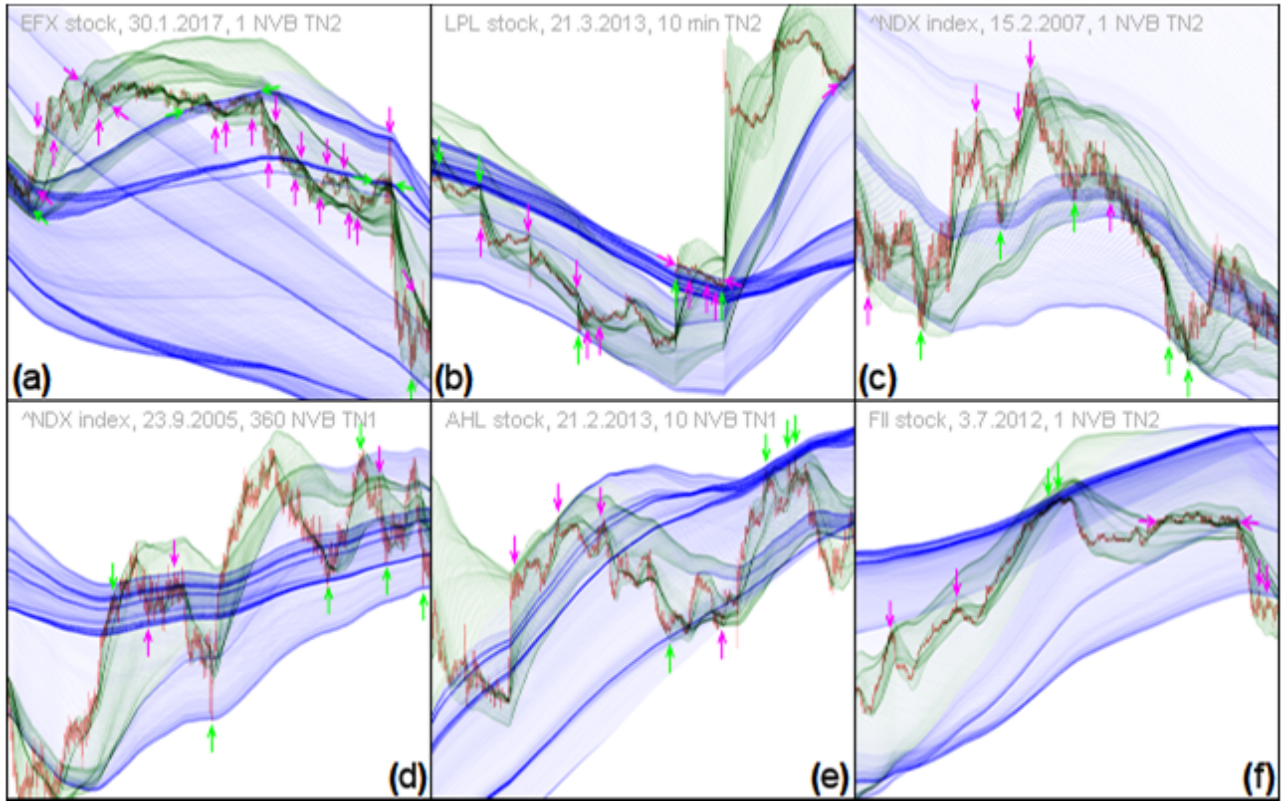

**FIG. 4. The three types of characteristic figures.** Networks showing the main interactions (thanks to green and pink arrows) of the price with a certain type of characteristic figure. The green arrows point to actions at a larger scale. (a) and (b) In these examples, the arrows point to the principal interactions with cords. (c) Here, the four bottom arrows point to interactions with envelopes. (d) The two green arrows at the bottom and the green arrow at the top point to interactions with envelopes. (e) The four topmost arrows point to interactions with boltropes. (f) The two topmost arrows point to interactions with boltropes. These interactions are fewer in number due to the fact that boltropes are less common among observable characteristic figures.

Figure 4(b) shows another example where cords are rarer, but very marked. This example shows how the same cord can lead to several interactions, even very far apart ones, as shown by the four green arrows around this thick cord. Note that the distant action at the bottom of the network is not due to a cord, but to a boltrope, which will be discussed later. One can observe the very strong actions of the thick cord and of the "emerging" cord, which provoke a jump followed by a landing on the latter. This kind of behavior is very classic. Numerous examples can be found in the appendices.

### *B.2. Envelopes*

Envelopes are rarer than cords and cannot be easily crossed for topological reasons. They act like a single magnetic pole by attraction/repulsion without the crossing of the characteristic figure. This type of characteristic figure is not very salient visually, but exerts no lesser action on the quantity, as seen in Fig. 4(c). Note that an envelope can be found topologically within the network, as is the case for the envelopes whose interactions are indicated by the downward-pointing arrows at the top. In contrast, the four upward arrows at the bottom point to interactions of envelopes with the price of the index, this time at the periphery of the network.

Fig. 4(d) illustrates well the force of envelopes, where top and bottom envelopes exert their actions at a relatively large scale (here, all the larger being that the "granularity" is high since each data point is averaged over the timeframe of a day). Note that envelopes tend to exert stronger actions when the network is tighter.

### *B.3. Boltropes*

Boltropes, which are both cords and envelopes, exert very powerful actions on the quantity, as shown by the chart in Fig. 4(e). This third type of characteristic figure is rarer. Boltropes are in a way super-envelopes. A boltrope, such as the very thick and ultra-dense blue boltrope on the top right, can only produce very strong and very precise interactions, as one can observe in the other examples in this article.

One can observe the remarkable boltrope in Fig. 4(f). Such a boltrope necessarily exerts its action at a non-local scale.

One can refer to the appendices to find numerous examples of cords, envelopes and boltropes. These examples have not been selected with any particular care; <u>characteristic figures are present on all the charts and repeatedly produce the same powerful and precise effects whatever the circumstances</u>. This makes TN a universal tool (refer to App. Q, R, and S).

### *C. Third Law (Topological Network)*

The third law can be expressed as: Characteristic figures are deformable and only shapes matter. Consequently, the prices or values per se do not bear any particular significance, unlike what is currently considered by most traders and financial analysts. Due to the fact that only the shapes matter, TN has a "topological" nature. The network and its characteristic figures continually deform as the quantity evolves within the network, both changing in an interdependent manner. Thus, the notion of absolute distances becomes meaningless. Only the relationship of the positions of the characteristic figures in relation to the quantity is meaningful. In other words, only relative distances have meaning. The absolute positions can however be used, but in a very limited way and only at a given time, when trying to make quantitative predictions, for example, predictions accompanied with prices, <u>using an extrapolation method</u>.

As we will see, the abandoning of the notion of absolute distance has consequences not only in the making of predictions, but also in the understanding of other aspects of the economy.

### D. Fourth Law (Scale of Attraction)

The fourth law can be expressed as: A characteristic figure interacts with the quantity in a manner that is proportional to the markedness of the characteristic figure.

As a general rule, the more pronounced a characteristic figure, the stronger the action exerted on the quantity. But what is remarkable is that even the least pronounced characteristic figures exert observable actions. The action of less pronounced characteristic figures are exerted at a shorter "distance", that is, more locally. In terms of the numerical representation of the quantity, this translates into an action within a shorter timeframe. This fundamental property is at the origin of the notion of self-similarity, which will be discussed later.

### D.1. Illustration of the Fourth Law

The example that follows illustrates well the relative strength of characteristic figures. Note that, on the charts, the characteristic figures, in particular, the green (color non-arbitrarily chosen, but this is not pertinent to this article) cords correspond to curves that are "shorter" (that is, based on moving regressions over fewer "points"). Empirically, the "shorter" characteristic figures (that is, corresponding to shorter curves) exert their action at a shorter spatial scale, that is, at a more local scale, data-wise. As it happens, the "shortest" (green) characteristic figures are generally less pronounced (dense), which is coherent with the idea that these characteristic figures are exerting a more local action, that is, not as strong. Looking at the charts in Fig. 5, one can see that the dense characteristic figures are also those that are the longest, and that act at a larger scale. The action of the long characteristic figures manifests as very strong interactions between them and the quantity. These interactions are shown using pink lines in Fig. 5. One can immediately observe how precise these interactions are. In contrast, the interactions of the less pronounced characteristic figures (which we will call secondary) with the quantity result in movements of the quantity at a much smaller scale. These interactions are shown here using green lines.

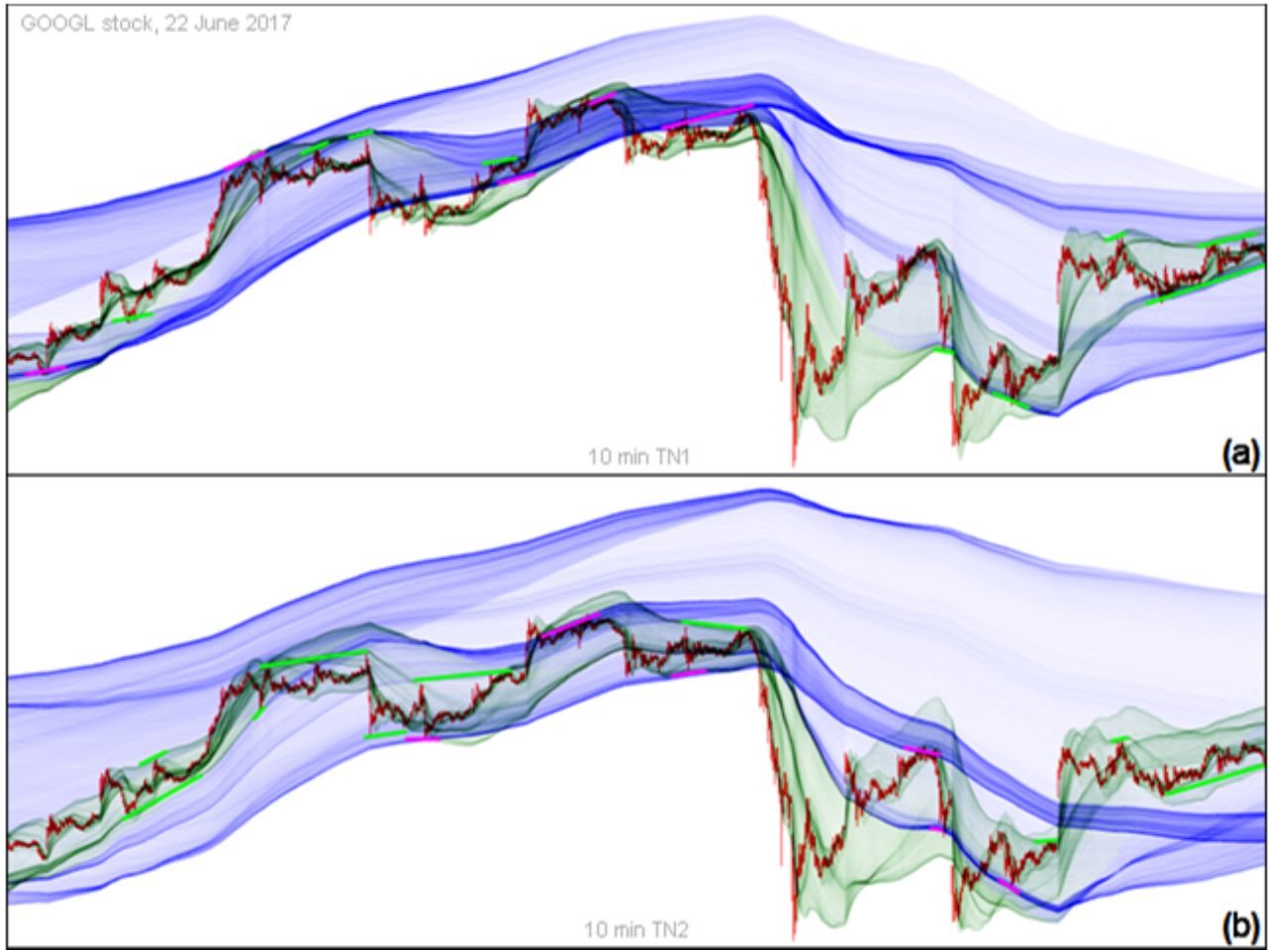

**FIG. 5. Network under the first two subtypes.** (a) Network under TN1 showing the actions of the main characteristic figures on the price. (b) The same network under TN2 allowing one to observe other characteristic figures accounting for other important price movements that were not the subject of interactions under TN1. This phenomenon is observed systematically. The pink and green lines represent the main interactions of the characteristic figures (with the price) exerting their actions at longer and shorter scales, respectively.

### E. Fifth Law (Relative Positions of Characteristic Figures and the Quantity)

The fifth law can be expressed as: The action of the characteristic figures depends on the positions of the characteristic figures in relation to the quantity. We will only briefly describe here the manner by which the characteristic figures exert their action, without going into much detail, since it would require an entire article itself.

The manner by which a characteristic figure attracts and then repels the quantity depends on:

a) the position of the quantity in relation to the characteristic figure (above or below it),

b) the positive or negative slope of the characteristic figure, and

c) the angle of the curve representing the quantity in relation to the characteristic figure (ranging from more or less perpendicular to more or less tangential).

In the example shown in Fig. 6(a), the price is above the bottom cord, the slope of the cord is negative, and the angle at which the price falls onto the cord is moderately open (45°). This results in a rebound (as in most cases), but of modest proportion. Likewise for the top cord.

In the example shown in Fig. 6(b), the price is also above the characteristic figures (envelope and boltrope, respectively); however, this time the slope is positive. Additionally, both angles formed by the price curve and the characteristic figures locally are more open. This leads to a rebound that is a lot sharper.

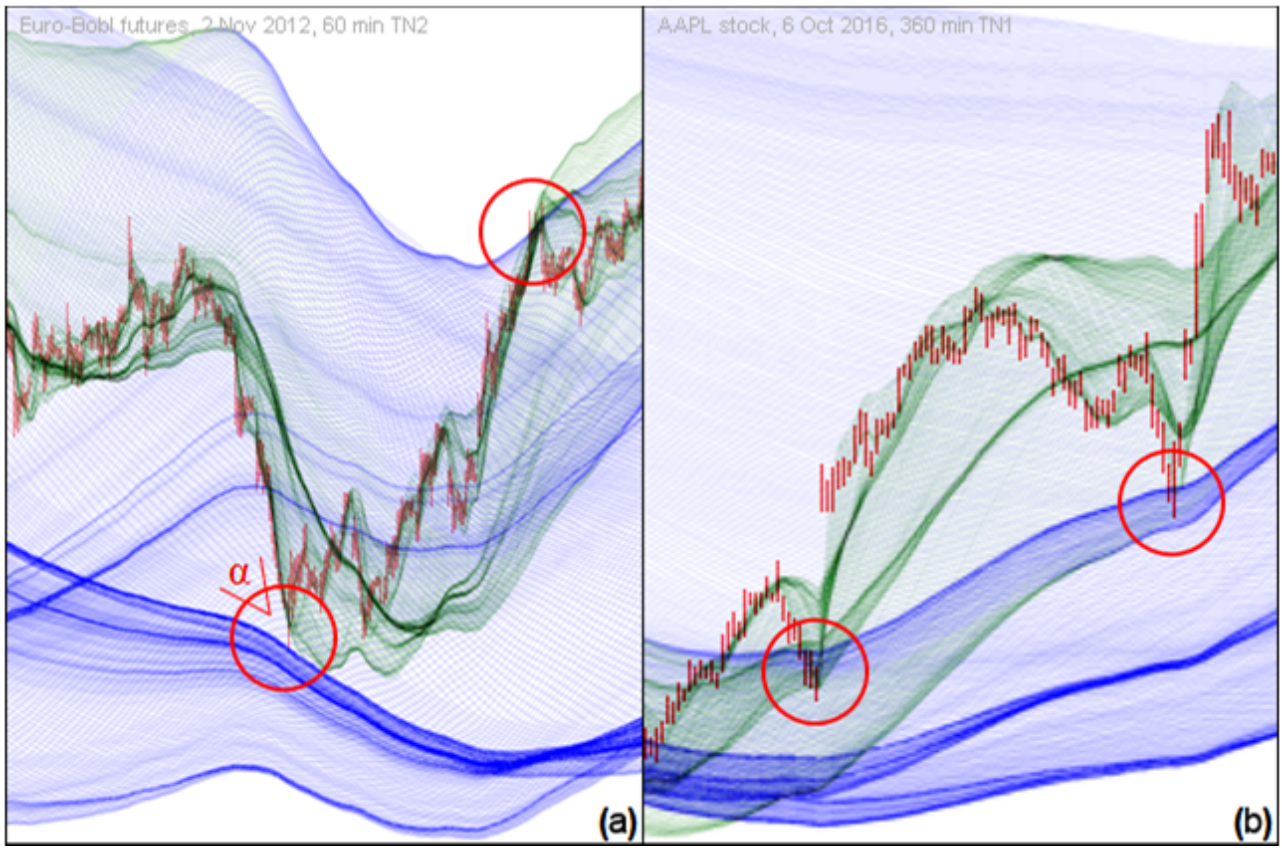

**FIG. 6. Illustration of the fifth law of the empirical theory.** (a) Chart showing rebounds (circled in red) of the price off top and bottom cords, under relatively closed angles. Refer to App. J for more details about this example. (b) Chart showing very violent rebounds off characteristic figures.

This is roughly how characteristic figures exert their action. Among the numerous scenarios, some are particularly evident. Serving as examples, two scenarios will be summarized as follows:

- Average price rising (in the given timeframe) + price above + cord with a positive slope = rebound occurs almost always
- Average price falling (in the given timeframe) + price below + cord with a negative slope = rebound occurs almost always

Note that attempts at quantifying absolutely (as opposed to relatively) the values of the slopes and relative angles are impossible due to the topological nature of TN.

### F. Sixth Law (Reoccurrence of Patterns)

The sixth law can be expressed as: The characteristic figures form well recognizable patterns which reoccur.

The previous section offered a glimpse at the notion of shapes. We will now look at the specific shapes that the characteristic figures/quantity pair produces within the topological networks. The property of identifiable shapes repeating themselves over and over again can be exploited in order to make predictions with relative ease, as we will see later. These identifiable shapes, which reoccur regularly, are called "patterns".

#### F.1. Illustration of the Patterns

Without making an inventory of all the patterns, since this is not the goal of the article, we will only look at a number of them to illustrate the concept.

Fig. 7(a) presents a classic pattern where a rapid price drop onto a thick bottom cord with a positive slope is followed by a sharp rebound. Without going into the details of this pattern, which is seen often, one can see that the four examples in Fig. 7(a) have a "family resemblance". One can also observe the extreme precision of the interaction (here, a rebound) between the cord and the price in each chart. We will return to the topic of the precision later on. The figures that follow show other examples of reoccurring patterns.

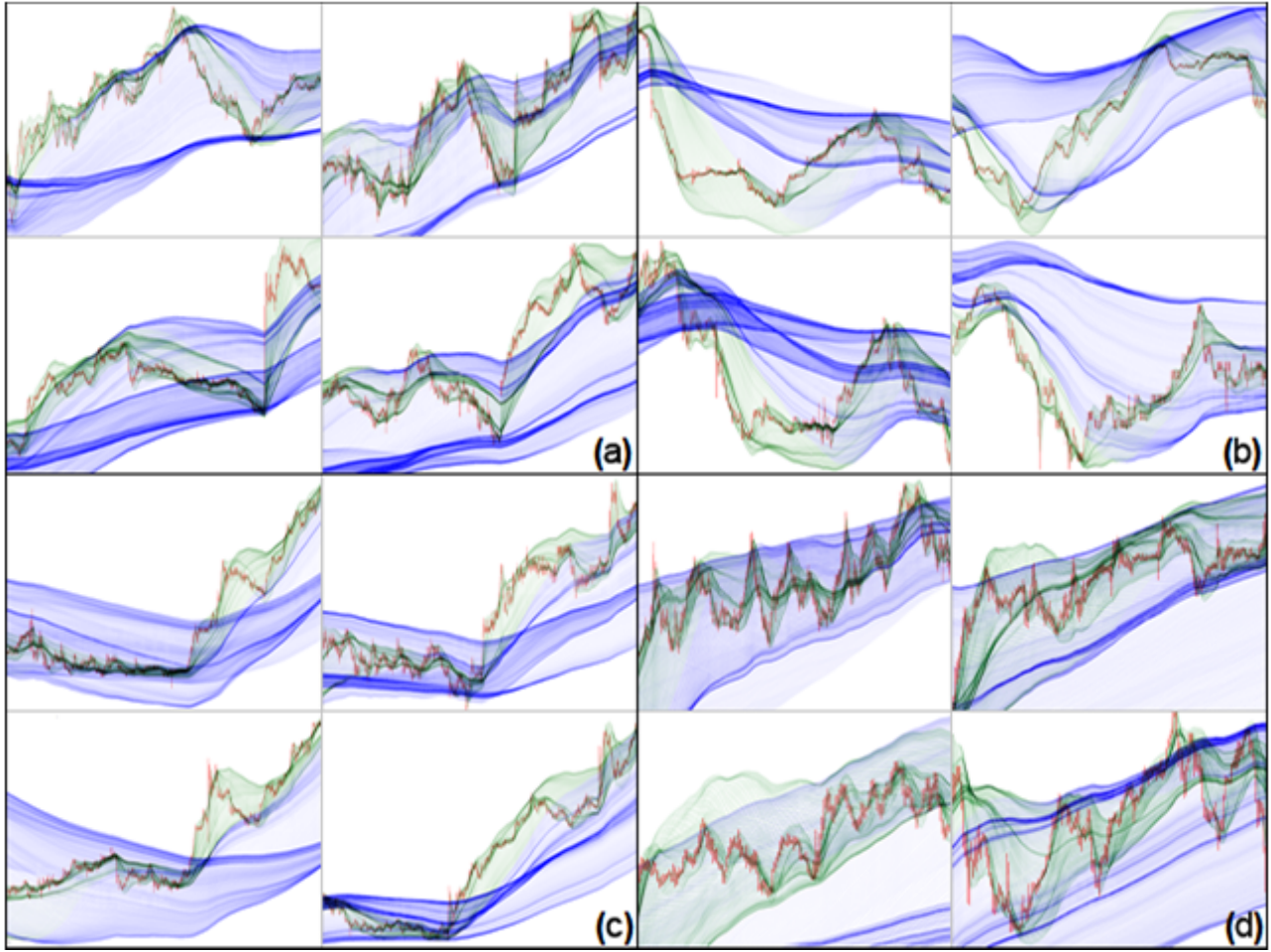

**FIG. 7. A small number of classic patterns frequently seen in the topological networks.** Larger versions of these examples, accompanied by other classic patterns, are found in App. D. (a) Pattern corresponding to a rising (positive slope) "bottom cord rebound". The rebounds are particularly sharp since the cords are pronounced and rare in these examples. (b) A "top cord rebound" (a top boltrope, to be precise). The rebounds are particularly sharp here too for the same reason as in (a). (c) "Buildup exit with return to emerging cord". This pattern is a big classic. (d) "Channel". Channels are, most of the time, rising or stationary (zero slope).

The pattern in Fig. 7(b) is the opposite, so to speak, of the pattern in the previous figure. This time, the rebound is preceded by a steady rise. The interactions with the marked cords occur with same degree of precision as in the previous example. The four examples here correspond to different resolutions (or different data representations of the quantity), which show that the same patterns are found whatever the granularity.

The pattern in Fig. 7(c) is a classic, classified under the name of "top buildup exit with return to emerging cord". This pattern can be seen several times a day when browsing stocks and leads to reliable and easy-to-make prediction. This type of ejection of the price is all the more violent since the pack of cords becomes more and more concentrated prior to the ejection. Note that pushing away from a bottom cord or a bottom pack of cords is not rare, and only reinforces the top buildup exit, as seen on the two charts on the right in Fig. 7(c).

The examples in Fig. 7(d) are given as a foretaste to the debunking of technical analysis, which is to come later. The oscillating upward movements are not underlain by hypothetical and magical "trend lines" (without the slightest causal principle, let alone a theory), but are explained by the presence of channels of parallel cords. These patterns are common and will perhaps one day allow one to put a definitive end to absurd charts covered with after-the-fact lines.

### *F.2.   The Importance of Patterns*

There is much more to say about patterns. We will see that identifying a pattern allows bypassing the detailed examination of the action of each characteristic figure on the quantity, so that predictions can be made more easily. Other examples of patterns are presented in App. D, but it is impossible here to make an exhaustive list. This glimpse into the notion of patterns, which are found in different resolutions, naturally introduces the notion of self-similarity.

### *G.  Seventh Law (Preservation of Topology Whatever the Representation)*

The seventh law can be expressed as: Topological configurations are preserved when changing the representation of the quantity.

To illustrate the topological nature of the tool, the most probative evidence is to observe how different numerical representations of a phenomenon (or quantity) result in compatible topological solutions. A visual demonstration of this should be less cryptic. Fig. 8(a) shows topological deformations of the topological network, and, in particular, of characteristic figures, when the type of data representing the price of a stock is changed. Without dwelling on the type of data used to generate the chart on the right, the type of data being normalized volume bars, which offer superior representativity of any phenomenon for which the principal value is accompanied by a weighting coefficient, such as the price of a stock and the corresponding volume exchanged. The same topological structures are identified by the use of colored dots. One can see that they are found on both charts, but they appear more or less deformed in relation to each other. What is remarkable is that the "same" topological structure can play different roles in relation to the quantity. The structures shown by the light blue and light green dots illustrate well this topological specificity. This phenomenon is undeniably counterintuitive, but confirmed empirically. Figuratively speaking, it is like a ball released from one point; it has the possibility to choose between several geodesics.

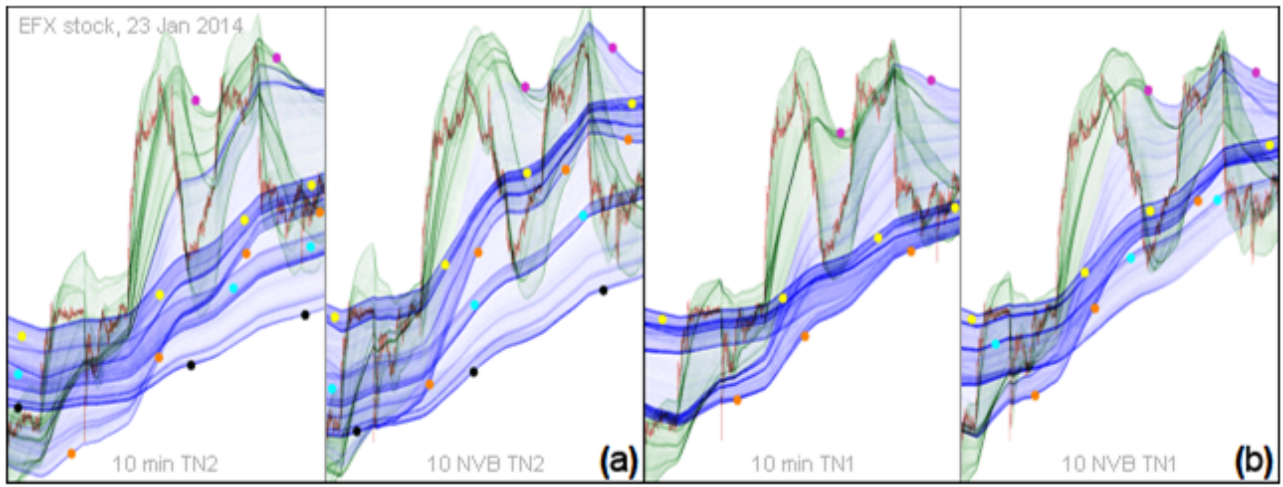

**FIG. 8. Conservation and deformation of the topology from one subtype to another and when changing the data representation of the quantity**. Larger versions of these charts can be found in App. E. (a) Charts of the network under TN2, based on the same data, 10-minute data in one case, and 10-NVB in the other. The same topological structures, identified by colored dots, are found, deformed, under different data representations. (b) They are also partly found under different subtypes, here under TN1.

In the examples shown in Figs 8(a) and 8(b), emphasis has been put on the major characteristic figures in common. Many other minor ones can be seen under closer examination.

We will return later to the notion of the topological network in order to deconstruct a certain number of well-engrained myths about the domain of "technical analysis".

### H. Eighth Law (Preservation of Topology Throughout Subtypes)

The eighth law can be expressed as: Topological configurations are preserved throughout the subtypes.

The example in Fig. 8(b) is interesting because it shows that, not only are the same topological structures found when the representation of the quantity is changed, but also when the regression order of the curves is changed.

That said, it is not worth dwelling on this law because, although interesting conceptually, it does not lead to anything easy to exploit. It would be indeed practically fastidious and useless to try to identify topological structures common to several subtypes. One can nonetheless examine them in Figs 8(a) and 8(b), where the colored dots allow identifying the common major topological structures, and use the charts of all five subtypes in Figs C1 to C5 (in App. C) to do this exercise mentally.

### I.   Ninth Law (Self-Similarity)

The ninth law can be expressed as: Whatever the granularity of the data, the characteristic figures and the patterns are identical.

TN is, within the limits imposed by the granularity of the data representing the considered quantity, unaffected by the resolution (that is, by the timeframe chosen to generate a network). For example, a network constructed using data whose granularity is 1 minute is indistinguishable from a network whose granularity is 100 times greater. In other words, the same patterns reoccur, whatever the resolution, from the lowest to the highest. Note that we are in no way talking about fractal behavior, but a scale invariance in relation to properties (effects and appearance of characteristic figures, patterns, and "modes" (not discussed here)) of TN. Note that these patterns and the self-similarity in TN are unrelated (they do not refer to the same objects) to the work of Mandelbrot [20] despite some common terminology.

### I.1.   Illustration of Self-Similarity

The phenomenon of self-similarity can be illustrated with the help of similar patterns seen in networks constructed using data of different granularities. For instance, the examples in Fig. 9(a) show a pattern called an "umbrella" in networks of different granularities. It is impossible to tell what the granularity is by looking at the network.

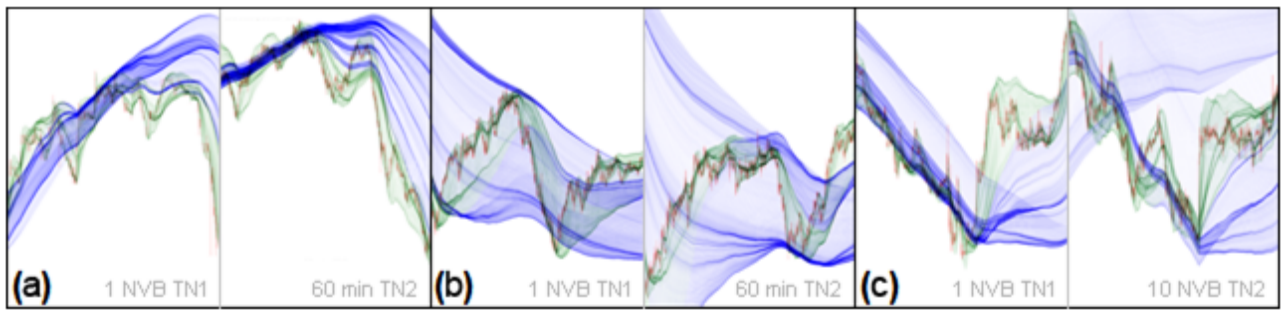

**FIG. 9. Notion of self-similarity through the observation of similar patterns in completely different resolutions.** Larger versions of these examples, accompanied by other examples, are present in App. F. (a) A very similar topological configuration (pattern), called an "umbrella", is present on two different charts corresponding to two different stocks in granularities (or "resolutions") that are very different (equivalent to 1 minute on the left, and 60 minutes on the right). (b) This pattern, called a "foot", comes from two instruments in the same granularities as in the previous example. (c) Here, the "fall brake"

pattern, followed by a trend reversal, as always, in the case of this pattern, is seen in granularities equivalent to 1 minute and to 10 minutes.

Fig. 9(b) shows two examples of a pattern, called a "foot", corresponding to very different granularities, considering that the chart on the left is of a granularity 60 times lower than the chart on the right. Note that the foot is a pattern that is observed in the case of trend reversals after a strong fall (within the timeframe considered).

Let us finish with another emblematic pattern, the "fall brake", to illustrate the identicalness of patterns whatever the resolution. The fall brake, of which Fig. 9(c) shows two nice examples, is characterized by a growing densification of the pack of cords that accompanies a very strong fall. When the pack becomes very dark, the fall stops abruptly and the price sets off on a rise.

### I.2. *The Practical Benefit of Self-Similarity*

The "self-similarity" of TN contributes to making predictions simpler since it is the same patterns that are encountered whatever the granularity. However, for certain quantities, it can happen that the self-similarity is not observed throughout all the granularities (usually in the lowest or highest granularities, but not necessarily). These "holes" in the granularities, characterized by the existence of poor-quality topological networks, and, thus, difficult to exploit, inform one about the quality of the signal, that is, the data representing the quantity (which other tools are not capable of doing).

## V. Predictions

Predictions made using TN are more or less complex and probabilistic in nature. The complexity is due to the semi-deterministic nature of the market and to the fact that the topological network comprises several network subtypes, each of which determines the future evolution of the quantity, and that the action of the characteristic figures occurs in all the timeframes concurrently (see App. R). It is by combining all the actions of the characteristic figures across the subtypes and the timeframes that one can make good predictions, both qualitatively and quantitatively.

### A. *"Semi-Deterministic" Nature of the Market*

It is obvious that the market is not deterministic since events that can influence the stock price of a company are themselves not deterministic. For example, a meteorite hitting the facilities of a company would likely result in a drop in the price of its shares. However, and this is where the notion of "semi-determinism" takes on its full meaning, TN teaches one something new about the nature of the market: events, from more or less predictable to totally unpredictable, do not affect the price in a totally unpredictable way. Quite the contrary, following such an event, the price respects the network. More precisely, soon after the "accident" (the event), the price returns under the influence of the network and starts interacting again with the first characteristic figures it meets. Therefore, "semi-determinism" should not be understood as "partial determinism", but as a very specific form of determinism, which incorporates very quickly "unexpected" events (see App. B).

### B. *Characteristic Figures Are Ahead of Time, a Specificity that Makes Predictions Possible*

Before going further, it must be said that all the other tools, beginning with "technical indicators", only follow the quantity, such as the price, with a certain temporal delay, which is basically proportional to the number of data points used to produce the tool or technical indicator. This is not the case for TN. The tool is not "lagging", on the contrary, it is "leading"; it anticipates the values taken on by the quantity over the course of its future evolution. Moreover, the more points used, the more the tool tends to be in advance. In the case of TN, the advance is proportional to the number of points used to generate the characteristic figure in question. A proportional lead in one case, that of TN, and a proportional lag in the other cases. This is a fundamental difference.

One should understand that the curves of the topological network themselves do, of course, lag; it is, in contrast, the characteristic figures, which form spontaneously, that are leading, even very much so. This is another fundamental difference (even though it is not technically speaking a difference since the other tools do not have anything other than their lines or curves, which is equivalent to what the topological network would be without its characteristic figures (which would be something totally useless)).

Of course, even though the characteristic figures are in advance, making predictions relies nonetheless on the extrapolation of the characteristic figures. This is an operation that is absent from the principles behind the other tools.

### C. *The Weaknesses of the Other Tools*

The goal of this article is not to attack the tools developed for stock market prediction; however, it is impossible not to mention these approaches given the fact that economists and analysts spontaneously tend to put TN in the "technical indicators" category, which, according to the current consensus among financial academic, cannot and do not work [21]. We will here only take a brief glance at the other tools in light of TN. We will focus here more on the tools that are not related to models and statistical methods because they are the tools that claim to be able to truly predict.

Certainly, TN relies on charts; however, this is not a sufficient reason for categorizing TN, and, thus, discrediting it. In fact, TN has nothing to do with technical analysis and is distinct from all the other tools used for stock market prediction, for obvious reasons, as we will see. The most obvious reason is that, with technical analysis and the like, "what you draw is what you get", whereas, in the case of TN, what one draws (the curves) is not what one gets. As mentioned earlier, if TN were only about the curves, it would have zero value, no matter how many there are ($0 \times N = 0$, to put it trivially).

To begin with, below, in no particular order, is a list of characteristics of the other prediction tools and their predictions.

1) The other tools never quantify the quality of the data used. To put it simply, data are "injected" and the tool indicates if the price will go up or down. TN, in contrast, allows quantifying the quality of the data. In certain cases, TN can reveal that the data are not or hardly exploitable according to appearance of the topological network obtained. Such information is precious. In particular, TN allows one to tell which resolutions (within the data from a given quantity, at a given time) can be exploited (see Figs 10(a) and 10(b)). This constitutes an even subtler level quantification of the quality of the data.

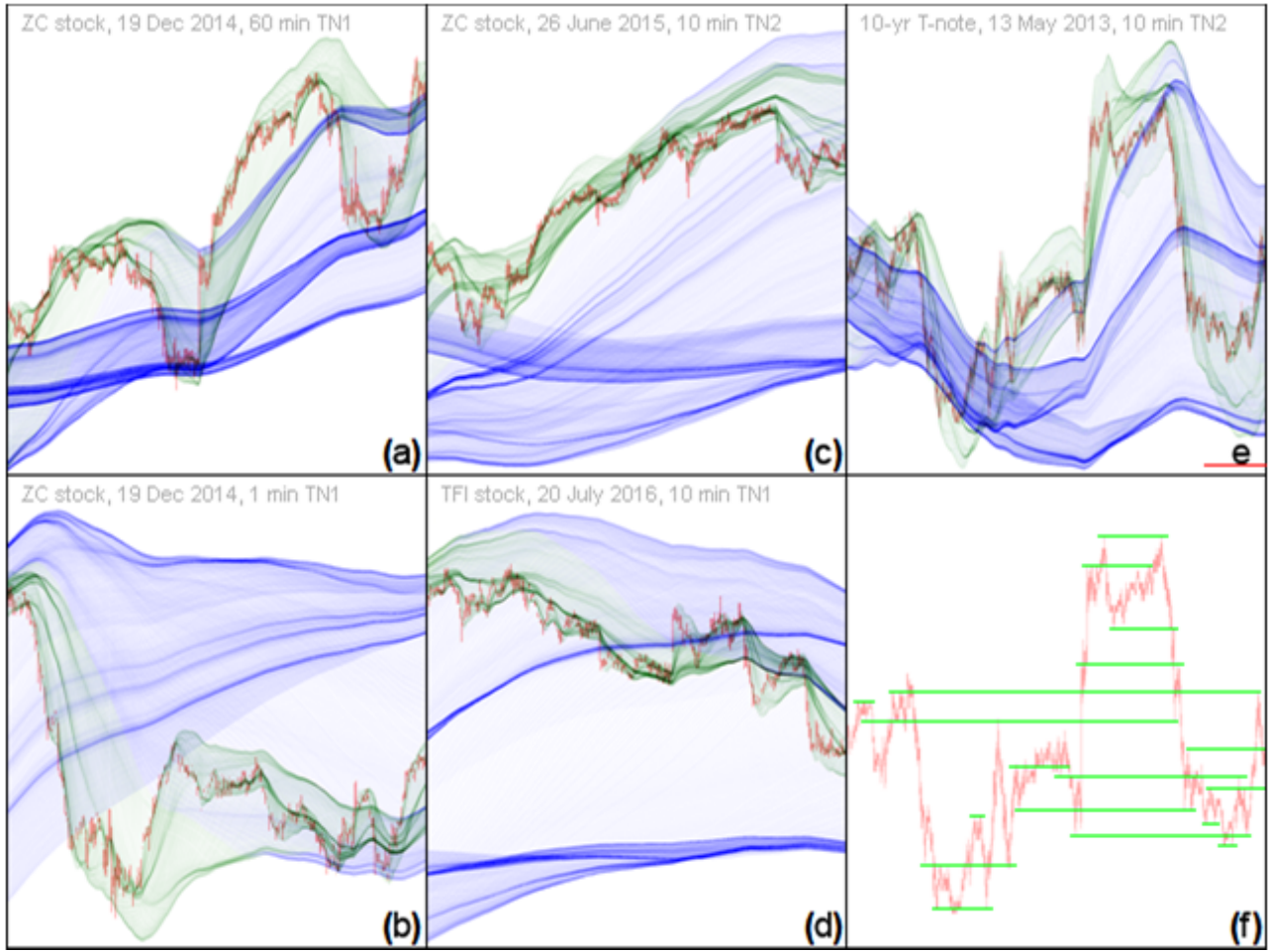

**FIG. 10. A mathematical tool satisfying a set of required criteria.** Refer to App. G for other examples and complementary information about these examples. (a) Network corresponding to the data of a stock in 60 minutes, at a given instant. This network is of optimal quality in terms of characteristic figures for providing predictions and understanding past price movements. (b) The same network, at the same instant, in 10 minutes, turns out to be of very poor quality in terms of characteristic figures. The meaning of this is simply that price movements in the 10-minute resolution are not determined by the characteristic figures appearing on this timeframe. By chance, here, the price movements happen to be very well determined by much "longer" characteristic figures (appearing only in higher resolutions). (c) Here, the probability of returning (and even quickly) to the cord is very high. This pattern is called "thin air" and is very common. (d) This network contrasts with the previous one since it is difficult to predict here whether the price will bounce off the bottom cord directly or not. (e) and (f) In light of TN, it becomes clear that "technical indicators" and other simplistic tools, such as "support lines", are fanciful mirages that are in reality nothing more than coincidences. The reason for trend reversals in price are explained by characteristic figures, as the top chart shows eloquently and unequivocally.

2) The other tools rarely accompany their predictions with probabilities. Predictions made using TN necessarily come with probabilities of actualization (see Figs 10(c) and 10(d)). Admittedly, it is subjective, but the subjectivity can be reduced to a minimum if the methodology used to make the prediction is good.

3) The various tools in use offer divergent predictions for the same data. This invalidates all or, at least, most of these tools.

4) Most of these tools rely on the use of parameters that are gratuitous or arbitrary. For example, one analyst will use a moving average of 13 days. Why not 17 or 41 days? In general, why would the mere fact of crossing such and such a curve, or exceeding one threshold or another when using a formula, make the price go up or down? If it were the case, one would already possess a very simple mathematical model (which would be widely known, given how long one has been looking for one). In reality, the reasons given, if any, for these crossings and the like are based in fantasy.

5) Most of the tools use a very limited number of data points in order to produce a prediction. Let us cite, for example, the famous Japanese candlesticks, which still have enthusiasts. The idea that, with the last 10 data points, or even less, one could predict something is completely naïve from a mathematically and physical standpoint, considering the complexity of the stock market phenomenon. Note that roughly as many as 10,000 data points, and never less than roughly 500, are needed in order to obtain a useable topological network. The numerical difference is striking. Of particular interest, in the case of TN, is that not only are the last 10,000 data points used conjointly, but the last 9,950 data points are also used conjointly, etc., down to the last 10 points, as explained in Section III.A, "Construction of the Topological Network". In other words, TN exploits, in the sense of extracting, a considerable amount of information contained in the data, which is not even remotely the case for the other tools.

6) It is interesting to note that the fact that these tools are ultra-simple, and used by a large number of traders, notably on markets that are actually zero-sum games, make the hypothesis that they are predictive logically absurd. Thus, it is not surprising that banks and firms specialized in trading are turning to high-frequency trading, which escapes this logical absurdity. TN escapes it too by the complexity of its analysis and its novelty. As a side consideration, admittedly, the widespread and proper use of TN, given that TN is very predictive, could not logically be unaccompanied by a change in the nature (behavior) of the quantity.

However, TN cannot stop working because TN is not a method or model, but a mathematical function that reveals the hidden order in data. If many people use TN, the hidden order will only increase. Note that the idea that, as soon as a method is publicly known, it ceases to work [22], is dubious for two reasons: first, there are no known methods or models [14] for predicting that have been proven to work, which makes it such that if such a method or model is publicly known it will cease to work (for the simple reason that it does not work in the first place); secondly, if such a method or model worked, became public, then, ceased to work as a hypothesized, it would be abandoned and would therefore start working again.

7) Other than the nice and naïve Japanese candlesticks, let us call attention to two other widely-used absurdities: trend lines (support and resistance trend lines) and gaps. TN allows clearly discrediting these two. Needless to say the other indicators, when confronted side-by-side with TN, reveal their senselessness. Refer to Figs 10(e) and 10(f) and App. B.

8) Indicators and methods are always very simple and easy to automate with the use of a computer program since they can be entirely formalized (which is clearly not the case with TN). Thus, in a few hours, anyone can implement any of these methods and use them to trade. Even with a relatively modest performance, for example, of 5% per month, given the numerous transactions that can be made, after ten years, at an exponential pace, a trader would make 350 times his original investment, and all this by just ensuring that his computer is turned on. Starting with 3 million dollars, he would be a billion dollars richer. We have never heard of such results.

9) Amusingly, but suspiciously, no tool is known to be better than any other. The only logical solution to this paradox is that none of these tools work.

Technical analysis and other tools are, therefore, nonsensical since they are arbitrary, without principle, and not validated empirically (see Figs 10(d) and 10(e)). These methods do not even comply with the criterion of falsifiability, so much so that the only remaining criterion consists in asking their proponents to prove their effectiveness by speculating on the market or by "backtesting" them, when applicable.

All these considerations do not prove that TN does work, but it proves that the other tools are quack science a priori. The proof of the effectiveness of TN is presented in this article and can be observed at any moment, past or present. In the case of TN, there are no arbitrary rules. The rules, which arise from an empirical theory, are derived from innumerable observable facts (the interactions).

Note that the fact that TN does not suffer from the aforementioned flaws is only a necessary condition. The property that makes the price bounce off (or stick to) a characteristic figure extremely precisely is a sufficient condition to, not only differentiate TN from all technical analysis and other tools, but also to make this tool fundamental to the comprehension of the stock market and the economy.

It would be a mistake, as we hope to have demonstrated, to see TN as a super-technical indicator. TN is of a different nature; it is underlain by an empirical physicomathematical theory that describes as well as probably possible the phenomena in question. In other words, TN is not 100 time more predictive than other technical indicators (since none are predictive), but it *is* predictive.

To distance itself from technical analysis, to which TN has no relation, we propose to name the examination of TN charts "topological analysis (examination or study)".

### D. A Criterion of Validity to Which the Other Tools Should Be Subjected

TN works indifferently with all kinds of data, provided they have a certain degree of intrinsic determinism, which can be summarized by the term "hidden order". Thus, data such as the unemployment rate, the consumer confidence index, water temperature, atmospheric pressure, etc., give rise to exploitable topological networks. We can postulate that all these quantities possess sufficiently similar structural properties at the data level, so that the phenomenon from which these quantities stem can be predicted with the use of the same predictive theory. The consequence of this is very important: this implies that a prediction theory that would fail to predict, for example, water temperature, would also fail to predict the price of a stock.

This postulate can be formulated as follows: "A predictive theory applicable to financial data should also pass the test of showing the capacity to predict other phenomena with "hidden order", such as meteorological and economical. It would be interesting to submit today's most advanced tools to the test.

### E. Human Analysis of Topological Networks

TN allows making predictions on a qualitative and quantitative basis, a property only shared with falsifiable scientific theories. The most natural approach consists in resorting to human analysis of the charts. The human brain lends itself all the better to this task that TN has a topological nature. It is well known that the human brain excels at pattern recognition. Anyone easily recognizes a familiar face that appears "deformed" due to the angle from which it is seen or due to changes produced by the passage of time. The forms and patterns in the topological network reoccur just like the same real-life changing faces or landscapes. The memorization of numerous forms and patterns viewed by a person, which takes place through a passive process, allows the recognition of shapes and patterns subsequently appearing before the eyes of the person who we will call the "analyst". As a past specific shape or pattern led to associated specific subsequent evolution of the quantity, if the same shape or pattern is later identified by the analyst, there is a high probability that the future evolution of the quantity will be the same as observed in the past with that particular shape or pattern. It is this phenomenon on which the human analysis of the topological network relies, which allows making very reliable predictions.

### *F. Two Complementary Approaches to Analyzing the Charts*

In addition to the above, one can analyze the action of each characteristic figures on the quantity at a finer level. By combining these two approaches, predictions can be made rapidly and with a high degree of precision and reliability. To understand this double approach better, it suffices to think of how a hand can be recognized. One can recognize it by analyzing the overall form and other global characteristics, but one can recognize it also by examining, at a smaller scale, the lines of the hand and the fingerprints. It is even possible to go much faster by performing a first discriminating pass to identify the overall shape of the hand, and then, by looking at the details of the hand. The earlier the recognition, the more the future evolution can be anticipated.

### *G. How to Predict – A Brief Presentation*

It is impossible to present in detail here the method by which predictions can be made. We will limit ourselves to presenting the general principles, using a few simple examples.

The method consists in the synthesis of the recognition of shapes, the analysis of the action of the characteristic figures, and the confrontation of charts of the same quantity (or same financial instrument) in different resolutions and under different subtypes. The synthesis is quite complex, but the human brain is particularly well adapted to this exercise, notably in terms of recognizing shapes. In any case, the principle is the same: the quantity (or the price) bounces from cord to cord (or, more rarely, sticks to one of them). This is observed all the time (See App. Q, R, and S). The exercise of predicting consists in anticipating the next rebound or rebounds off characteristic figures, knowing that the latter must be extrapolated, since the topological network continues to construct itself, as the quantity keeps evolving in a semi-deterministic manner. In Fig. 11(a), one can clearly see how the price bounces from cord to cord. The difficulty is in anticipating the price movements to come, that is, identifying which cords the price will go to next.

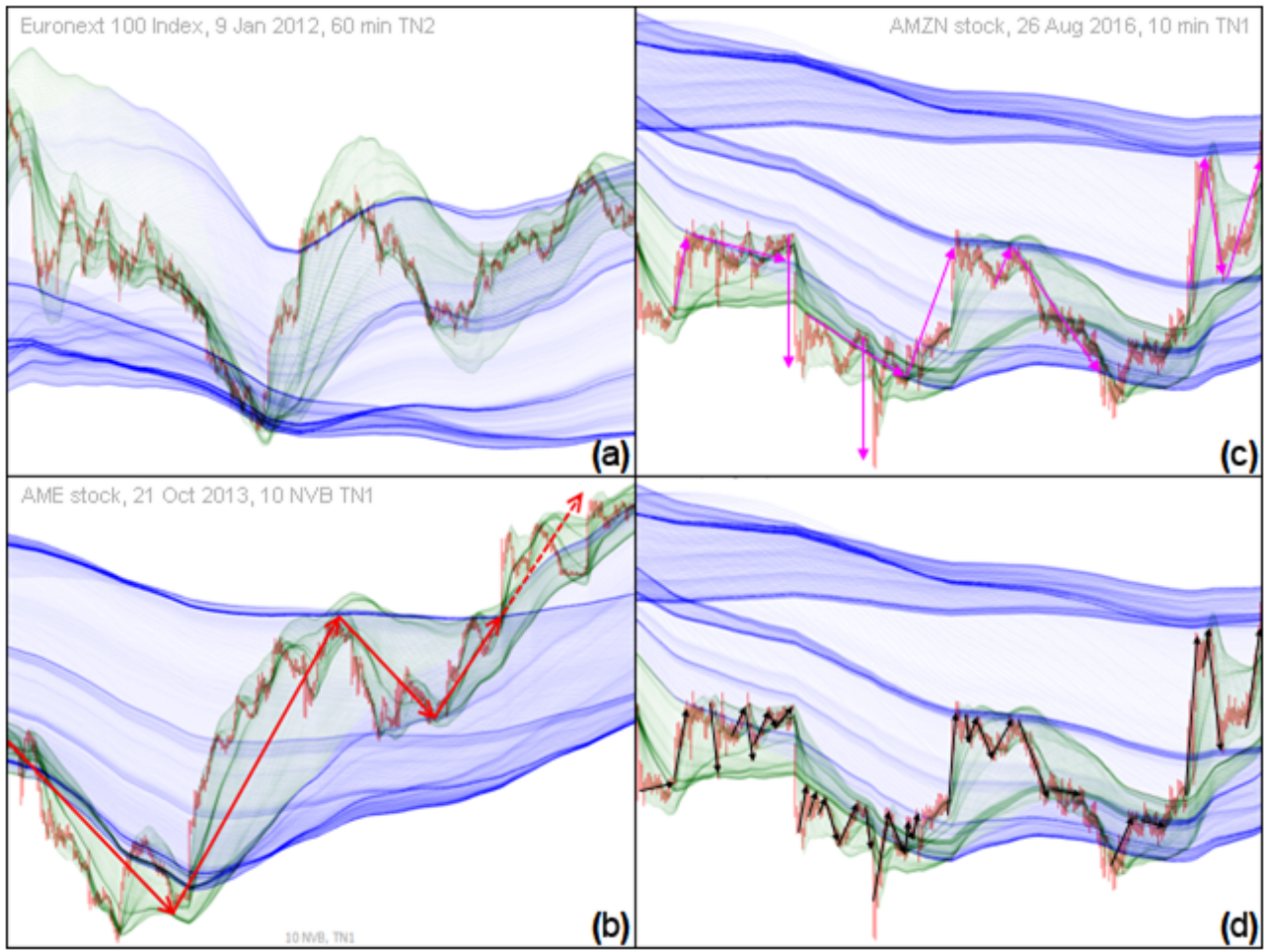

**FIG. 11. The possibility of predicting is based on the simple and fundamental faculty of the quantity to bounce from one characteristic figure to another.** Larger versions of the charts in this figure can be found in App. H. (a) On this chart corresponding to an index, the cords are rare and pronounced. This leads to easy predictions. (b) Major cords (often "longer") must be distinguished from minor cords (with a more "local" action). The major cords determine the larger price movements (see red arrows). (c) The pink arrows show the effects of each of the major cords on the price. It is at this level that predictions are made, in relation to each subtype and resolution. (d) It is nonetheless interesting, even fascinating, to observe that even the smallest price movements are explained by minor cords (at this level of resolution). One sees nevertheless that certain price movements are not explained by the cords that are present. This only illustrates the fact that one should use several subtypes and several resolutions.

While one can easily see on any chart, like on the previous chart of a well-known stock in 10 minutes (Fig. 11(c)), that the price bounces from cord to cord (and, more rarely, oscillates around a cord), one can also observe that:

1) Certain price movements do not land on a cord, or other characteristic figure. It is as if characteristic figures were missing on the chart.

2) For every case, it is impossible to come up with just one scenario based on a single chart (however, in some cases, depending on the instrument and the market conditions, using just a single chart, it is relatively easy to find the right scenario).

Regarding point 1), the fact that certain local extrema do not interact with any characteristic figures <u>on a given chart</u> leads to two comments:

a) It is possible that the future price movement will not go to a characteristic figure present on the chart in considered.

b) The local extrema interact with characteristic figures present on the same chart under one or several different subtypes, or present on the chart of the same instrument at a higher granularity (that is, in a higher resolution).

Point b) is fundamental, and it is this notion that is exploited for making predictions. We will return to this notion in what follows. It is point a) that justifies the simultaneous use of several charts for a given instrument at the same instant, corresponding to different subtypes and to different resolutions, in order to make reliable predictions qualitatively and quantitatively.

Point 2) justifies as well the simultaneous use of several charts, in order to decide between two or more scenarios.

If the scenarios represented by the black arrows in Fig. 11(d) were not obvious to predict at the time they occurred, a smaller number of more global scenarios were more certain. They are represented by pink arrows on the chart in Fig. 11(c). We will see later how the simultaneous use of several charts helps in making a prediction.

### *H. Predictions Are Sometimes Extremely Easy to Make*

Fig 11(b) presents a textbook case, since the interactions between the price and the characteristic figures are simple, rare, strong, and very precise. Additionally, we are dealing with one of the most classic scenarios, which allows an experienced analyst to anticipate the future evolution of the price relatively easily. The scenario can be broken down into four big movements (represented by the arrows in red), each one solidly pushing away from the main characteristic figures. The last dotted arrow represents the price movement that any analyst, even a beginner, would have made at the time when the price just went through the boltrope (at the top, in blue). It is a very classic pattern, already presented, called "top buildup exit". The exit is always sudden and systematically accompanied by a "return of the price to an emerging cord". Refer to Fig. 8(c) to better understand and mentally photograph this pattern.

The chart in Fig. 11(b) is a good example of when a single chart suffices for making a prediction without making a mistake; however, it is not recommended not to confront one's conclusions with the examination of other related charts. For example, it could be that a very strong cord, or other characteristic figure, which is absent on a seemingly simple chart, but present on another related chart, in particular in a higher resolution, is strongly attracting the price, weakening or even invalidating the analyst's prediction.

### *I. Simultaneous Use of Several Resolutions*

We have seen that use of several subtypes allows one to account for certain local extrema. There are some cases where these extrema, especially the less local ones, are explained by the presence of much longer characteristic figures. These appear on charts corresponding to higher resolutions. Fig. 12 illustrates well this phenomenon. See App. K for other examples.

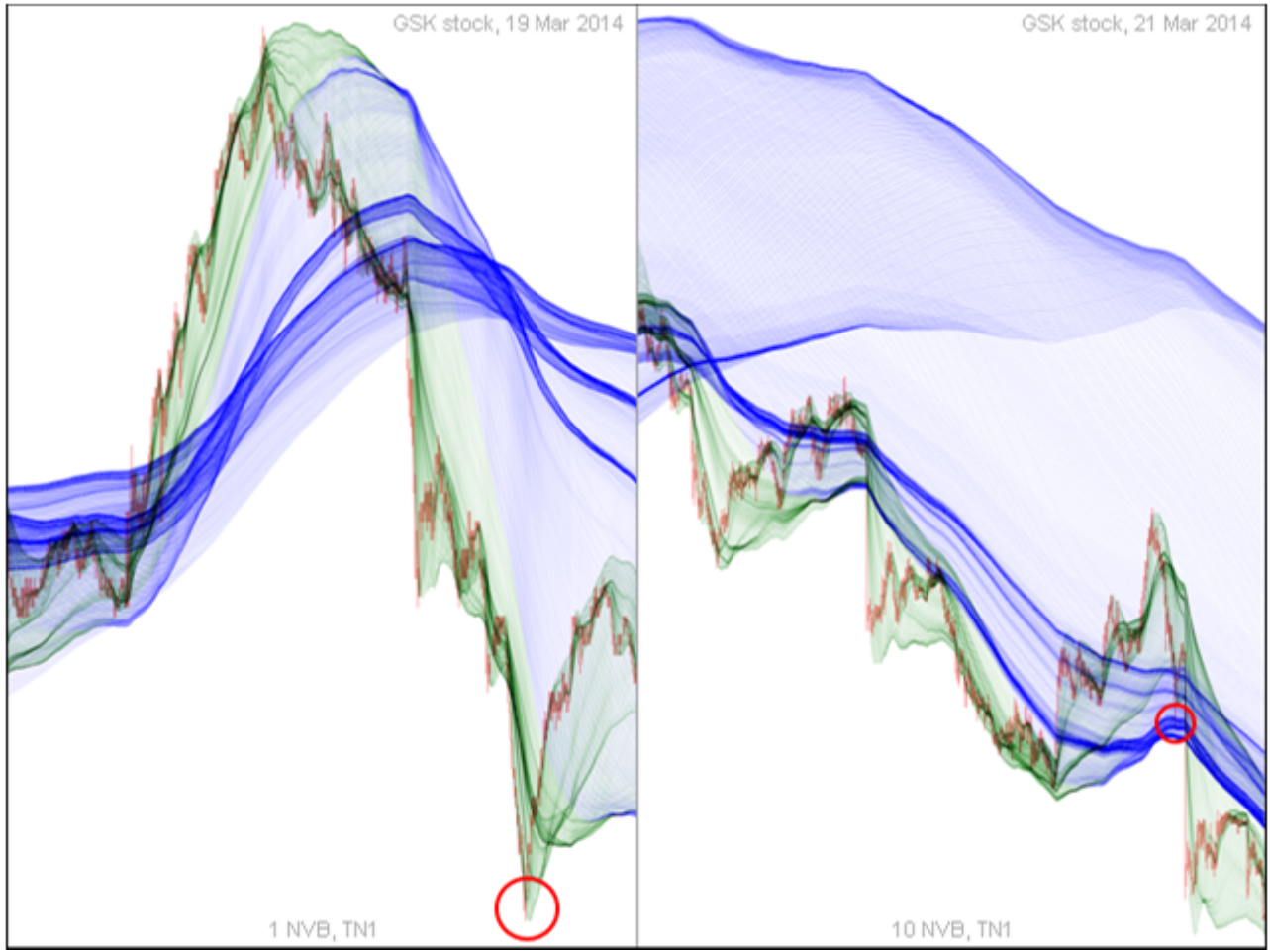

**FIG. 12. Conjoint use of a higher resolution to account for or predict a significant trend reversal.** The extremum circled in red is outside of the scope of the characteristic figures of the topological network on the chart on the left. However, a longer characteristic figure explains this trend reversal or allows (when in real-time) to anticipate it. Thus, the value of looking at the topological network under a high resolution.

### J. Why It Is Naïve to Put Forward Future Prices

Indicating a fixed, absolute price in a prediction, as a large number of analysts do, is nonsensical, in light of what TN teaches one, since a prediction consists in forecasting the next interaction of the price with a characteristic figure, such as a cord. And since characteristic figures rarely have a zero slope (as in the case of the charts shown in Figs 10(c) and 10(d)), indicating a future price implies the extrapolation of the slope, and, furthermore, knowing when precisely the interaction will take place. Indeed, as more time passes (or as more exchanges take place), the more the ordinate of the characteristic figure deviates from that of its past abscissa (as in the case of the chart in Fig 4(a), with the two major interactions with the slanted characteristic figures). Of course, a price range can be given, with more or less precision (that is, narrowness of the bracket), but the easiest predictions consist in indicating the characteristic figure(s) with which the price will interact next. The predictions remain quantitative nonetheless, not in a directly measurable manner, but in an indirect manner relying on the determination of the abscissa of the prolonged characteristic figures at the future instant considered.

### *K. Predictions – A Summary*

Take the top cord with nearly no slope in Figs 11(c) and 11(d), the closer the price moved toward it, the more certain it was to rebound off it. That which could have invalidated a rebound (that is, provoked a crossing of the cord) would have been the presence of a more massive cord (absent at this granularity, but that could have been present at a higher granularity). But this was not the case here.

We will not detail the rest of the scenario since the principle is the same. One can observe that characteristic figures that are less marked produce movements at a smaller scale, always with the same precision. This observation allows one to realize immediately that the examination of concurrent charts of other granularities allows vastly improving the quality of predictions. More precisely, higher granularities inform one about the large movements to come, whereas, the lower granularities inform one about movements at a more local level. By combining the knowledge of small and large price movements, the quality of the prediction is improved both qualitatively and quantitatively. In certain cases, a single granularity can suffice for predicting the future evolution of the price with a high degree of confidence, as in the case of Fig. 13; however, it would be a lie to say that this is true most of the time. Confrontation of charts of different granularities, but also under different subtypes, is often necessary for improving the quality of predictions. This is where the experience of the analyst comes in, to make the synthesis of local and global (large scale) actions. The game of chess offers a good analogy, where the local and the global represent tactics and strategy. It is the delicate synthesis of the two that allows one to be a good chess player.

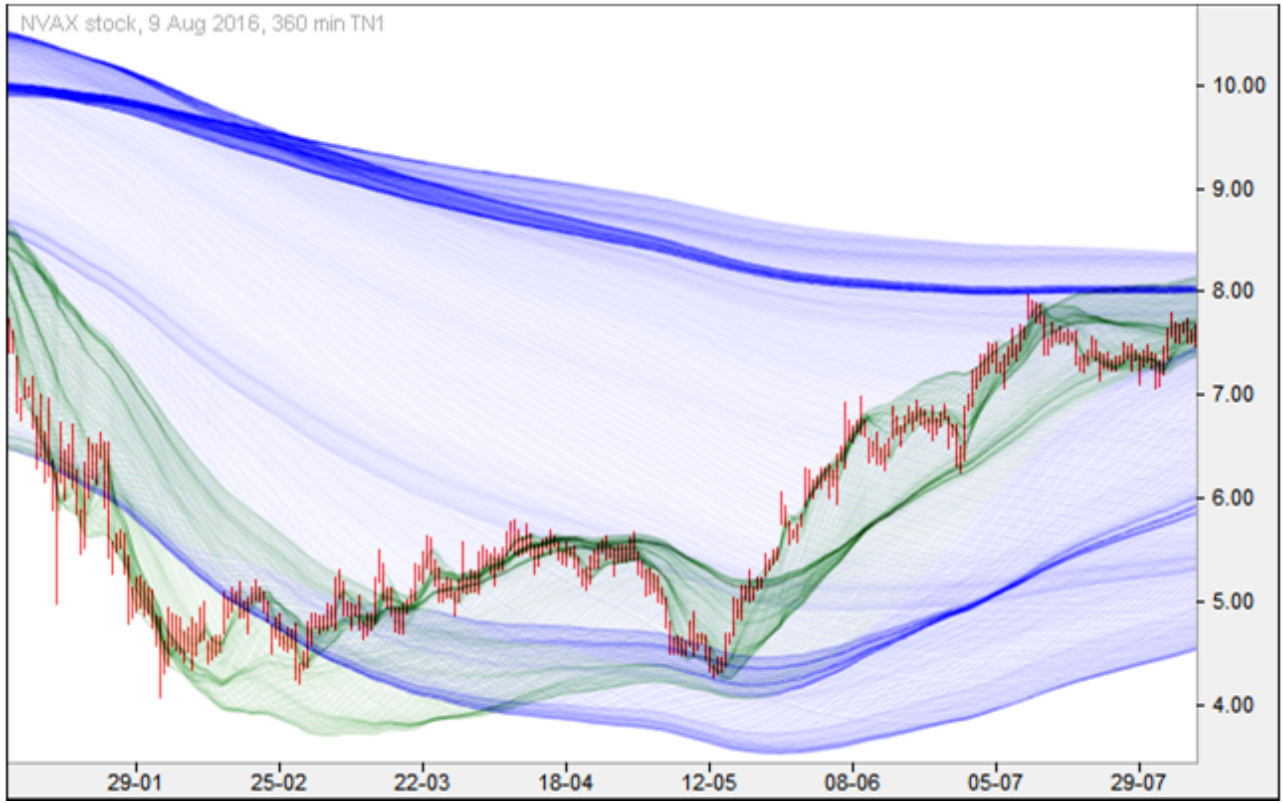

**FIG. 13. Predicting can be child's play in certain cases.** When there is a small number of cords, distant (vertically) from one another, as on this chart of a network using 360-minute data and under TN1, predicting is easy. For example, at the end of April, it was almost certain that the price was going to bounce off the bottom cord and then travel back to the top one.

## VI. On the Difficulty of Convincing the Non-Scientific Community that TN Is Not an Illusion

### A. *Demonstration by "Reductio Ad Absurdum"*

For reasons that escape the author, certain people (such as traders) still believe that the characteristic figures do not interact with the price, in other words, that they are located where they are by chance. The best way to refute this belief is provided by the simple test that follows: spatial shifting of the network (and hence, of the characteristic figures). As can be seen on Fig. 14, the smallest shift eliminates dozens or hundreds of interactions, from the strongest to the most subtle, leaving only a few (false) fortuitous interactions. The demonstration is trivial. <u>If TN were an illusory phenomenon, shifting the characteristic figures or not would not modify, on average, the number of interactions.</u> Since this is blatantly opposite to what is being observed using any chart, then, the characteristic figures interact with the quantity, and, therefore, the tool is genuine. Amusingly, if one can easily tell that the network has been shifted, then, it clearly suggests that characteristic figures interact with the price. Other examples can be found in App. N.

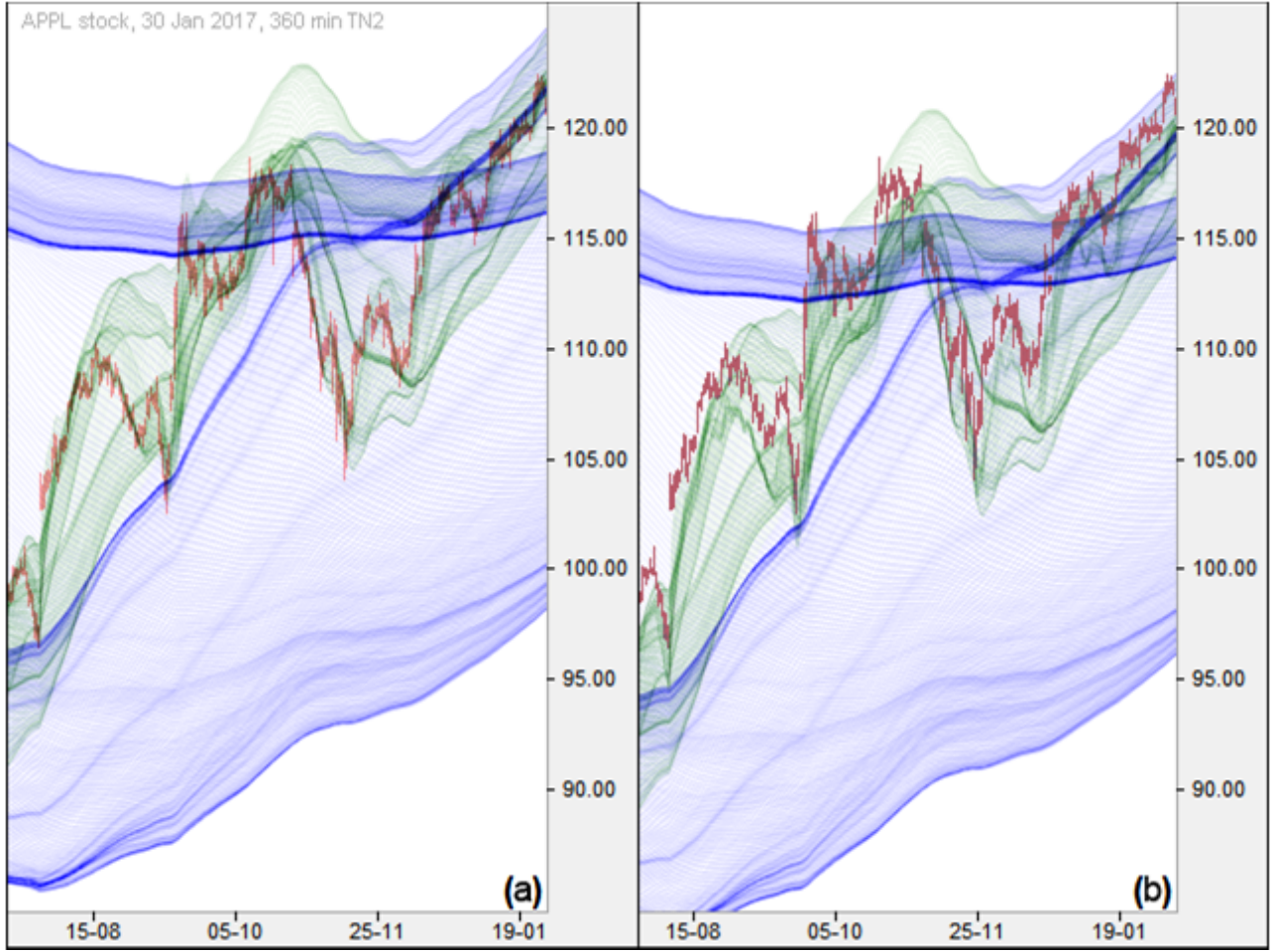

**FIG. 14. Demonstration by "reductio ad absurdum" with the help of a random chart.** Larger versions of these two images can be found in App. N. (a) A TN chart in 360 minutes under subtype TN2. One can observe the number of interactions between the local extrema and the characteristic figures. More than 20 of them can be counted. (b) A very slight vertical shifting of the network (in one direction or the other), here, arbitrarily upward, eliminates all interactions except for a few (new) interactions (around 5), which can be explained by pure chance, given the large number of characteristic figures and of local extrema. It is clear that if the shift had been bigger, the number of fortuitous interactions would have been even smaller. See App. N for other even more obvious examples.

### B. Invitation to Do a Simple Test

One may be quick to point out, and rightly so, that the price movements were spotted after the fact. However, if one places a sheet of paper over any chart and moves it horizontally from left to right, one will notice that the characteristic figures do obviously anticipate the price. The price, during the course of its evolution over time, is "guided" by the characteristic figures, from the smallest to the most salient ones. The more salient they are, the more global (at a larger scale) the action is.

### C. *The Expectations of Investors Is Incompatible with the Nature of the Market*

A large number of individuals in the field of finance, to whom TN was briefly presented, were very skeptical about predictions that were made. The reason seems edifying: accustomed to very simple and black-and-white predictions ("target price of $103.50", "entry at $96.24", etc.), the individuals saw in the predictions made using TN (which are complex, include several scenarios based on the characteristic figures present, and include probabilities of actualization) predictions that looked convoluted and imprecise, compared to the ones they were used to. Indeed, how can one be more precise than "target price of $103.50"? What they failed to realize is that the predictions that they are used to may be very precise, but that they are simply false, or, more precisely, senseless, and, therefore, worthless. We were even asked to predict the price of a stock in exactly one week's time (see App. M), to prove that TN works. This is of course absurd and in contradiction with the very nature of the phenomenon that constitutes the price of a stock, as demonstrated throughout this article.

## VII. Discussion and Conclusions

### A. *Highlights*

TN is not an econo-mathematical curiosity. TN brings to light a fundamental property of the market: there exists hidden order in the price. TN also reveals hidden order in other types of data. In reality, the versatility of TN is precisely due to the existence of hidden order much more present than imagined in many types of data or physical quantities. The topological network acts, in relation to the price, the economic indicator, or, more generally, the quantity considered, like the gravitational field in relation to massive bodies.

TN is some kind of a mathematical lemon squeezer that expresses all the substance, down to the testa of the seeds. This contrasts with the current hypothesis that the evolution of the price is Markovian, that is, where past prices would be irrelevant to its future evolution. If we take one slice of network (corresponding to one data point of the time series visible on the chart), using the $n_k$ parameters used for the TN5 charts presented here, TN exploits 1,919,328 (7,500 + 7,481

+ … + 9) values, hence the lemon squeezer image. A graphical representation of such a slice can be found in Fig. O1 in App. O. If we take a visible network of 1000 bars (that is, data points), the total number of values used (data points used and reused, but each time in a different manner) reaches 1,919,328,000 values. To summarize, almost 2 billion values are exploited for the visible part of a network under TN5. This contributes to explaining the formation (emergence) of new objects, the characteristic figures. The fact that such clean and delicate networks emerge from such a considerable amount of processed values suggests that the tool greatly concentrates and discriminates in order to reveal the hidden order in the data. It is the extreme sensitivity of the tool to the slightest order in the data (order never revealed by other approaches) that allows talking about a physical law behind stock market prices.

One of the principal contributions of TN is that one now knows the "reason" behind every (or almost every) price movement. Why does the price reach 45.71 and then, falls to 44.89 in Fig. 1? TN tells us the reason: it is because cords are there. One should not confuse characteristic figures, which are present well before price movements, with trend lines that are placed in an ad hoc manner (according to principles that are based more on magic than anything else) in order to supposedly account for what is observed after the fact. Each local extremum interacts with (by bouncing off or sticking to) a characteristic figure of the network. Not only is this completely novel, but could one have imagined that such a theory existed? It is even fascinating to observe that even the slightest price movements can be explained. They are caused by characteristics figures acting at a more local scale.

Another principal contribution of TN is the notion of the "topological nature" of the price. Predictions cannot be price-based (another reason for throwing out the window all the "metrical" tools) since they rely on objects that keep deforming over time. More precisely, the predicted value of the price is only an approximate translation, valid only at a given instant, into a measurable unit, of the interaction between the expected future price and the projected characteristic figure responsible for the interaction. This instant and this price are completely anecdotal since 1) it is not the time that determines the phenomenon, but the transactions (which can happen at any time), and since 2) the topological objects keep deforming quite freely (thus, the ordinate (that is, the price) also keeps changing). The precise price of an interaction corresponds simply to where the characteristic figure and the quantity meet. It has no other significance. TN teaches one to stop placing any particular importance on the value of the price (and other values of quantities). This is enough of a reason to explain why fundamental analysis cannot be used to predict stock prices.

Even though TN allows making very reliable and precise qualitative and quantitative predictions, this does not mean that they are trivial (such as "the price will be so much at such-and-such a date"). On the contrary, they are complex and are accompanied by probabilities. The topological (as opposed to metrical) nature of TN should lead one to abandon the widespread practice of numerical predictions (only characteristic figures matter, not the values of the price). This is reinforced by the semi-deterministic nature of the market, as revealed by TN.

One can also hope that TN will indirectly allow finally putting an end to all sorts of fanciful tools and methods that are gratuitous, without any scientific basis, deluding and tricking people.

TN requires one to revise one's understanding of the stock market and a certain number of preconceived ideas. First, it needs to be realized that each price movement, even at the most local scale, is determined by one or several characteristic figures that drive the price. If one thinks again about the image of the "drunken walker" used earlier, now that one has these characteristic figures, it is as if there were walls in his way, making him change directions or turn around as he hits them. This is no longer a random walk; this is a semi-deterministic walk. Secondly, one should be aware that a price movement lasting several days is determined by a cord requiring, for example, three years of data. Lastly, we should mention that TN is a good example of the phenomenon of emergence since, from a collection of individual mathematical objects, emerges something new, very powerful, in no way comparable to the informational sum of each constituting object, that is, each curve.

### B. *The Efficient Market Hypotheses Is False*

We will briefly examine Efficient Market Hypothesis (EMH), in light of TN, since it is so closely related to the random walk hypothesis and since it is the reigning paradigm.

The EMH is a theory that claims that security prices "reflect" the available information, with the important consequence that it would be impossible on average to outperform the market. This theory has the peculiarity of existing under three forms: "weak", "semi-strong" and "strong". In the first form, prices would reflect historical prices only. In the second form, prices would, in addition to the first form, adjust quickly to new public information. In the third form, prices would, in addition to the second, adjust quickly to "insider" information. Although some researchers believe that the EMH is false, or in part false [3], and others believe it is largely true [23, 24, 21], the research has only been able to scratch the EMH, by finding in some selected data evidence of a lack of randomness. This theory has been considered more or less valid for the reason that it has not been invalidated by the statistical models in use. TN shows that the EMH is not valid.

First, TN teaches us that it is not the price at instant $t$ that "reflects" past prices, but it is the characteristic figures at instant $t$ that reflect past prices. As for the price, it does not reflect characteristic figures; it is attracted/repelled by characteristic figures.

Second, the price at instant $t$ does not adjust to new information. New information makes the price move (what is referred to as "adjustment of the price" in the EMH) in whatever direction unrelated to the characteristic figures, and, afterwards, the new price "adjusts" to the characteristic figures. In other words, the price will adjust to the network. The network, on the contrary, adjusts very little to the news. The mathematical reason for this is that the network is constructed using many data points, and the prices affected by the news represent only a few data points. Therefore, over the long term, a piece of news (unless dramatic) has very little effect on the evolution of the price since the characteristic figures do not change much because of that news. The "adjustment of the price" as in the EMH is actually very anecdotal and the value of the price right after the new information does not have any meaning. The real adjustment is the movement of the price, following that first anecdotal movement, caused by the characteristic figures. For example, if there is a marked cord below the price at instant $t$, and a piece of news is made public to the traders who decide to act upon it, making the price, say, go up, then, afterwards, the price will not continue to go up, but will go down because of that cord will still be there (only very marginally modified by the price movement caused by the news). If the news produces a large price change, then, the adjustment, consecutive to that initial price change and caused by the characteristic figures, can be particularly large (see App. B).

Third, the EMH treats historical prices and news as information, which would cause the price to adjust to it. Historical prices and news produce totally different effects on the price, as shown by TN. Therefore, combining past prices and news into a single "information" category, as the EMH does, has no pertinence.

Lastly, TN shows that predicting future prices can be done well, which invalidates the direct consequence of the EMH, which actually proves by reductio ad absurdum that the EMH hypothesis is false.

### C. The Market Is Semi-Deterministic

One can say that there are two large groups of researchers: those who consider that news and psychology are economic factors determining the price (or the market as a whole), in other words, that the price (or the market) would be the result of a causal mechanism, while not possessing a predictive/explicative theory, and those, the econophysicists, who consider that the price (or the market as a whole) behaves in a largely random manner, in other words, that the price (or the market) would be unpredictable.

TN invalidates these two hypotheses; however, it would be absurd to believe that certain events are not capable of influencing the price. To this effect, one only needs to think of Black Monday, September 11, 2001, or, more recently, Brexit. An econophysicist sees the price as a random walk perturbed by some external forces [12]. They would cause fluctuations sometimes far beyond what a random walk probabilistically allows. For an econophysicist, a stock market crash is seen as a "tsunami", that is, an outside force [12]. TN shows that the tsunami is not an outside force; the tsunami is the consequence of the past prices themselves, through the characteristic figures. The outside force is simply the news, which can kick the price in one direction or another. But, it is the appearance of the network, that is, the presence, positions, and shapes of the characteristic figures that will determine the next price fluctuations. And, as indeed observed, there is absolutely no probabilistic limit to the amplitude of these fluctuations, especially when there are no strong cords that surround the price.

How then is it possible that the price be influenced at the same time by TN and by events? This is precisely because of the semi-deterministic nature of the market (or price). What happens is that, when an important unexpected event (of any kind) occurs, the price "jumps" (or lands) onto a characteristic figure that is more or less distant, depending on the importance of the event. But, in any case, the price conforms to the topological network and to the characteristic figures, which, moreover, continue to evolve, integrating the new prices posterior to the event. In the absence of significant events, the market, through TN, is largely deterministic.

Because the market is semi-deterministic, the predictions made using TN are more or less complex and probabilistic in nature (while remaining precise) because it is impossible to know in advance if events will tend to push the price in one direction or another, even modestly.

Samuelson [2] starts his 1965 article by saying, "In competitive markets there is a buyer for every seller. If one could be sure that a price will rise, it would have already risen", asking himself if this statement is correct. Let us immediately point out, as an aside, that Samuelson's statement is suspicious in itself, semantically speaking. Samuelson is working under the hypothesis that the evolution of the price is random. Yet, he says that the future price (the statement says "price will rise") "would have already risen", which is opposite to the idea of a random evolution of the price. Therefore, the statement contradicts the hypothesis. For Samuelson's statement to be assessed as correct or not, one would first need to ensure that the quantity (the price) does not behave like a predictable physical phenomenon. Let us take the weather. If we are sure that the temperature at a location will rise, it has not already risen, obviously. It may sound like a trivial counter analogy, but only because we now have a method to decently predict it and because there are no known ways (at least for now) to influence it. But suppose that the price behaves like a predictable physical phenomenon, the statement would then be: "If one could be sure that a price will rise, it would not have already risen", it will rise. Since TN shows that the price behaves like a predictable physical phenomenon, therefore, this last statement is the correct one. Now, one could and should ask oneself whether this last statement will still hold when TN is known to the public and used by buyers and sellers? We would be in a situation analogous to weather that one could act upon. But would that make the weather forecasting method no longer valid? No. In the case of TN, the law should remain even more unaffected because it is not a method, but closer to a usual physical law. The worst case scenario would be if all the buyers and sellers would use TN optimally, which is very unrealistic (because, even in the case where all the buyers and sellers were using solely TN, some would buy and others would sell, they would all use different timeframes, and some would analyze well and some poorly). This would likely cause the price to stagnate (which, in passing, would not invalidate TN since TN would be able to predict the stagnation). This would make some people abandon the use of TN, therefore, making the price move again. So, even if the TN information is publicly known, based on TN, one can say "if one has good reasons to think that the price will rise, it will likely rise".

### D. What the Tool Brings in Practice

Other than the possibility of predicting with great precision the future price of a financial instrument or the future value of an economic indicator, TN allows differentiating very easily between "good" and "bad" instruments (at a given instant or in general), that is, those which can give rise to predictions and those which cannot, due to the absence of easily exploitable topological networks (that are lacking characteristic figures or that are too messy). Note that this possibility is new to the understanding of the stock market since no other tool or theory today seems capable of distinguishing between good and bad instruments, not in terms of expected gains, but in terms of "signal quality" (which is unrelated to the richness of the data representing the instruments in question).

TN allows one to judge when an instrument is not, at a given instant, predictable. This is notably the case during what we call "stationary mode" (see Fig. L2 in App. L), which is a temporary oscillating network configuration that cannot be missed. We can say that the stationary modes illustrate the notion of the semi-determinism of the price, here toward temporary indeterminism. Note that when a stationary mode ends, the price goes looking for a characteristic figure, which shows that the indetermination of the stationary mode is only partial (see Fig. G6 in App. G).

TN can be used to detect price anomalies, for examples, errors in the data or possible price manipulations, based on an abnormal appearance of the topological network. Such an abnormal appearance can correspond to unusual topological structures (groups of characteristic figures within the network), or to local extrema that do not interact with characteristic figures, or to breaks in the topological network. We can say that such abnormal networks are rare, and most "accidents" are due to well identified causes (announcements of financial results, major news, etc.).

### E. Economics Is a Science After All

If there is still a doubt about the scientific nature or not of economics, TN should contribute to tilting the balance toward the hard sciences. Since the market, through the prices of instruments and the values of economic indicators, conforms, in a surprisingly powerful manner, to an empirical theory that has a purely mathematical nature, economics is clearly a science. We regret contradicting Louis Bachelier, the father of mathematical finance, who, in his 1900 thesis [1], posits that it is impossible to mathematically predict the stock market and that the dynamics of the stock market will never be a hard science. That the economy is actually a hard science should please the mathematicians and physicists working in the fields of economy and finance. One can even affirm now that non-scientific propositions (such as "investors are being cautious", "the market is getting nervous", "investors are optimistic this week", "the market reacted well to this news", etc.) should be definitively abandoned. Upon examination of TN charts, it seems obvious that the "psychology of the market" is an unfounded illusion.

TN also allows proving, in an indirect and empirical manner, but very convincingly, a number of things, and, in particular, answering the questions: "Can the price of a financial instrument be distinguished from a random walk?", "Are financial time series reversible?", "Are (nontradable) economic indicators (such as the unemployment rate) of the same nature as (tradable) financial instruments, and, in particular, can they be predicted?", "Can crashes be predicted?", "Are all financial instruments equally suited to TN?", "Can the data of one instrument be distinguished from that of another?", and "What effects does news have on topological networks?" These questions cannot be covered in this already lengthy article.

### F. And What if the Physicomathematical Law of the Market Had Been Found?

There is more to the physicomathematical law than the mathematics of the network itself (network based on regressions of order $D$) in that the law rests not on the mathematical curves, but on the complex actions of formations (the characteristic figures) that appear spontaneously within the network (emergence). Since the appearing of characteristic figures and their effects cannot be mathematically theorized (nor mathematically conjectured or deducted), this physicomathematical law falls into the "empirical theory" category. For this reason, the proof is also empirical. To this effect, there is a wide range of evidence that validates the theory (exposed in this article), and, needless to say, TN has never been invalidated by observation (meaning that the characteristic figures always appear and always interact with the quantity). Refer to App. Q, R, and S for a formal indirect proof of the validity of the theory. The fact that the proof is indirect (as opposed to a direct mathematical demonstration, which is impossible) does not weaken TN. Note that researchers have tried to prove the randomness of the evolution of the price also using an indirect demonstration. If such proof had been found (provided that the evolution of the price was actually random), it would have not weakened the proof.

Consequently, it seems difficult to refute that TN embodies the physicomathematical law of the market (or of the price). Even if some reject the claim (although proof would need to be given), it seems difficult to deny that TN forces one to change one's comprehension of part of the economy. Specifically, to those who would maintain the position that the physicomathematical law of the market has not been found, they will have to explain what TN is, why characteristic figures form spontaneously, and why the latter attract and repel the price systematically and with such precision. Considering the consistency and breadth of the evidence, to reject or ignore the phenomenon would be rejecting what is observable, and therefore be a form of scientific denial.

TN is a very powerful tool, maybe as powerful as it may exist for the quantities considered. It allows, with the theory, understanding market prices well, taking into account the fact that the market is not fully deterministic. One cannot exclude the possibility that another family of curves could work also, which would incidentally reveal that the market is even a bit more deterministic than TN shows. If so, the empirical theory would probably remain unchanged. TN brings proof that the market is much more deterministic than previously thought. For now, TN seems to be an isolated mathematical object. Time will tell if TN can be included in a broader mathematical theory.

Beyond economy, the fact that the physical, concrete world (such as the market) is so well described, not to say governed, by pure mathematics (the tool), as opposed to equations or models with specially chosen constants or parameters, in other words, physical laws, is quite fascinating.

To summarize, we believe to have proven that the market is not at all random; it is, on the contrary, semi-deterministic, towards a rather strong determinism. This has been achieved thanks to a new mathematical function (the TN tool) that is extremely sensitive to the slightest order in data (hardly or not at all detected by statistical methods) and that has the unique property of engendering emergent mathematical objects, the "characteristic figures" (among them, the "cords") that happen to drive the price (and other quantities), by successive rebounds and adhesions. The tool outputs a "topological network" which is relatively complex to analyze, as it exists under multiple subtypes. Nonetheless, predicting the price can be done relatively easily and reliably, thanks to an empirical framework consisting of empirical laws and by putting to use the existence of reoccurring topological patterns specific to TN. As a consequence, the random hypothesis is proven false, now that we have a physical law that describes and predicts the evolution of security prices.

## Conflict of Interest Disclosure Statement

Regarding competing interests, the author is an independent researcher (not paid by any institution, research center, university, or company) and has a patent and software program in relation to the tool.

# Appendices

## Appendix A: Where, Why, and How the Market Turns Around

      No other tool today allows knowing why and how the price of a financial instrument (for example, a stock) turns around. Two other examples (Figs A1 and A2) are presented here, again focusing on 4 spots where the price turns around (4 local extrema), as in Fig. 1. One can identify without difficulty other local extrema that are explained by the presence of characteristic figures in these two examples, or in any other chart present in this article.

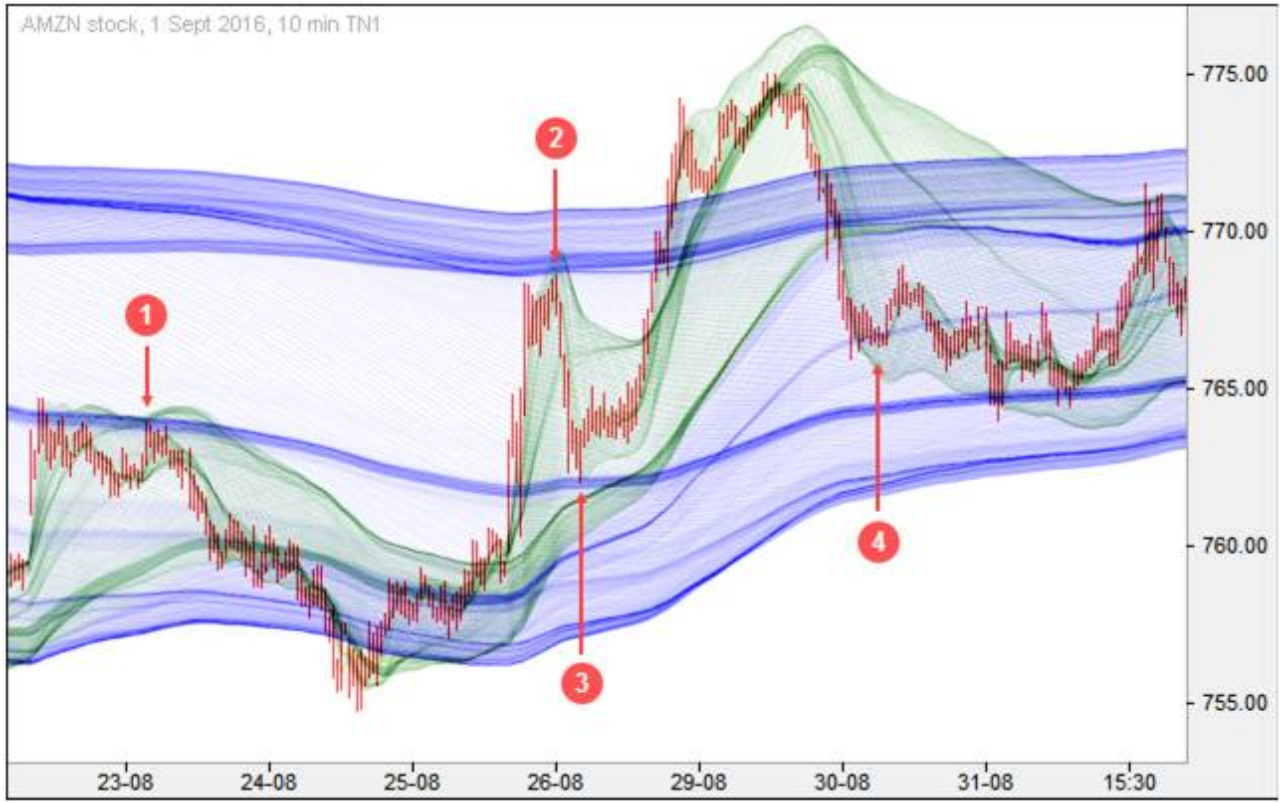

**Fig. A1. At the origin of price turnarounds: the characteristic figures.** Four local extrema have been arbitrarily chosen on this topological network under TN1 to illustrate the fact that, thanks to TN, we now know the cause of these price turnarounds (local extrema). One can see that both the smallest and largest price turnarounds are subject to interactions (rebounds) with a characteristic figure.

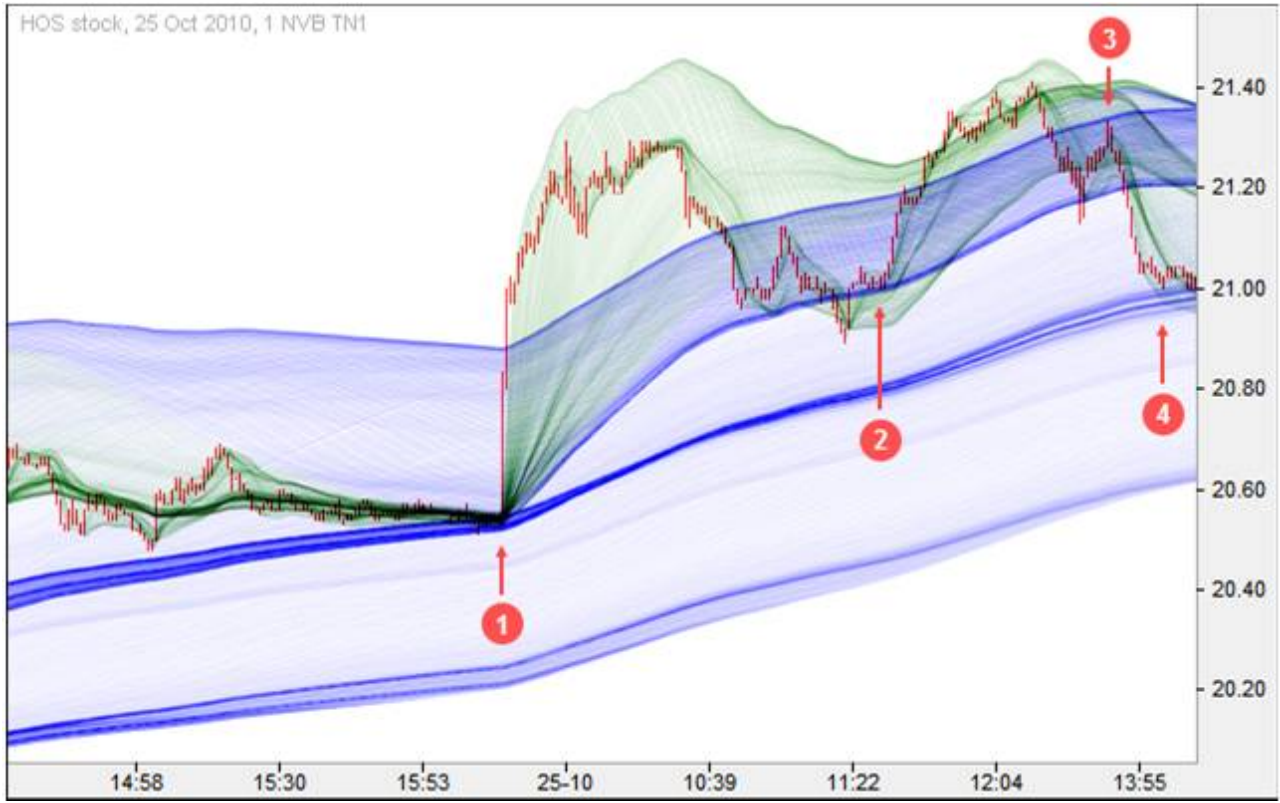

**Fig. A2. A nice example that is particularly simple and striking.** Note also the precision with which the price turns around, exactly where the cords and boltrope are found, in this network in a smaller resolution and, here too, under the TN1 subtype.

One may wonder why some price turnarounds do not interact with any characteristic figures. For example, on Fig. A2 below, some local extrema are "in the middle of nowhere". Figs A3 to A5 provide the answer and supply the interactions for most of the extrema that are without interactions in Fig. A2. In other words, all of the turnarounds/extrema are explained thanks to TN. Refer to Section II.F, "Curve Subtypes", in the article in order to understand better the figures below.

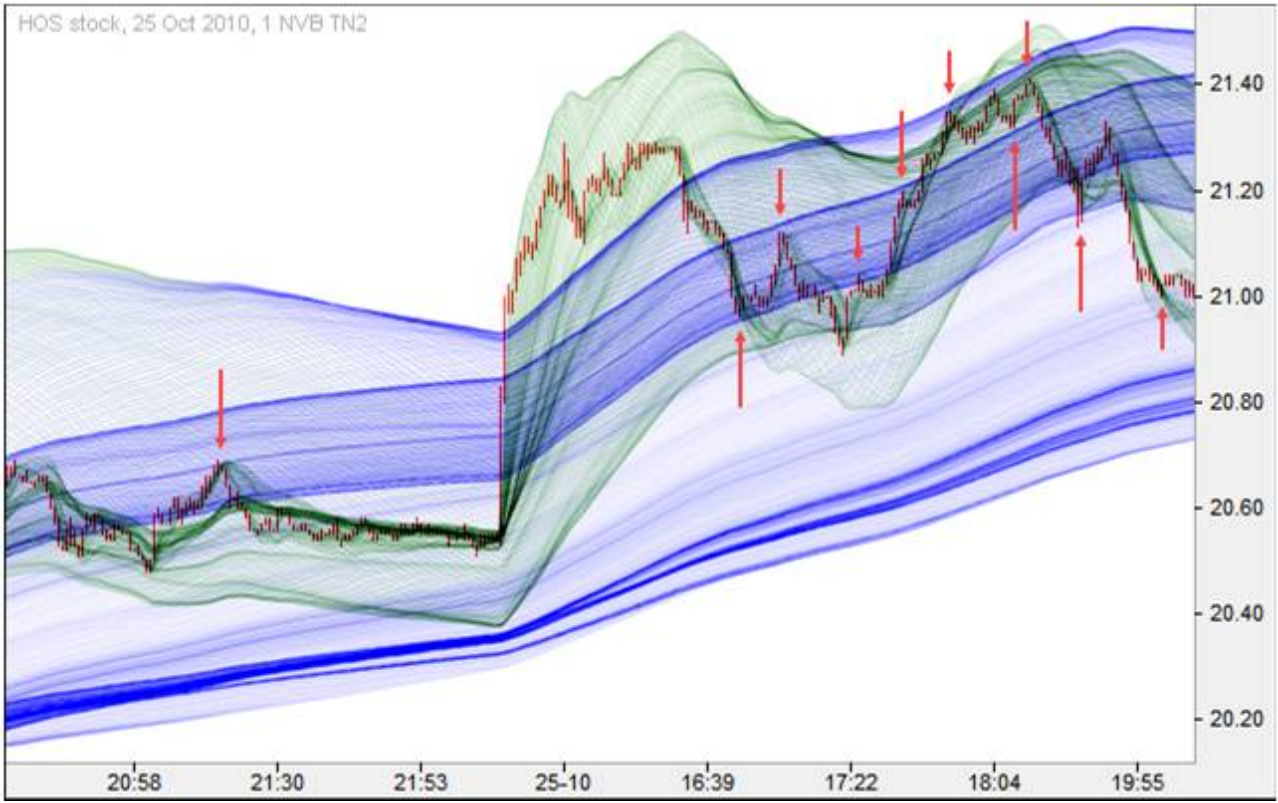

**Fig. A3. Other interactions present on the network under TN2.** Notable local extrema tend to be seen under several subtypes. Here, subtype TN2 reveals a dozen more interactions.

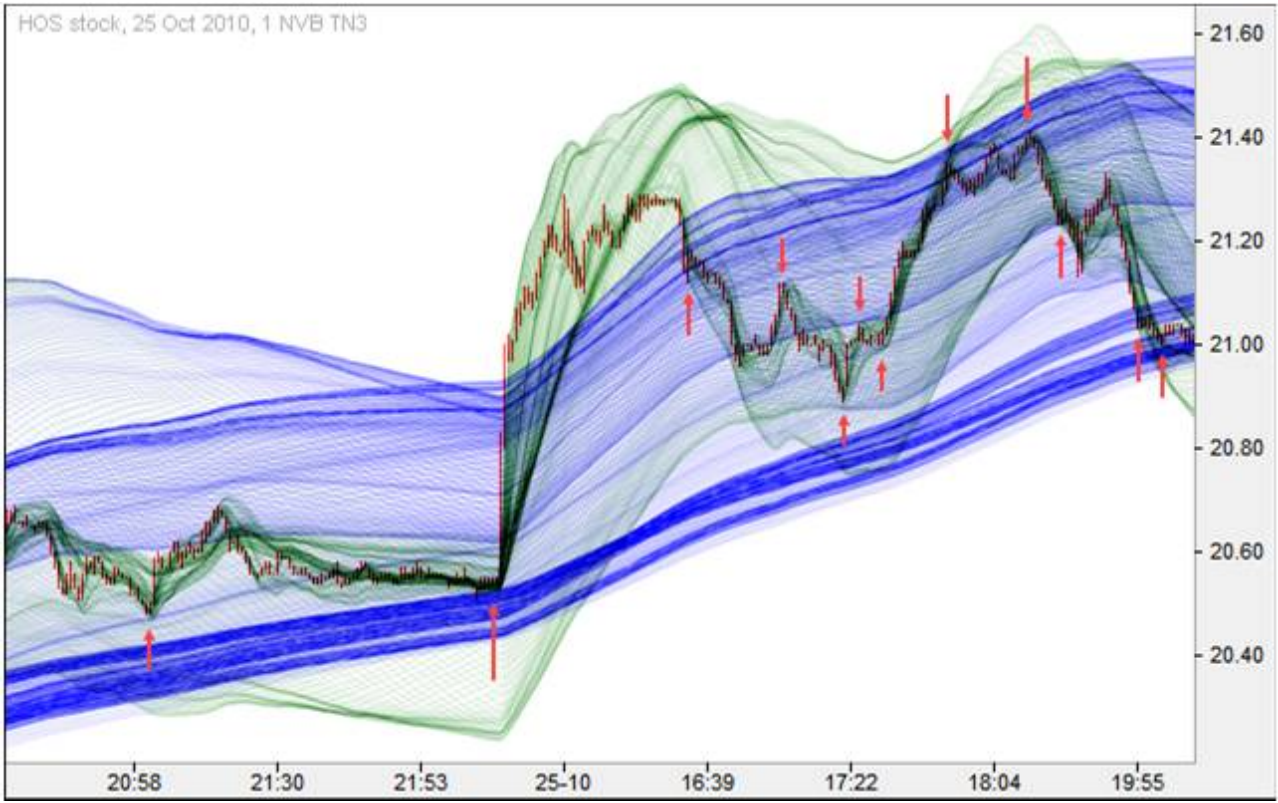

**Fig. A4. New interactions under TN3 and some repeated interactions under at least two subtypes.** It is interesting to note that certain interactions are present under both TN1 (Fig. A2) and TN2 (Fig. A3). The most salient of these repeated interactions is found at the bottom of the network and is identified by the number (1) on Fig. A2.

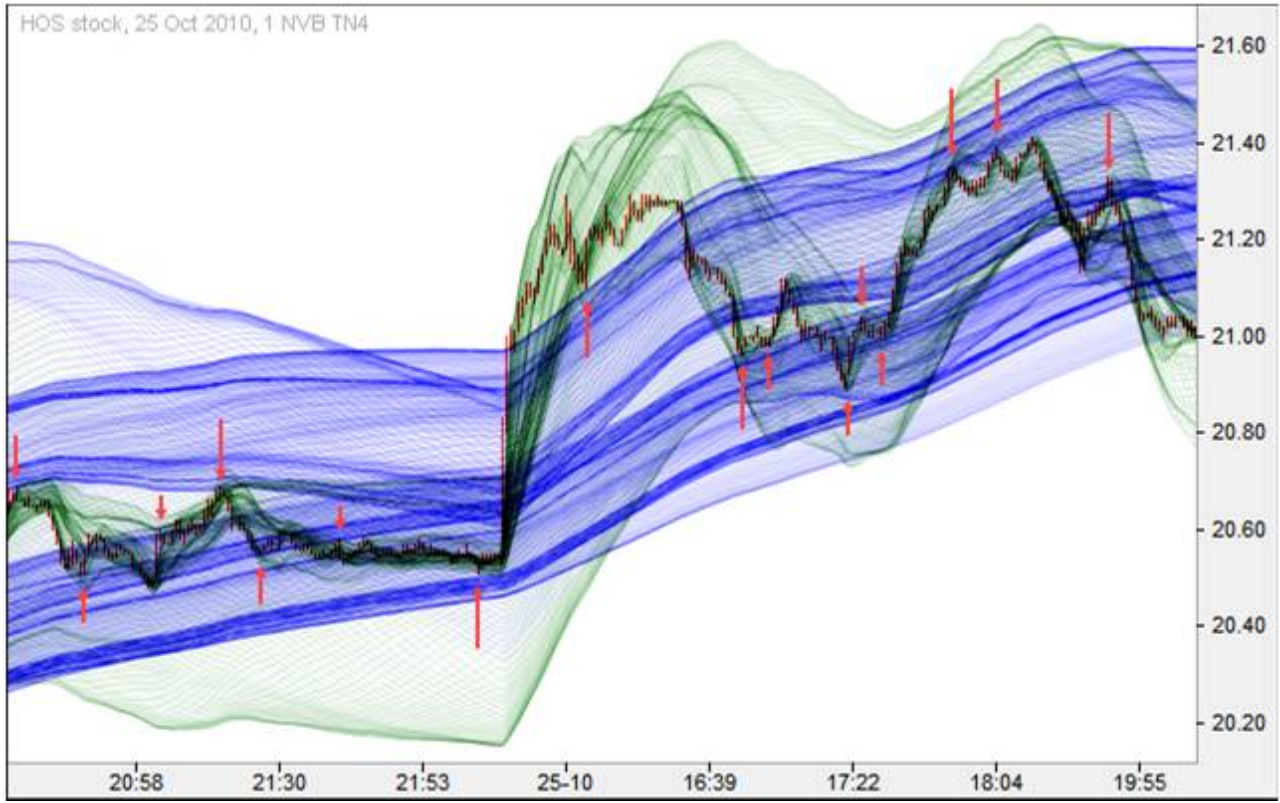

**Fig. A5. Confirmation of many interactions in this network under TN4.** The higher subtypes reveal more characteristic figures. Under higher subtypes, certain interactions tend to repeat themselves, especially the most major ones, that is, corresponding to local extrema of larger amplitude.

## Appendix B: How "Accidents" Land on Characteristic Figures

It is commonly accepted that accidents of all kinds (unexpected news, various financial results, financial scandals, etc.) produce completely unpredictable effects. This is actually not true; the price very quickly goes back to the characteristic figures, no matter the timeframe (or granularity) used. This reinforces the affirmation that TN has never been invalidated by observation.

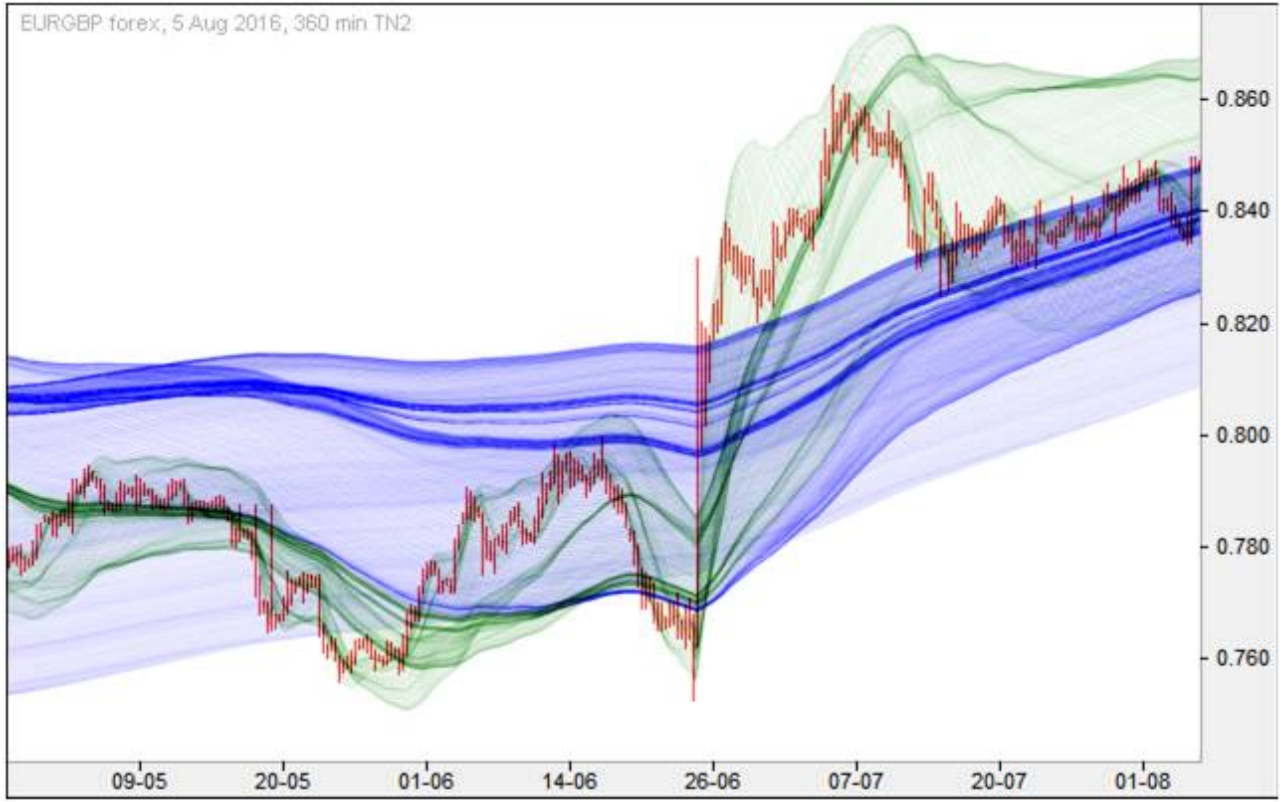

**Fig. B1. Brexit, as seen through TN.** It is remarkable how the Euro-Sterling exchange rate fell, very quickly, back onto the cords at the top of the network. One can even see how quickly the rise found support on a green "emerging" cord.

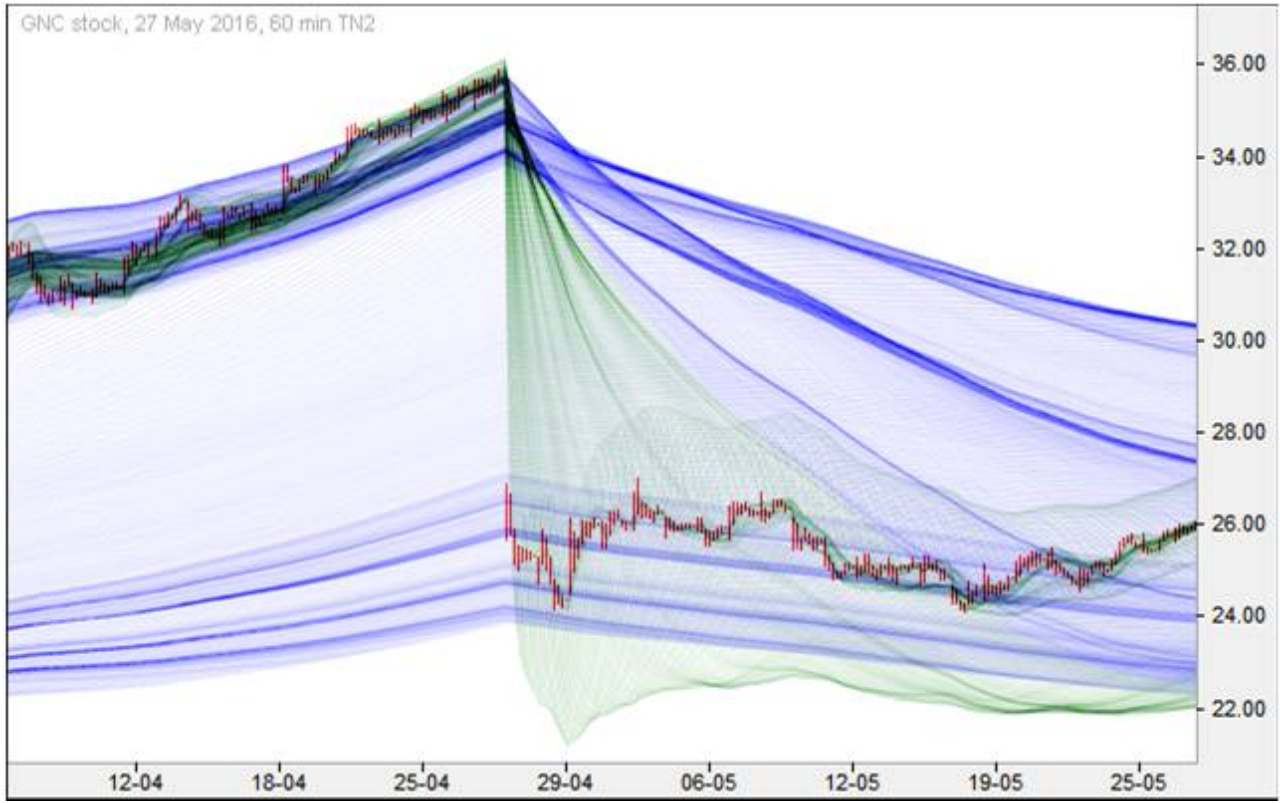

**Fig. B2. Fascinating landing on cords after a major price drop.** On April 28, 2016, this stock immediately dropped after the release of quarterly results. Not only did the evolution of the price following the drop fully interact with the bottom cords, but the drop itself landed on a cord. Note that the existence of cords at the very bottom makes it so that, even in the absence of this news, a fall onto these cords would have been very likely. Generally, even major news cannot prevent the price from being attracted by characteristic figures.

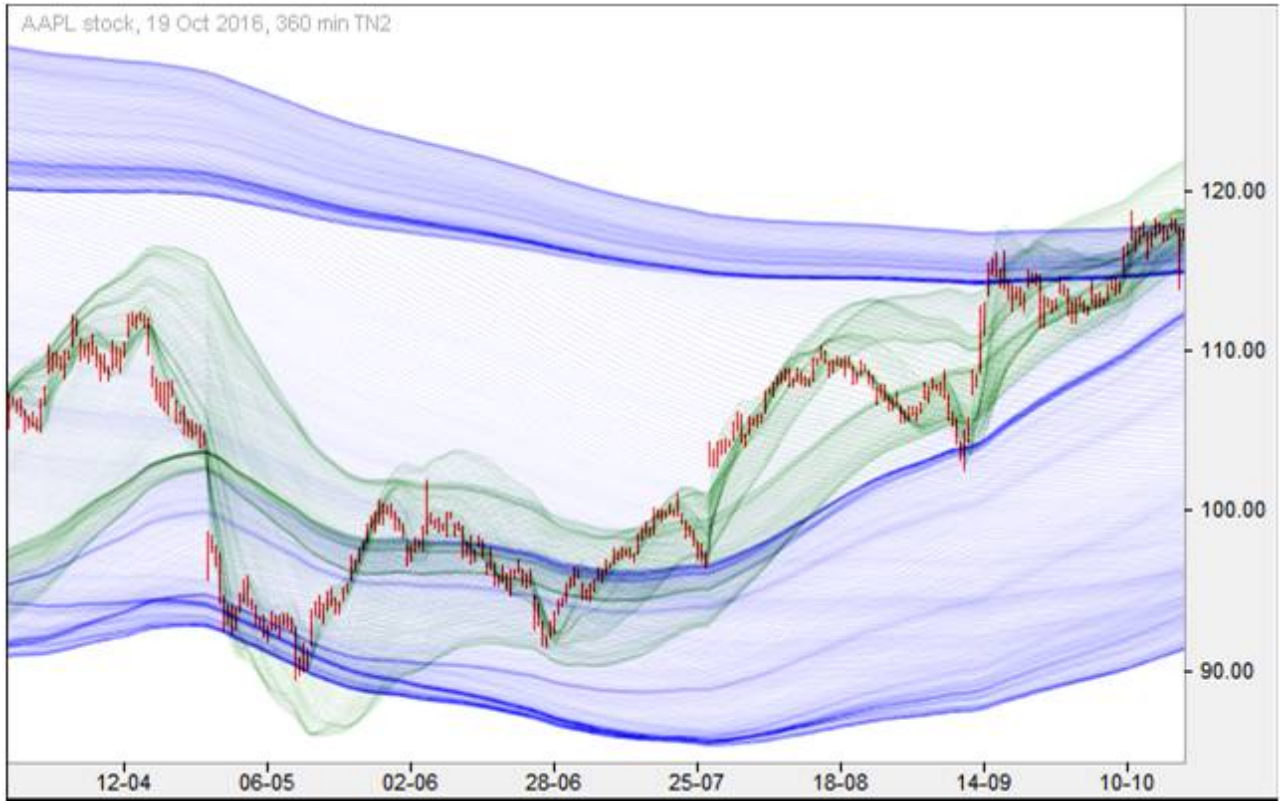

**Fig. B3. Moderately important news has little effect on the network.** On April 26, 2016, Apple Inc. announced disappointing news after climbing for 13 years. The next day, at market open, the stock plummeted. One can see how the price landed on the bottom cords. This temporary fall, which would have likely occurred in any case in the absence of this news (because of the presence of those blue bottom cords), was of no consequence, as the subsequent evolution of the price, guided by the cords, shows. This illustrates the weak effect of moderate news in relation to the network.

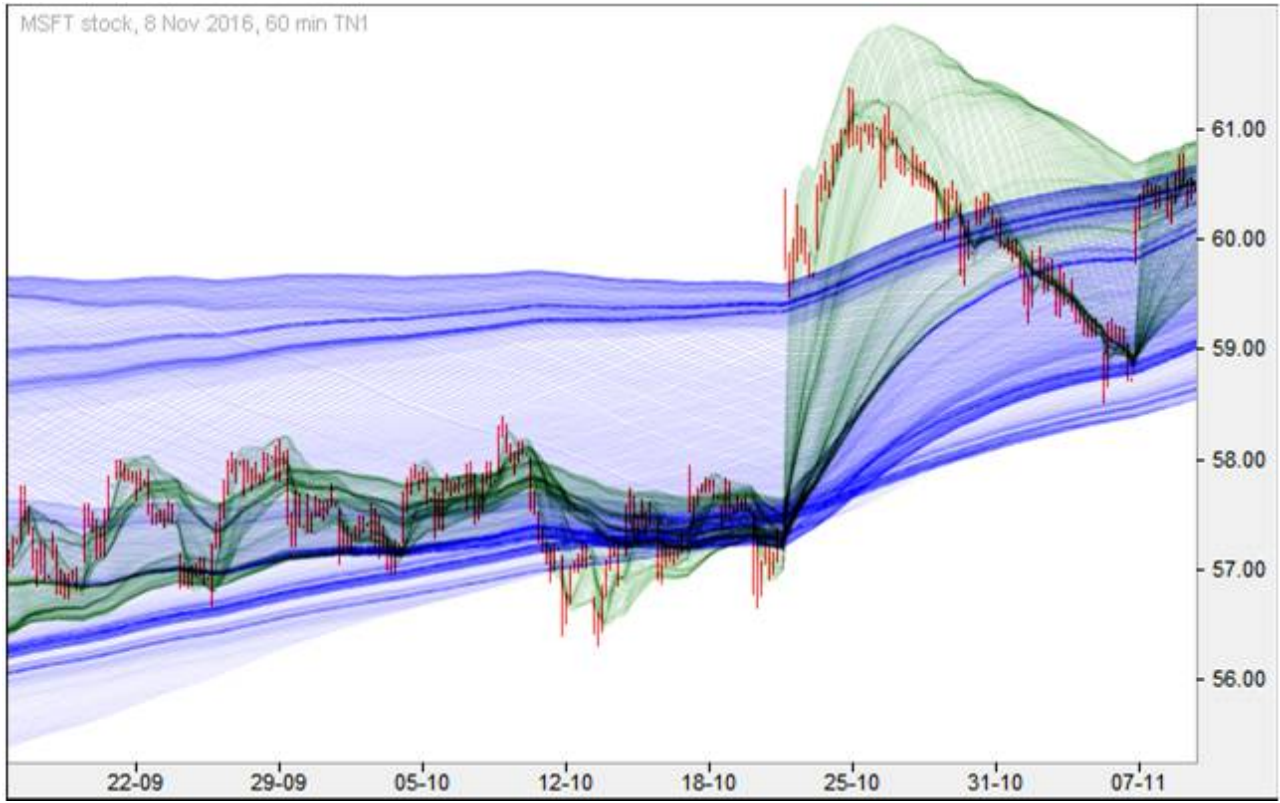

**Fig. B4. Moderately important news always respects the network.** On October 20, 2016, Microsoft Corporation announced higher-than-expected quarterly revenue. The next day, at market open, the stock jumped, landed precisely on the boltrope, before finally falling back onto the thick blue emerging cord. Note that the presence of the strong blue condensation preceding the news was conducive to making the price move violently in one direction or another. Under all circumstances, things would have returned to normal given the shape of the network at the time. One can see a second jump, at the beginning of November, right into the same top cords.

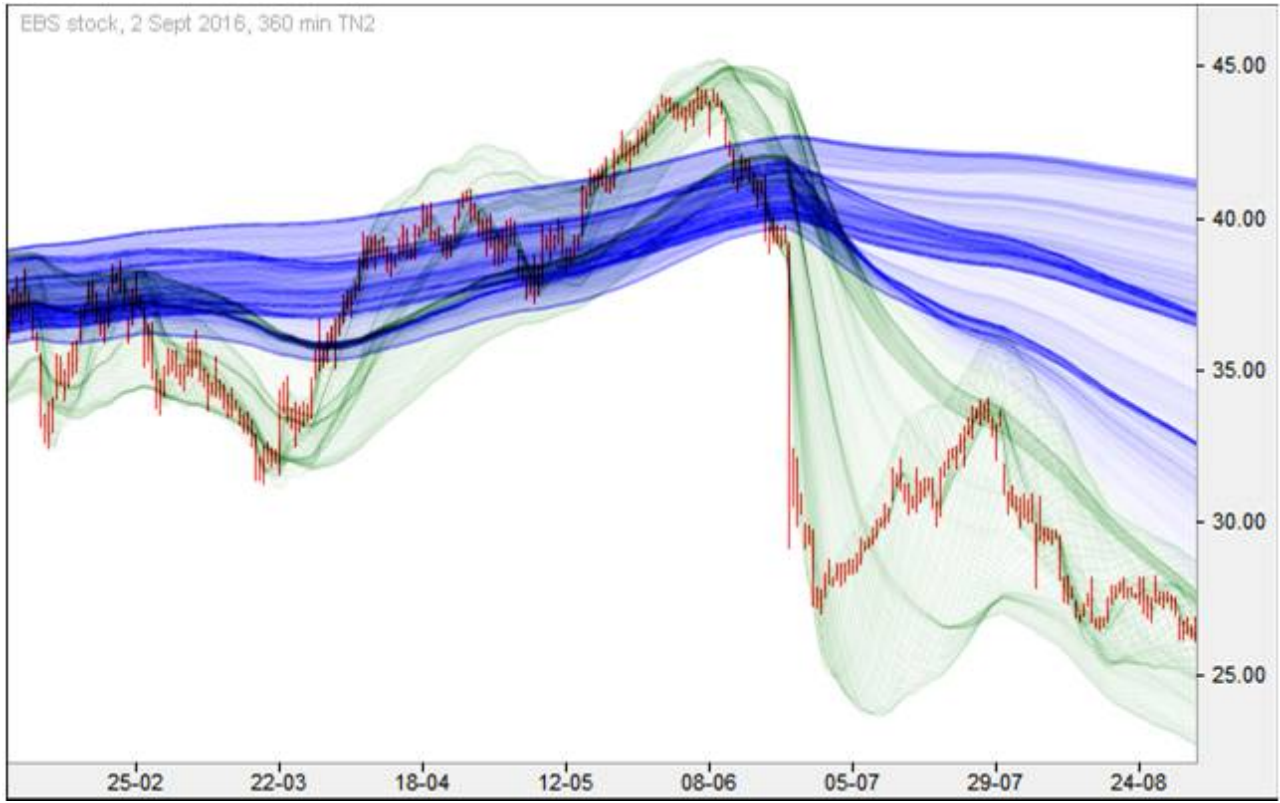

**Fig. B5. Even after drastic drops, the price goes rapidly looking for a cord.** On June 22, 2016, Emergent BioSolutions dropped after a commercial decline related to vaccine sales. Note the very "dangerous" position of the price underneath the very thick cords just prior to this news. The action of such a topological configuration is always a violent nosedive. This illustrates the usefulness of TN in relation to upcoming news.

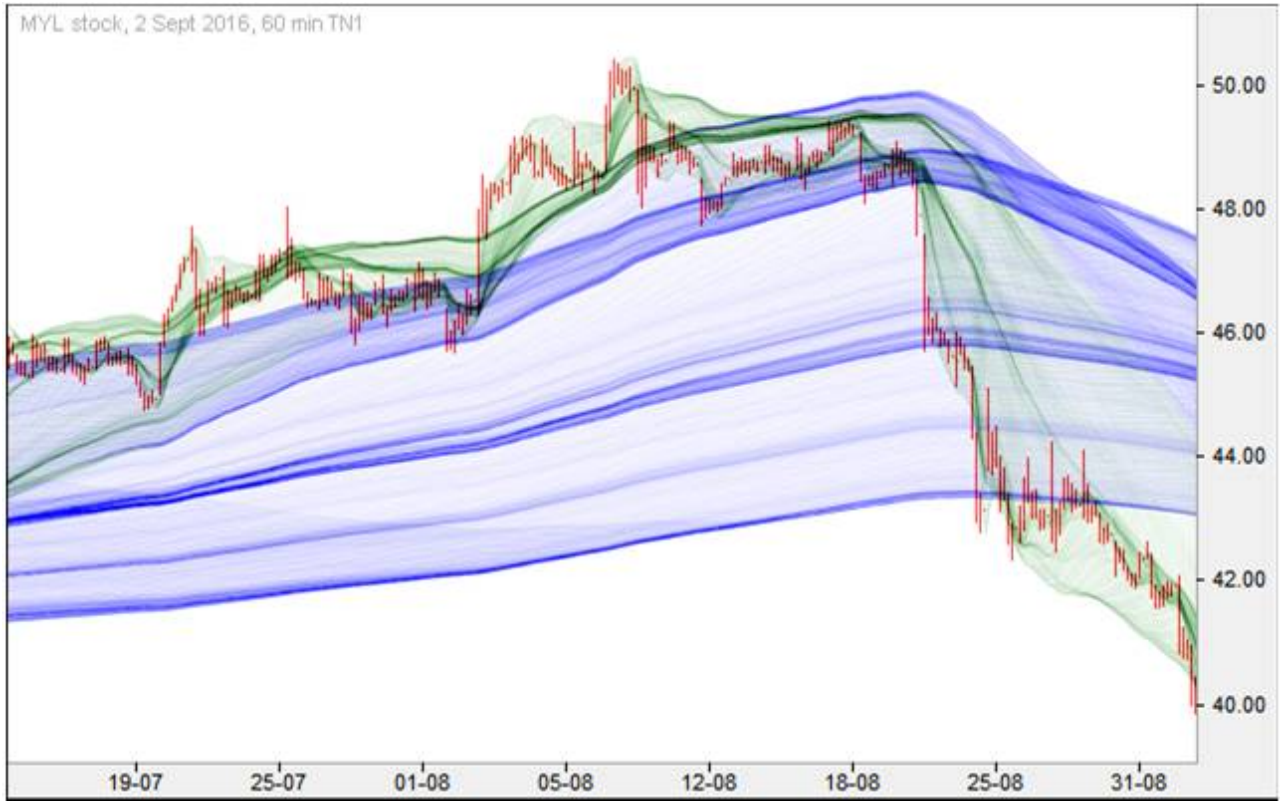

**Fig. B6. News of little importance can produce marked effects when the network shows an unstable topological configuration.** On August 22, 2016, Mylan is being asked to explain the price hike on the EpiPen. This is followed by the price tumbling from cord to cord, as if falling down the stairs. This type of minor news, which normally should not produce such an effect, produced here a very strong effect because of the presence of the distant bottom cords in the network. The way the price paused for a moment on the first major cord, before continuing its fall to the other bottom cord, is, although common, always fascinating.

## Appendix C: Curve Subtypes

The topological network comes in different regression orders, hence the term network subtype. Figs C1 to C5 present the charts from Fig. 3 in larger dimensions so that the characteristic figures and their effects under each subtype can be observed and compared more easily.

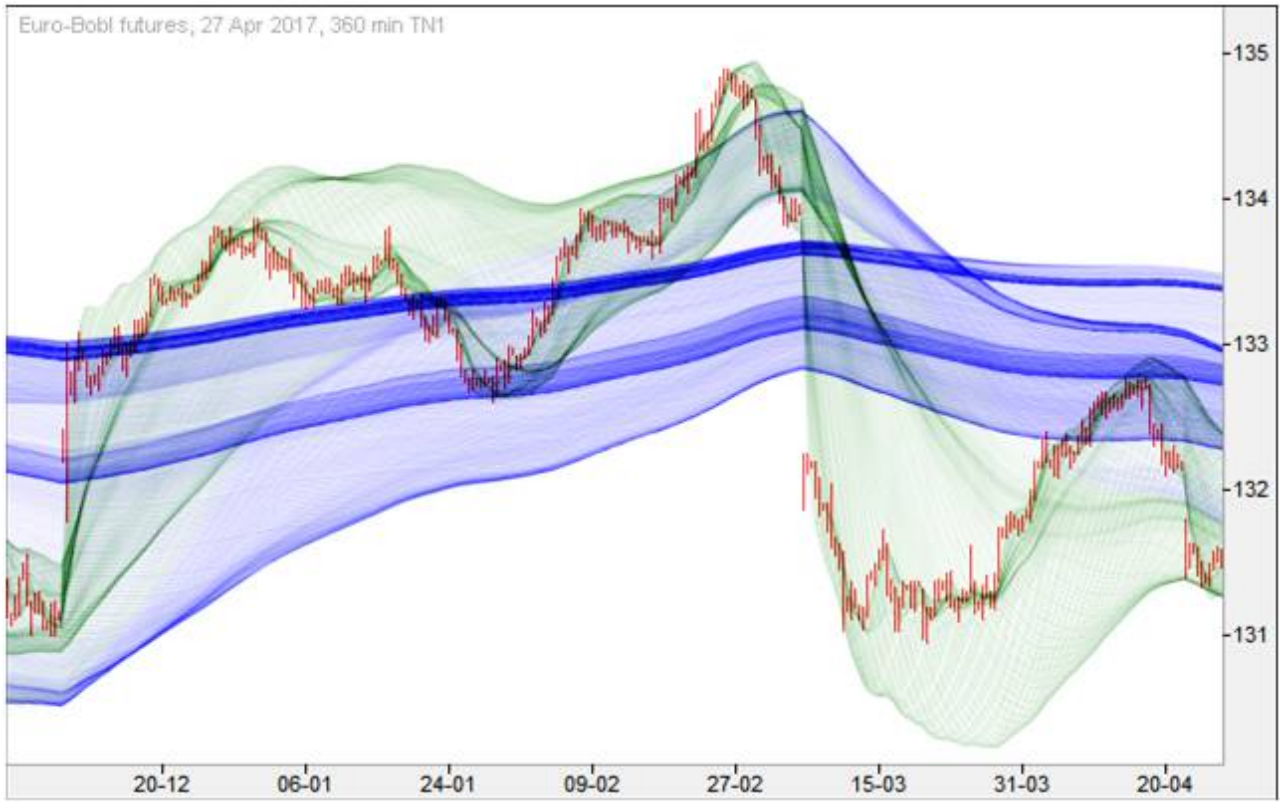

**Fig. C1. The topological network under TN1 of the Euro-Bobl futures.** Network corresponding to the Euro-Bobl future (futures of the medium-term German bond) in 360-minute resolution, April 27, 2017. The subtype TN1 reveals numerous and extremely precise rebounds off the few very marked cords present under TN1.

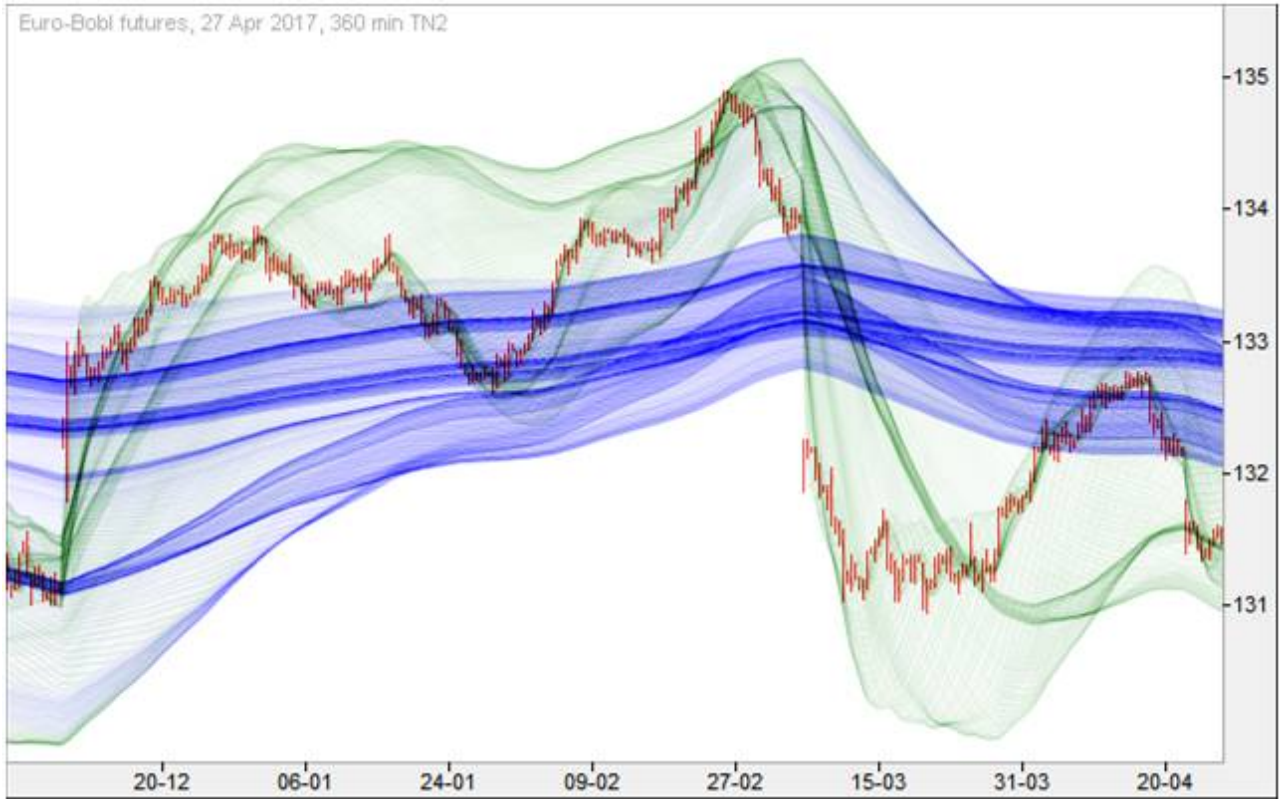

**Fig. C2. Same network under subtype TN2.** Under TN2, some of the extrema that did not have any interactions under TN1 interact with cords present under TN2. In particular, the first bottom local extremum (on the far left) and the highest local extremum (around 135) are explained by the network under TN2.

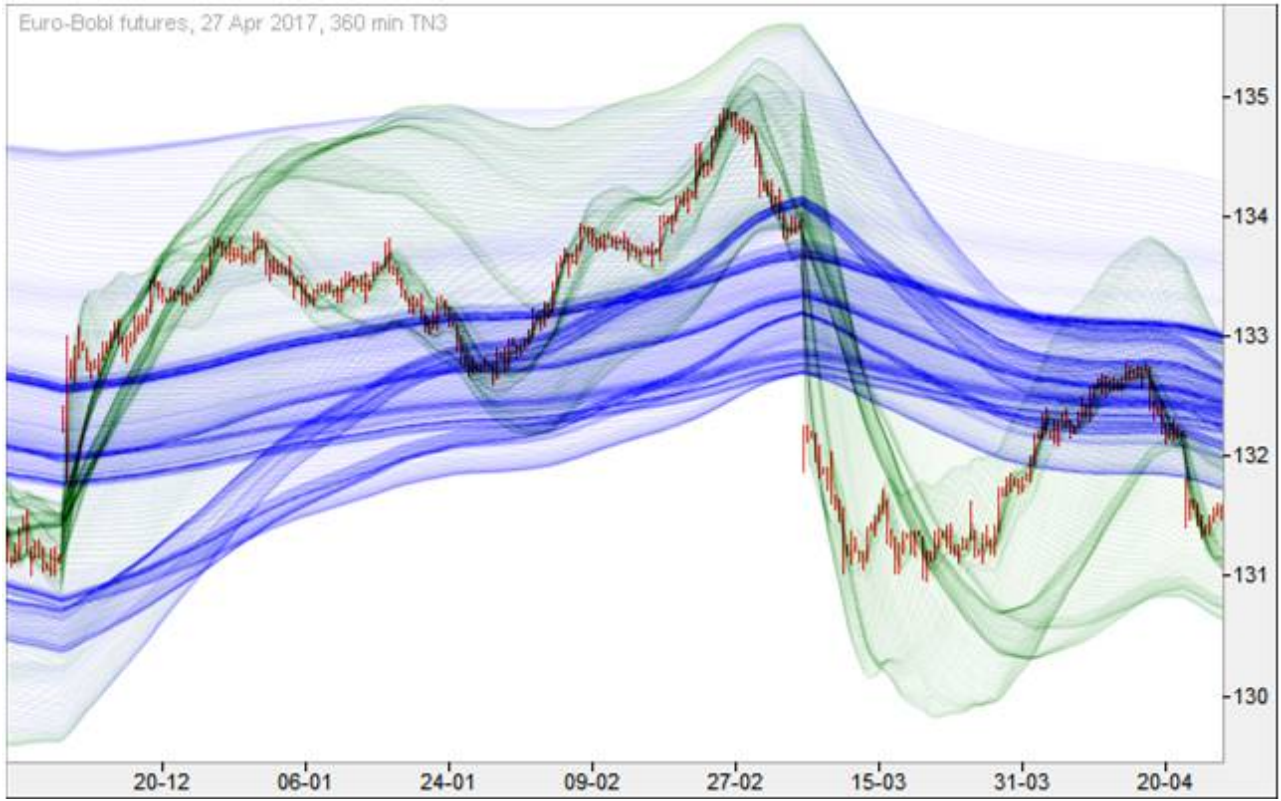

**Fig. C3. Same network under subtype TN3.** New extrema are accounted for, as well as a certain number of extrema which were already present under TN1 and TN2.

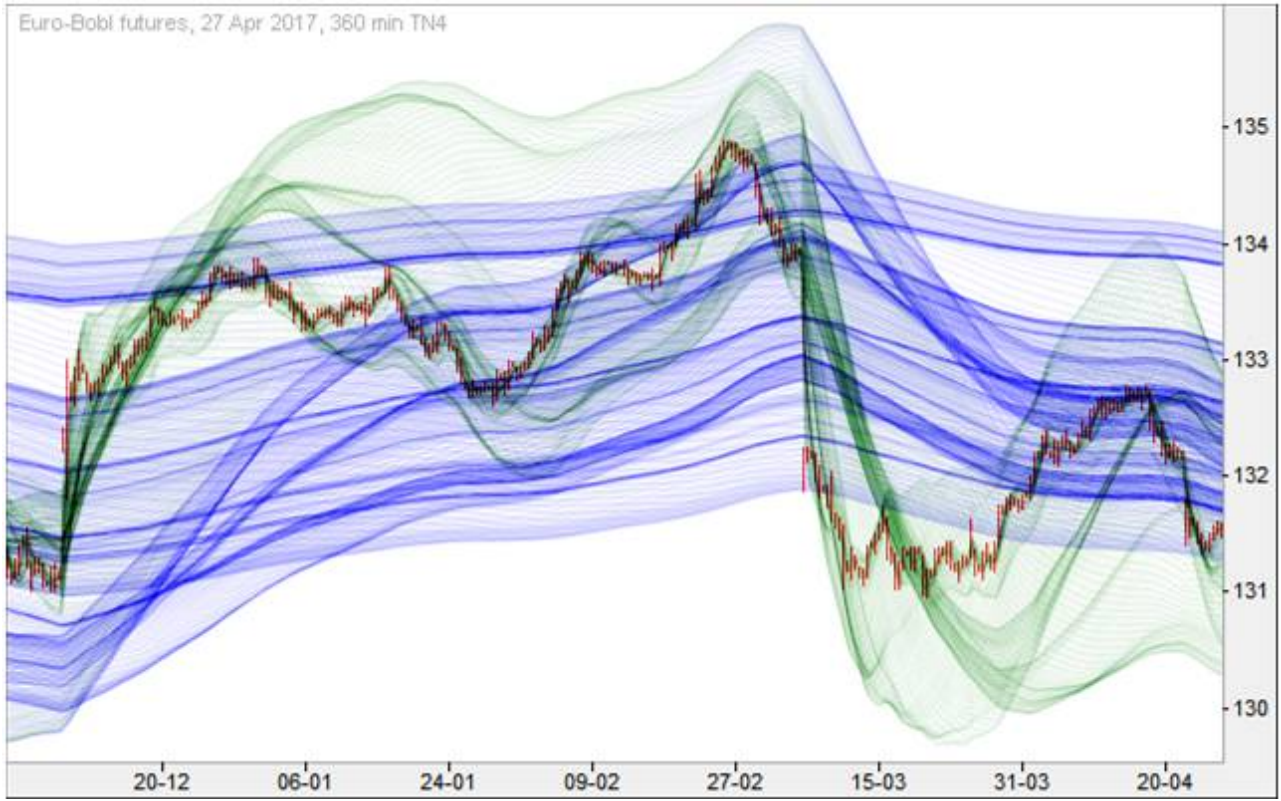

**Fig. C4. Same network under subtype TN4.** The same comment applies to TN4. Note the number of major interactions under TN4.

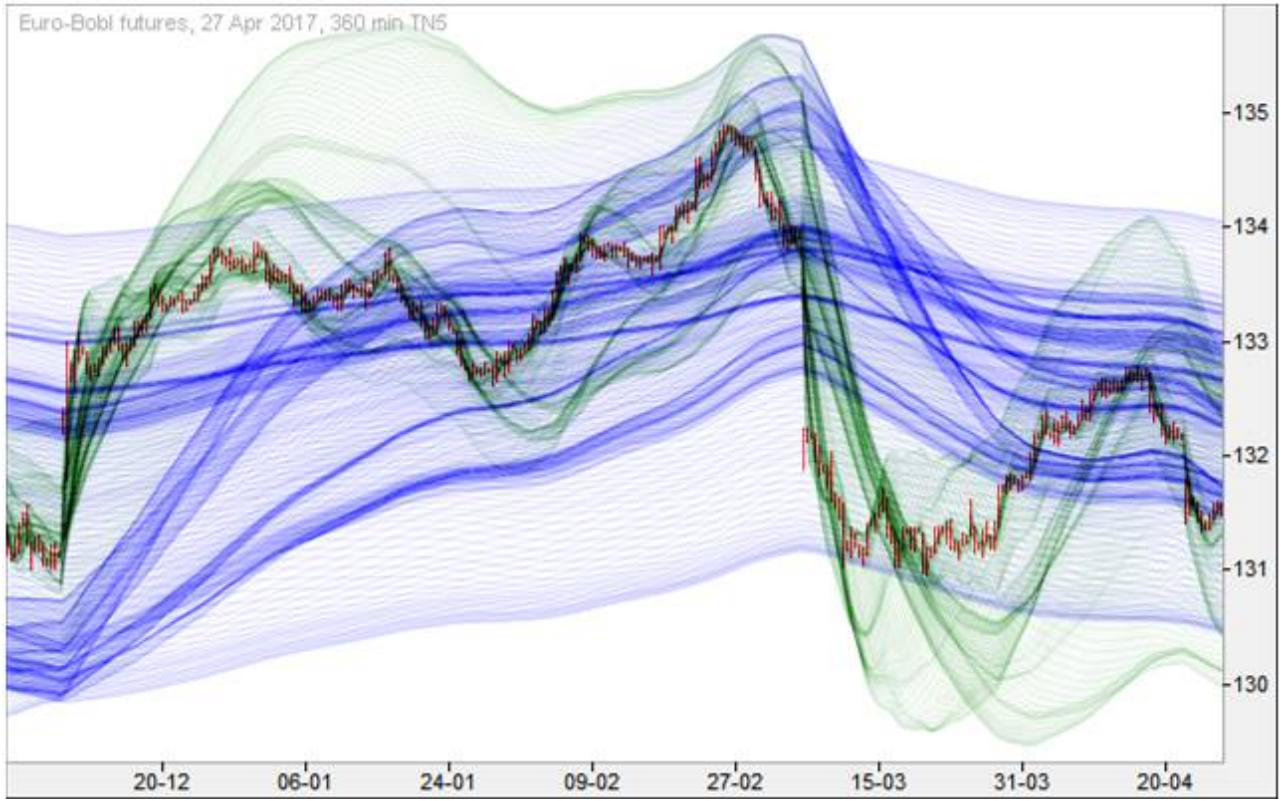

**Fig. C5. Same network under subtype TN5.** The same comment applies to TN5. This time, one can observe, among other things, interactions with the lower extrema, around March 15, 2017.

## Appendix D: Patterns

The patterns (topological configurations) presented in Fig. 7 are reproduced here in larger dimensions. Other classic patterns are also presented in this appendix. Note that, in what follows, the patterns are described using a figurative style. The purpose of this is to simplify the description of the shapes that reoccur regularly. Note that the examination of these patterns here is done without the help of other longer resolutions, which would make the analysis and the possibility of predicting even more precise. Certain specificities would be understood better by doing so. However, for pedagogical purposes, we will limit ourselves here to a single chart.

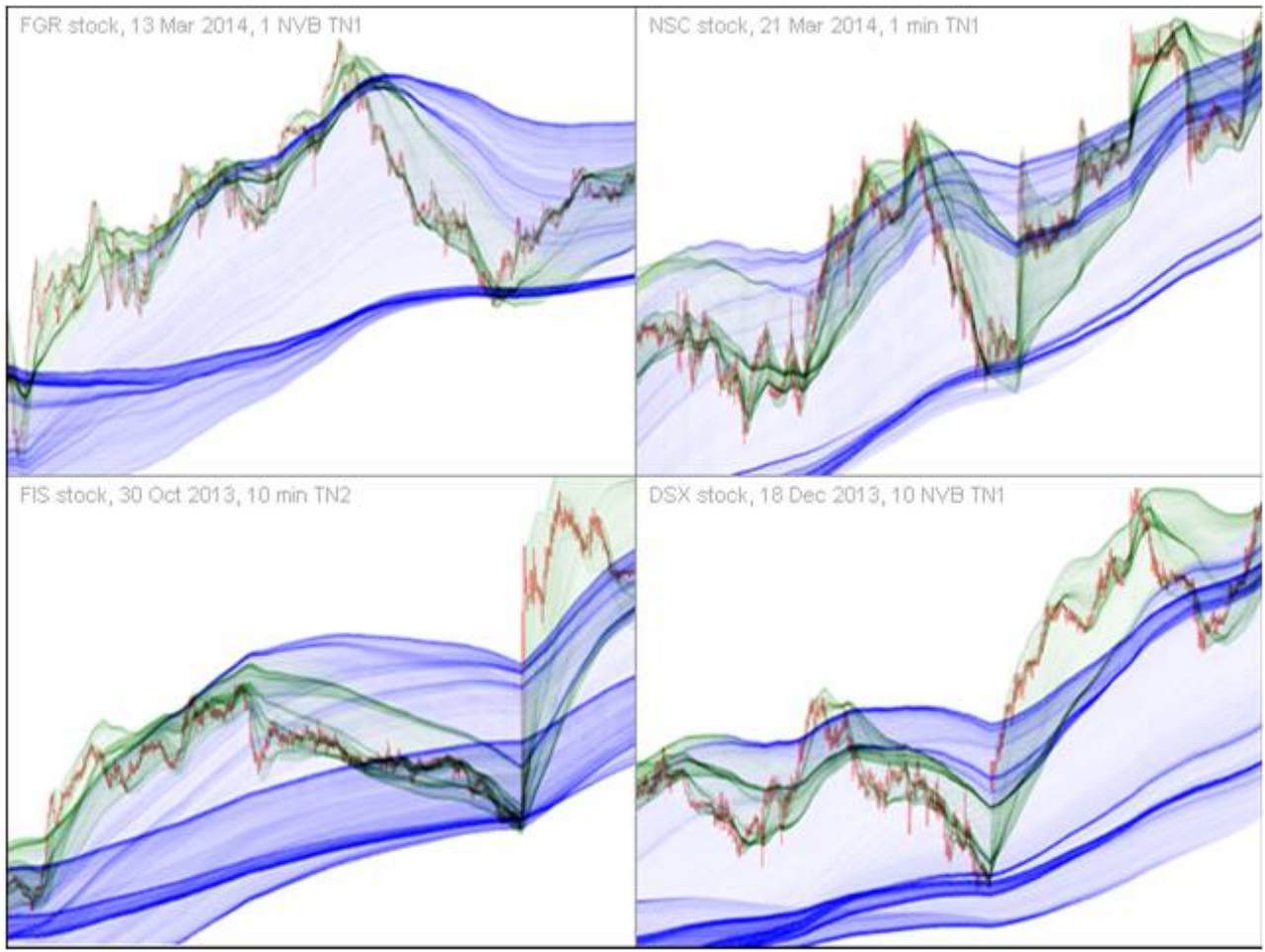

**Fig. D1. Pattern corresponding to a rising (positive slope) "bottom cord rebound".** A classic which is easy-to-spot in advance due to the presence of a very marked bottom cord. The further away from the bottom cord the quantity is and the fewer the cords between the two there are, the stronger the rebound will be. The most extreme case is a rise "in thin air" (ascent without the presence of cords). It is always followed by a violent fall back onto the cord underneath, which makes predictions particularly easy to make in this case. The nature of the slope of the bottom cord is also very important and is discussed in Section III.F.1, "Illustration of the Patterns".

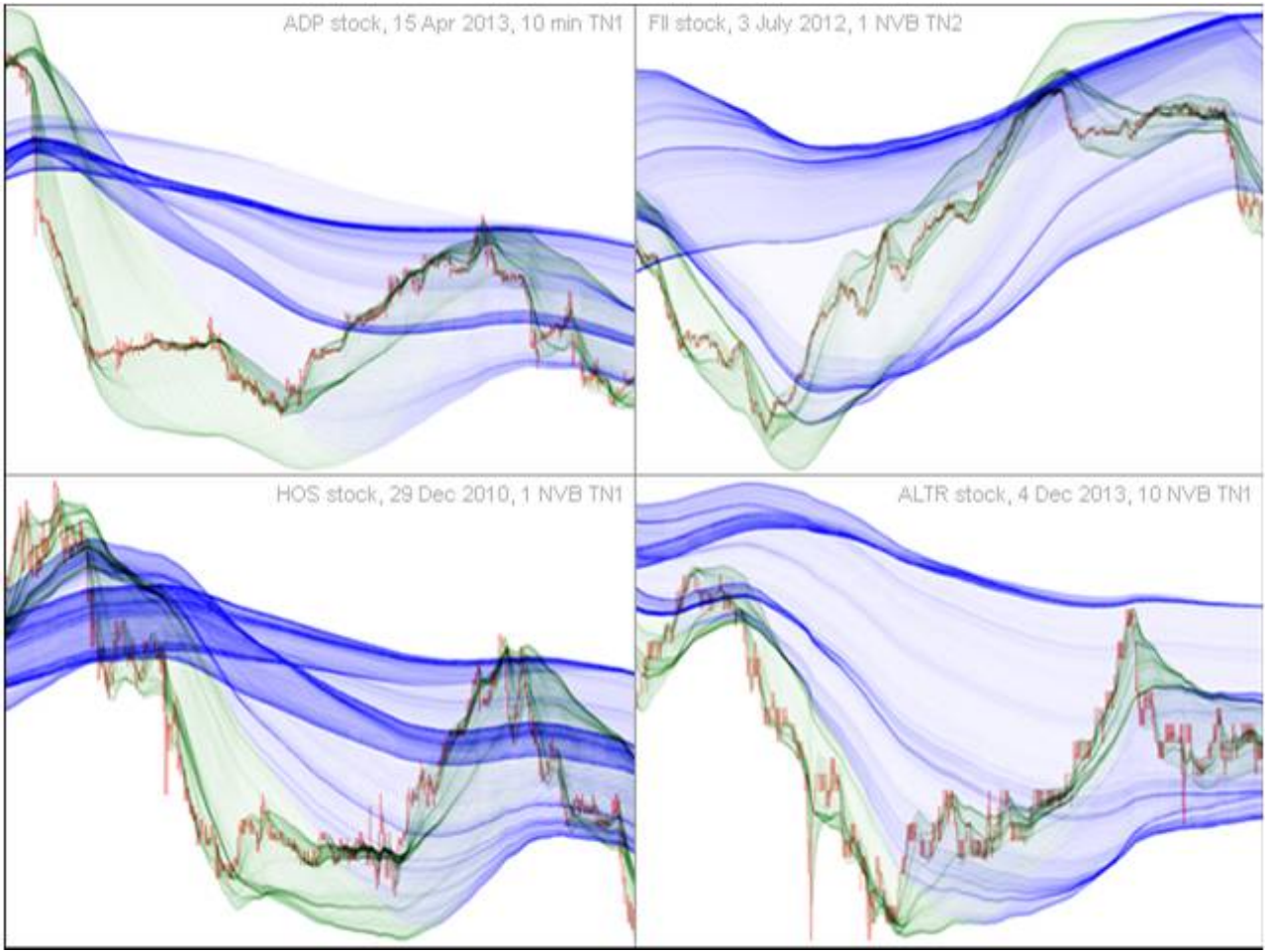

**Fig. D2. Pattern corresponding to a "top cord rebound".** After a rapid fall, and in particular, "in thin air", the price goes back up to hit and bounce off a long top cord. The "umbrella" pattern (see Fig. D9), for example, is almost always followed by a (negative slope) top cord rebound, called, in this particular case, a "false recovery" (see Fig. D10). This is usually followed by a fairly violent "fall resumption".

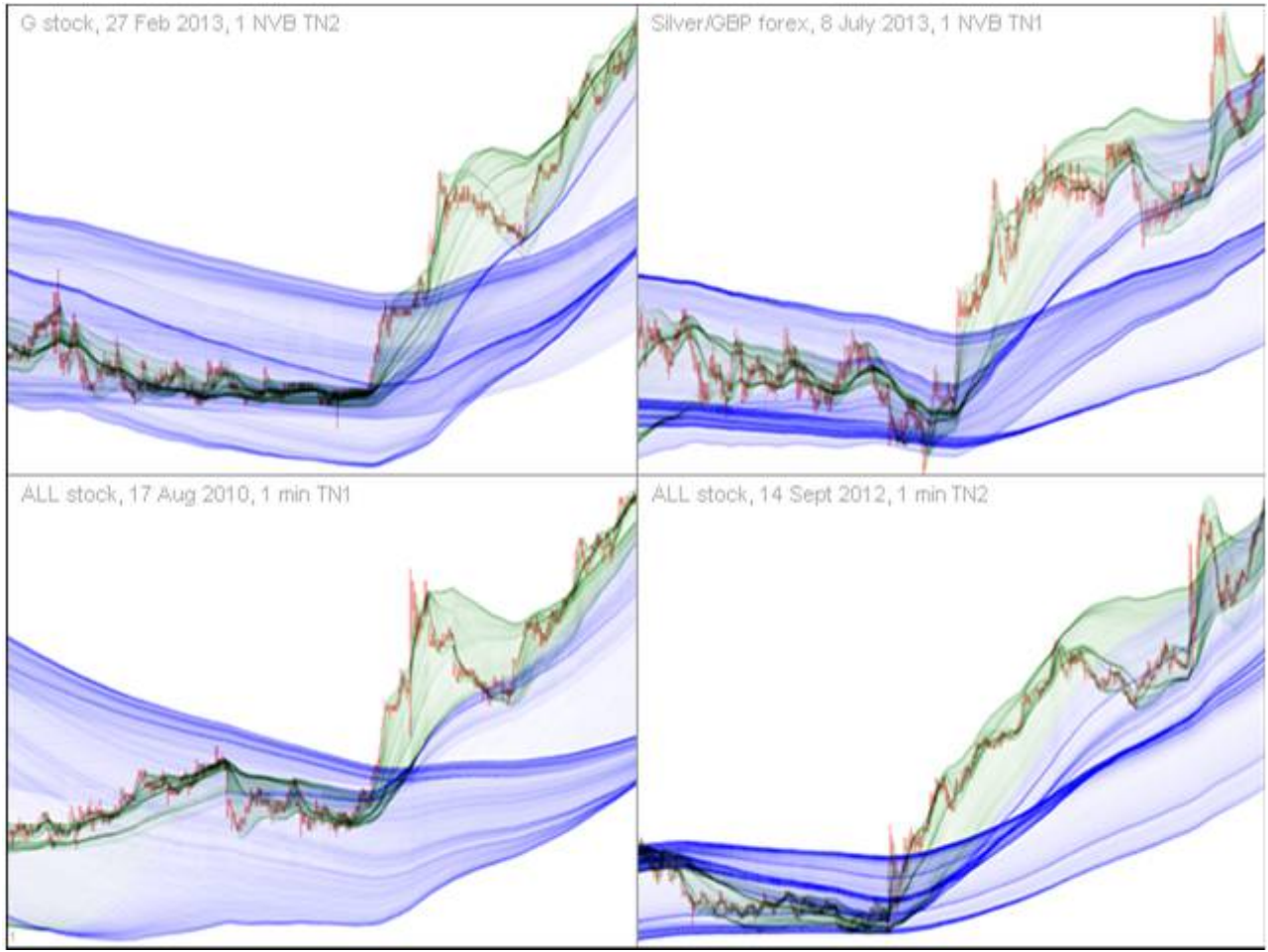

**Fig. D3. Pattern corresponding to a "buildup exit with a return to emerging cord".** After spending some time in stationary mode (characterized here by a buildup around a pack of dense cords), the price eventually shoots out. It first hits the boltrope at the top, then, pushes away from it to start a rise "in thin air". This rise almost always ends in a return to an emerging cord, stemming from the initial pack.

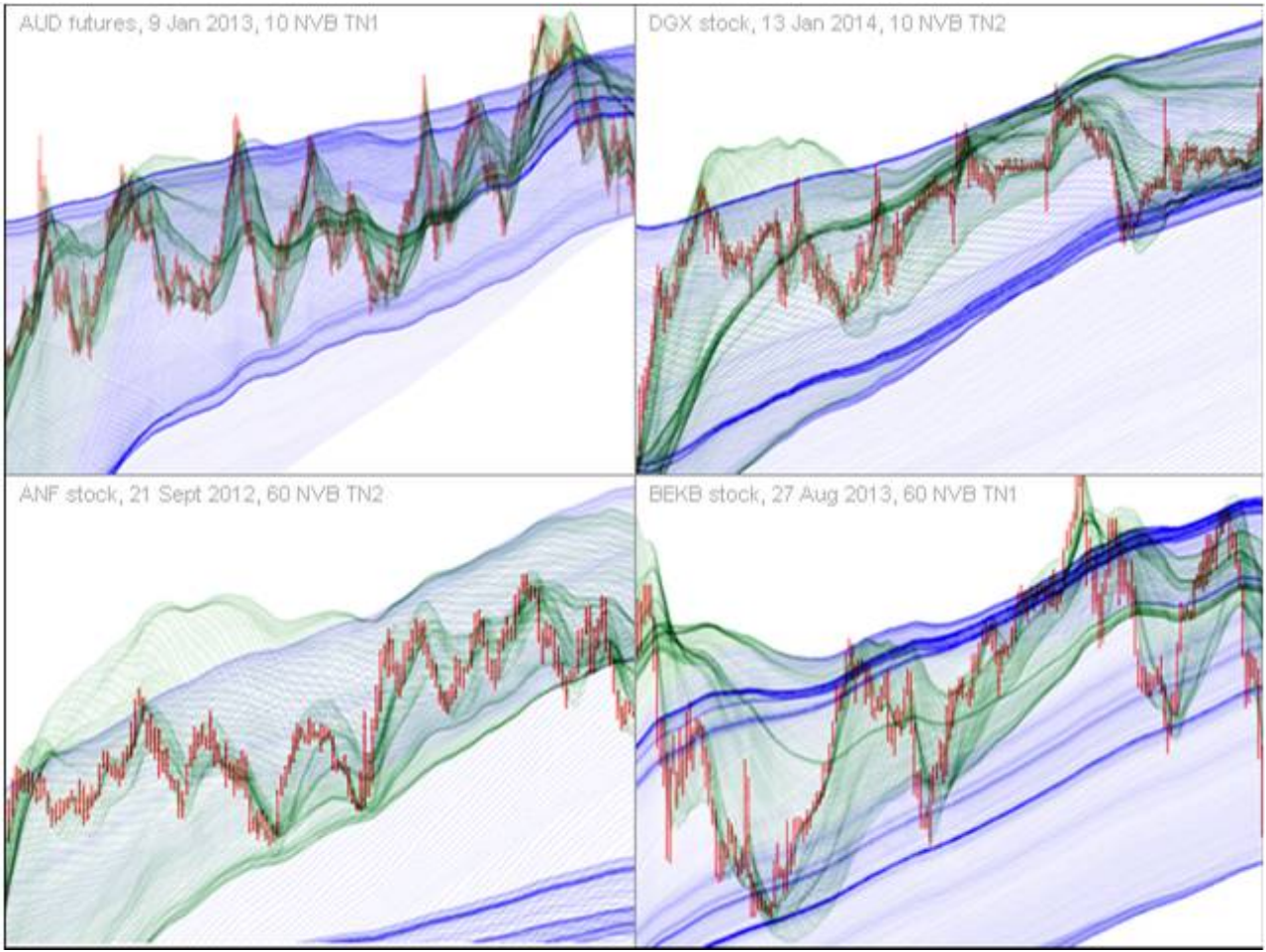

**Fig. D4. Pattern called a "channel".** A channel is a local stationary mode determined by parallel cords or boltropes. This usually ends in a very violent drop or a rise resumption.

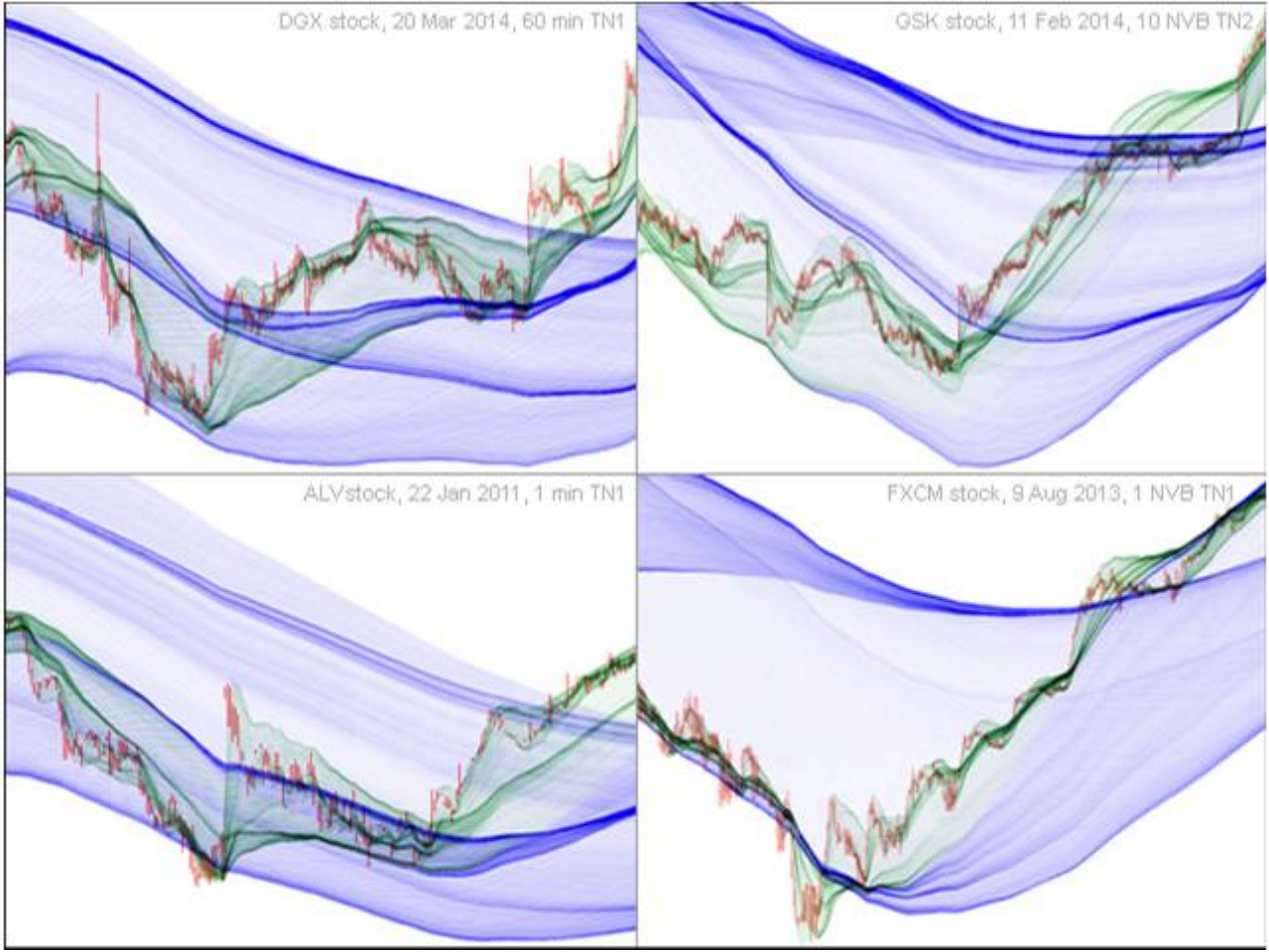

**Fig. D5. Two "upward trend reversal" patterns.** The two charts on the left show examples of "feet". A foot is a trend reversal which pushes off a cord before a rise. The two examples on the right are trend reversals with buildup exits (characterized in one case by marked sticking to the blue boltrope). These trend reversal charts all have in common the presence of a buildup exit where the rise is always marked by an acceleration, as explained in Fig. D3.

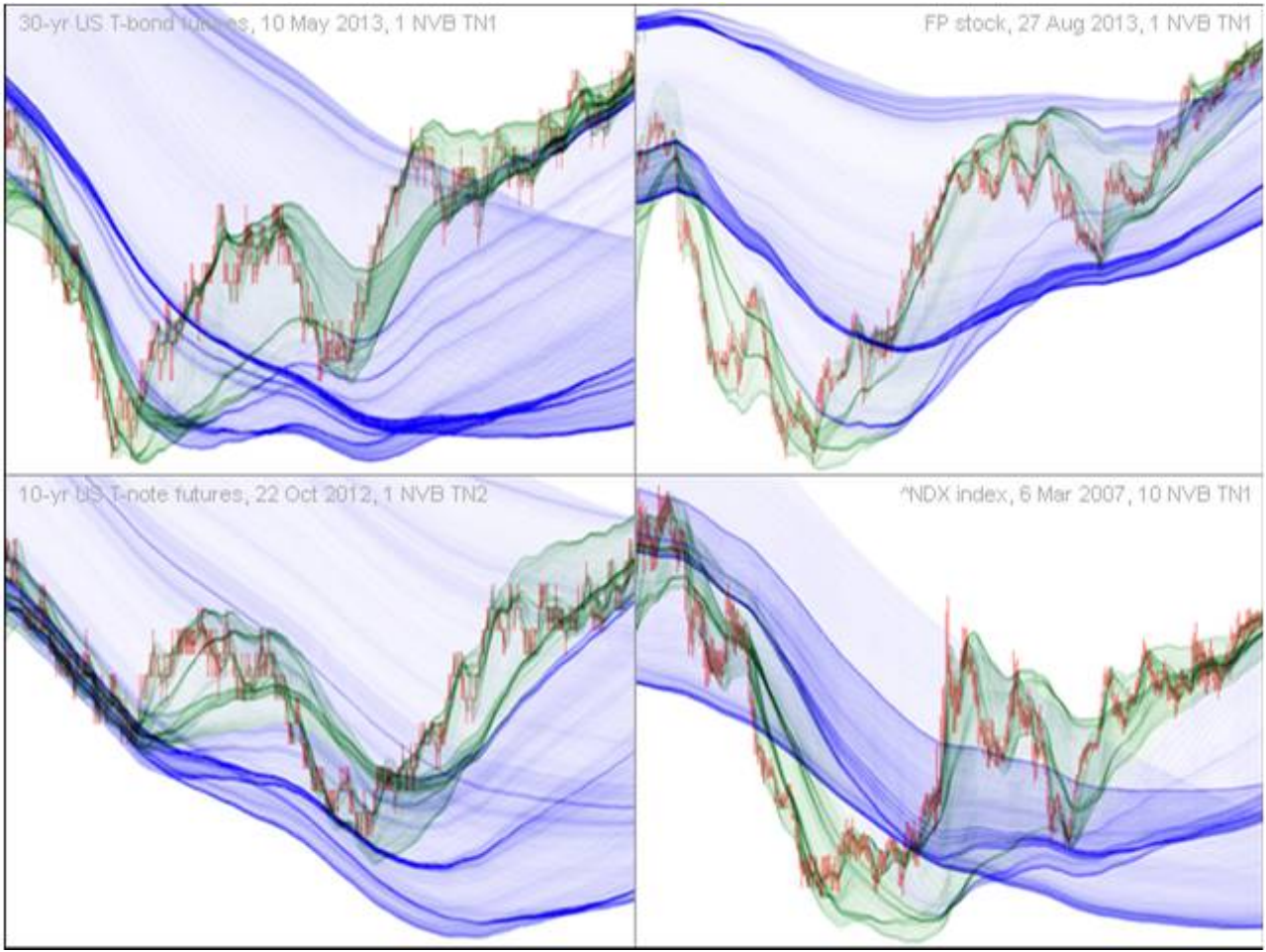

**Fig. D6. Pattern called a "foot".** These are four more typical examples of feet. Note the way the price pushes away from a strong blue cord. This pattern is very "safe", which allows early anticipation of an upward reversal. Before even pushing away from the cord, one can expect a future rebound.

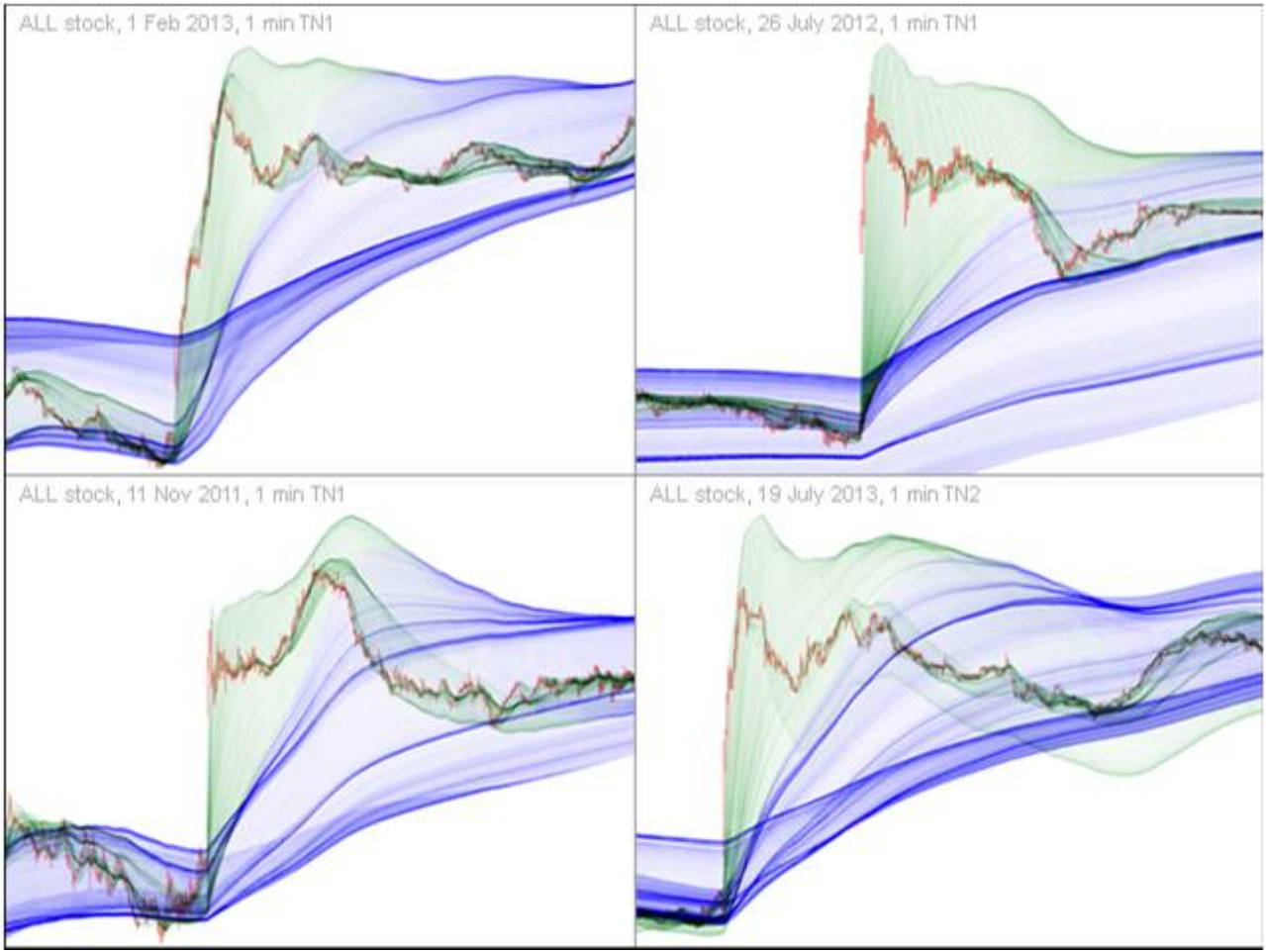

**Fig. D7. "Rise in thin air with return to emerging cord".** As seen in Figs D3 and D5, buildup exits are violent, and sometimes explosive, as seen in these four examples. These rises in thin air are almost always accompanied by a more or less quick return, to an emerging cord. This is very classic pattern.

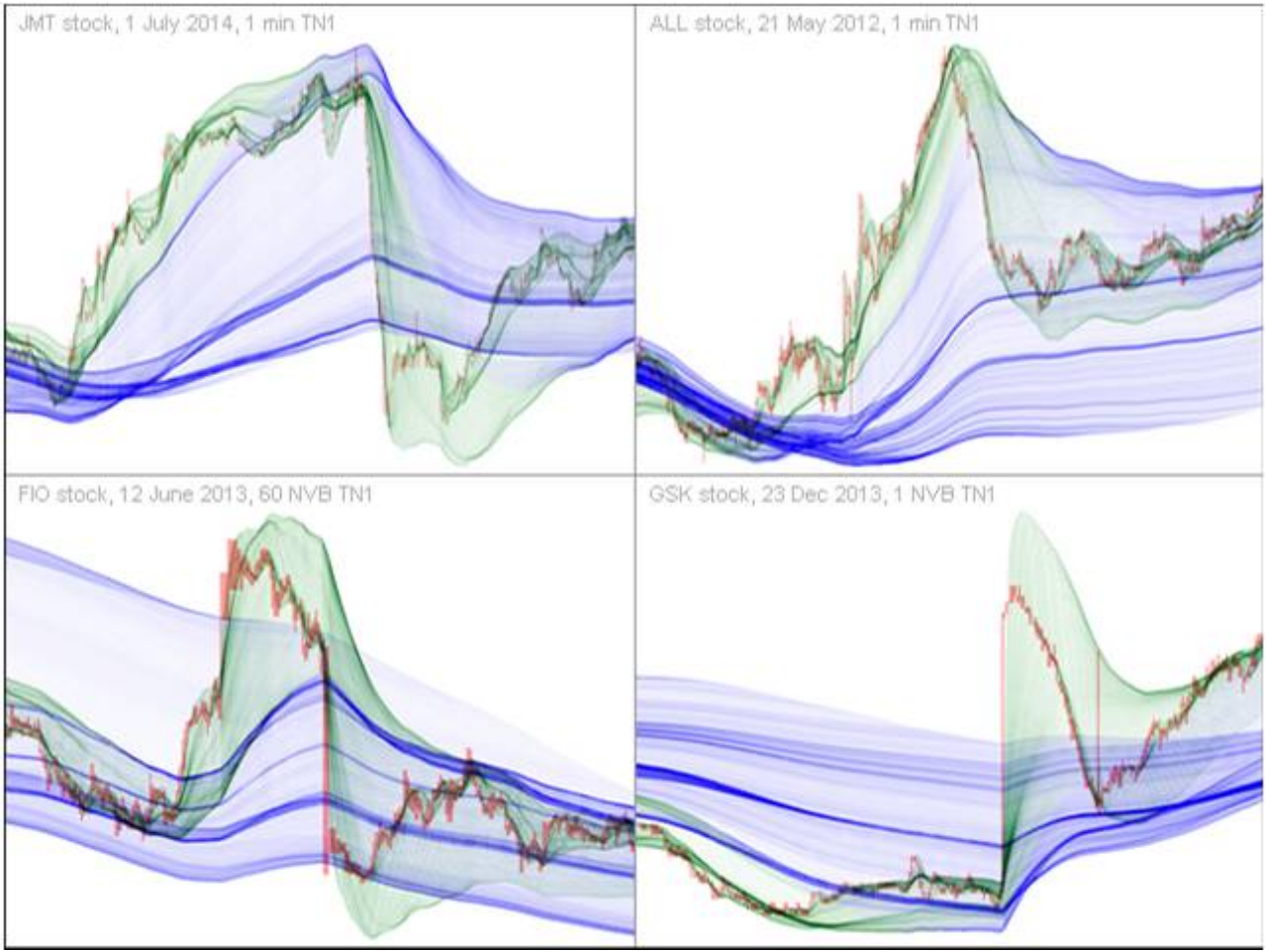

**Fig. D8. "Rise in thin air" and "fall in thin air".** Rises in thin air, especially sudden ones, are followed by falls in thin air, sometimes very violent. The emerging cord is sometimes "missed" (and, thus, crossed), as is the case of the two examples on the left. In any case, the price always ends up bouncing off a cord.

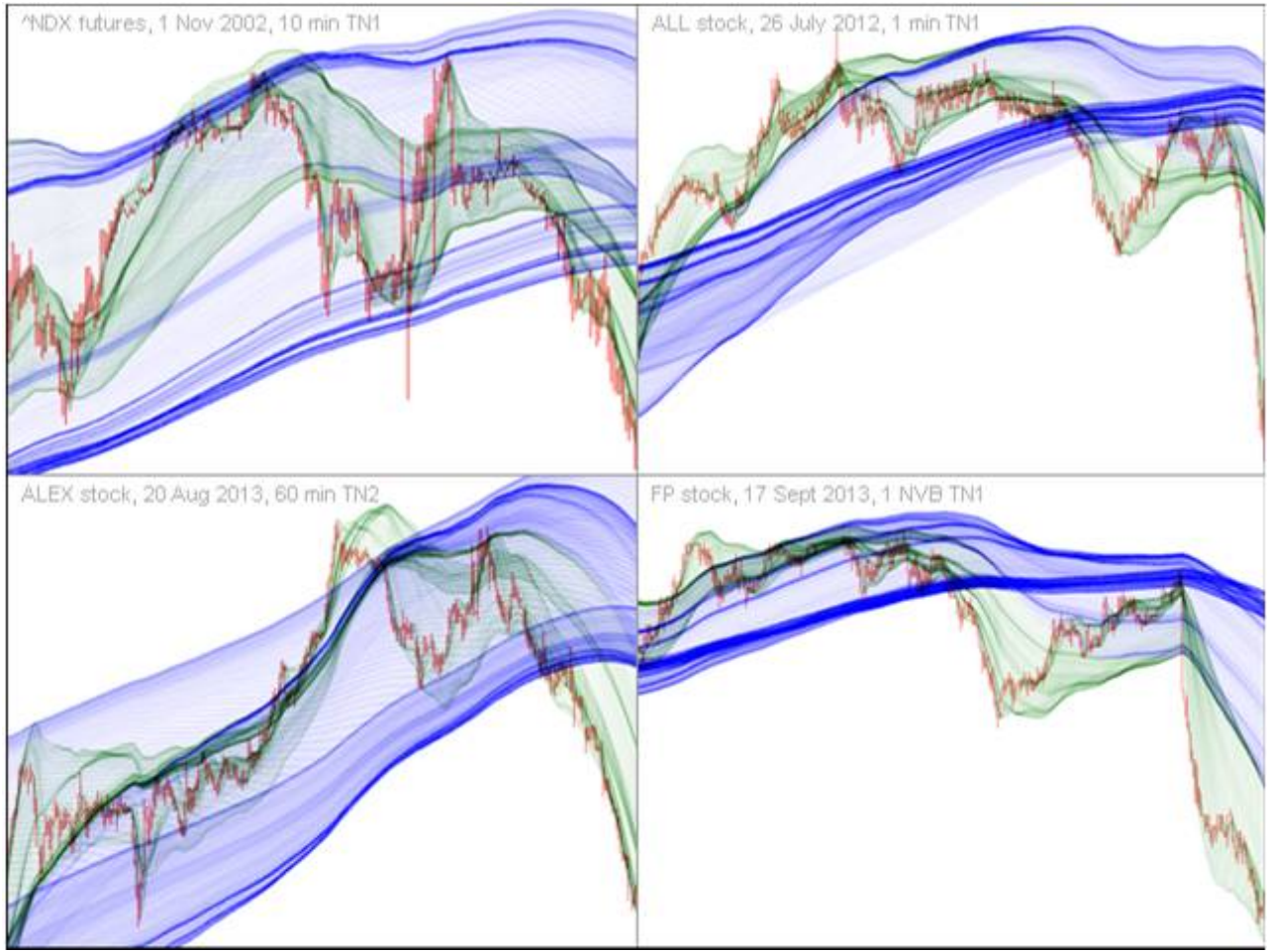

**Fig. D9. "Umbrella".** As its name suggests, the umbrella is characterized by the existence of one or several strong cords above a last local extremum at the end of a long rise. After "hitting the ceiling", so to speak, the price begins a very violent fall. After passing under an ultra-dense cord, and having bumped into it, as in the two examples on the right, a Niagara-Falls-like plunge is guaranteed.

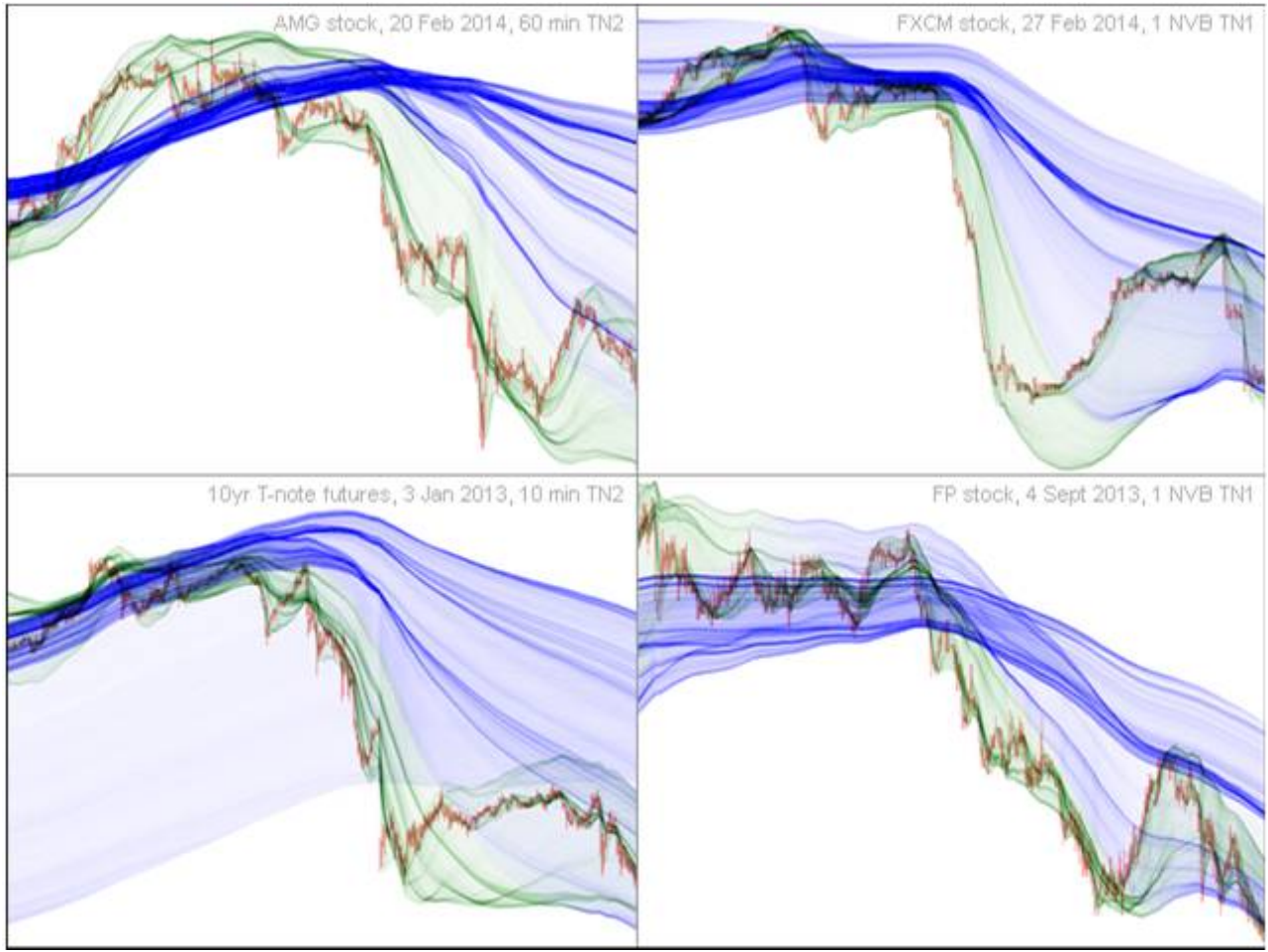

**Fig. D10. "False recovery".** Following the plunge of an umbrella, the price rises sharply (a phenomenon which has already been observed by everyone). What TN reveals is that this rise stops as soon as the price meets the first marked emerging cord. After bouncing off this cord, either the price continues to fall even more, or a new scenario takes place, depending on the characteristic figures present.

**Appendix E: Conservation of Topology**

    The notion of conservation of topology when changing the mode of representation of the data and when changing the subtype is admittedly delicate, but it is interesting to observe how the apparently same characteristic figures "deform" during these two types of changes. We will limit ourselves here to reproducing larger versions of the images in Fig. 8. One can easily study this phenomenon in other charts, as, for example, in the images presenting the five subtypes in App. C.

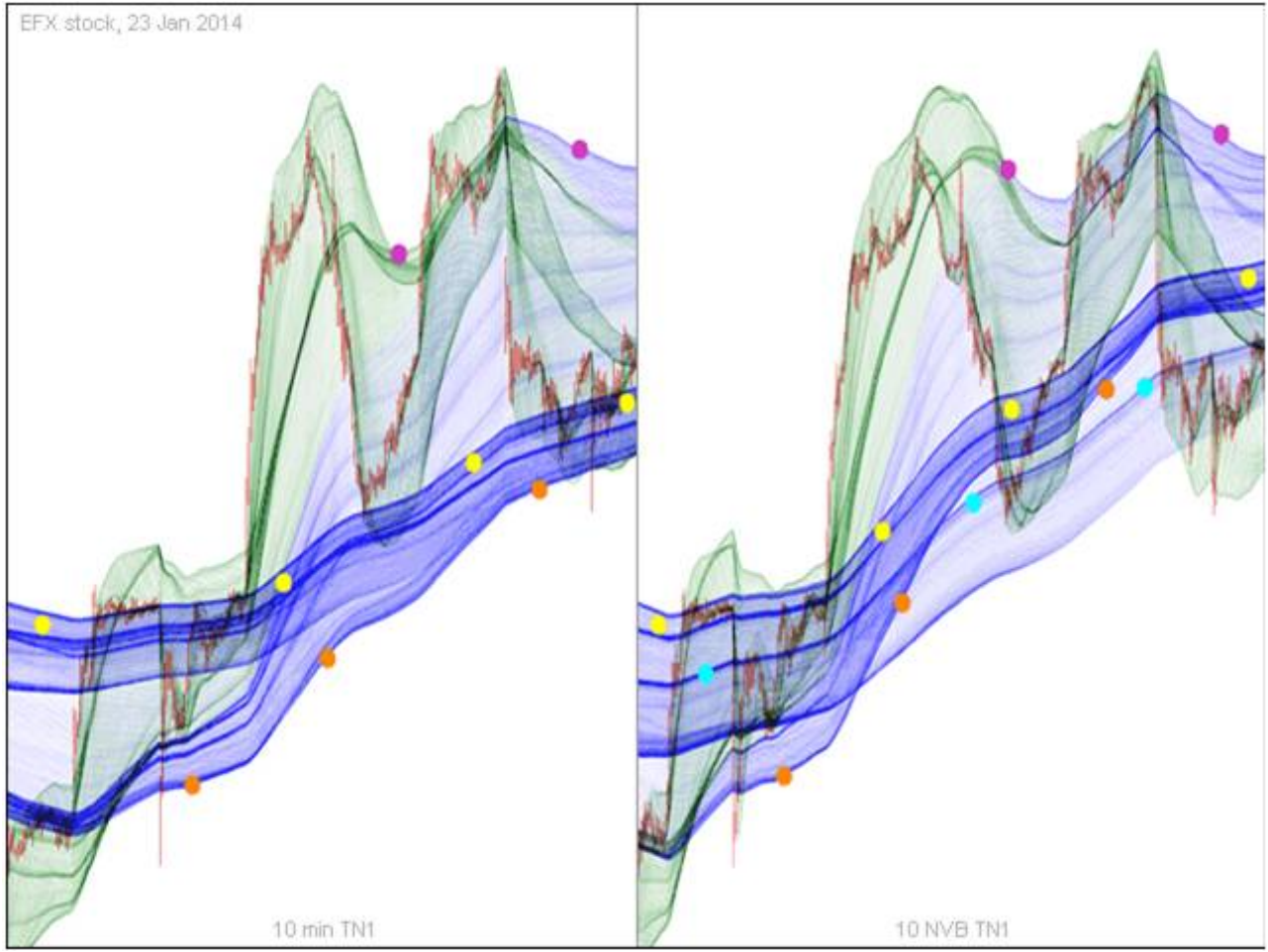

**Fig. E1. Conservation and deformation of the topology of the network under TN1 when changing the data representation.**

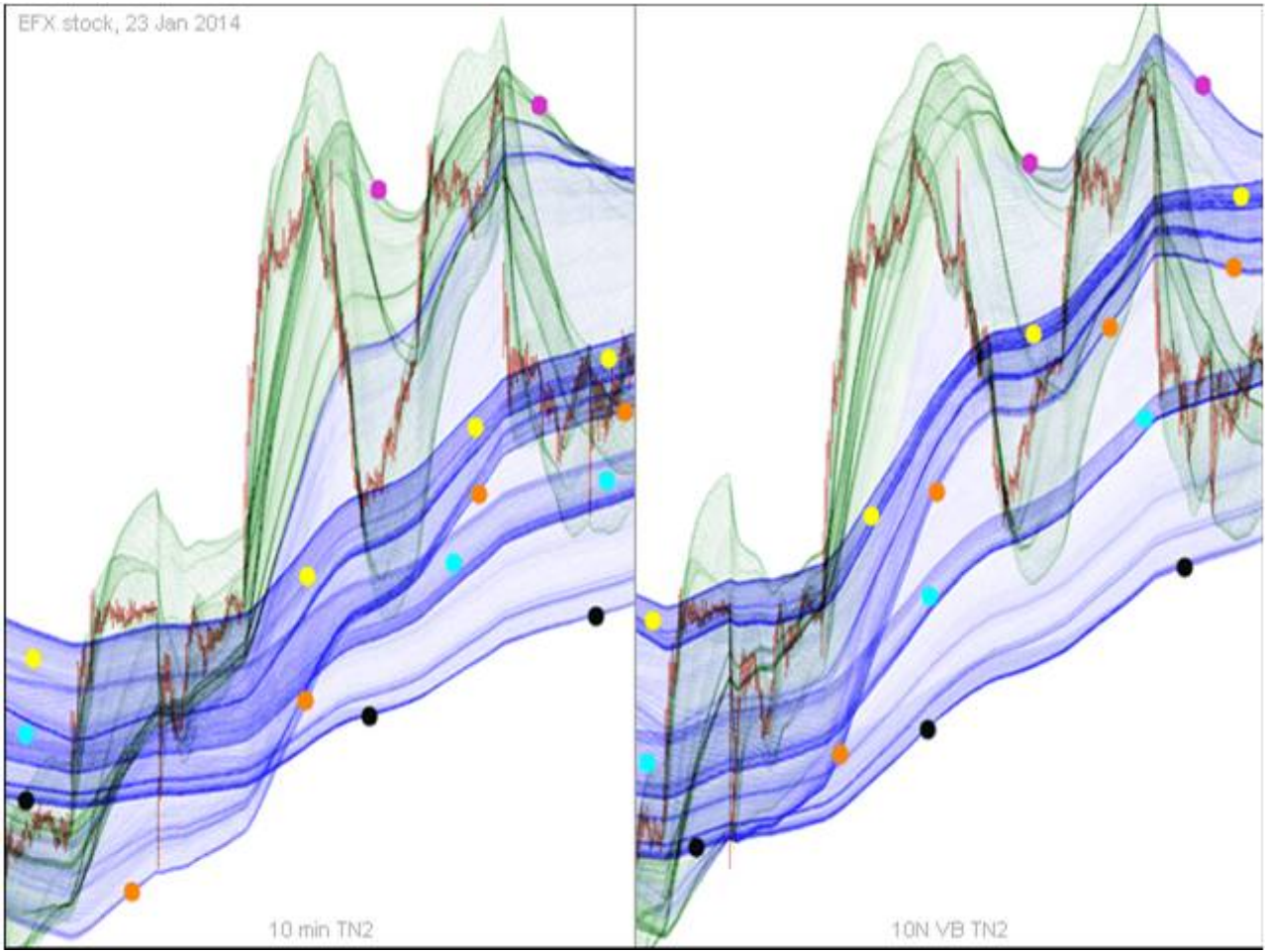

**Fig. E2.** Conservation and deformation of the topology of the network under TN2 when changing the data representation.

## Appendix F: Self-Similarity

This appendix presents larger versions of the charts from Fig. 9. The notion of self-similarity simply illustrates the fact that the same patterns are found in a totally identical manner in completely different timeframes (or granularities), at, of course, different instants. App. D. also illustrates this phenomenon.

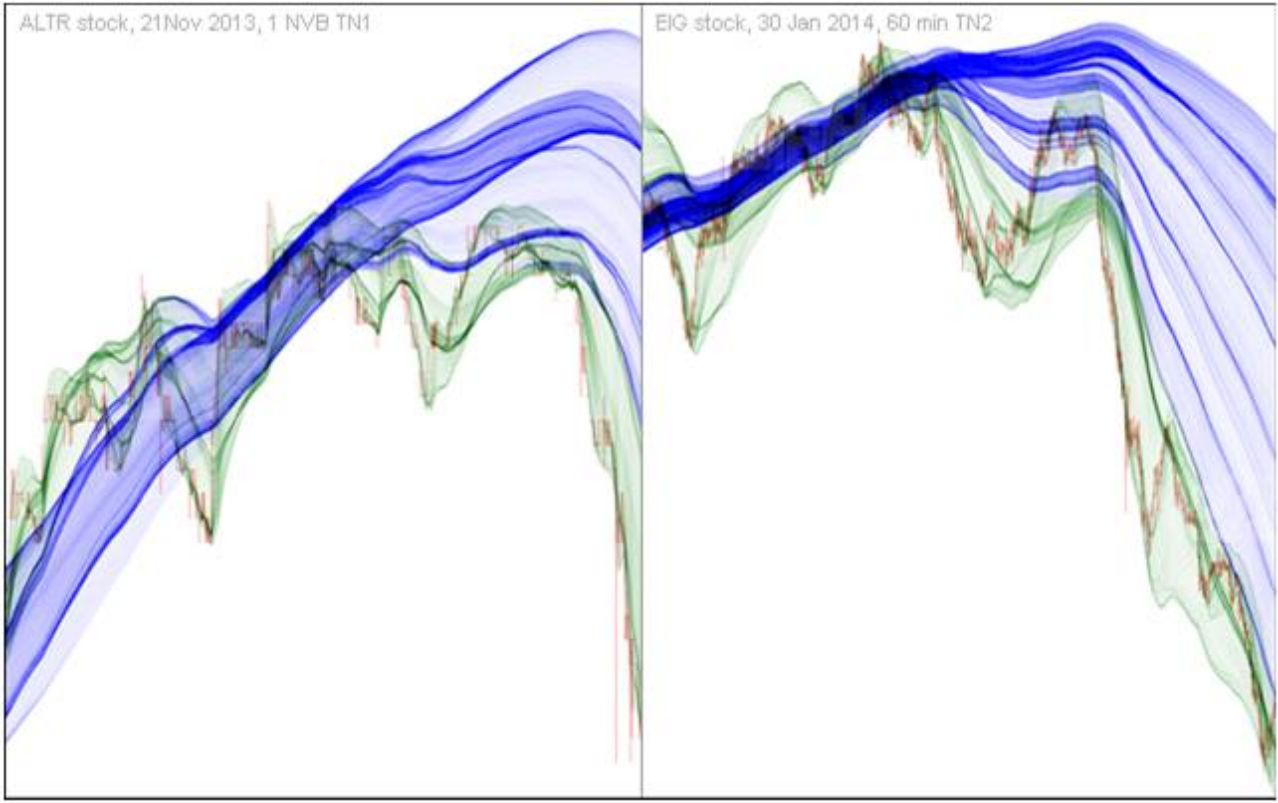

**Fig. F1. Two umbrellas in very different granularities.** Note the large number of cords in the case of the chart on the right. This is due to the fact that the subtype is higher.

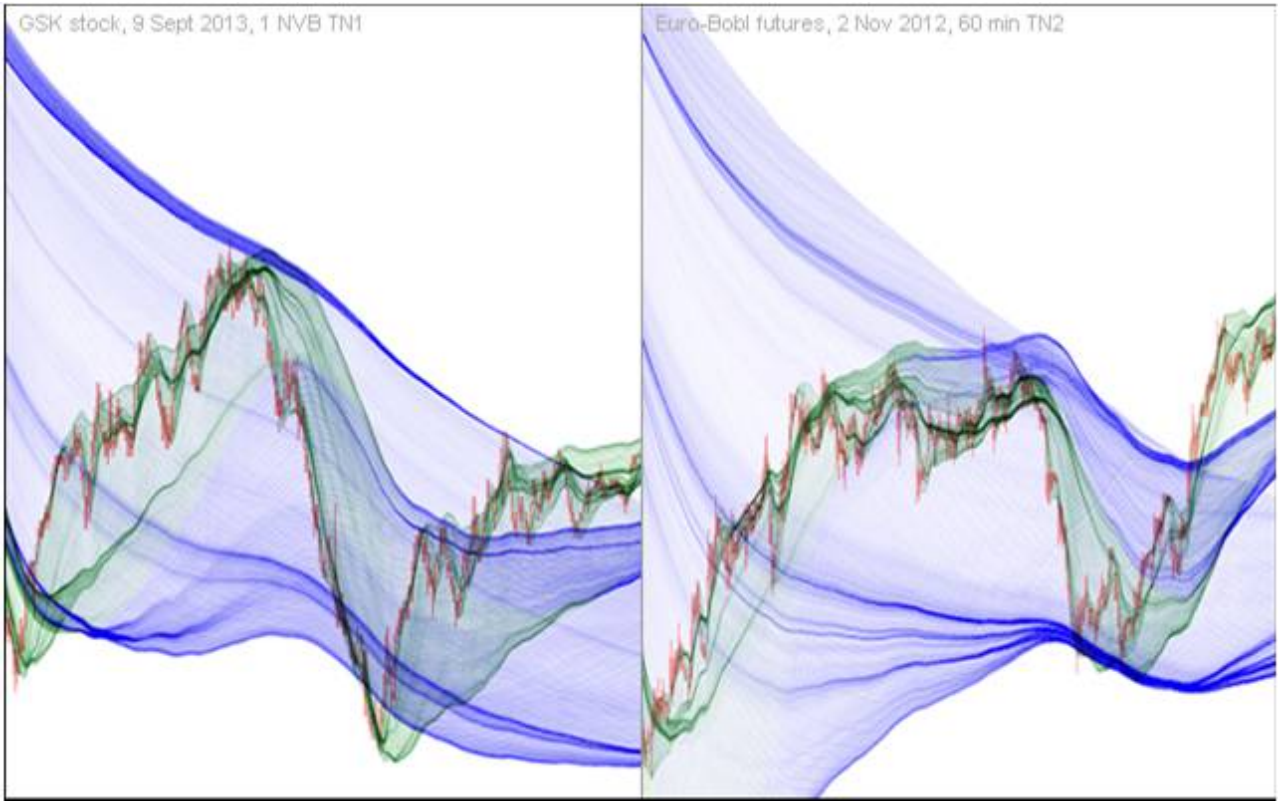

**Fig. F2. Top cord rebound followed by a foot.** In the same ratio of granularities as in Fig. F1, one can see here very similar topological configurations. Refer to Fig. J1 of App. J for more details about the chart on the right.

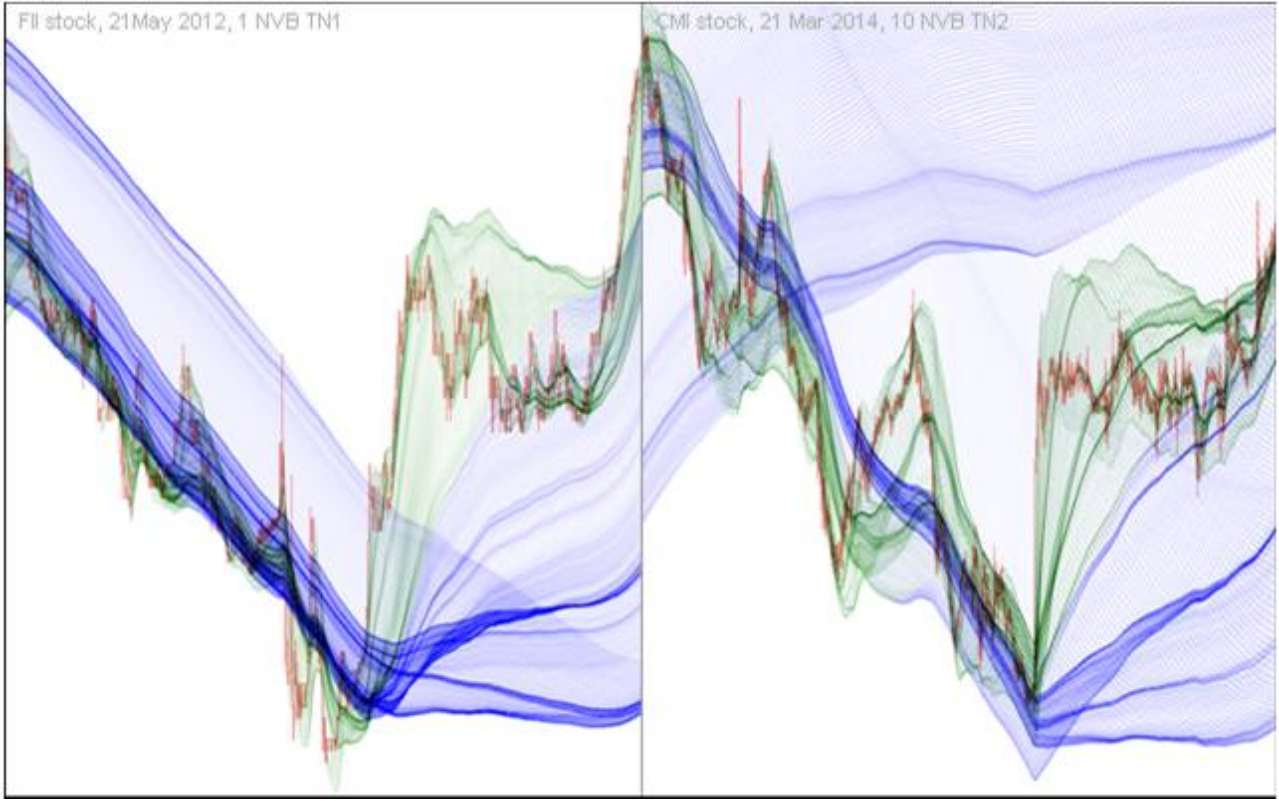

**Fig. F2. Fall brake and rise in thin air.** Fall brakes are characterized by a progressive densification of a pack of cords accompanying a fall (or more rarely, a rise). These patterns allow anticipating imminent violent trend reversals. Again, they can be seen in all granularities.

# Appendix G: The Possibility to Predict More or Less Easily Depending on the Granularity

What other mathematical tool is capable of quantifying the predictability of the quantity depending on the timeframe? This phenomenon can be observed to varying degrees whatever the quantity being considered.

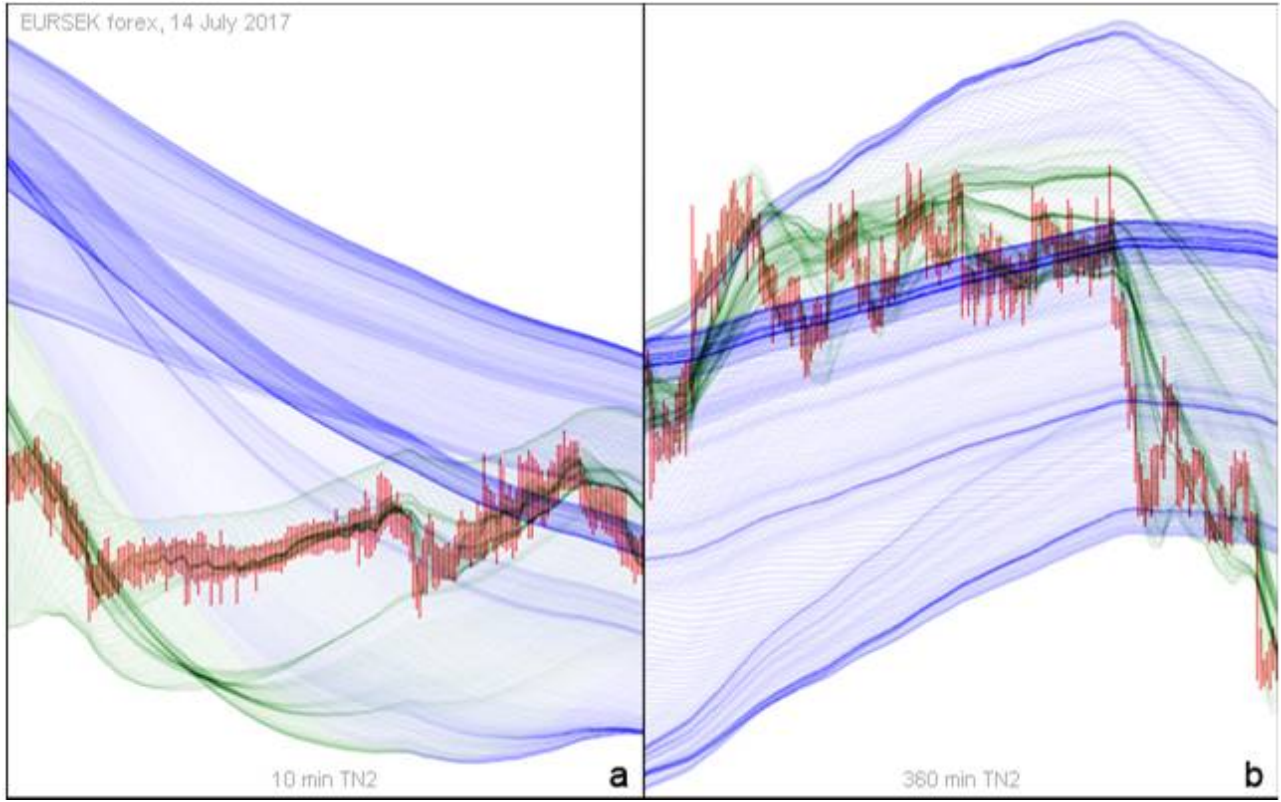

**Fig. G1. Resolutions that are of uneven quality for a given instrument at the same instant.** While the network of this exchange rate in 10 minutes is of mediocre quality in terms of characteristic figures, it is excellent and simple to analyze in 360 minutes, at the same instant.

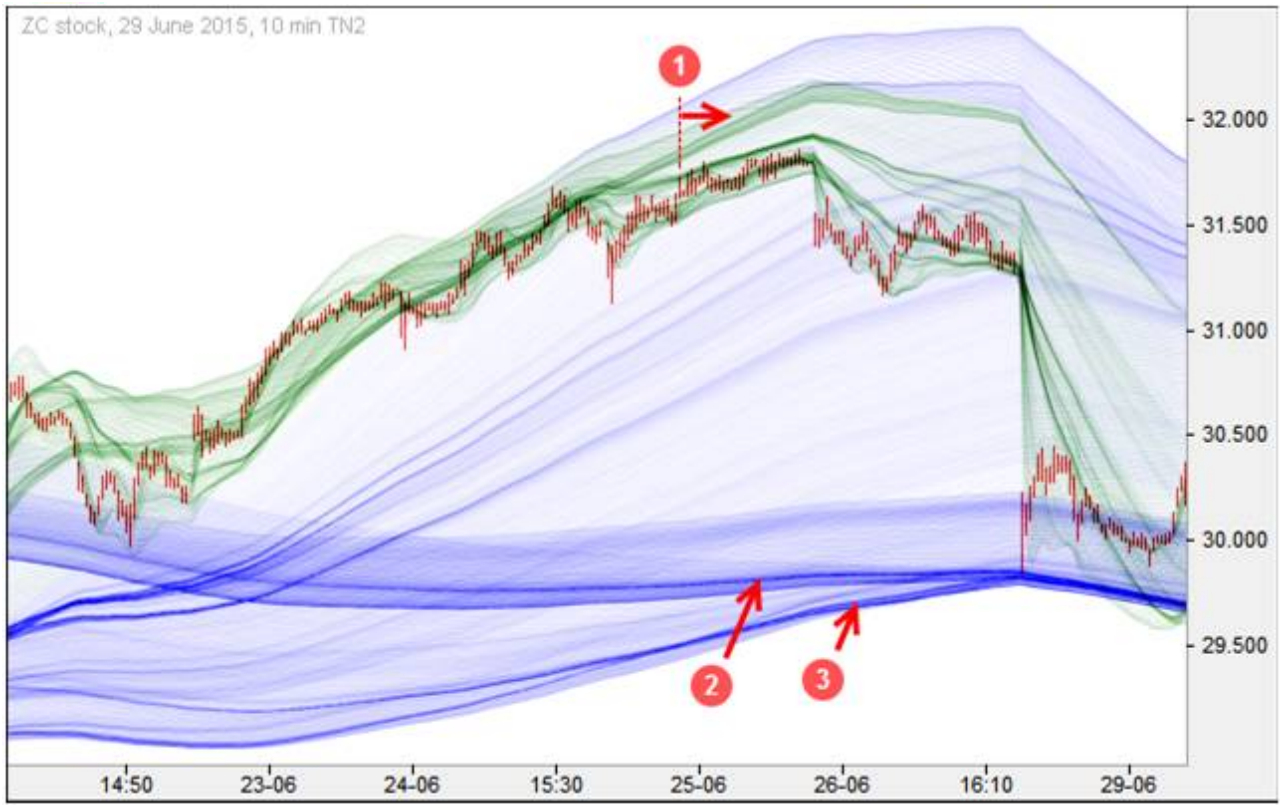

**Fig. G2. Easy prediction in a particular resolution.** This example corresponds to the subsequent evolution of the chart from Fig. 10c. Starting from point (1), it is fairly evident, even without confronting it to a higher resolution, that the price will meet (or, here, return to, to be precise) cord (2), or, possibly, cord (3). This is the fate of rises in thin air.

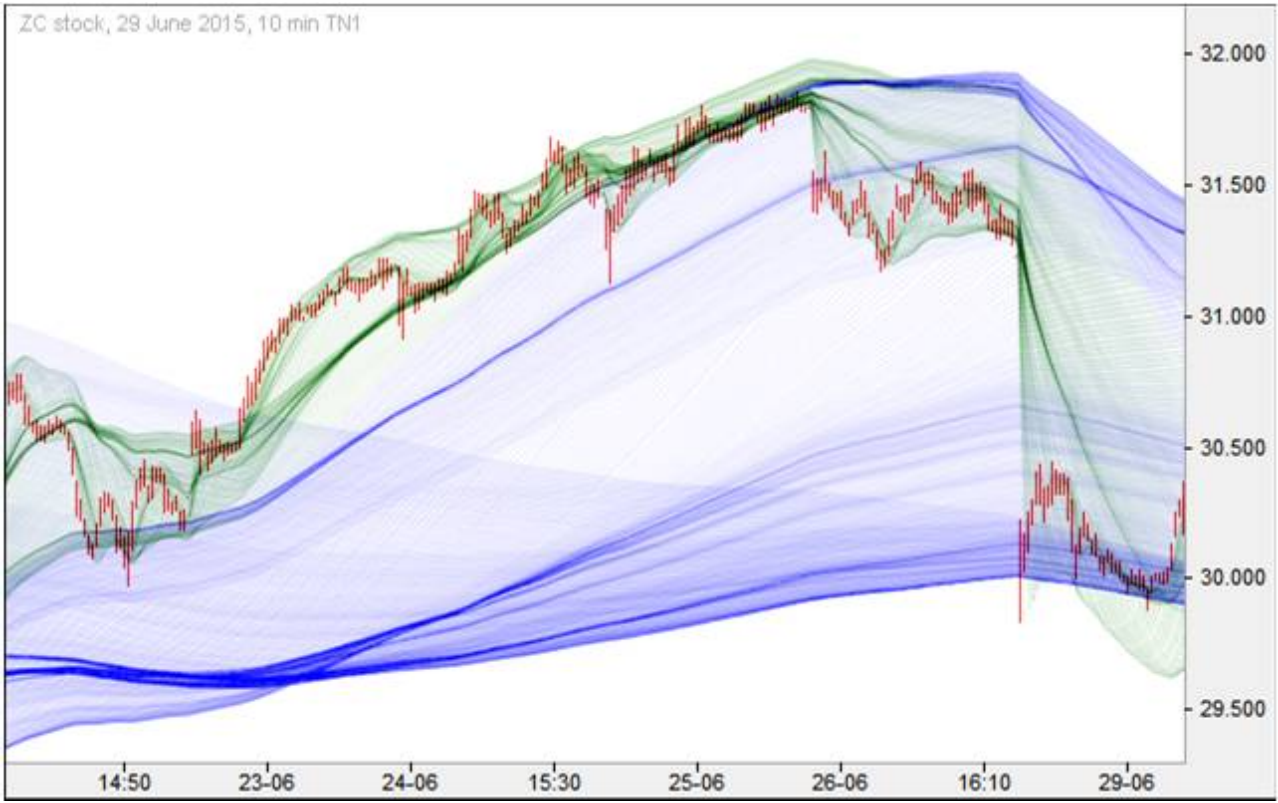

**Fig. G3. Confirmation of the easy prediction in one resolution with the use of another subtype.** One can observe here that the drop takes place after a "rise brake". The drop lands nicely on a boltrope present under TN2.

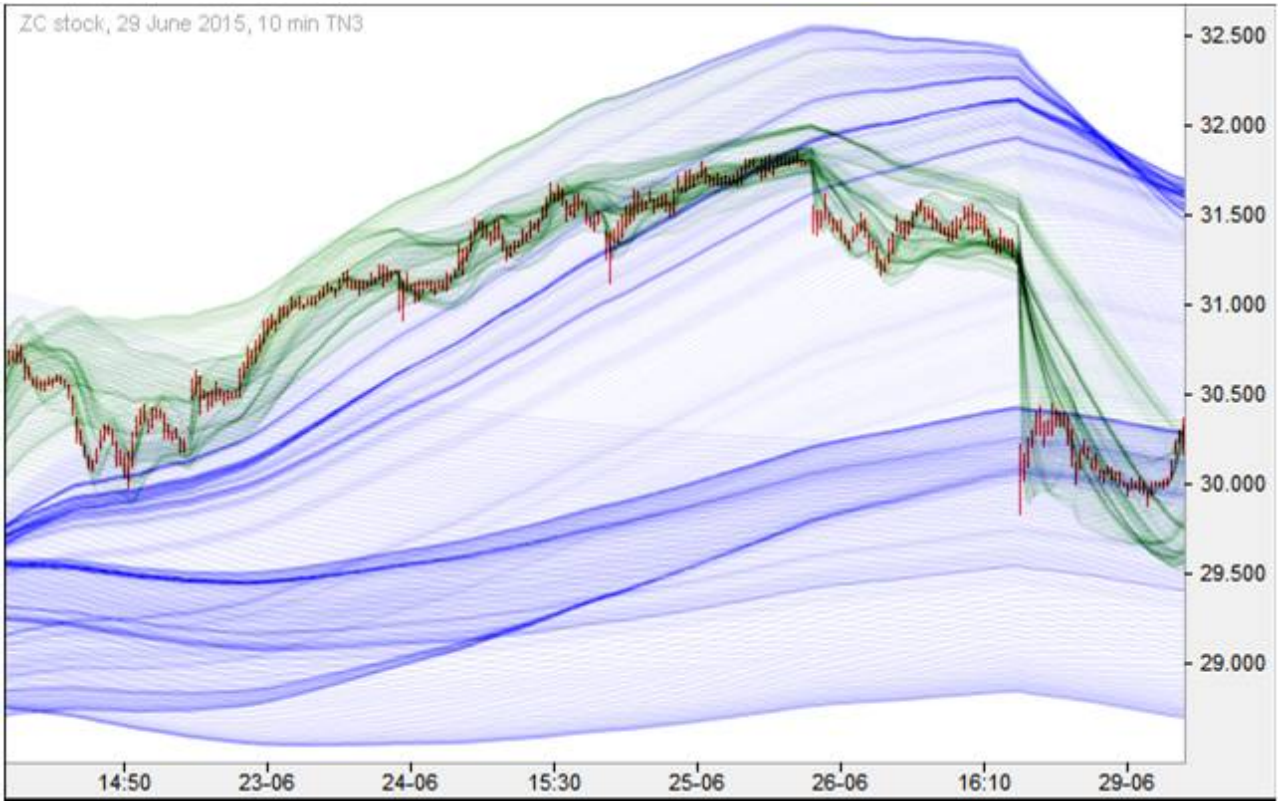

**Fig. G4. Help predicting thanks to a higher subtype.** The landing points are clearly present in the network under subtype TN3.

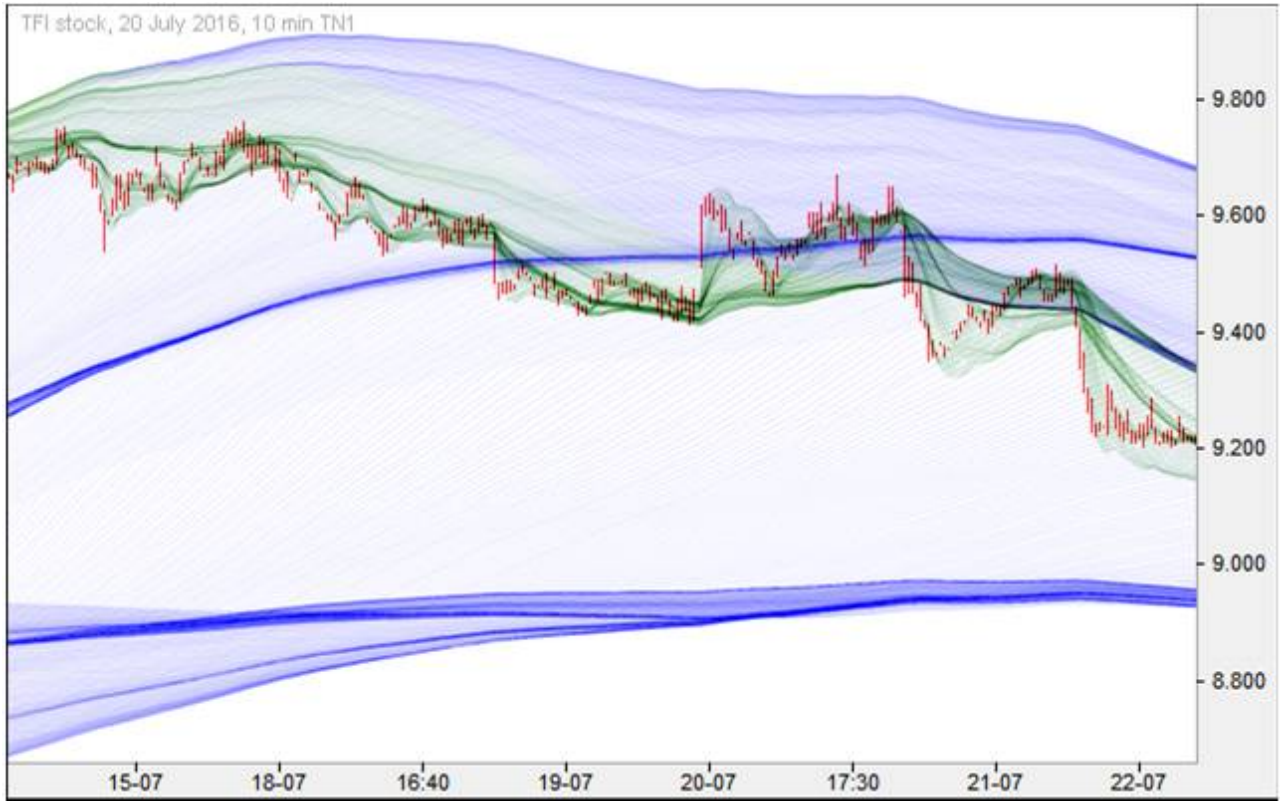

**Fig. G5. A resolution that is difficult to exploit at a given instant does not remain like that forever.**
This chart is the one from Fig. 10d. The next figure allows one to observe that the qualification of the granularity for a given instrument depends on the instant.

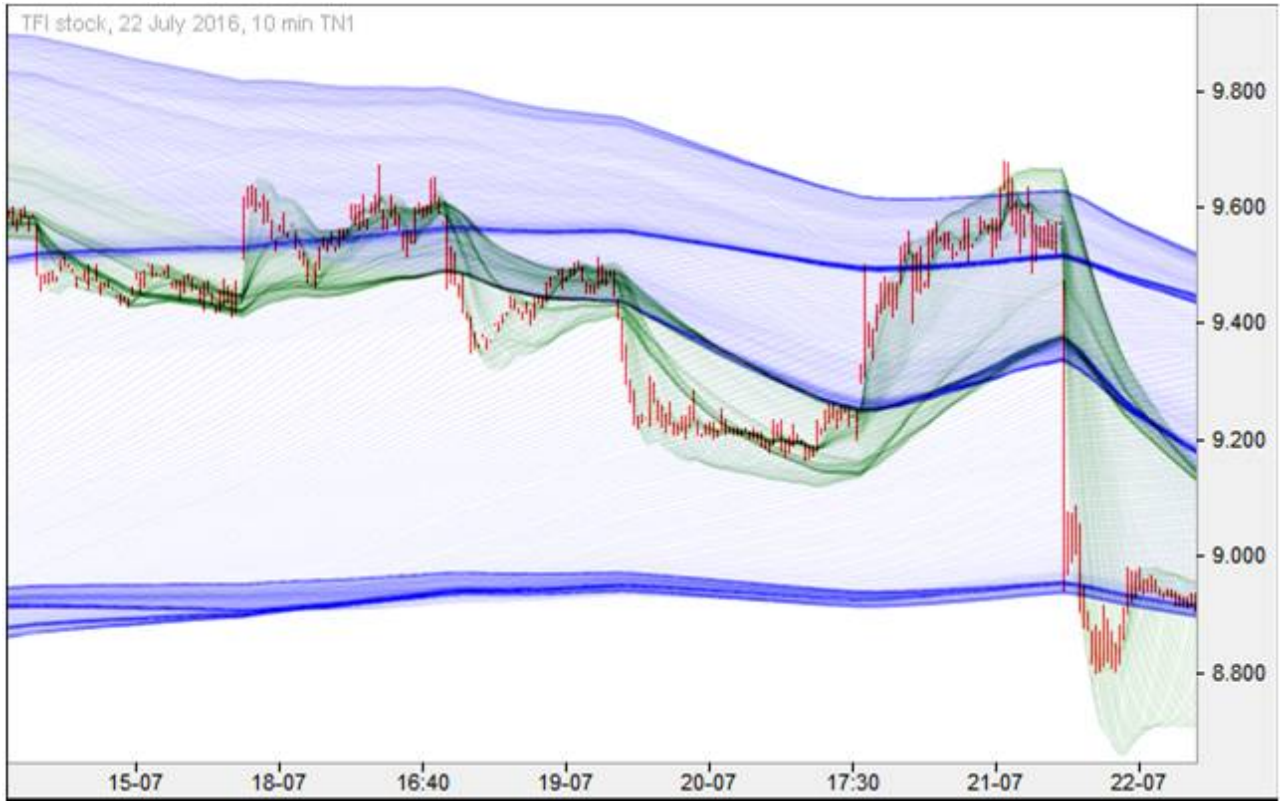

**Fig. G6. The same resolution at an ulterior instant.** Shortly after the instant of the network in Fig. G5, the price starts bouncing from characteristic figure to characteristic figure in a marked manner. Thus, one changes from a stationary mode, in relation to that particular resolution and subtype at that instant, to a mode that is perfectly described by the characteristic figures present.

## Appendix H: How to Predict

This appendix shows larger versions of the charts in Fig. 11. To limit the number of examples used, the reader is invited to perform the exercise presented below with any chart.

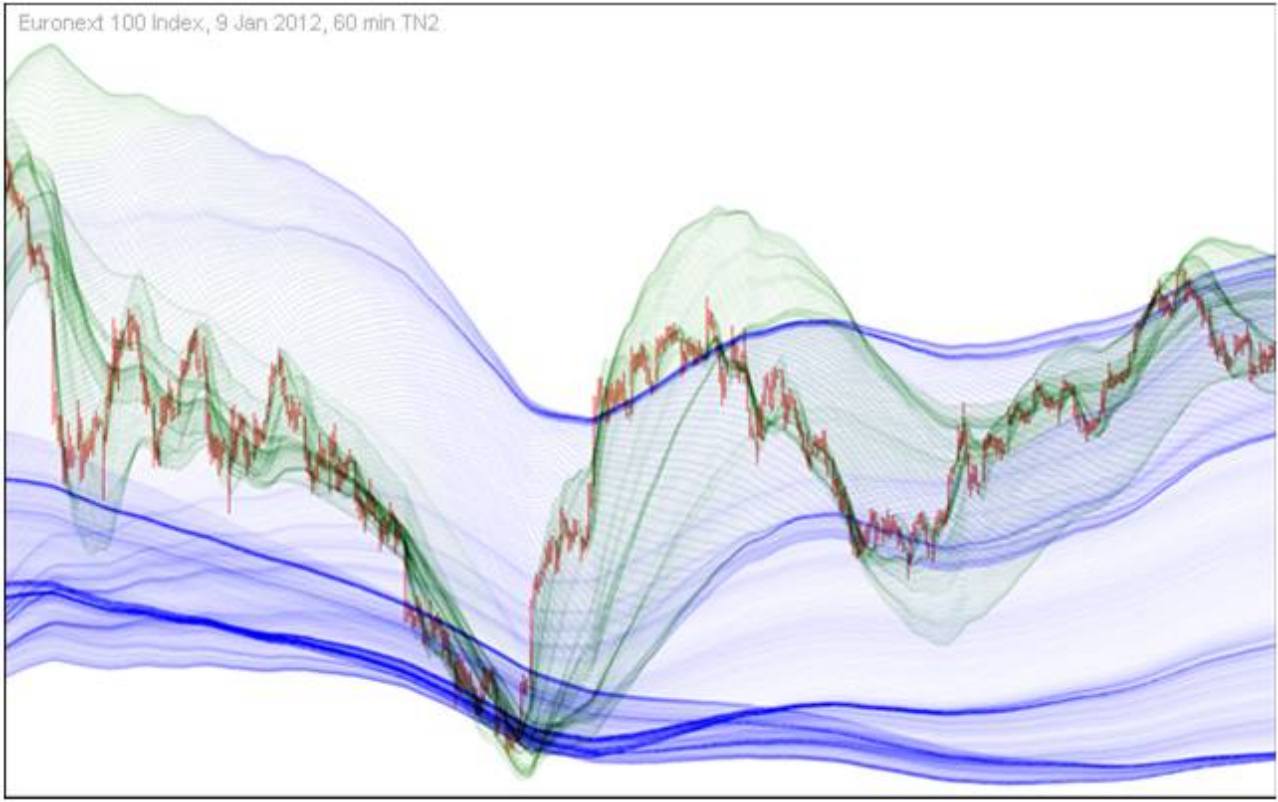

**Fig. H1. The price or the quantity bounces from characteristic figure to characteristic figure.** One can see on the network corresponding to a stock index numerous interactions, and, in particular, those that correspond to larger scale local extrema. Predicting consists in anticipating the next rebound off one of the characteristic figures already present.

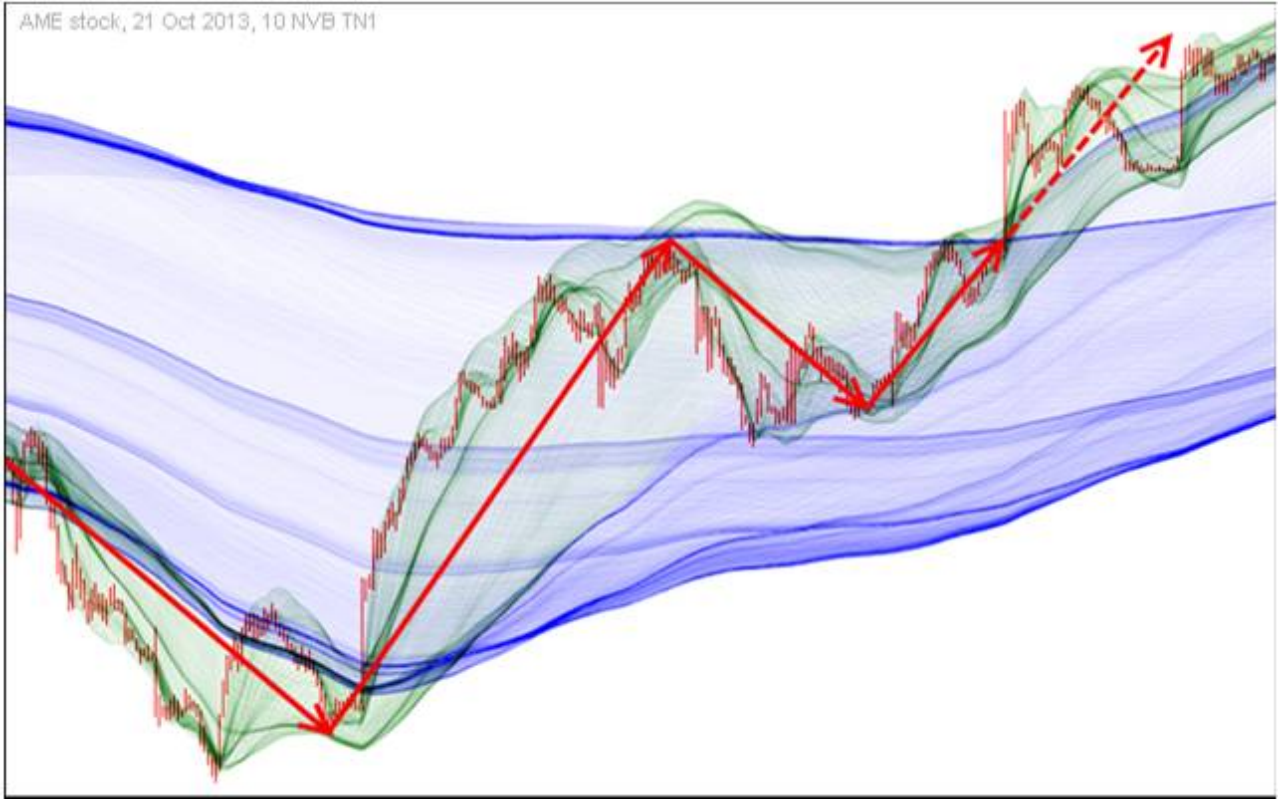

**Fig. H2. From the general to the particular.** Practically each characteristic figure present on the topological network interacts with the price. However, when making predictions, it is about determining (from a network of a given resolution) the next movement of sufficiently large amplitude. For example, on this chart, these movements of large amplitude are indicated by red arrows. In order to predict reliably smaller price movements, lower resolutions need to be used.

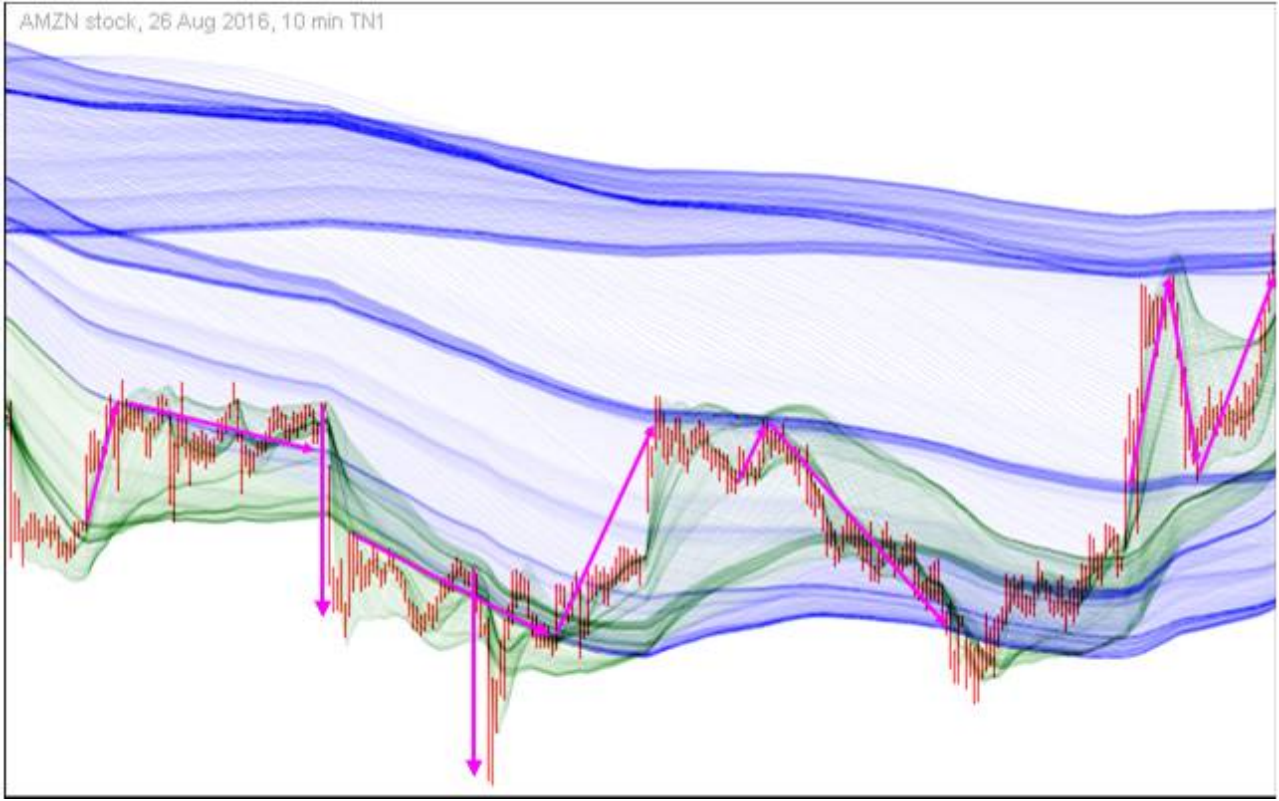

**Fig. H3. The global (large scale) actions of the main characteristic figures.** Characteristic figures attract and repel the price or quantity, as if they were dipole magnets. The pink arrows represent the effects of characteristic figures in relation to price movements of sufficiently large amplitude. Notice the movements where the price follows or temporarily sticks to (or oscillates around) a cord. This is when the cord attracts the quantity as much as it repels it, so to speak. One can observe this phenomenon on most of the charts.

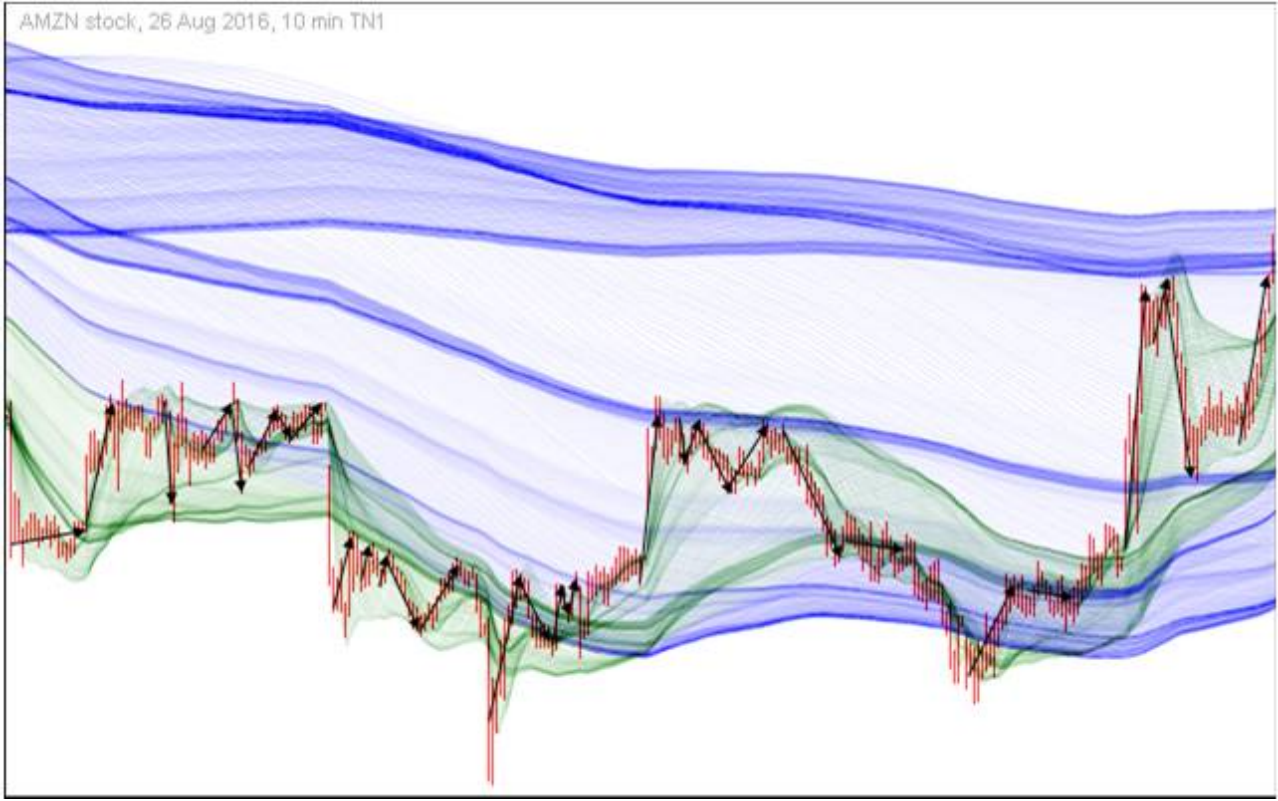

**Fig. H4. The precise action of the main characteristic figures.** Here is what a more detailed examination of the effects of the characteristic figures of the chart in Fig. H3 looks like.

**Appendix I: Examples Among Millions**

The examples presented in this article are in no way ad hoc cases, but are charts chosen at random, to illustrate a particular aspect of TN. If one has the opportunity to look at more TN charts, one can see that characteristic figures, patterns, etc., are present on every chart. Alternatively, one can refer to App. Q, R, and S for a more formal proof.

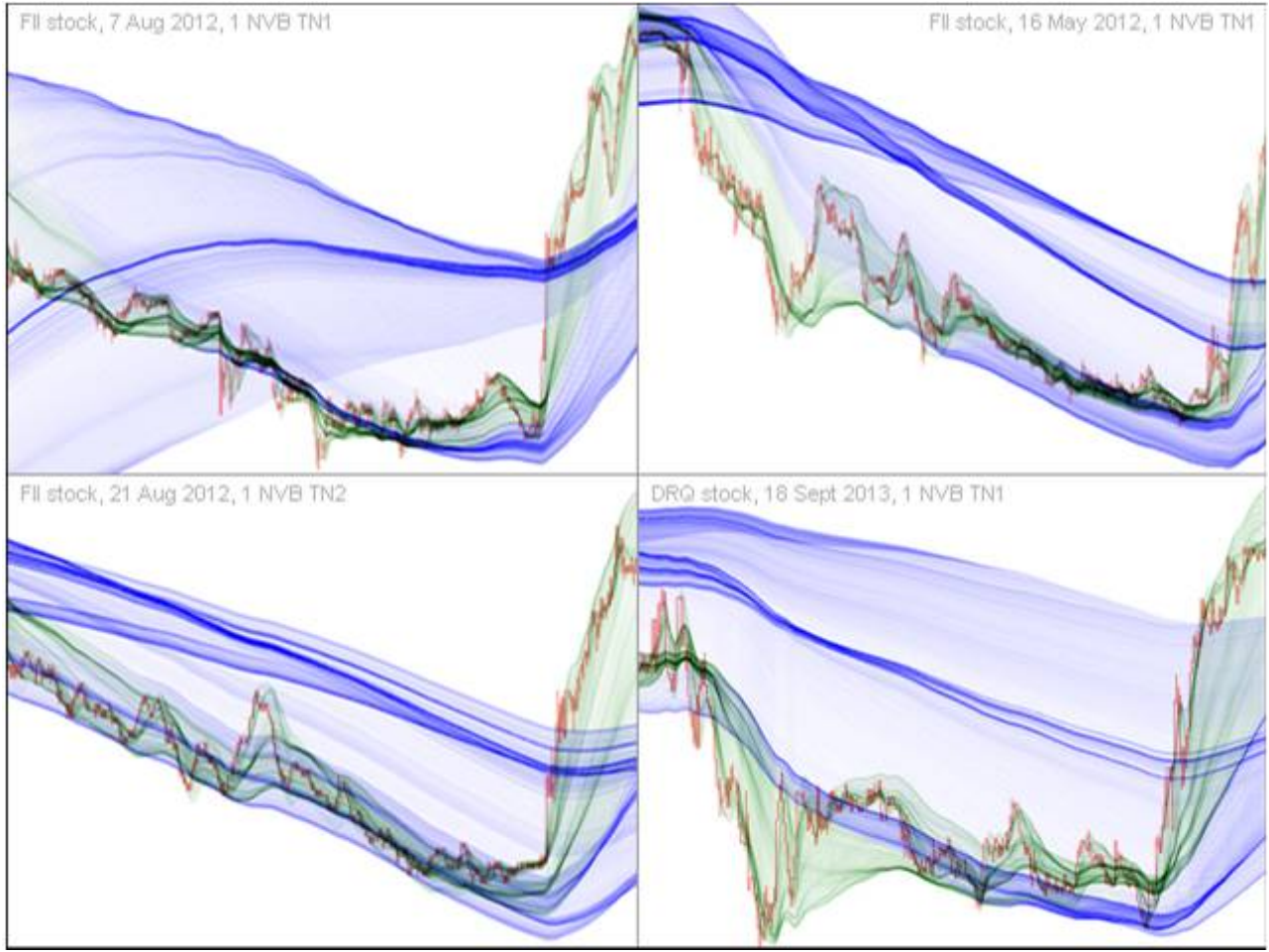

**Fig. I1. "Fall brake" followed by an "exit buildup".** These patterns repeat themselves so often that the same two patterns appear three times in a row in the same instrument, in the same resolution, over a 3-month period, as seen in this figure. The fourth chart purposely shows another instrument to illustrate how even a combination of two patterns can repeat itself very similarly at any time and in any instrument.

## Appendix J: Use of Several Subtypes for Making Predictions

One may have wondered earlier why one notable local extrema happens to be "in the air", in Fig. 6a (reproduced below in Fig. J1a). The reason is provided by the network under TN2, as can be seen in Fig. J1b.

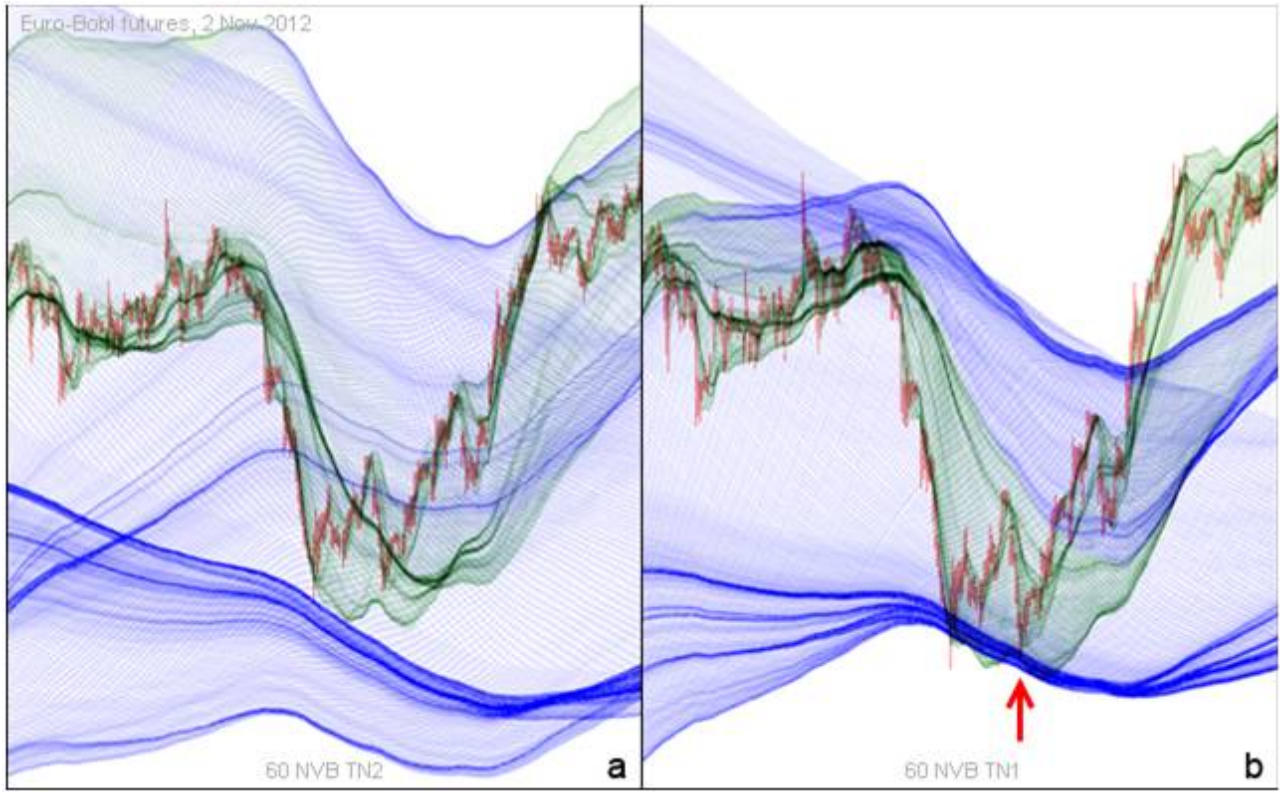

**Fig. J1. The network under the first two subtypes.** Each subtype provides a certain amount of information in terms of characteristic figures and interactions. For example, the two bottom local extrema (and also even the small extrema between the two) do not show any interactions under TN2. However, under TN1, an extremely dense cord accounts for this extremum.

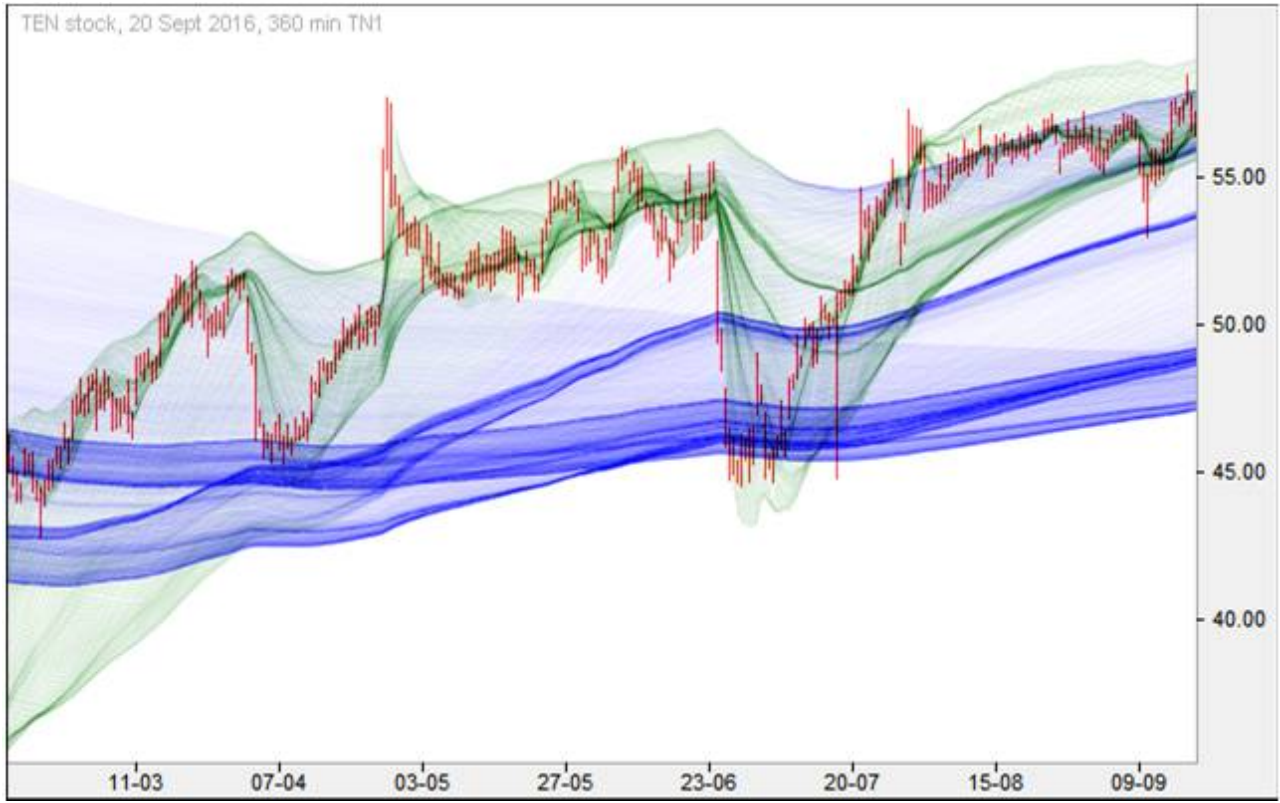

**Fig. J2.** An example where there are numerous characteristic figures in the lower part of the network under TN1.

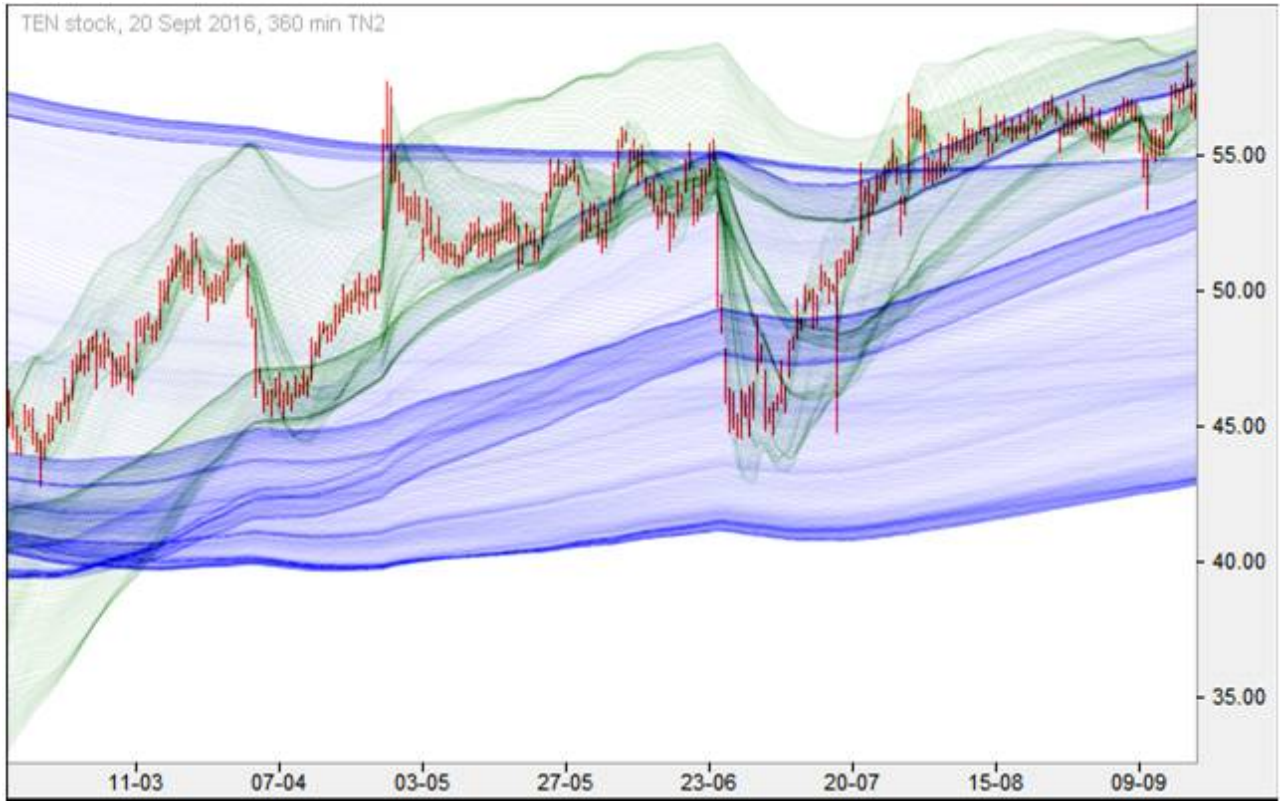

**Fig. J3. The same example under TN2.** Some characteristic figures are present, this time, in the upper part of the chart, which help to make predictions.

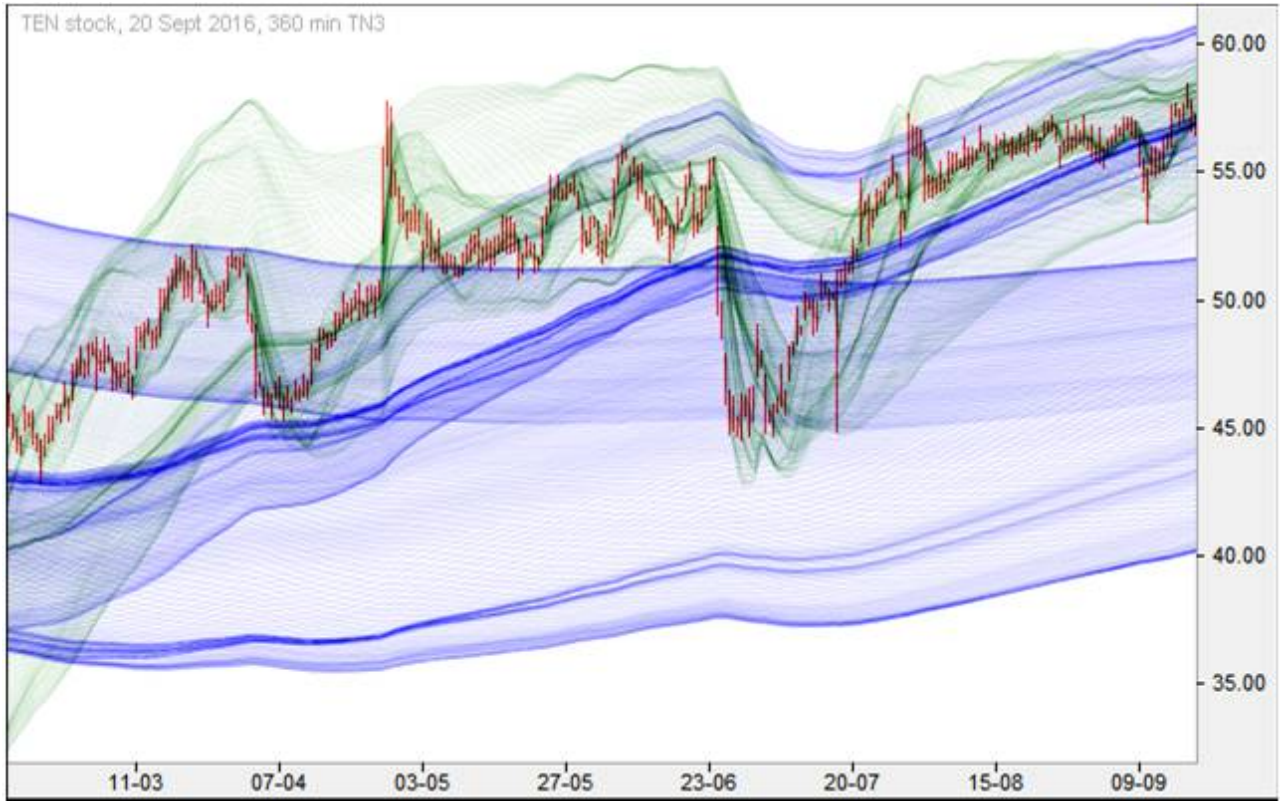

**Fig. J4. The same example under TN3.** Some characteristic figures are present, again in the upper part of the network, providing even more elements for making predictions.

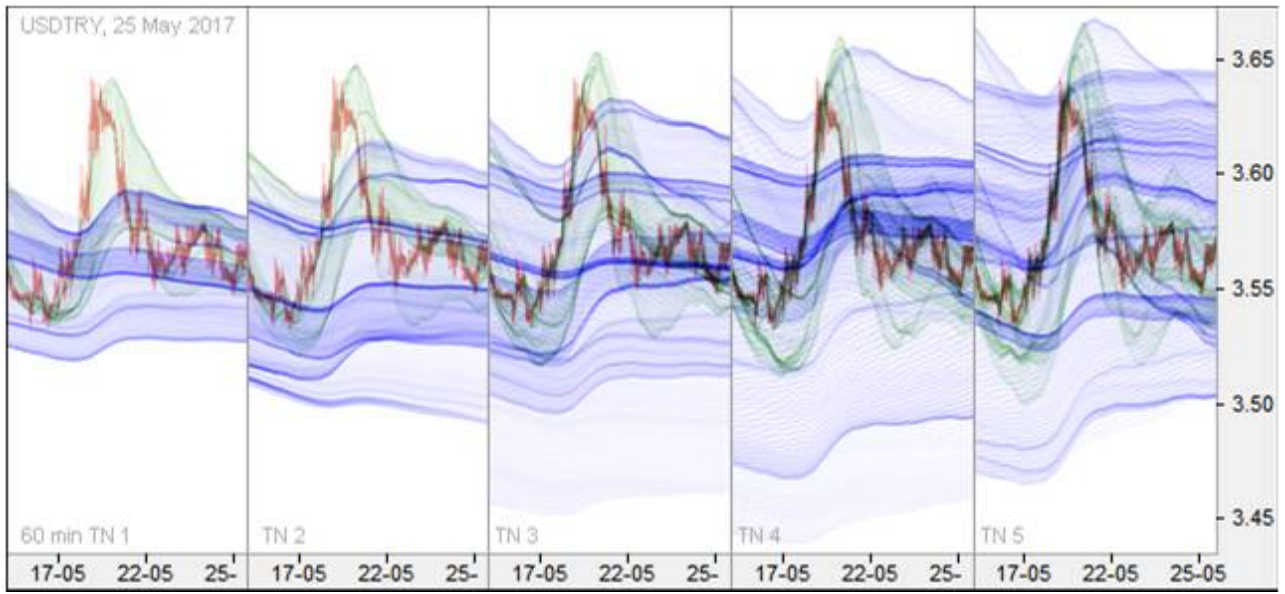

**Fig. J5. Another example illustrating the usefulness of higher subtypes.** The ordinate of the local maximum "in thin air" could not have been known in advance under TN1. On the other hand, under TN4 and TN5, the top envelopes allow knowing the ordinate reached by this violent price movement.

**Appendix K: Use of Several Resolutions for Perfecting Predictions**

It is rare for a local extremum not to interact with any of the characteristic figures present on the ensemble of the subtypes or any of the characteristic figures appearing in higher resolutions, that is, longer characteristic figures. Fig. K1 shows examples of local extrema that can only be understood with the help of networks in higher resolutions. In the case of example b, it is interesting to note that the cord which accounts for the local minima (circled in red) is particularly long considering that the ratio of the resolutions is 60:1.

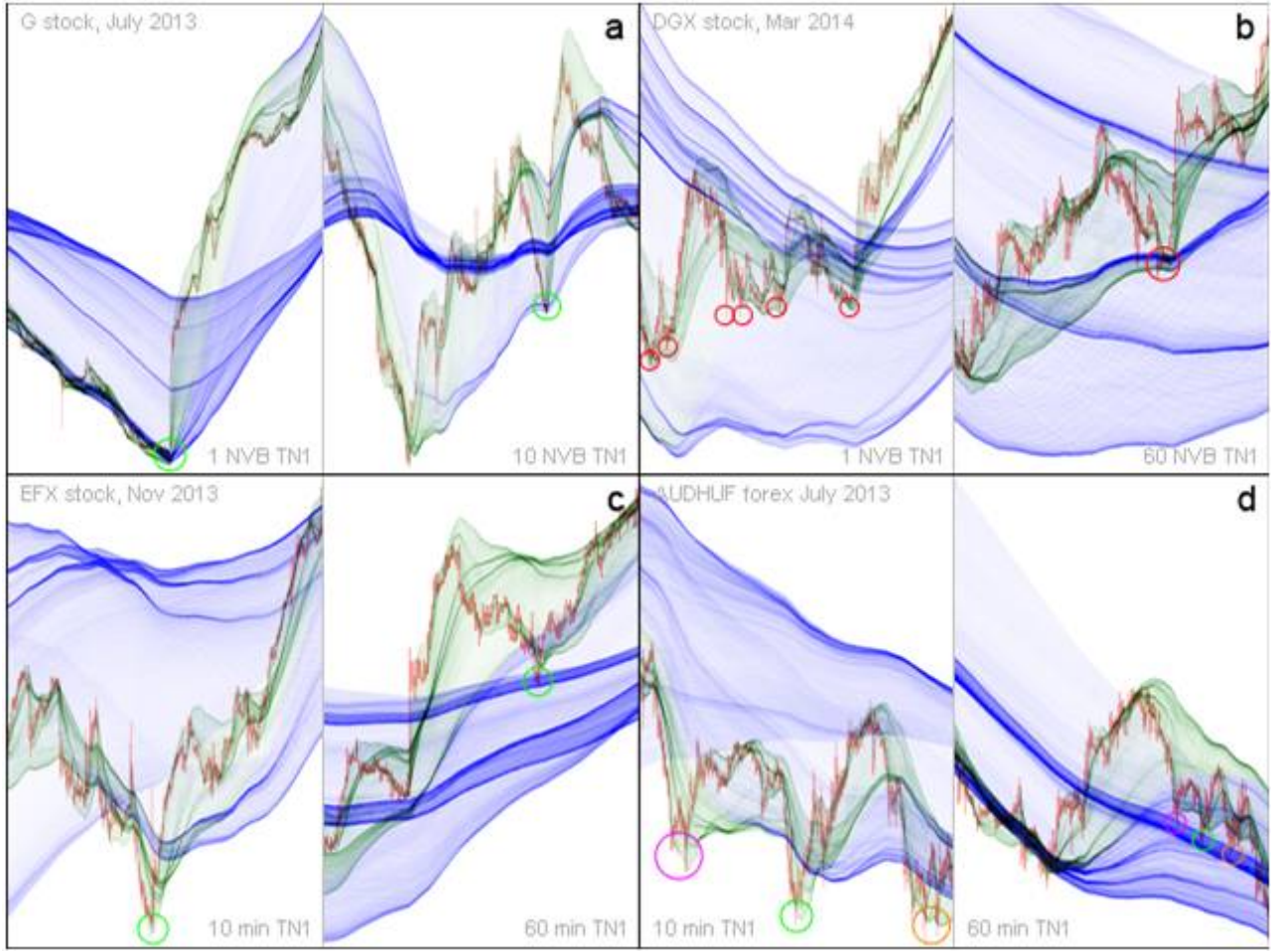

**Fig. K1. Notable local extrema are explained and are predicted with the help of higher resolutions.** When characteristic figures are sparse in a topological network or when the quantity is largely outside of all cords, it is necessary to turn to the network in higher resolutions. Observe the circles of the same color in each of the four pair of examples to understand the phenomenon.

**Appendix L: The Impossibility of Predicting in Certain Cases**

There are cases where it is impossible to make predictions. Fortunately, these cases are rare, particularly with stock market data. When it is the case, the reason is one of the following: data producing a topological network of poor quality; lack of marked characteristic figures; or future movements that cannot be ascertained due to the configuration of characteristic figures being too "symmetrical".

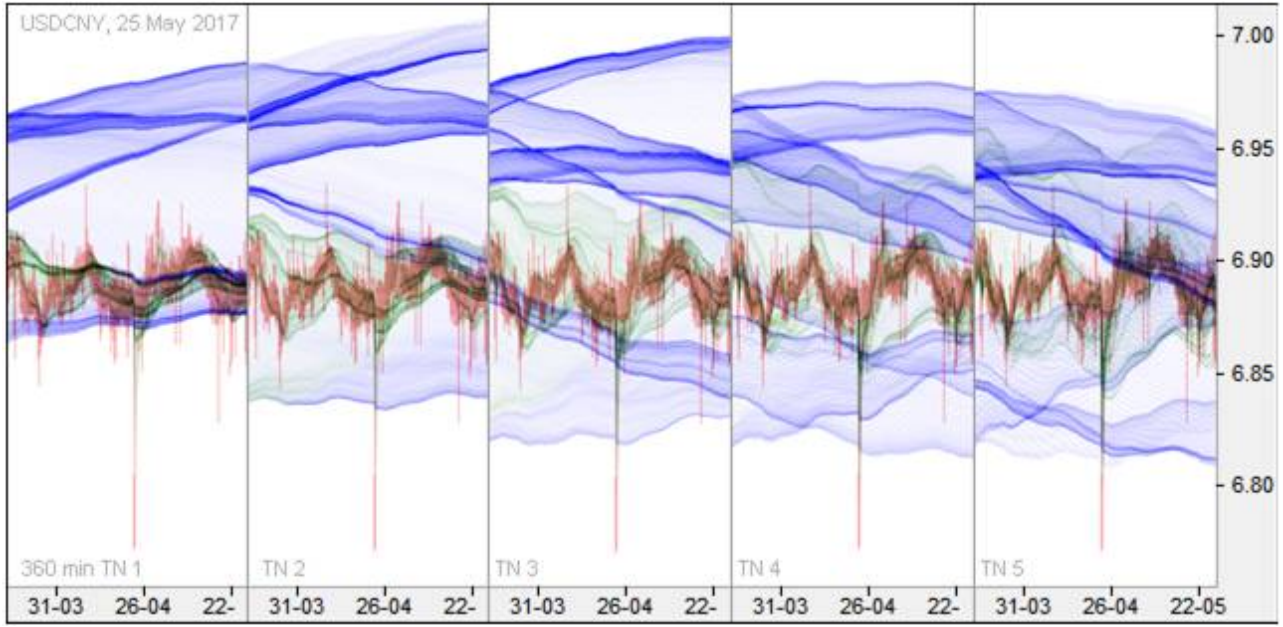

**Fig. L1. Even the consultation of all the subtypes does not allow predicting, here.** Some characteristic figures are present, but they exert little action on the price.

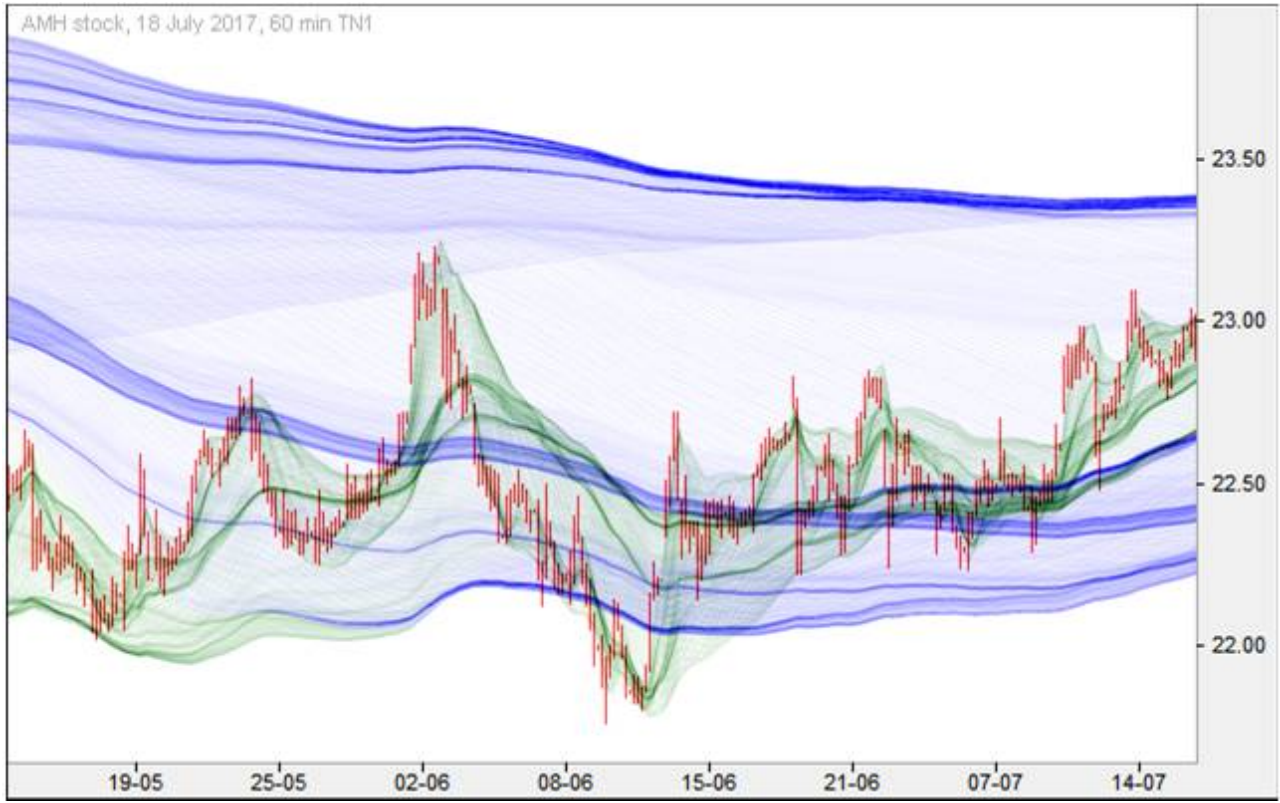

**Fig. L2. A case that cannot be ascertained.** The presence of cords, above and below the price, fairly symmetrically, does not allow predicting the next price movement in this network exhibiting a "stationary mode". Nevertheless, two scenarios with numerical values can be put forward, something that already provides valuable information.

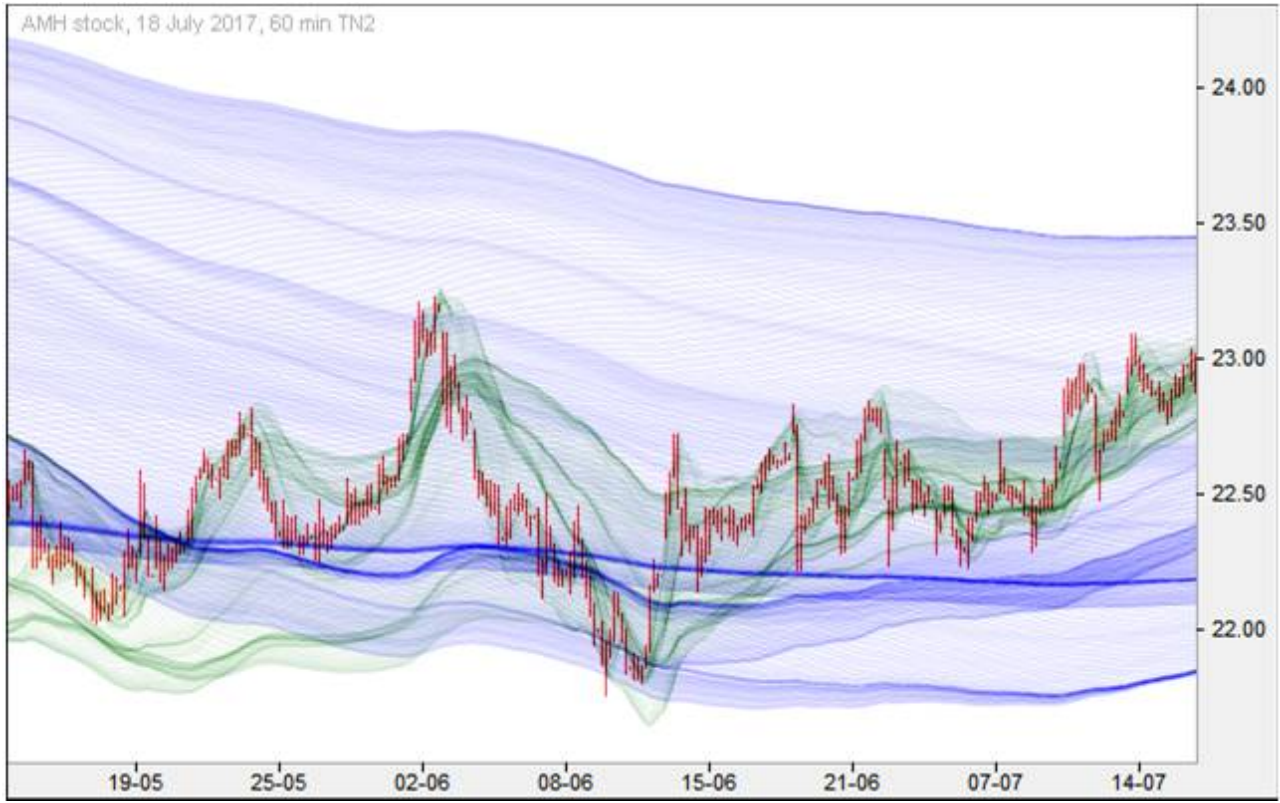

**Fig. L3. Here, under subtype TN2, it is even worse.** Besides the symmetry of the network, there is a lack of characteristic figures. The examination of a higher resolution is indispensible here.

# Appendix M: On the Difficulty in Convincing the Non-Scientific Community that TN is Not an Illusion

As a side note (to see how difficult or not it would be for people to understand this counterintuitive theory), on September 9, 2016, the author sent several traders the chart of this stock (which ended where the vertical line is), accompanied with explanations about the tool and a very simple prediction: the price will make its way to the blue cord (indicated by the pink arrow). This is precisely what happened (after a small "umbrella"). Note that this prediction was easy due to the presence of that very thick cord.

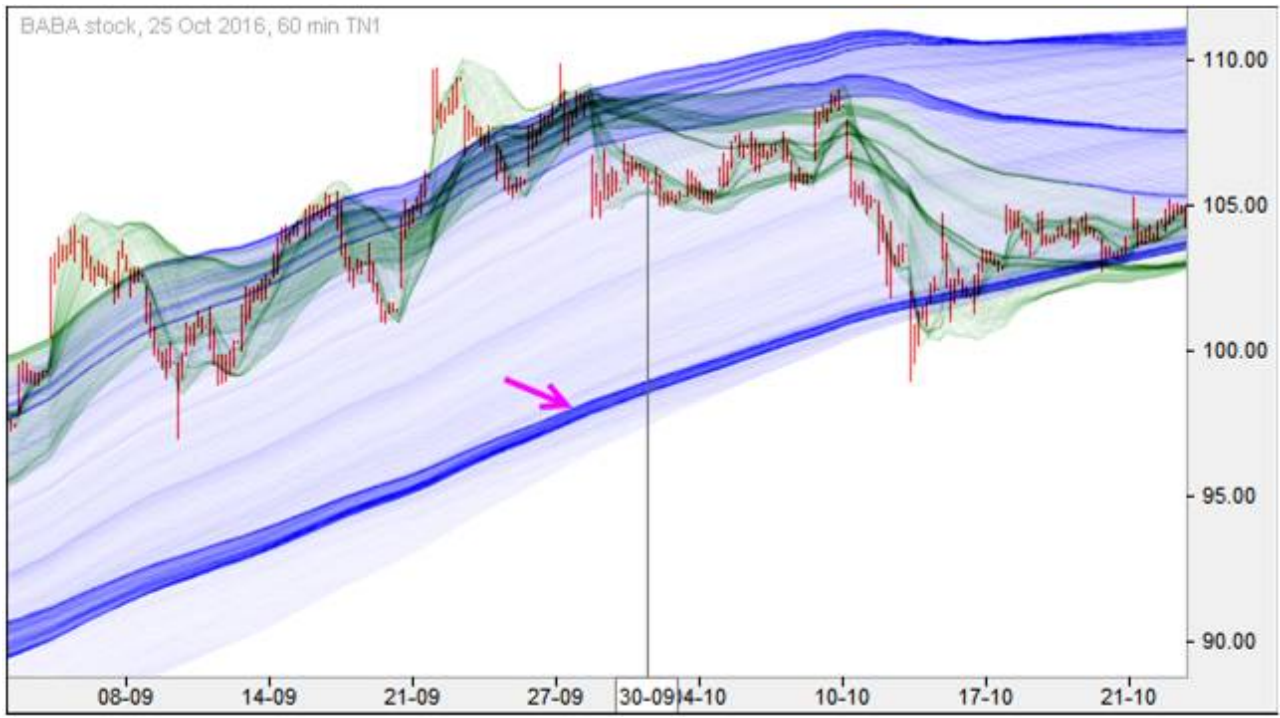

**Fig. M1. An easy-to-make prediction.** For the first time, we now have a mathematical tool that allows predicting, but the public (accustomed to hearing other things, no matter their validity, or lack thereof) does not realize its importance and predictive power.

Although it is objectively astonishing to observe three weeks later that the price indeed went to sit gently on the cord in question (what other tool is able of such a prediction?), there was very little reaction from the traders, and nothing suggesting an understanding of the tool. Yet, is it not difficult to understand how TN works and what the implications are. Anecdotally, the author was once told, "What I want to know is what the value of the stock will be in exactly one week." One may also like to know what the temperature will be in six months and one day, but some desires only pertain to dreams. Not being able to know this temperature would not invalidate the best weather forecasting model because of the degree of indeterminism in the phenomenon. To be able to predict the value of the price in three weeks would actually imply that one is capable of predicting the value of the price in two weeks, but also, in one week, one day, one hour, etc. In brief, it would imply that the price is deterministic, which is in contradiction with the semi-deterministic nature of the price, as TN teaches one. People intuitively know this, but continue to want to obtain "tips" of a deterministic nature. This is absurd for the reasons above. TN brings a great degree of richness regarding predictions, and predictions clearly subtler than "the price will be 104.30 euros in three weeks". People now need to move from deterministic predictions, which are false and unfounded (what is the use of predicting that the price will be worth 113.47 in 15 days if this prediction comes out of nowhere?), to real predictions, valid, but complex, in conformity with the semi-deterministic nature of the market.

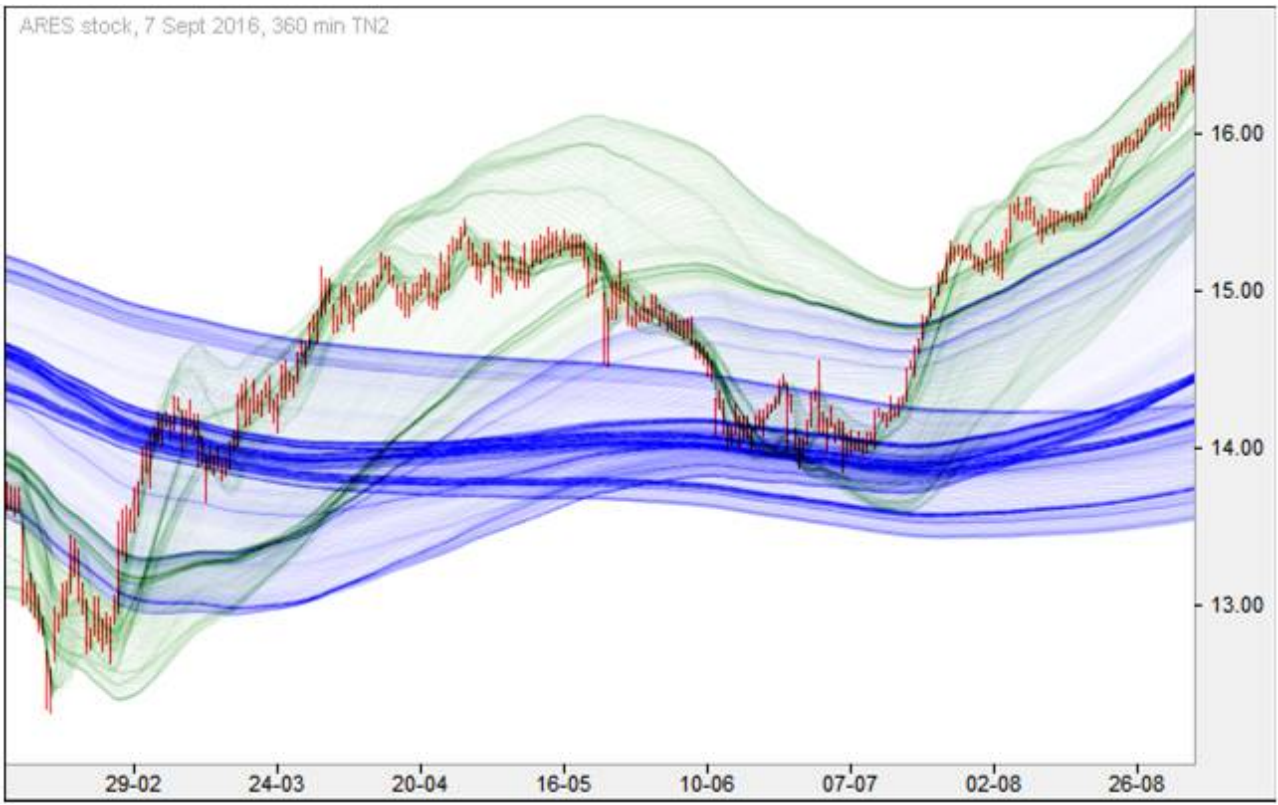

**Fig. M2. TN is not that difficult.** It is simply about spotting the cords and understanding that the price lands and bounces off them.

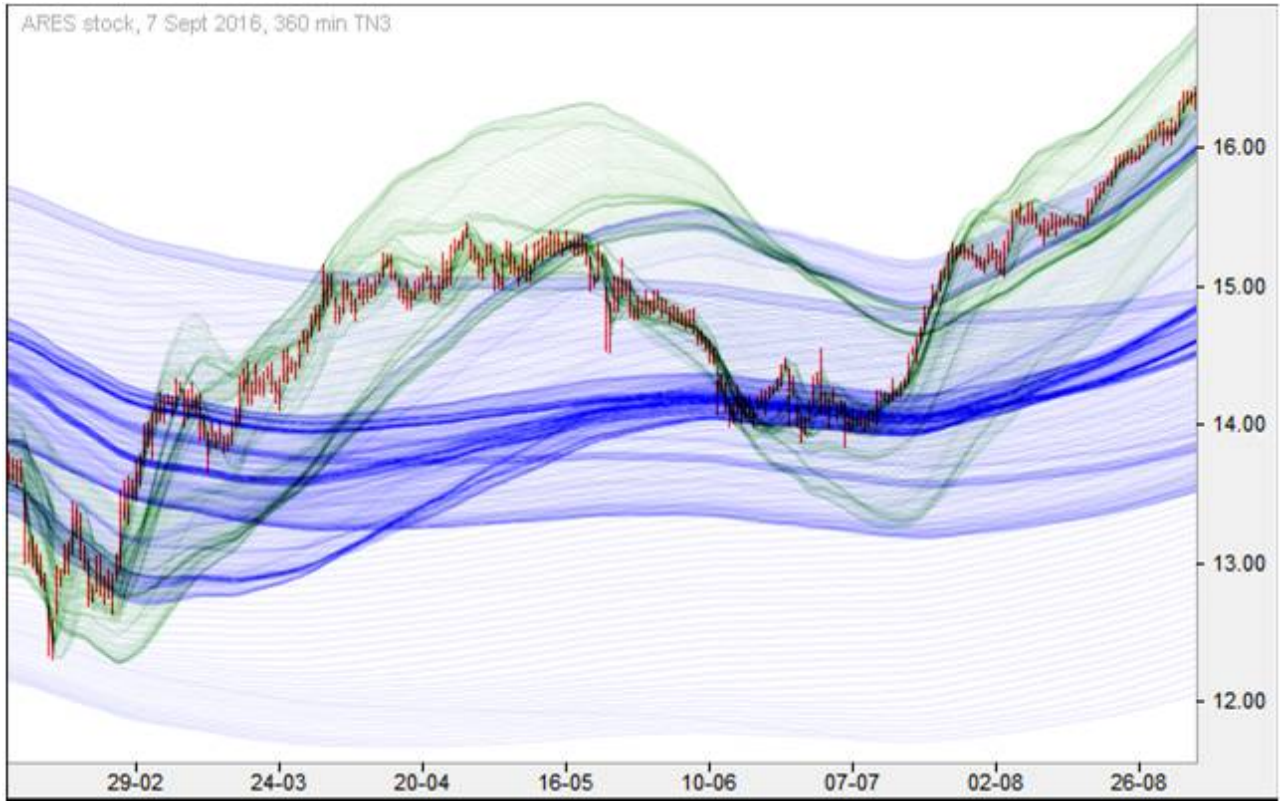

**Fig. M3. It is even less difficult, considering how the information can be crossed-checked and enriched.** Suppose one suspected that the price would turn around when it reached the large blue cord in TN2, after looking at the TN3 chart, one would have been certain.

**Appendix N: Demonstration by "Reductio Ad Absurdum"**

The demonstration in this appendix is intended for people who think that the interactions are fortuitous (despite their systematic presence and extreme precision in each of the millions of topological networks). The argumentation can be found in the body of the article in the section by the same name. We will limit ourselves here to providing, based on the Fig. 14 example, examples of shifting in both directions, and under another subtype.

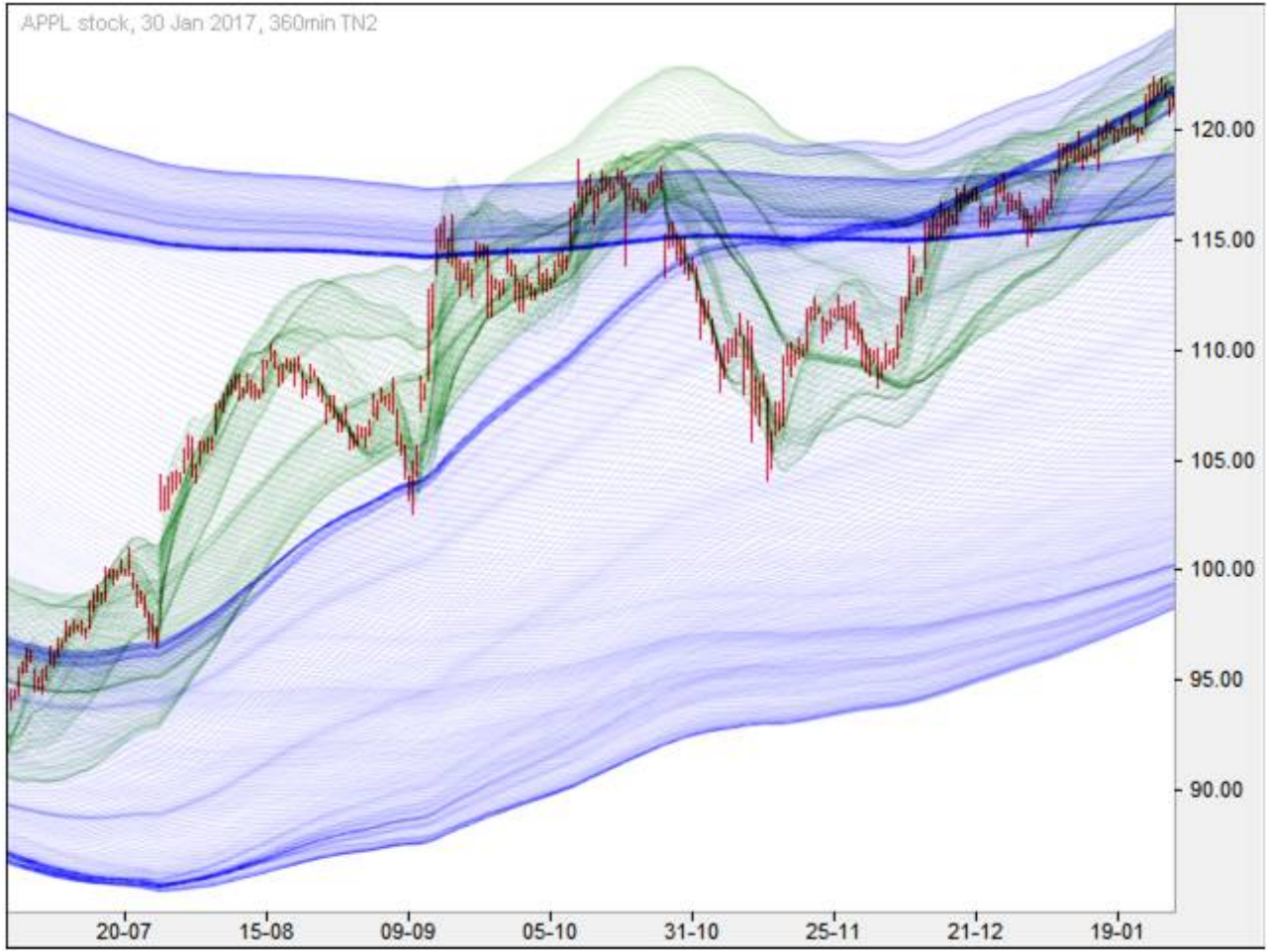

**Fig. N1. The topological network of a stock under TN2 in a long resolution.** Observe the characteristic figures and their interactions with the price. The interactions are extremely precise.

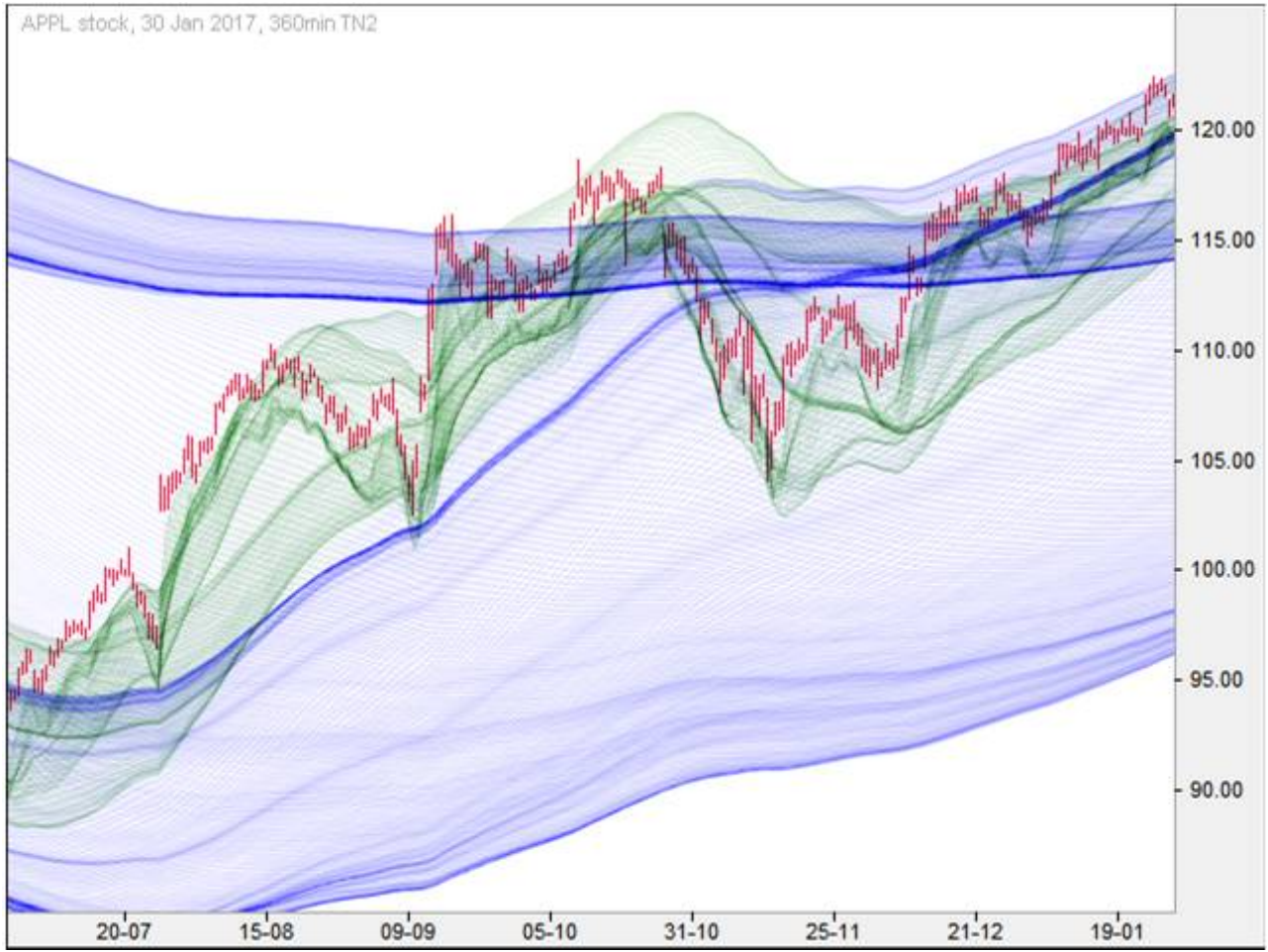

**Fig. N1. The same network with a small shifting of the price toward the top.** Only a few (fake) interactions, due purely to chance, are found. Note in particular, the absence of interactions with the most notable local extrema, for obvious probabilistic reasons.

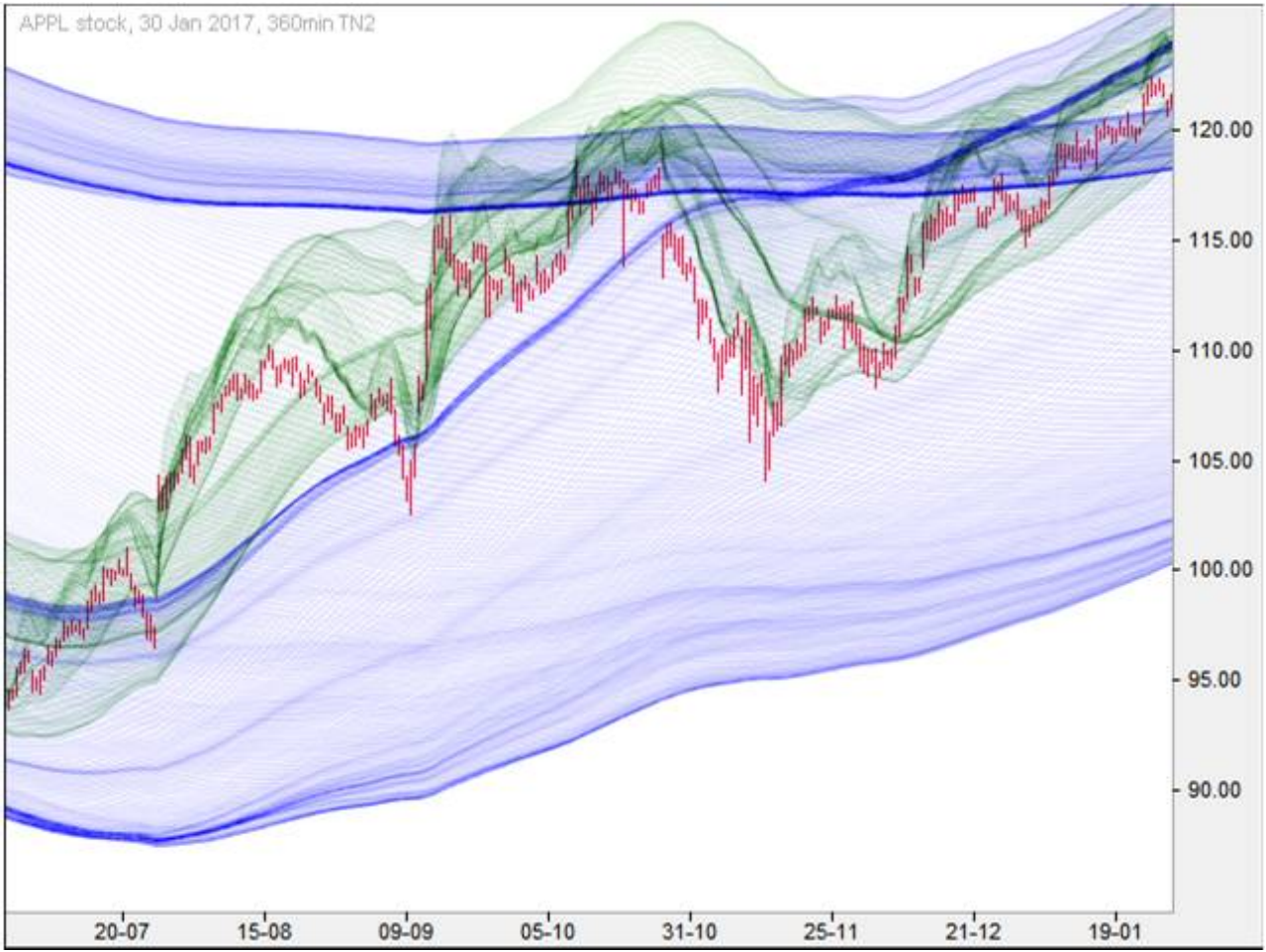

**Fig. N3. The same network with a small shifting of the price toward the bottom.** In particular, when one visually follows the movements of the price, one no longer sees any rebounds from one characteristic figure to another. Cf. App. H.

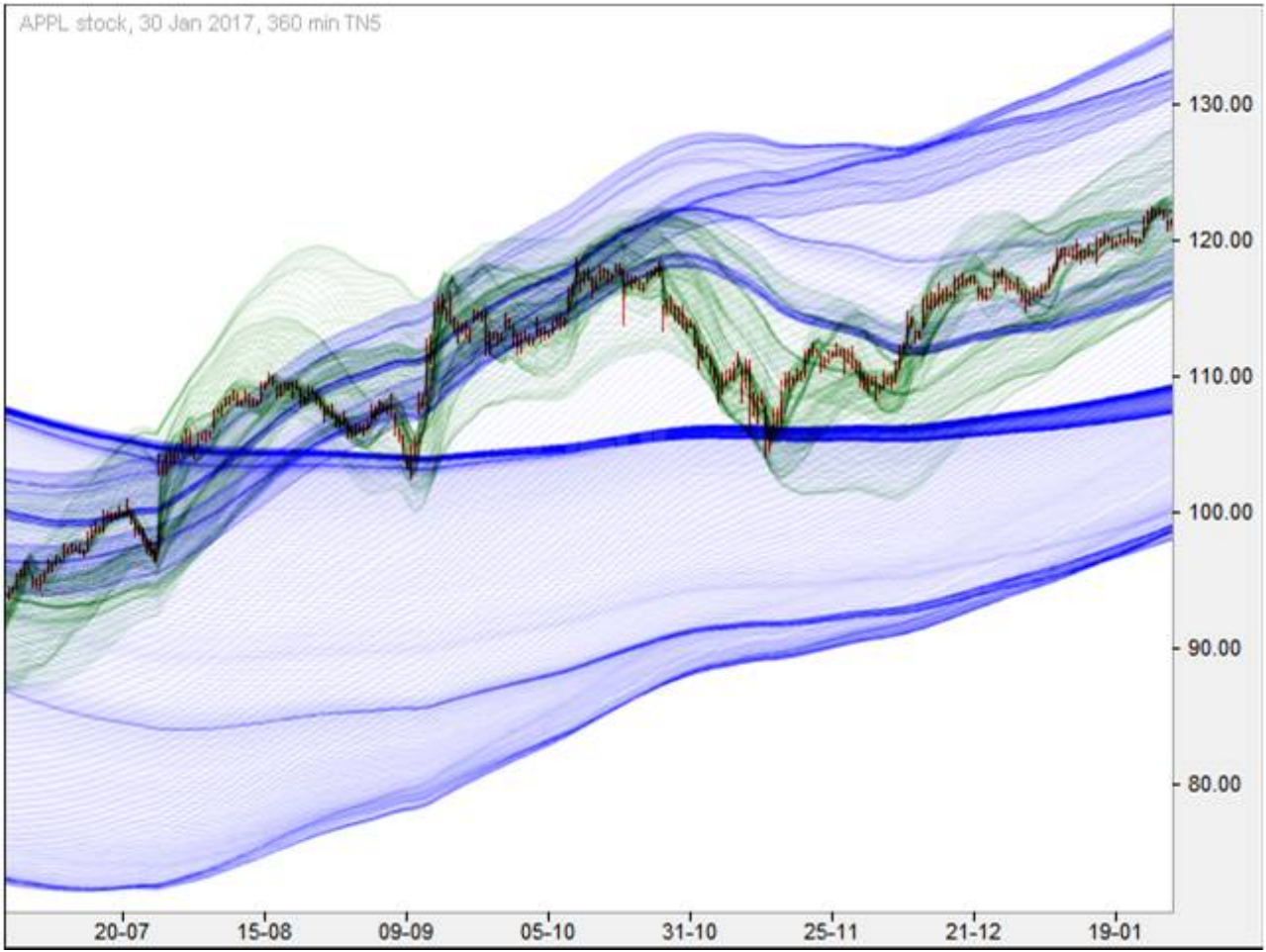

**Fig. N4. The test is even more powerful since it can be done on all the subtypes with the same shifting.** Here, the same network is shown under TN5.

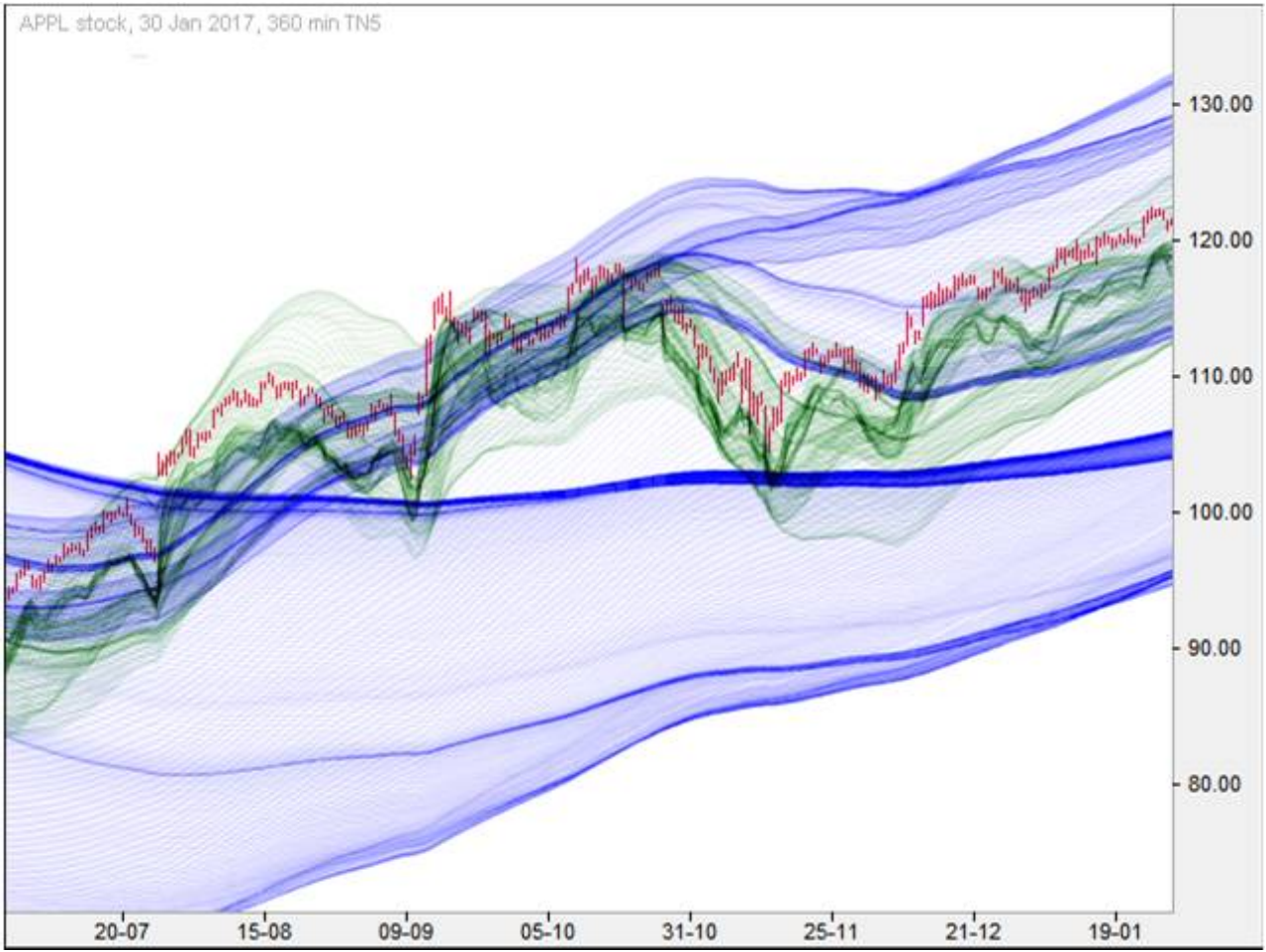

**Fig. N5. The same shifting toward the top, under TN5.** The rare interactions observed are fortuitous. Taking a closer look, one can see that they are absurd (they do not fit into a course of interactions).

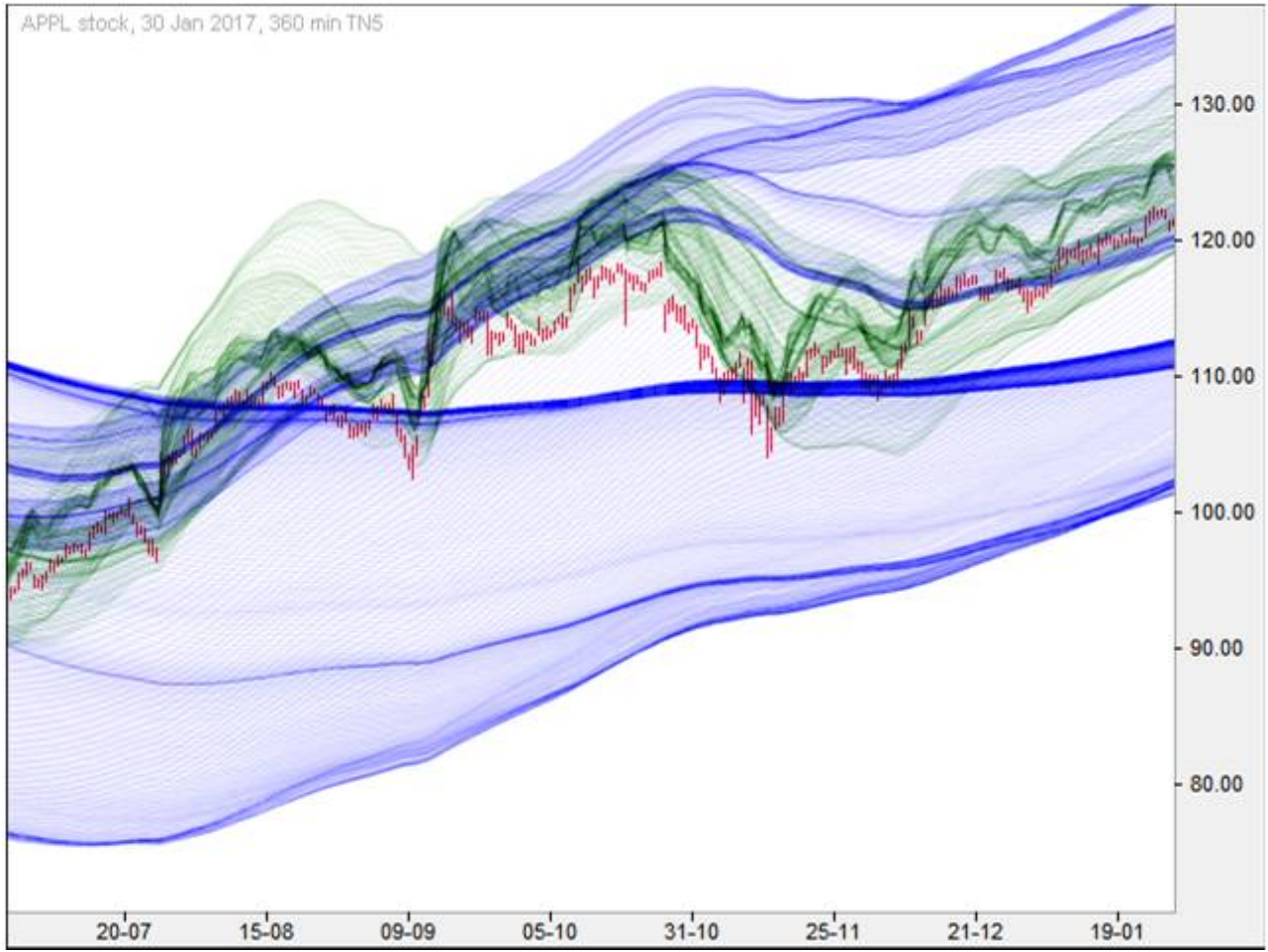

**Fig. N6. The same shifting toward the bottom, under TN5.** The principal local extrema are almost always outside of the cords, which is in opposition to what is observed in the absence of shifting. In addition, the probability that one local extrema interacts concomitantly with a characteristic figure under several subtypes is close to zero, which proves the point one more time.

## Appendix O: The Non-Pertinence of In- and Out-of-Sample Testing

We have been asked about in- and out-of-sample testing. This is irrelevant (not applicable) to TN. TN is not a model. It is not "fitted". With TN, one injects a certain amount of historical data from a time series and the corresponding topological network is generated. By injecting more data, one just gets a wider topological network (covering a longer time period). It does not improve the prediction because the number of data points used for the network is the same whatever the "slice" of network considered. In Fig. O1b, one can see two zoomed-in slices of the network from the chart in Fig. O1a. Each slice, as any other slice from the network, uses the last 1500 data points (up to the data point corresponding to the slice). A slice consists of the ordinate of, here, 500 curves. The thick blue or green lines in Fig. O1b correspond to cords or characteristic figures.

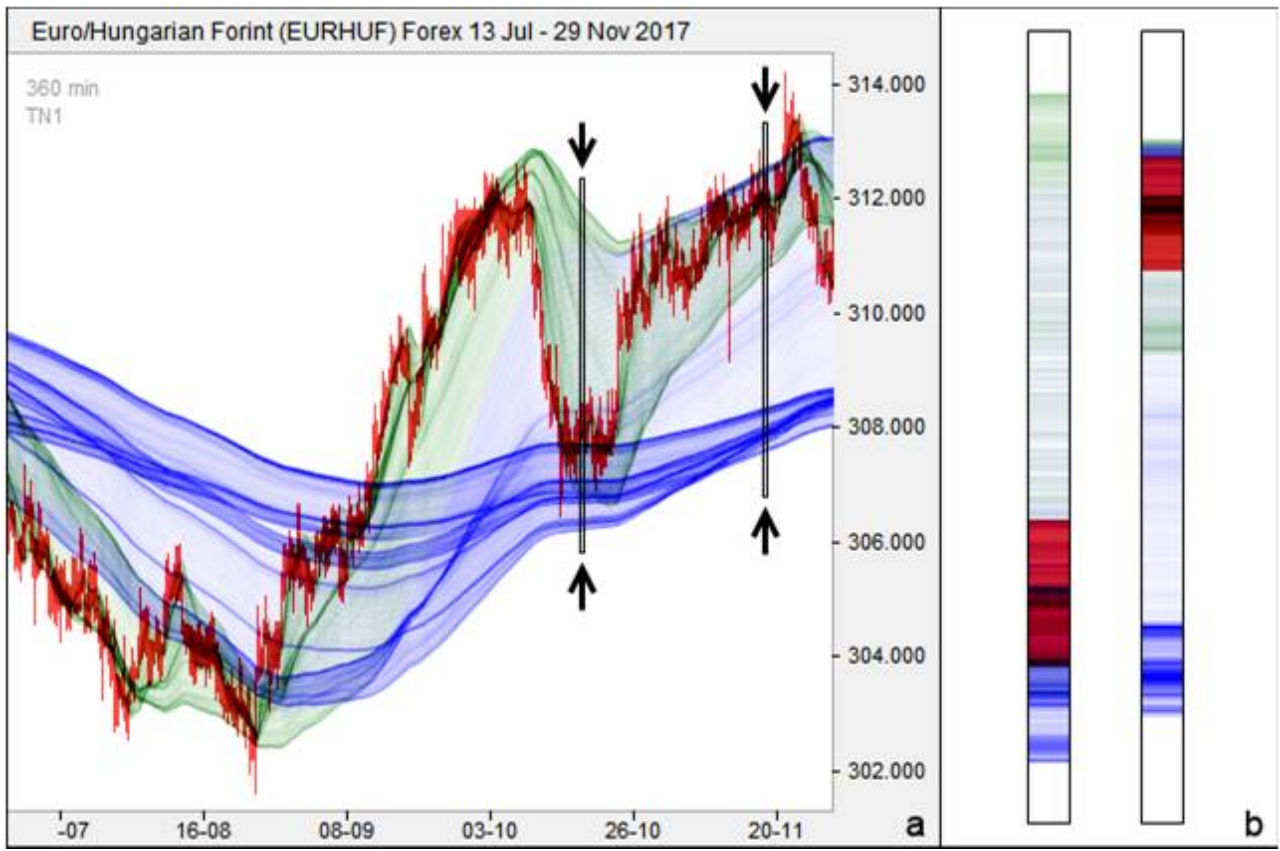

**Fig. O1. A network corresponding to a forex instrument in a large granularity.** a) Two slices have been selected to better understand what a network consists of. b) The two corresponding slices are magnified. Each one contains 500 portions of curves (points). Each slice requires here 1500 data points.

Experience leads us to choose between 1500 and 7500 data points for the various subtypes for a network slice. With these numbers of data points, one gets rich and clean networks that can be easily exploited. With too few data points, many relevant cords will be missing. With too many data points, very long, out-of-context cords will appear. For these very long out-of-context cords, one should generate the network from the same time series using a larger granularity. This way, one better understands the relationship between these long cords and the price. Without a sufficiently large network to observe, the cords are out of context. It is a bit like relying on a slice such as in Fig. O1b to analyze and predict the price. Refer to Figs 12 and K1 to understand why the generation of a network in a larger granularity is more helpful than increasing the number of data points of a network. In these figures, the network in a larger granularity provides the longer cords that are needed in order to account for some of the extrema present in the smaller granularity network.

## Appendix P: No, the Cords Do Not Join the Extrema

Some academics still believe that the network is just "curves joining extrema" despite the total falsehood of this affirmation.

This affirmation reveals three things:
a) they confuse curves and cords (and other characteristic figures),
b) they observe that the price and the cords do interact, and
c) they refuse to accept the idea that the price is governed by the cords.

Regarding a), it is the curves that are mathematically generated, not the cords (and other characteristic figures). Curves do not join anything; curves are moving regressions of order $D$. As for cords, they are not mathematically generated; they emerge spontaneously. They have nothing to do with curves. They are not even parallel curves; they are complex intersections between curves. An elementary observation of the charts shows that.

Regarding b), the interactions are so obvious that they are not denied, thus, the affirmation. Refer to App. Q, nonetheless.

Regarding c), it can only hold if the affirmation were true. But since the proposition is trivially false, one can logically only admit that the price is governed by cords.

We should normally not have to prove that something false is false, but let us provide two proofs for those to whom it is not mathematically obvious that curves do not join extrema.

1) Let us point out again that when new data points are added, the past network does not change. This obvious fact discredits the idea that cords (let alone the curves since they do not even join anything) joining extrema. How could an existing cord go join the next extremum when one does not know where the next extremum will be (by definition)? To do so, the cords would have to all run after the price as it keeps evolving, which is clearly not what is observed, besides the fact that the mathematics behind the network does no such thing.

2) The interactions are between the price and the characteristic figures. When using a larger granularity than the exchange itself, what interacts with the characteristic figures is usually the $H$ ($M+d$) or $L$ ($M-d$) value (refer to Section II.A, "Construction of the Topological Network"); however, the $H$ and $L$ values are never injected into the mathematical tool. It is the $M$ value that is injected. This also discredits the idea that the cords join the extrema. How could something that does not even know the $H$ or $L$ value join the $H$ or $L$ value?

## Appendix Q: Proof by Simultaneity

Proving that TN embodies a physical law that governs security prices amounts to proving that the characteristic figures drive the price, that is, that the price bounces from one characteristic figure to another. However, it is impossible to prove formally (with a direct mathematical approach) that the price bounces from one characteristic figure to another because the characteristic figures are not mathematically engendered by a function (they therefore do not have a mathematical expression that could be used for a direct proof). Refer to Section II.A, "Construction of the Topological Network". Recall once again that characteristic figures emerge spontaneously from the dense network of curves, which, in contrast, are mathematically engendered by a function. While it is not possible to prove directly that the property (the price bouncing from one characteristic figure to another) is true, it is possible to prove it indirectly, especially since it is probabilistically evident.

We will try to prove it a fortiori. Suppose, to begin with, that the property is fortuitous (by chance). Under this hypothesis, one can calculate the probability of occurrence that a given chart "qualifies", that is, is such that about half of the local extrema are located precisely on (within, at the scale of a full-size image in this article, say, two millimeters of, as opposed to further than two millimeters from) the characteristic figures. This probability is delicate to calculate due to the complexity of the topological network, but it is, in any case, very small (see App. N, "Demonstration by "Reductio Ad Absurdum"", to see how fortuitous rebounds look). Let us call $\varepsilon$ the average probability of charts qualifying by chance. This probability is intuitively inferior to 1/1000. If one has thousands or millions of charts, one can a priori find by chance examples of qualifying charts. However, to reach such a number of charts, one would need to examine the charts of all the instruments over their entire data history, from all the markets, in all the granularities, and under all the sub-types.

To put an end to the idea that the examples in this article have been selected as described above (by scouring all these charts), let us use a simple protocol, doing the exact opposite, that is, showing charts without the possibility of selection. We will restrict ourselves to a single date, for example, the date at which this appendix was written, June 6, 2018. We will also restrict ourselves to the 100 most traded instruments of the NYSE starting with the letter "A", say, in 10-minute resolution. Under the hypothesis that the property is fortuitous, one should not find even a single qualifying chart, considering $\varepsilon$, which is evidently much smaller than 1/100. Yet, all the charts (except a small number due to data producing poor quality networks (see App. L, "The Impossibility of Predicting in Certain Cases")) qualify, that is, exhibit numerous and precise rebounds. This proves that the characteristic figures drive the price. Below are 92 qualifying charts out of the 100 (scaled down to 50%).

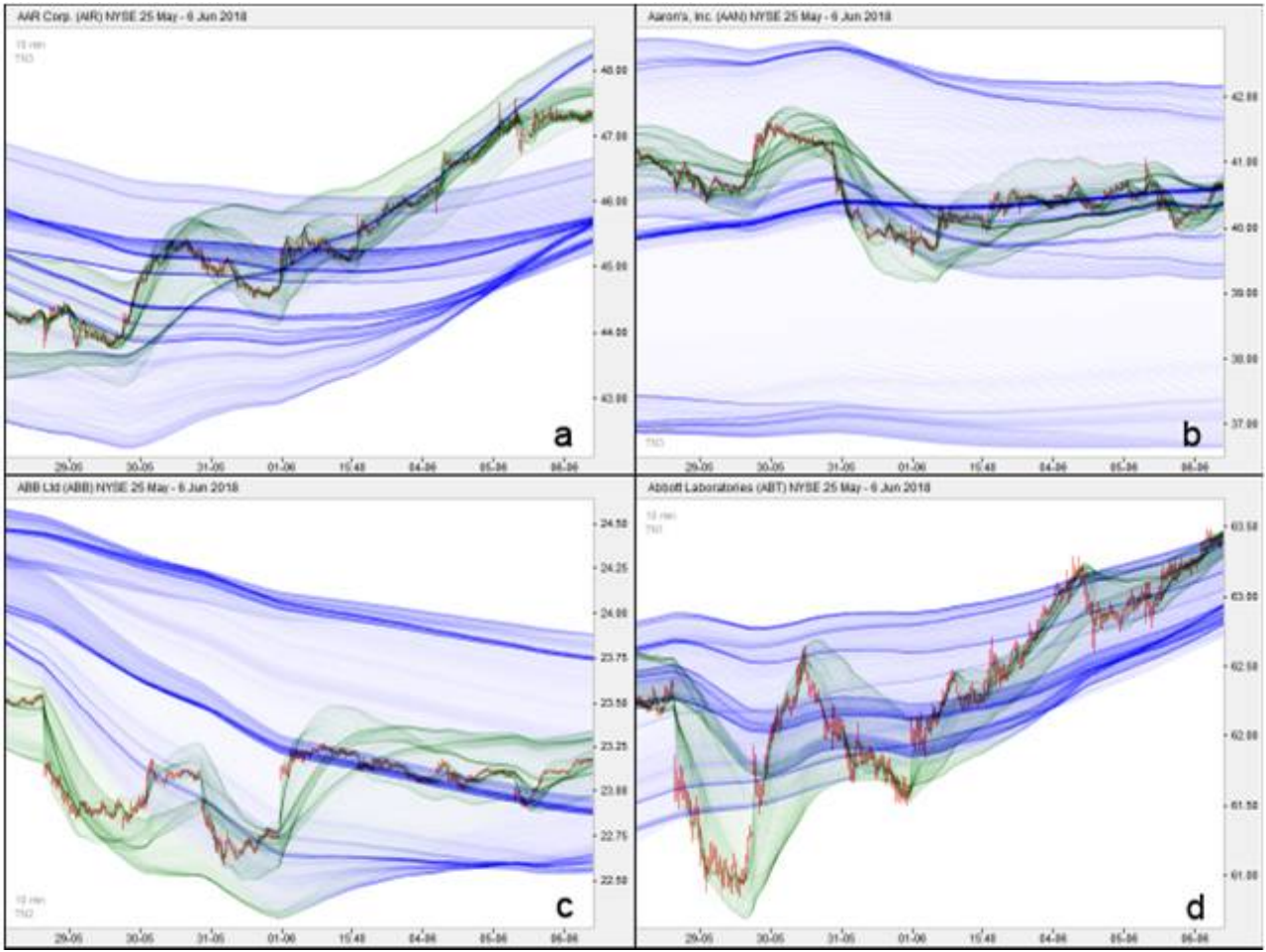

**Fig. Q1. Four "A" NYSE stocks in 10-minute at the same instant, June 6, 2018 16:00.** a) AAR, TN3. b) Aaron's, TN3. c) ABB, TN2. d) Abbott Laboratories, TN1.

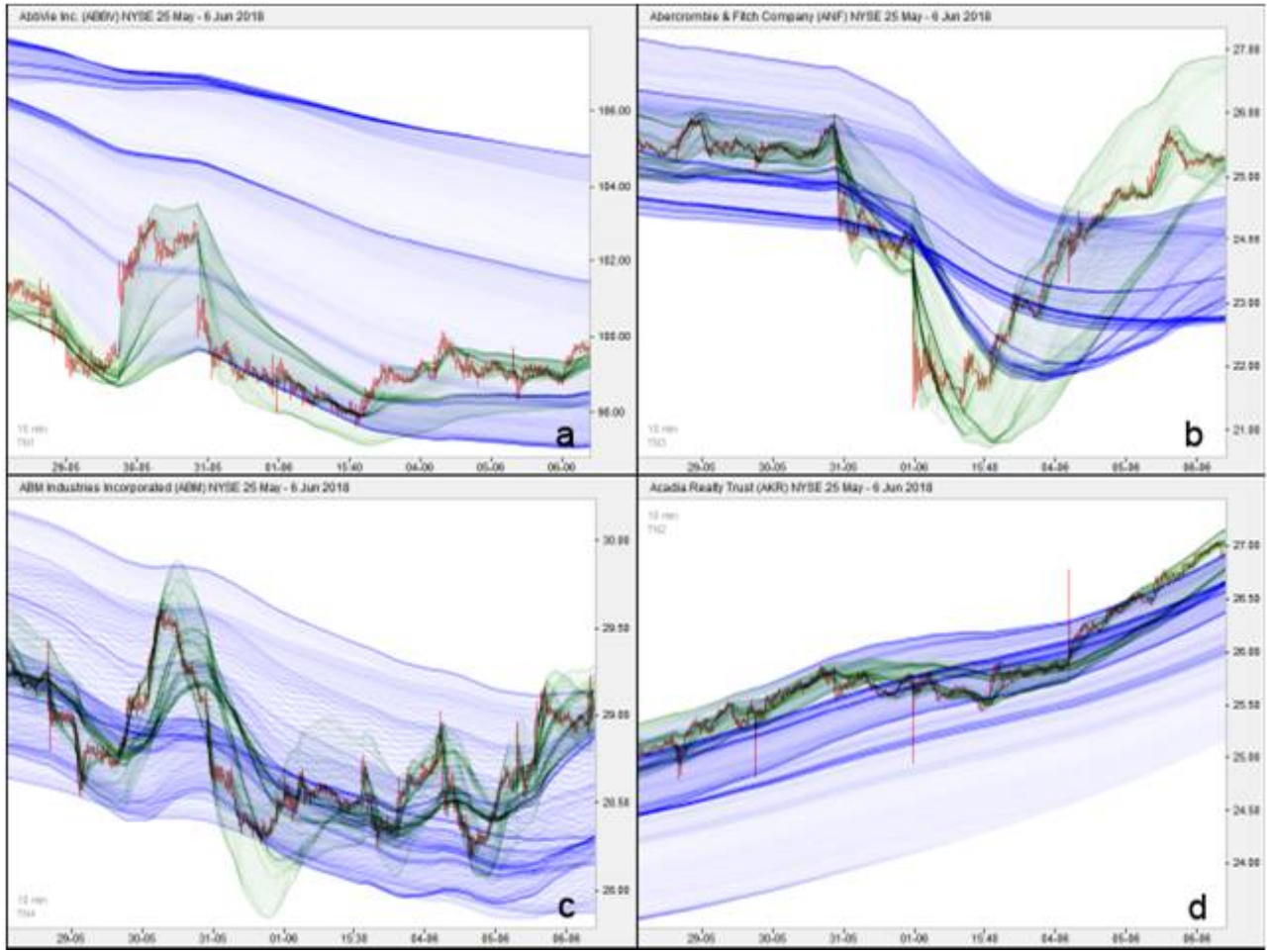

**Fig. Q2. Four "A" NYSE stocks in 10-minute at the same instant, June 6, 2018 16:00.** a) AbbVie, TN1. b) Abercrombie & Fitch, TN3. c) ABM Industries, TN4. d) Acadia Realty Trust, TN2.

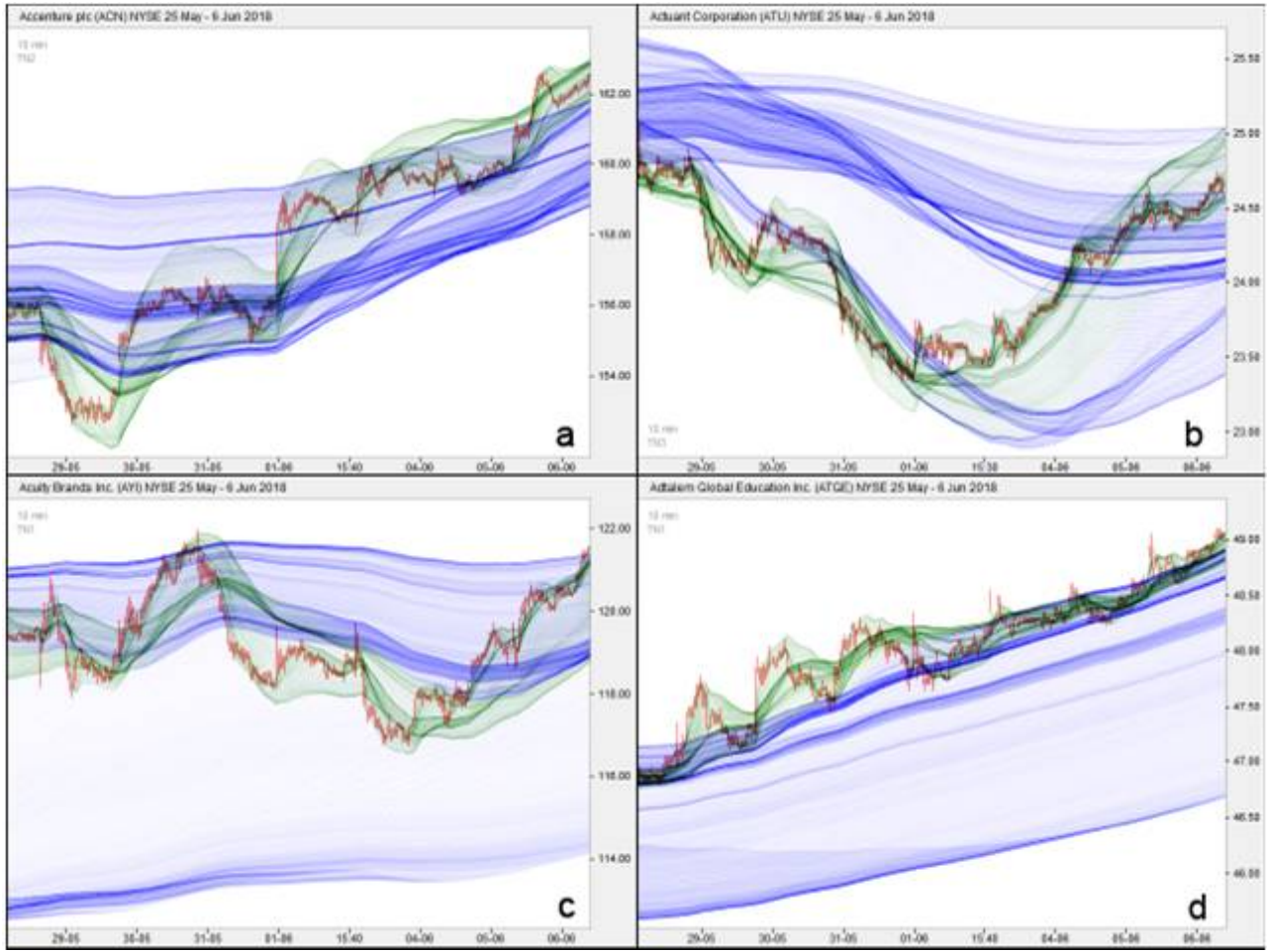

**Fig. Q3. Four "A" NYSE stocks in 10-minute at the same instant, June 6, 2018 16:00.** a) Accenture, TN2. b) Actuant, TN3. c) Acuity Brands, TN1. d) Adtalem Global Education, TN1.

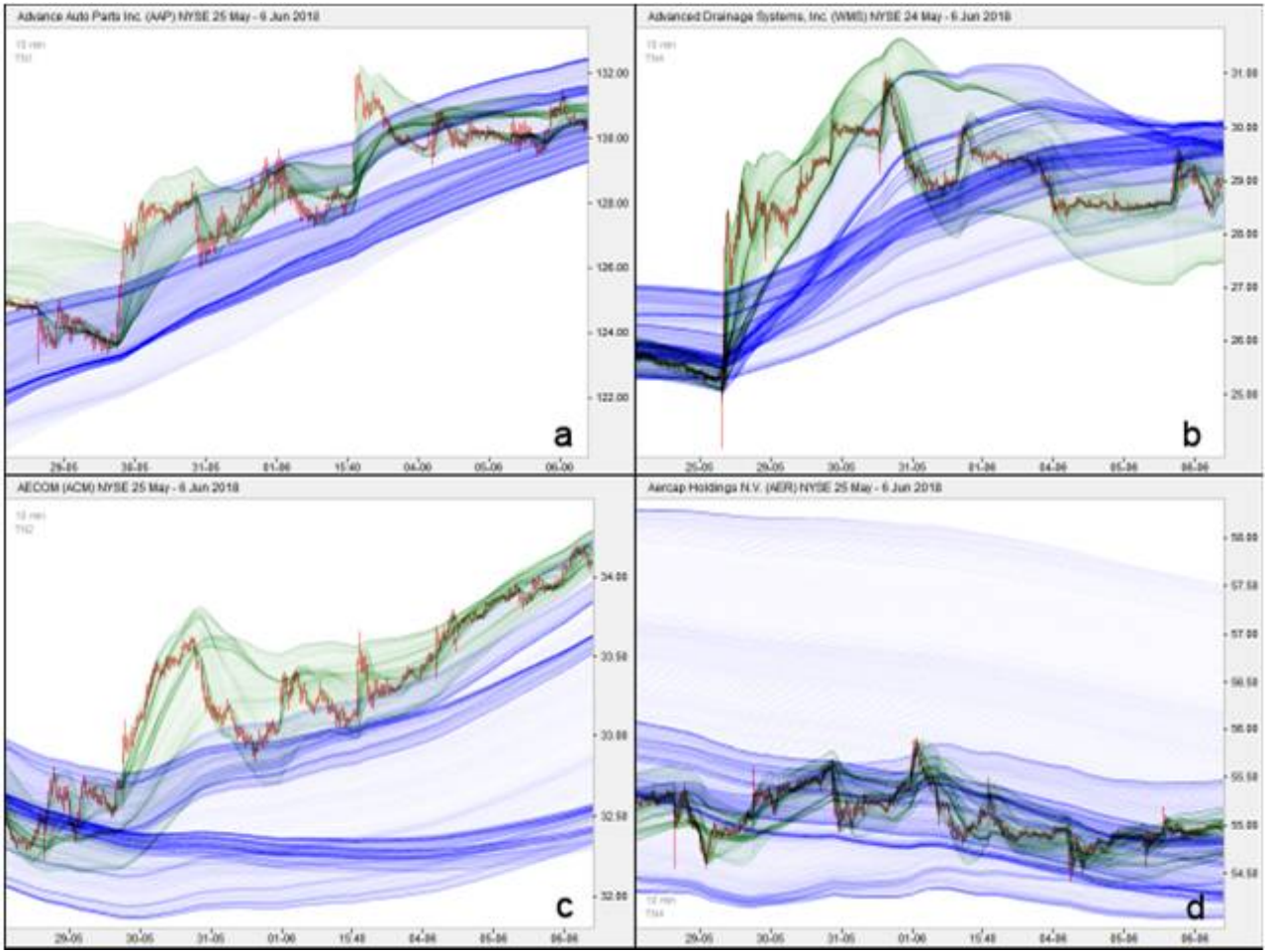

**Fig. Q4. Four "A" NYSE stocks in 10-minute at the same instant, June 6, 2018 16:00.** a) Advance Auto Parts, TN1. b) Advanced Drainage Systems, TN4. c) AECOM, TN2. d) Aercap Holdings, TN4.

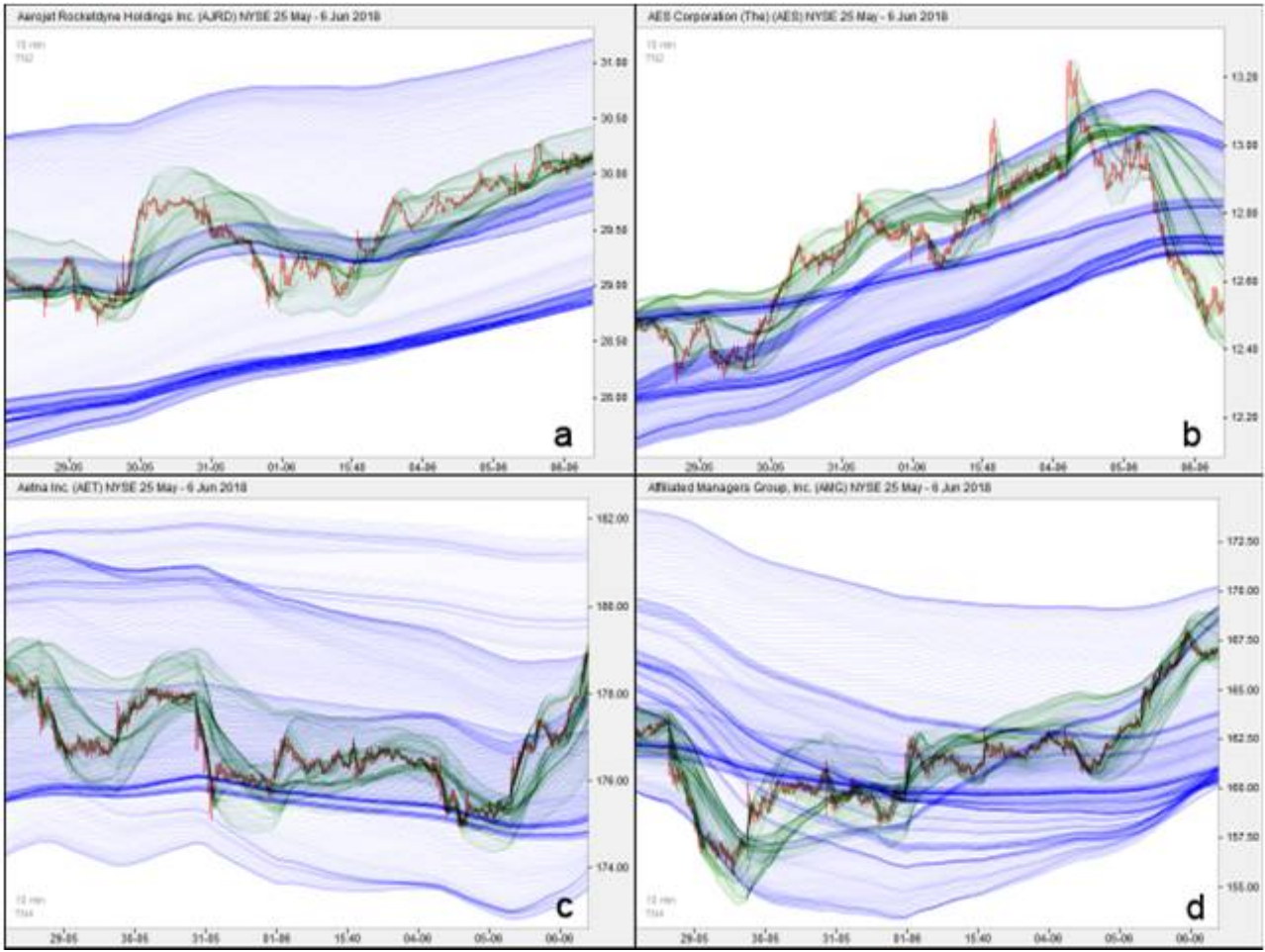

**Fig. Q5. Four "A" NYSE stocks in 10-minute at the same instant, June 6, 2018 16:00.** a) Aerojet Rocketdyne Holdings, TN2. b) AES, TN2. c) Aetna, TN4. d) Affiliated Managers Group, TN4.

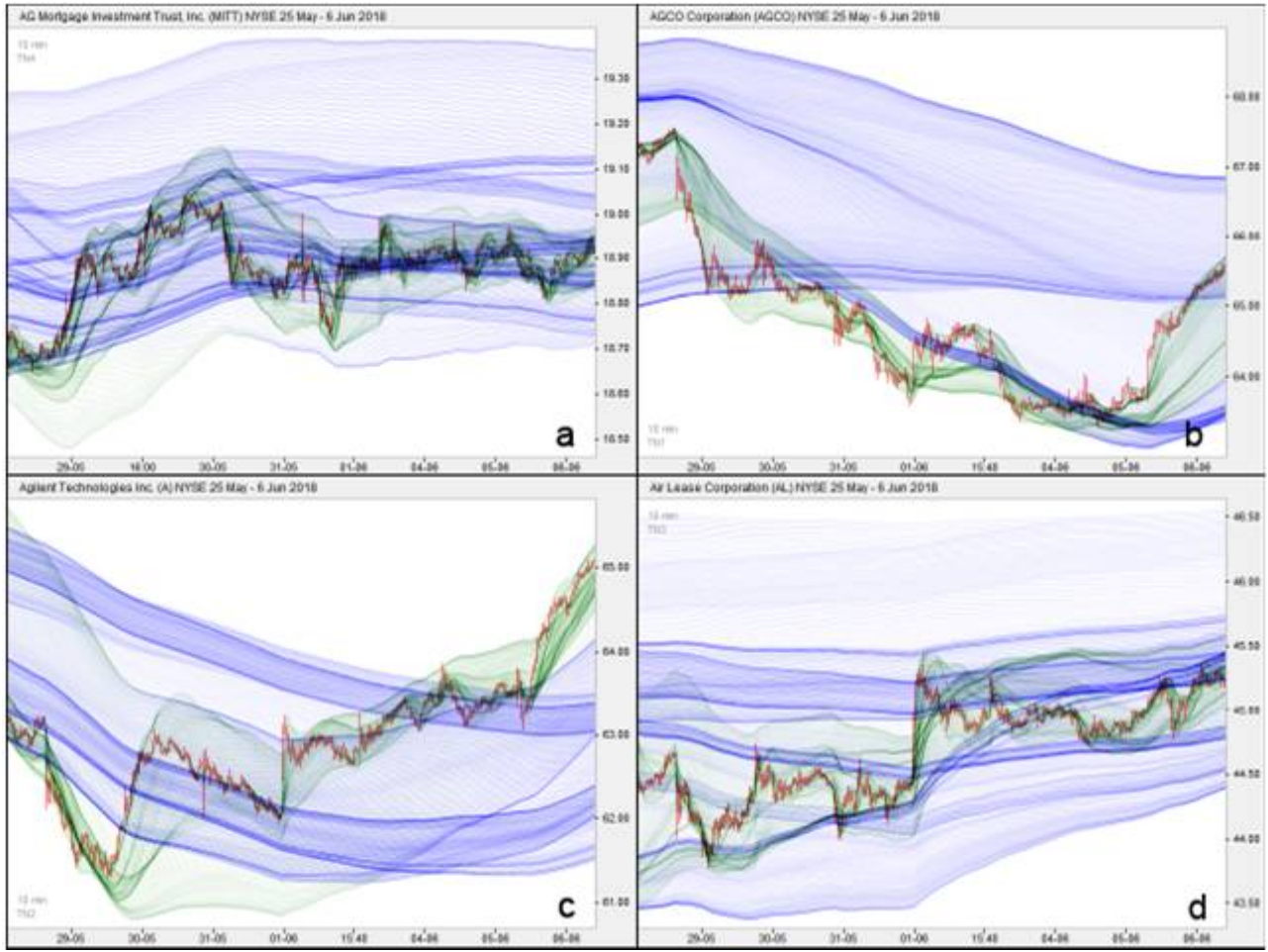

**Fig. Q6. Four "A" NYSE stocks in 10-minute at the same instant, June 6, 2018 16:00.** a) AG Mortgage Investment Trust, TN4. b) AGCO, TN1. c) Agilent Technologies, TN2. d) Air Lease, TN3.

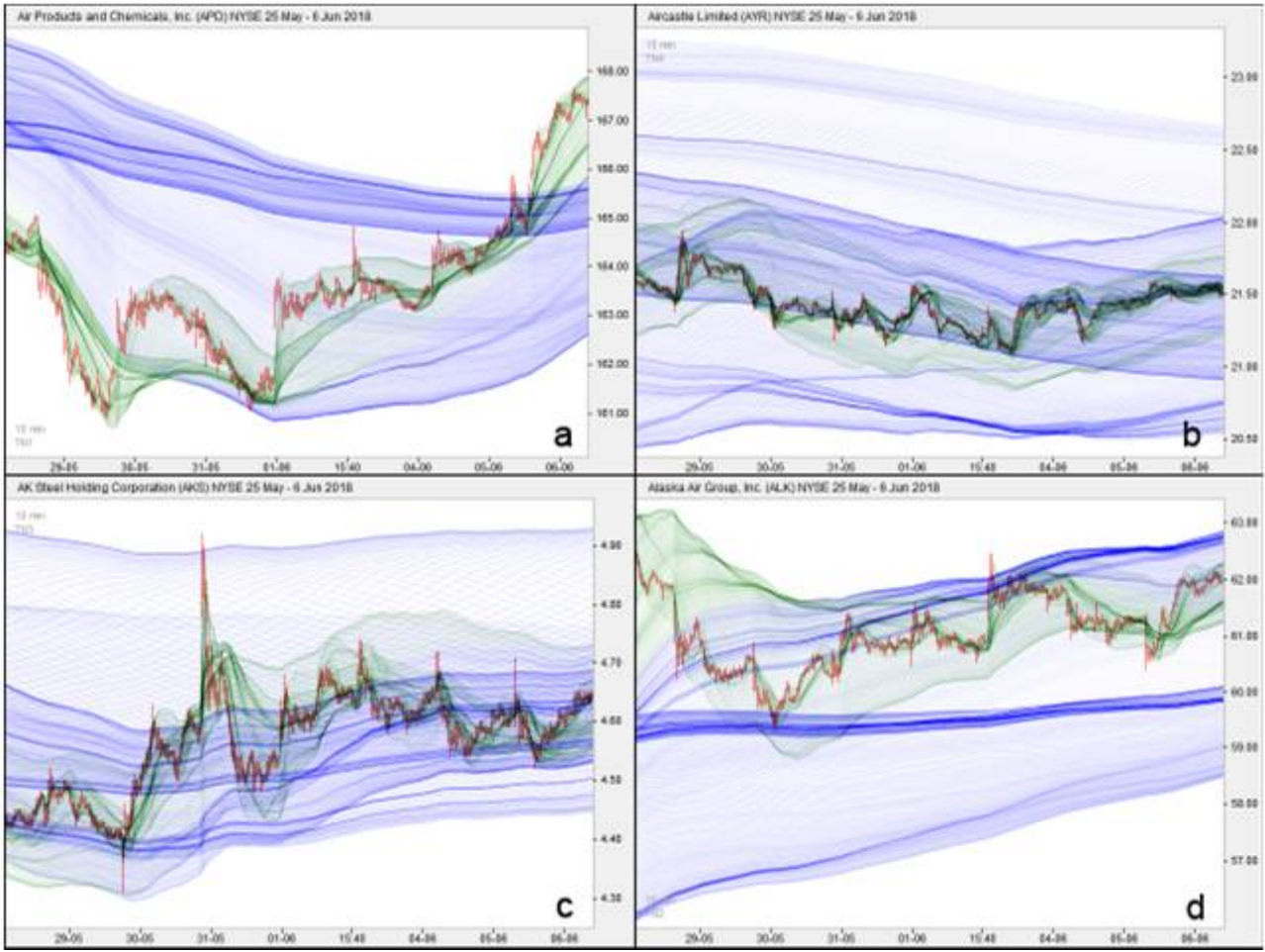

**Fig. Q7. Four "A" NYSE stocks in 10-minute at the same instant, June 6, 2018 16:00.** a) Air Products and Chemicals, TN1. b) Aircastle, TN4. c) AK Steel Holding, TN3. d) Alaska Air Group, TN2.

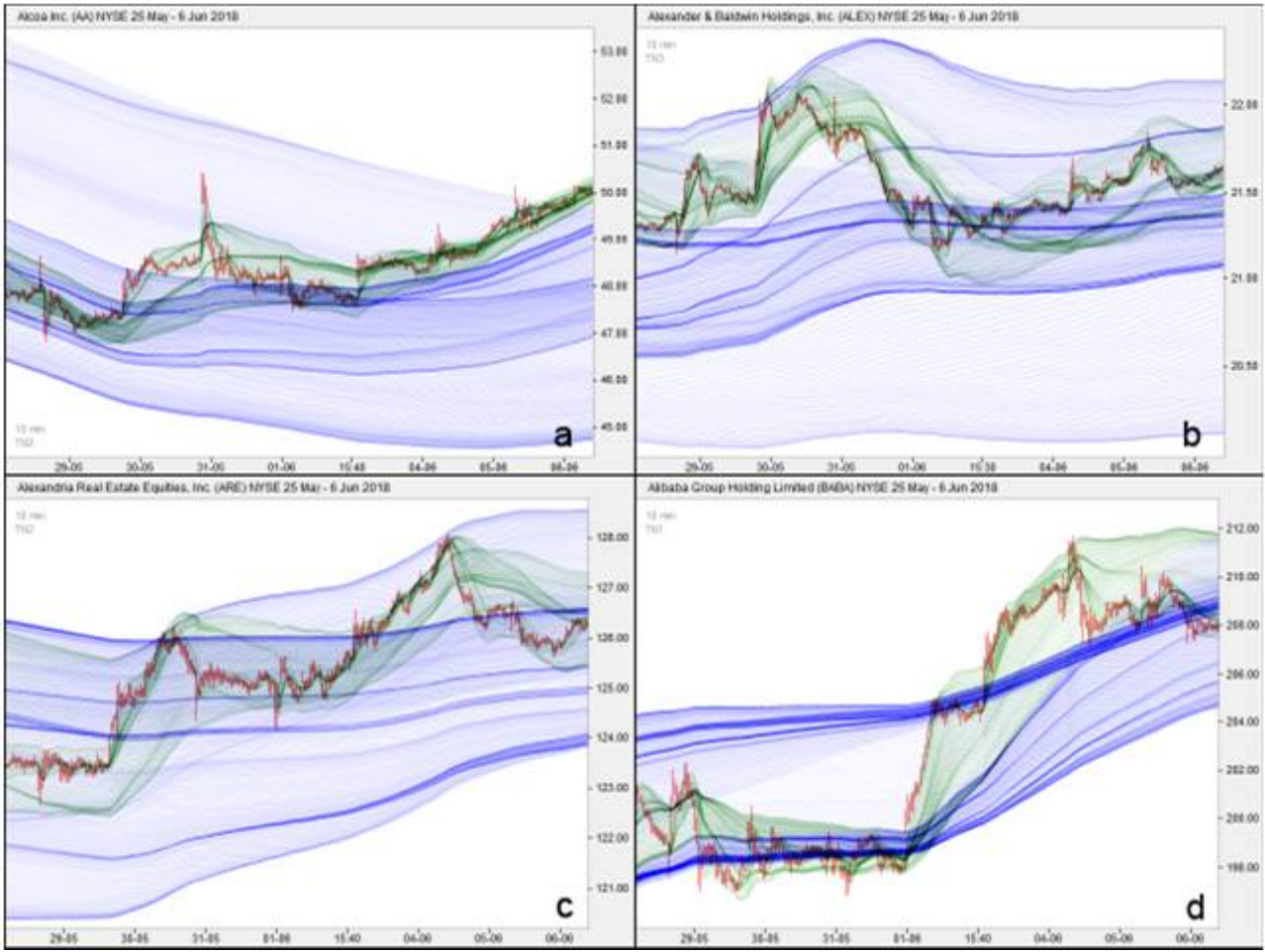

**Fig. Q8. Four "A" NYSE stocks in 10-minute at the same instant, June 6, 2018 16:00.** a) Alcoa, TN2. b) Alexander & Baldwin Holdings, TN3. c) Alexandria Real Estate Equities, TN2. d) Alibaba Group Holding, TN1.

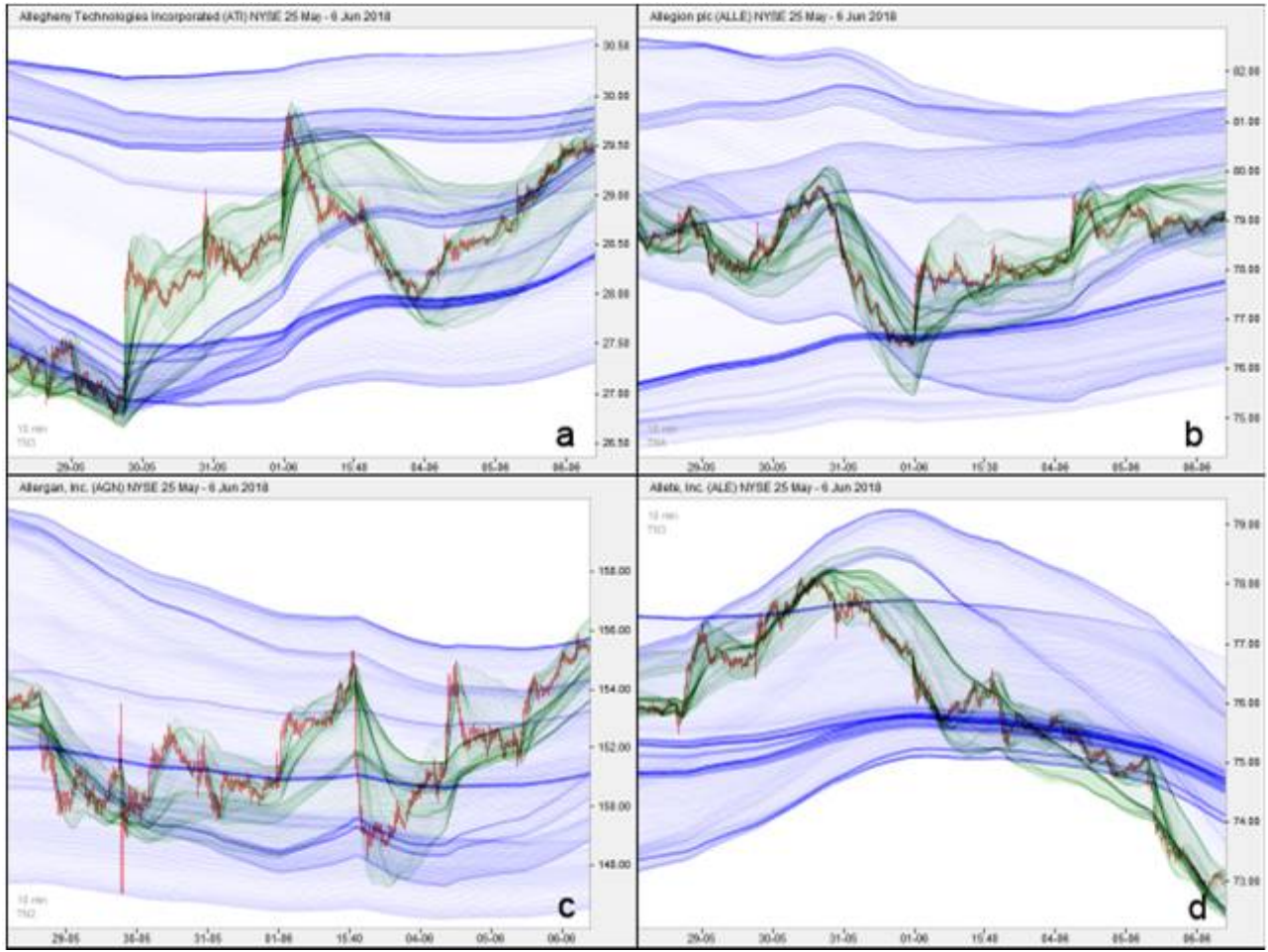

**Fig. Q9. Four "A" NYSE stocks in 10-minute at the same instant, June 6, 2018 16:00.** a) Allegheny Technologies, TN3. b) Allegion, TN4. c) Allergan, TN2. d) Allete, TN3.

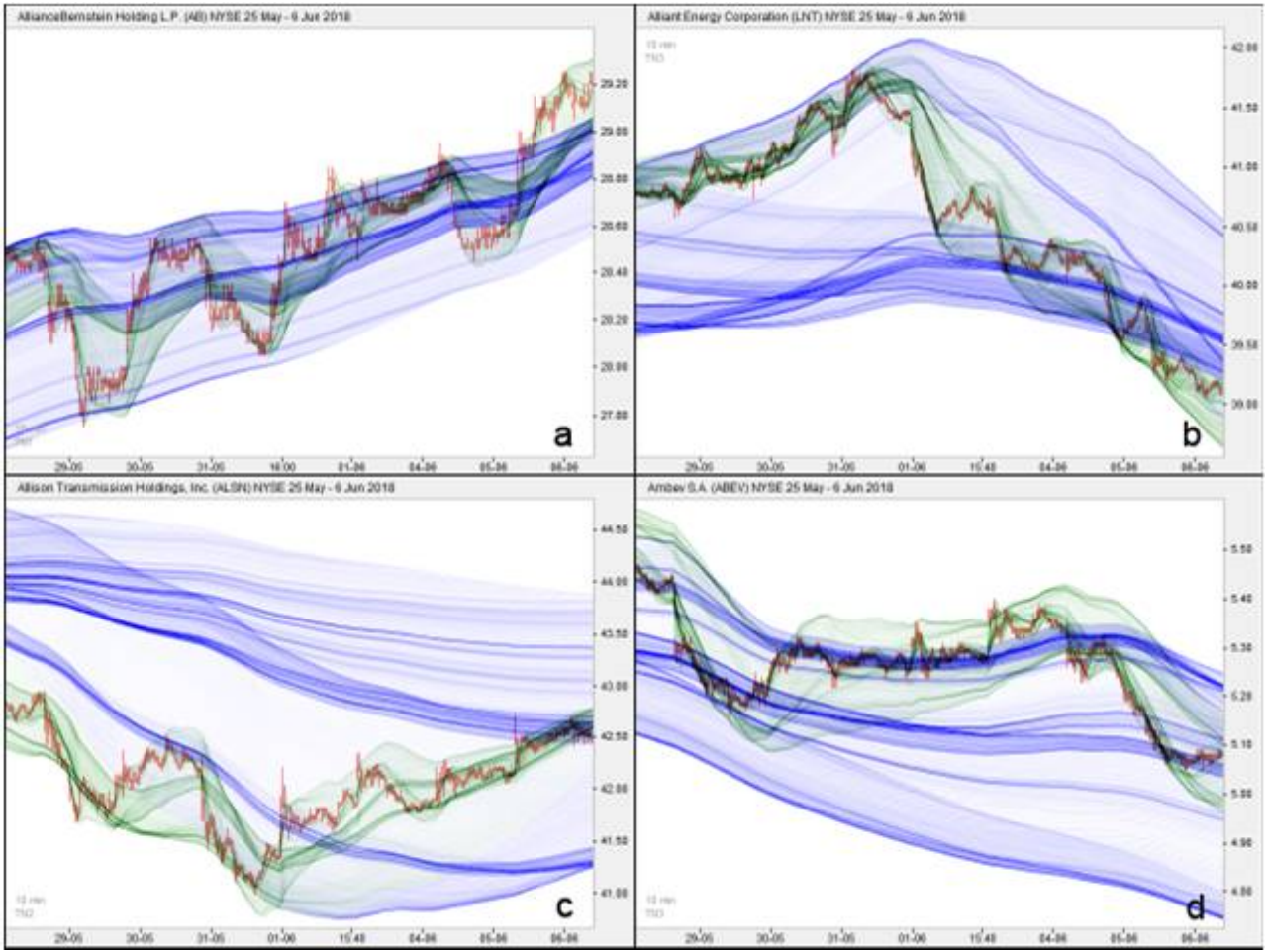

**Fig. Q10. Four "A" NYSE stocks in 10-minute at the same instant, June 6, 2018 16:00.** a) AllianceBernstein Holding, TN1. b) Alliant Energy, TN3. c) Allison Transmission Holdings, TN2. d) Ambev, TN3.

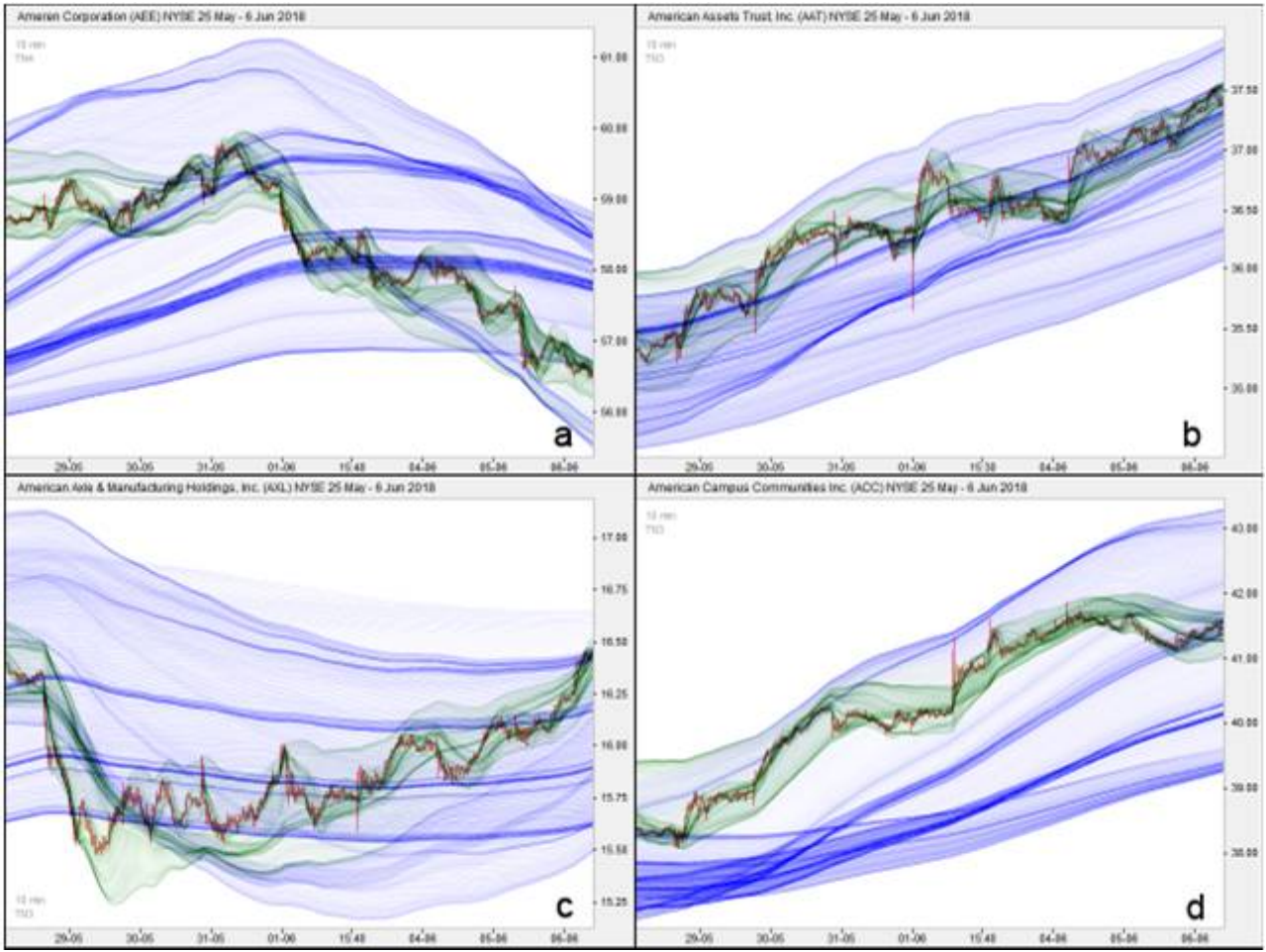

**Fig. Q11. Four "A" NYSE stocks in 10-minute at the same instant, June 6, 2018 16:00.** a) Ameren Corporation, TN4. b) American Assets Trust, TN3. c) American Axle & Manufacturing Holdings, TN3. d) American Campus Communities, TN3.

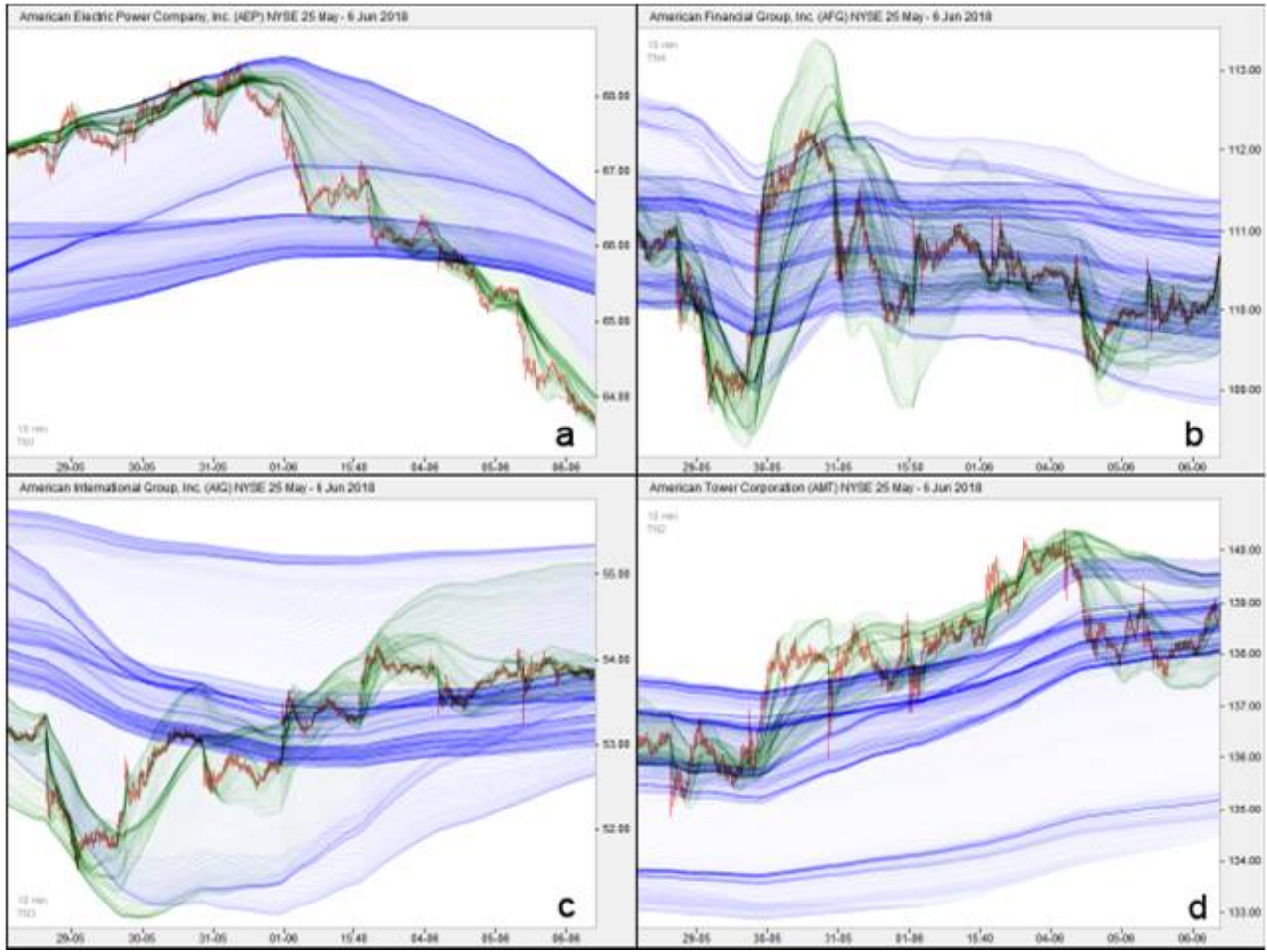

**Fig. Q12. Four "A" NYSE stocks in 10-minute at the same instant, June 6, 2018 16:00.** a) American Electric Power Company, TN1. b) American Financial Group, TN4. c) American International Group, TN3. d) American Tower Corporation, TN2.

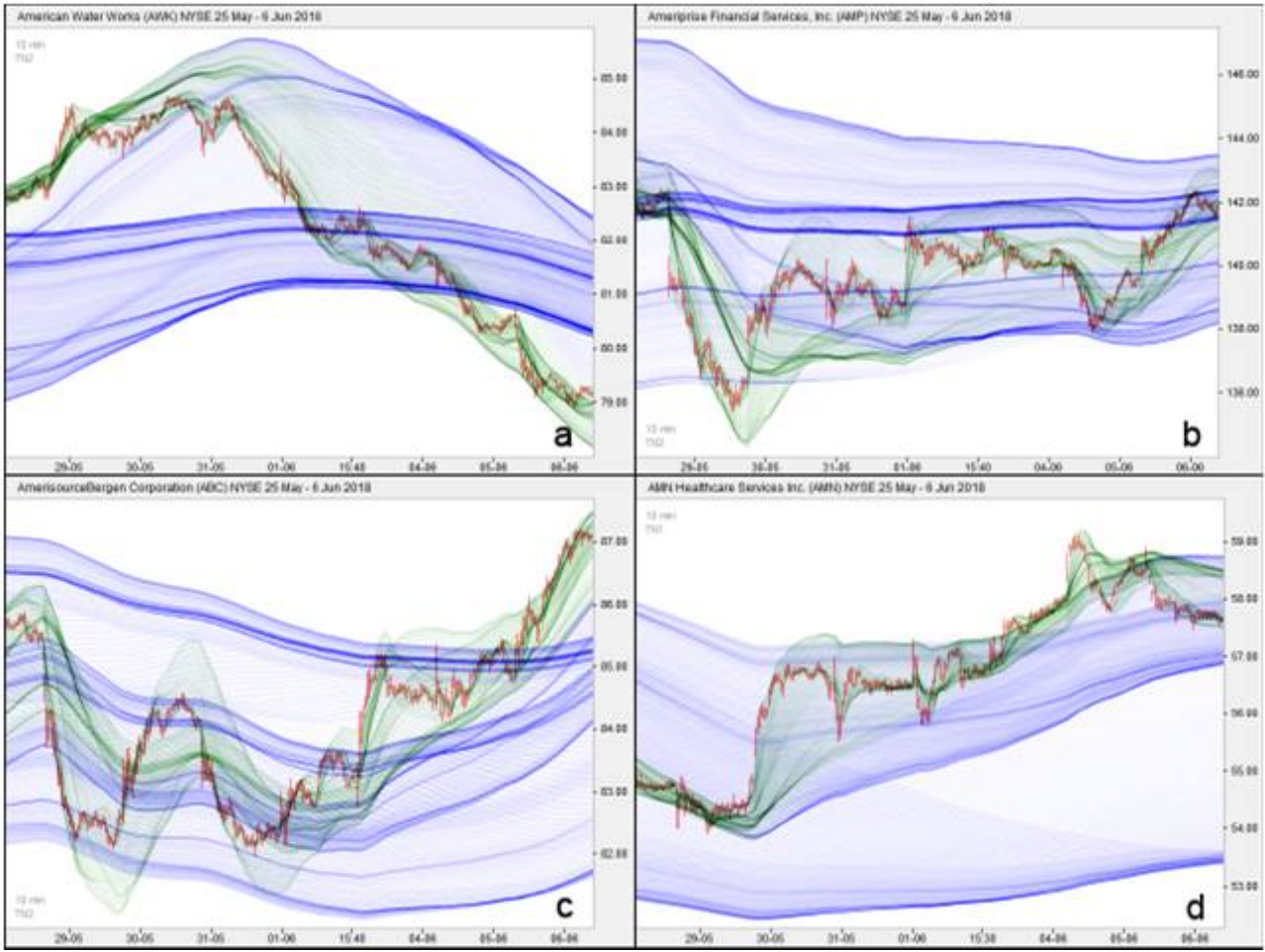

**Fig. Q13. Four "A" NYSE stocks in 10-minute at the same instant, June 6, 2018 16:00.** a) American Water Works, TN2. b) Ameriprise Financial Services, TN2. c) AmerisourceBergen, TN2. d) AMN Healthcare Services, TN1.

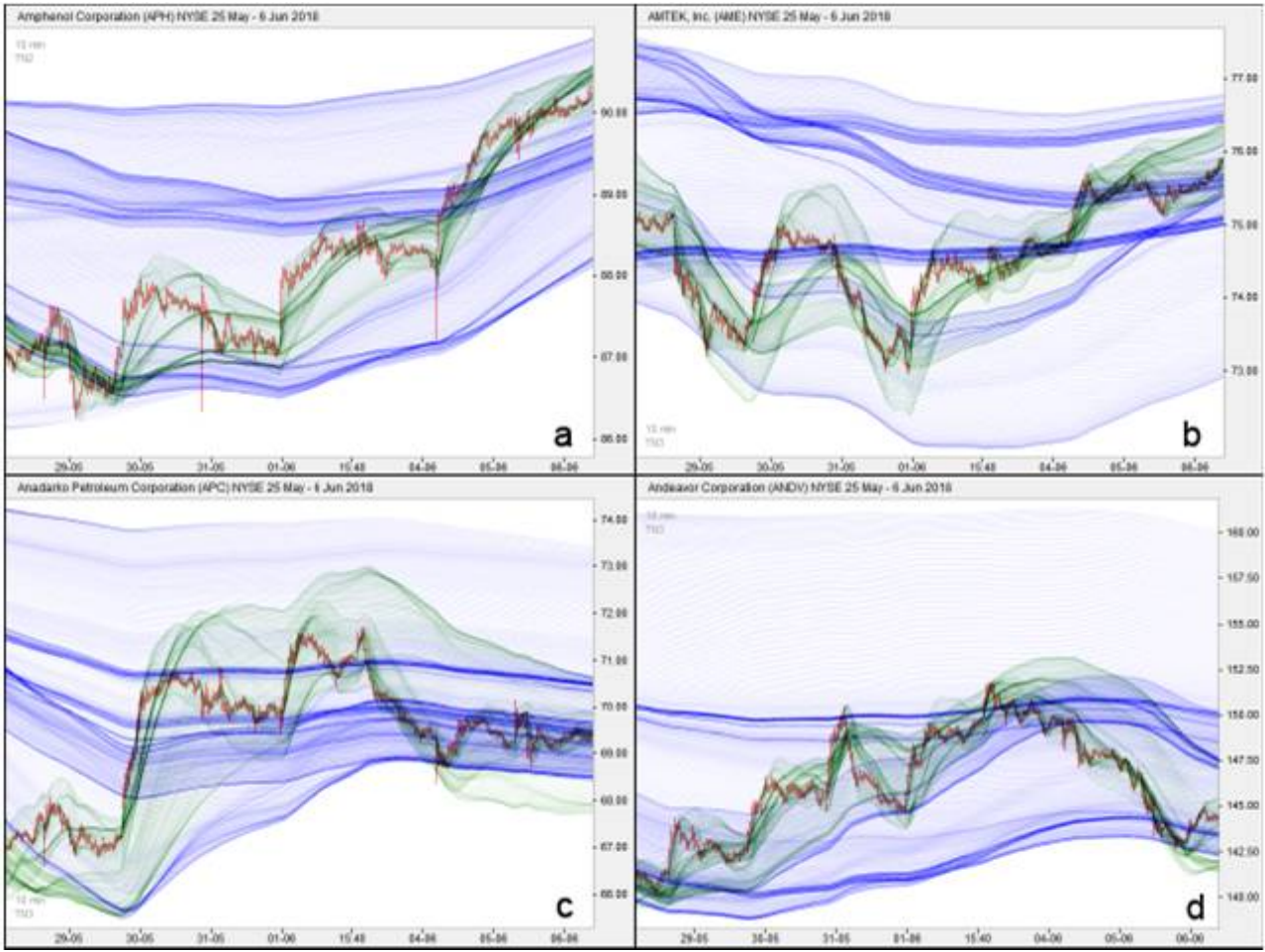

**Fig. Q14. Four "A" NYSE stocks in 10-minute at the same instant, June 6, 2018 16:00.** a) Amphenol, TN2. b) AMTEK, TN3. c) Anadarko Petroleum, TN3. d) Andeavor Corporation, TN3.

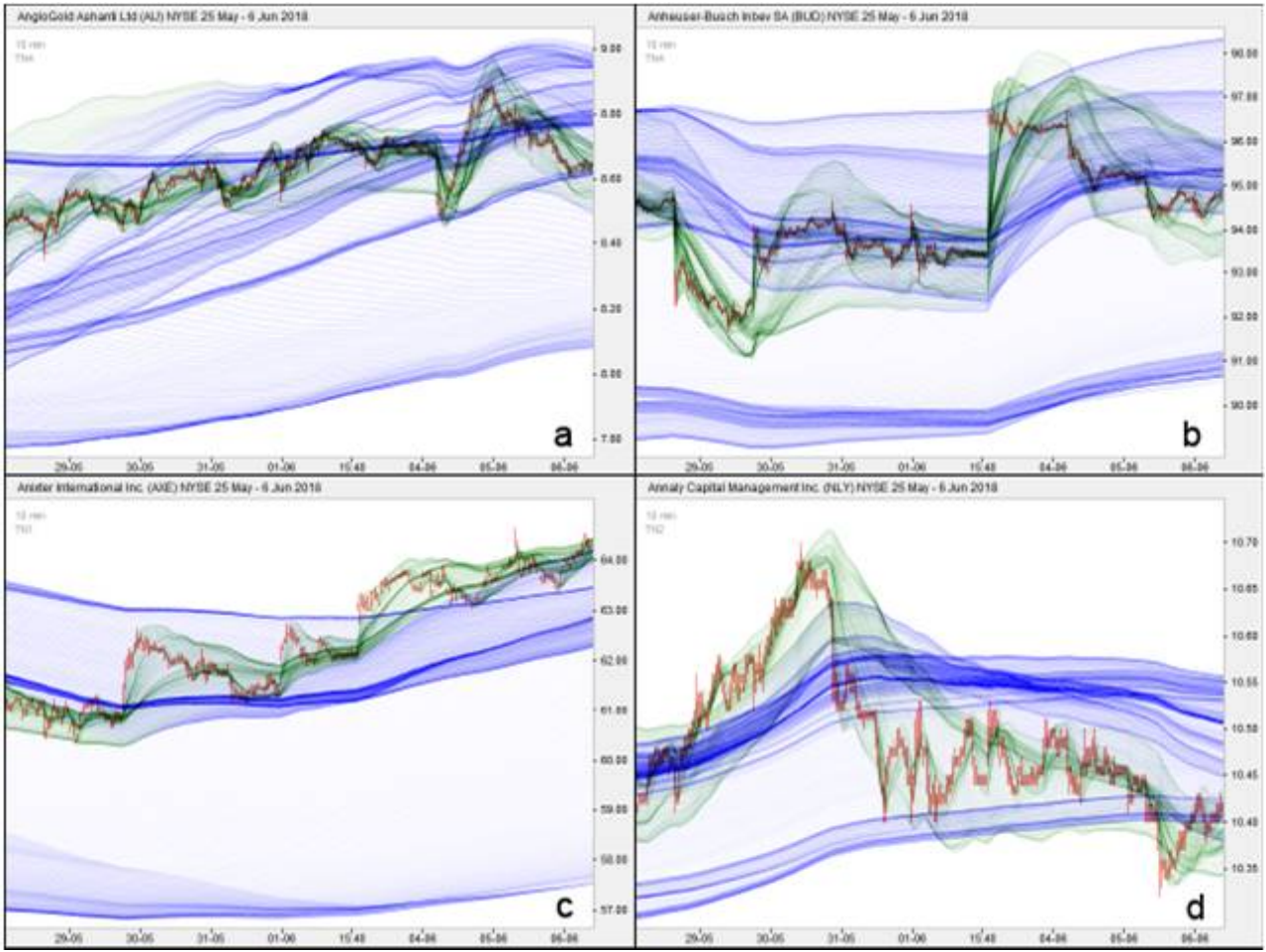

**Fig. Q15. Four "A" NYSE stocks in 10-minute at the same instant, June 6, 2018 16:00.** a) AngloGold Ashanti, TN4. b) Anheuser-Busch Inbev, TN1. c) Anixter International, TN1. d) Annaly Capital Management, TN2.

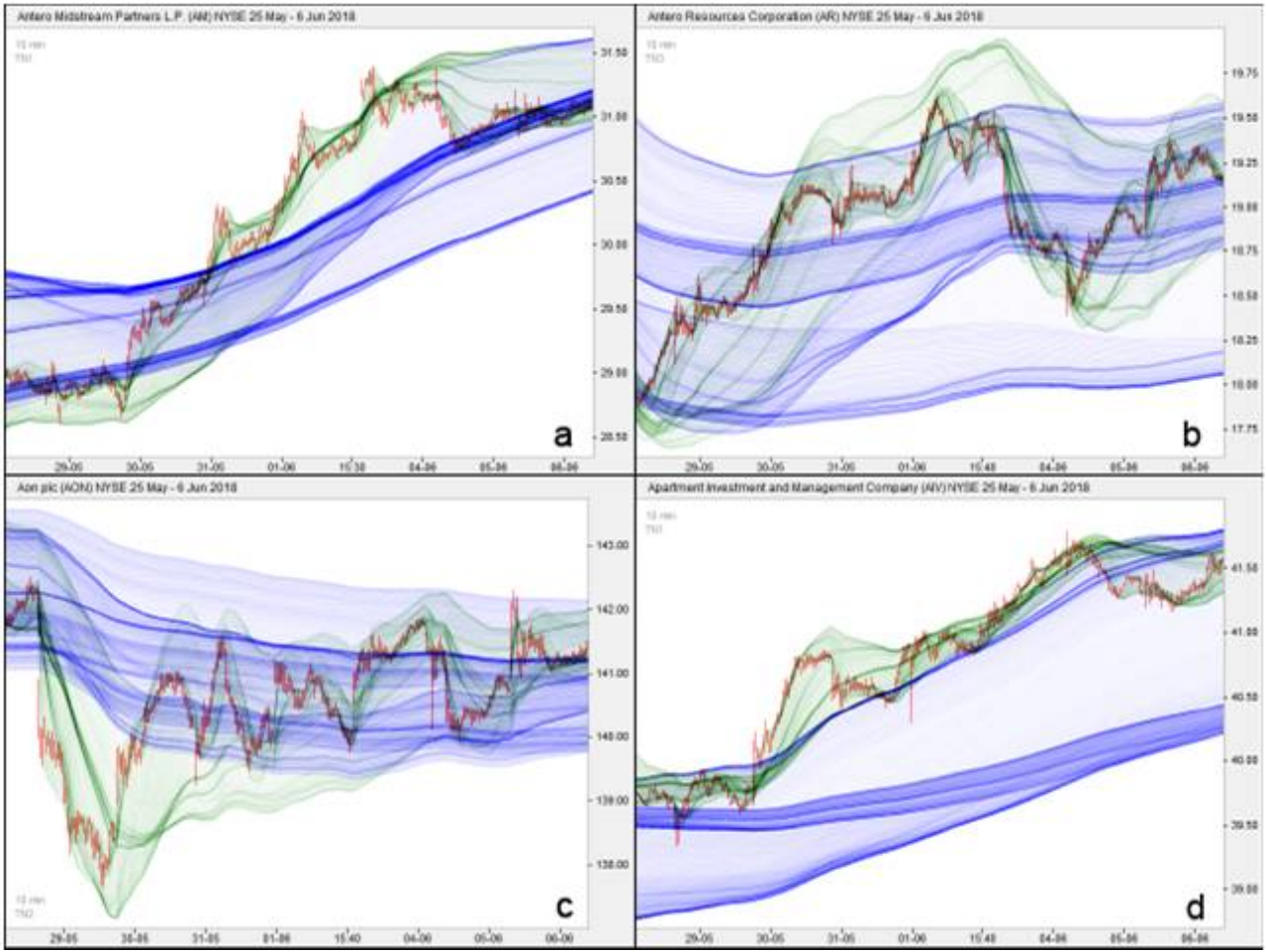

**Fig. Q16. Four "A" NYSE stocks in 10-minute at the same instant, June 6, 2018 16:00.** a) Antero Midstream Partners, TN1. b) Antero Resources, TN3. c) Aon, TN2. d) Apartment Investment and Management, TN1.

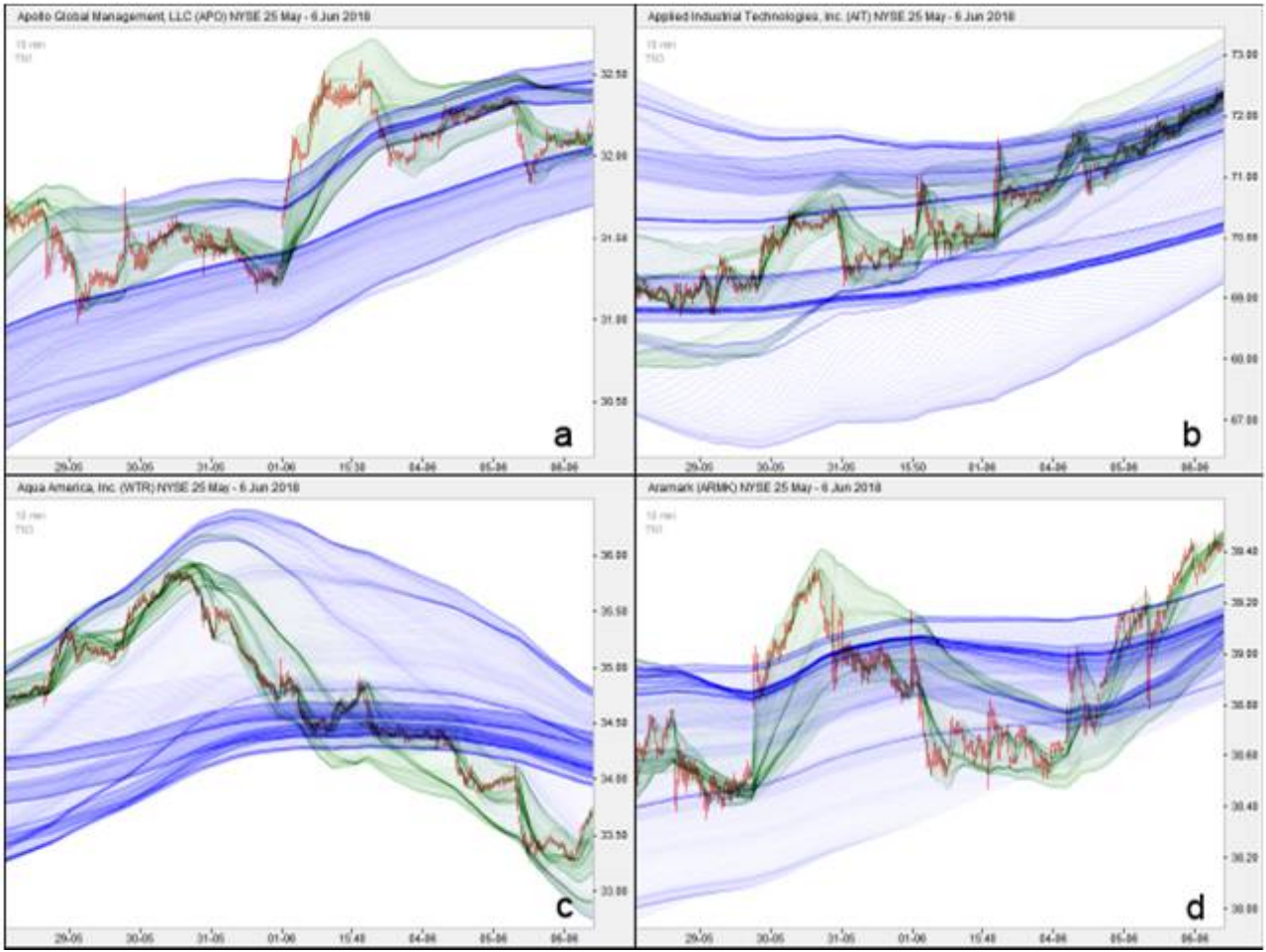

**Fig. Q17. Four "A" NYSE stocks in 10-minute at the same instant, June 6, 2018 16:00.** a) Apollo Global Management, TN1. b) Applied Industrial Technologies, TN3. c) Aqua America, TN3. d) Aramark, TN1.

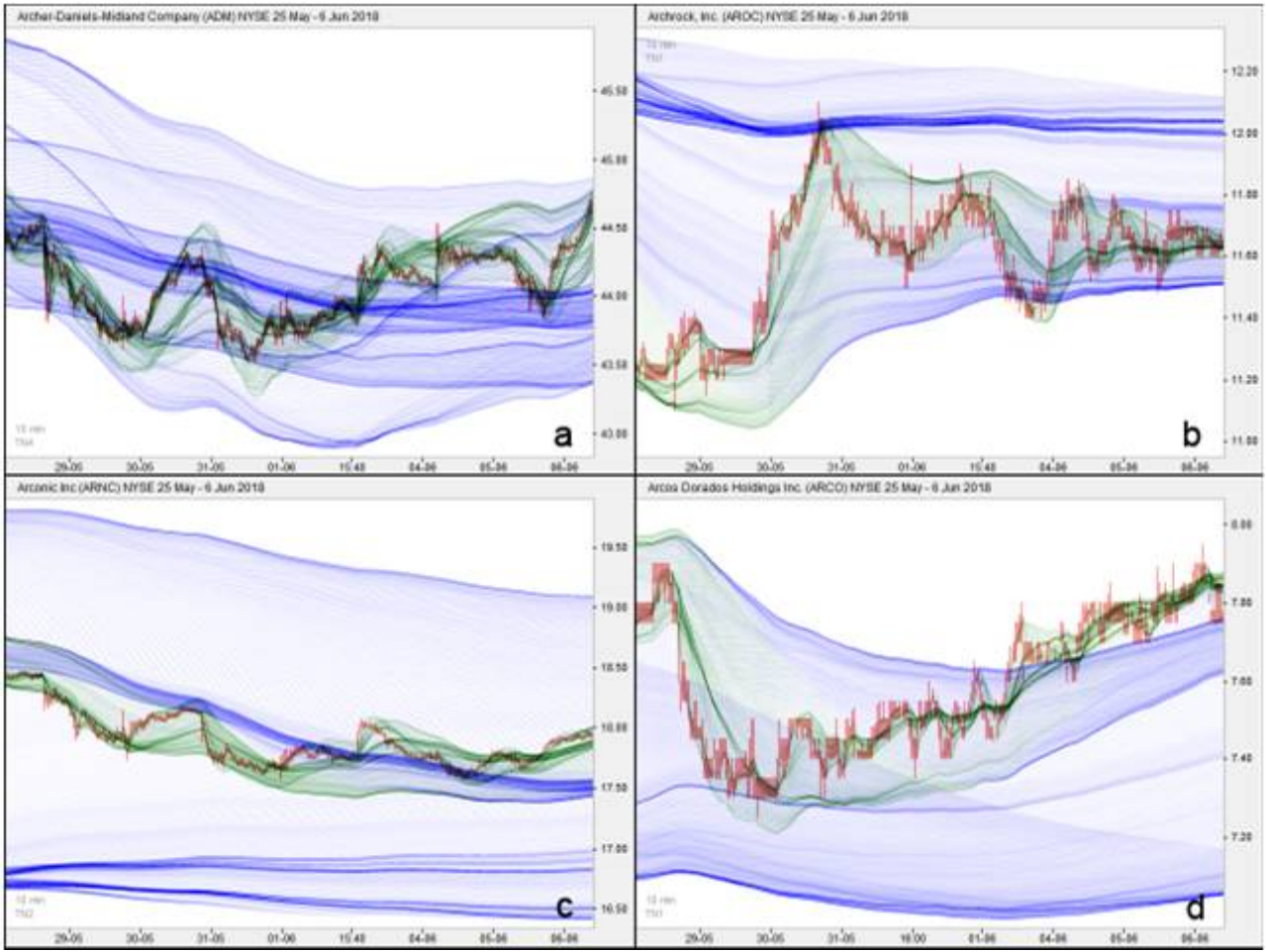

**Fig. Q18. Four "A" NYSE stocks in 10-minute at the same instant, June 6, 2018 16:00.** a) Archer-Daniels-Midland, TN4. b) Archrock, TN1. c) Arconic, TN2. d) Arcos Dorados Holdings, TN1.

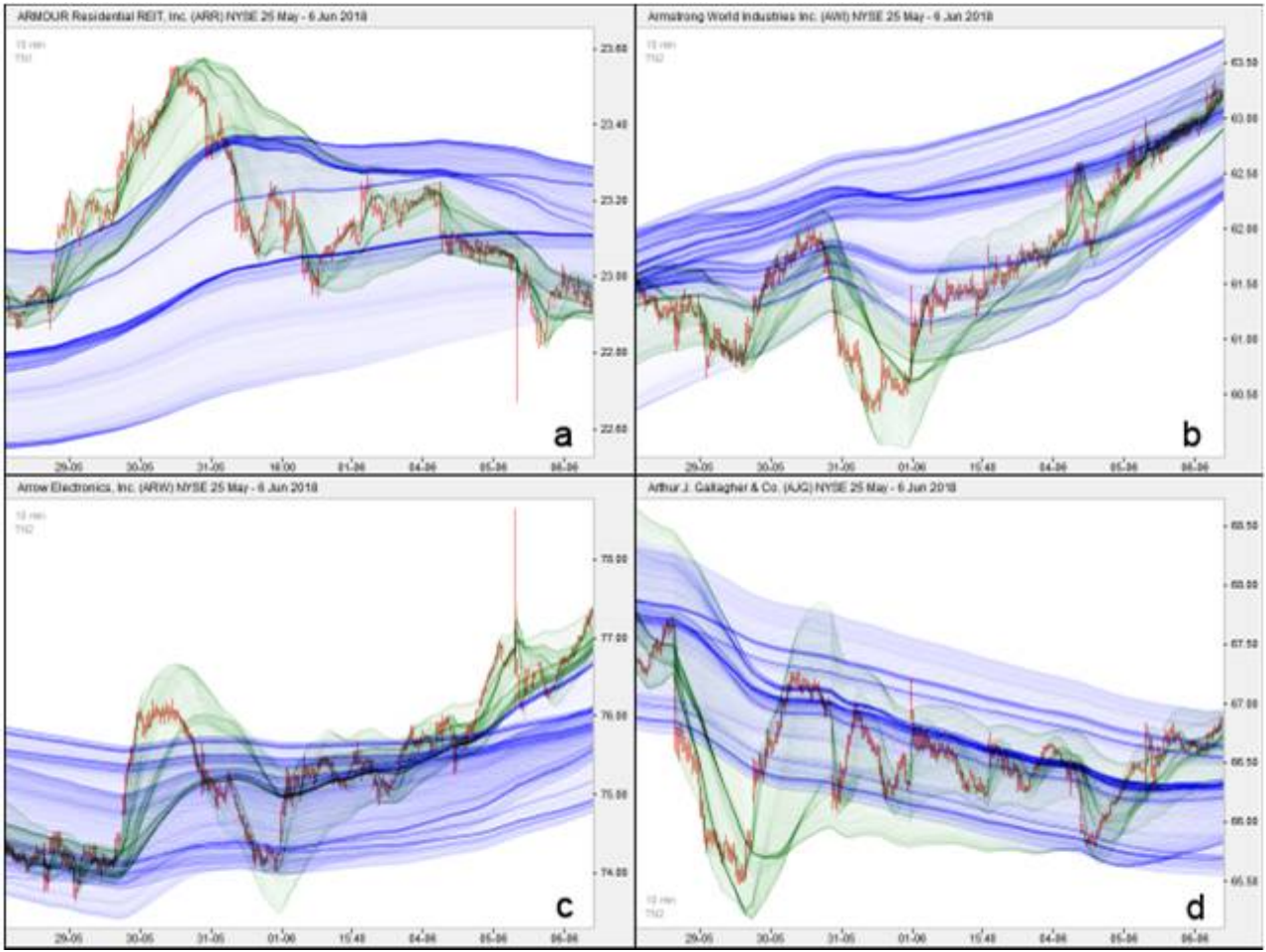

**Fig. Q19. Four "A" NYSE stocks in 10-minute at the same instant, June 6, 2018 16:00.** a) ARMOUR Residential REIT, TN1. b) Armstrong World Industries, TN2. c) Arrow Electronics, TN2. d) Arthur J. Gallagher & Co., TN2.

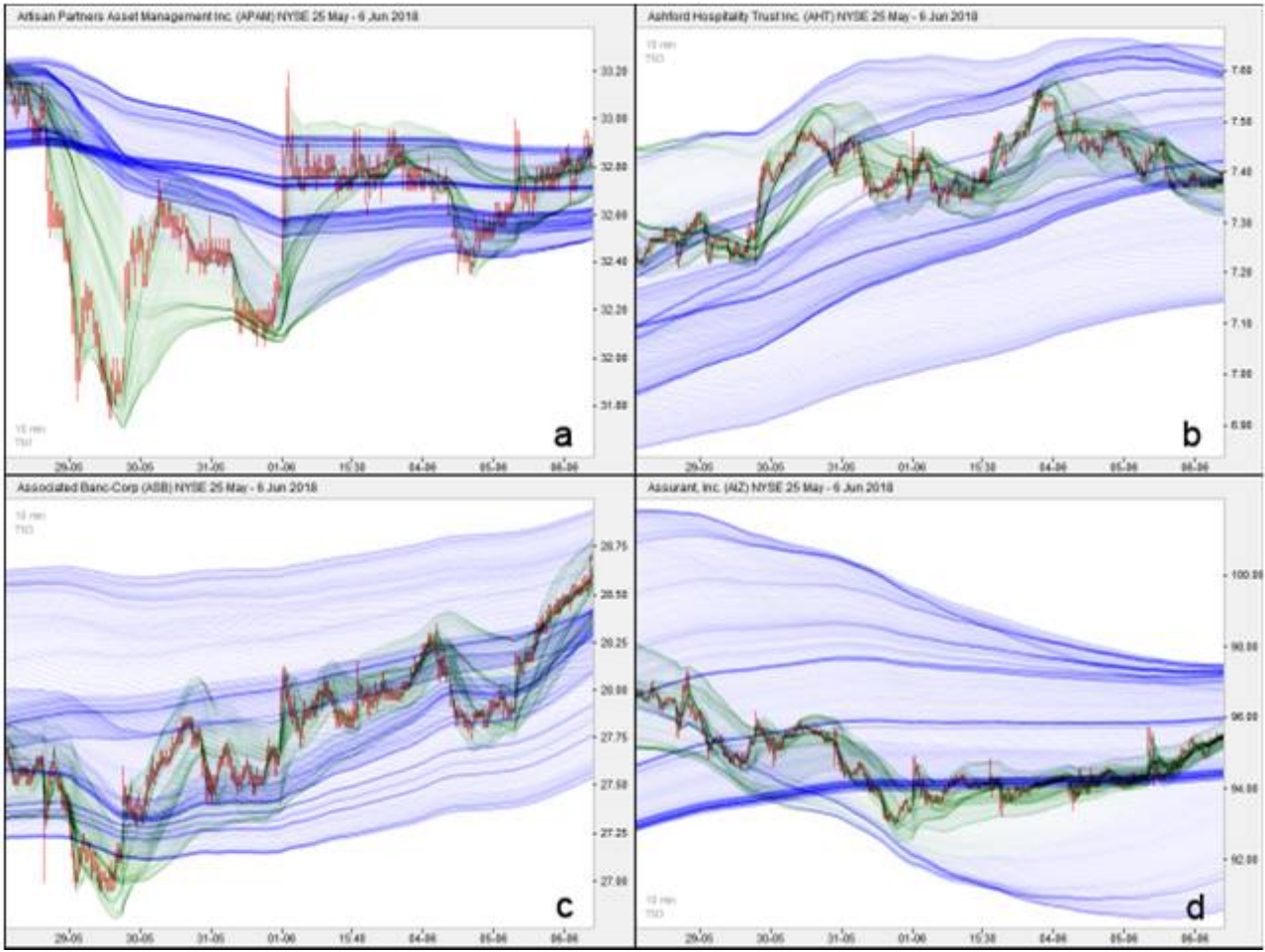

**Fig. Q20. Four "A" NYSE stocks in 10-minute at the same instant, June 6, 2018 16:00.** a) Artisan Partners Asset Management, TN1. b) Ashford Hospitality Trust, TN3. c) Associated Banc-Corp, TN3. d) Assurant, TN3.

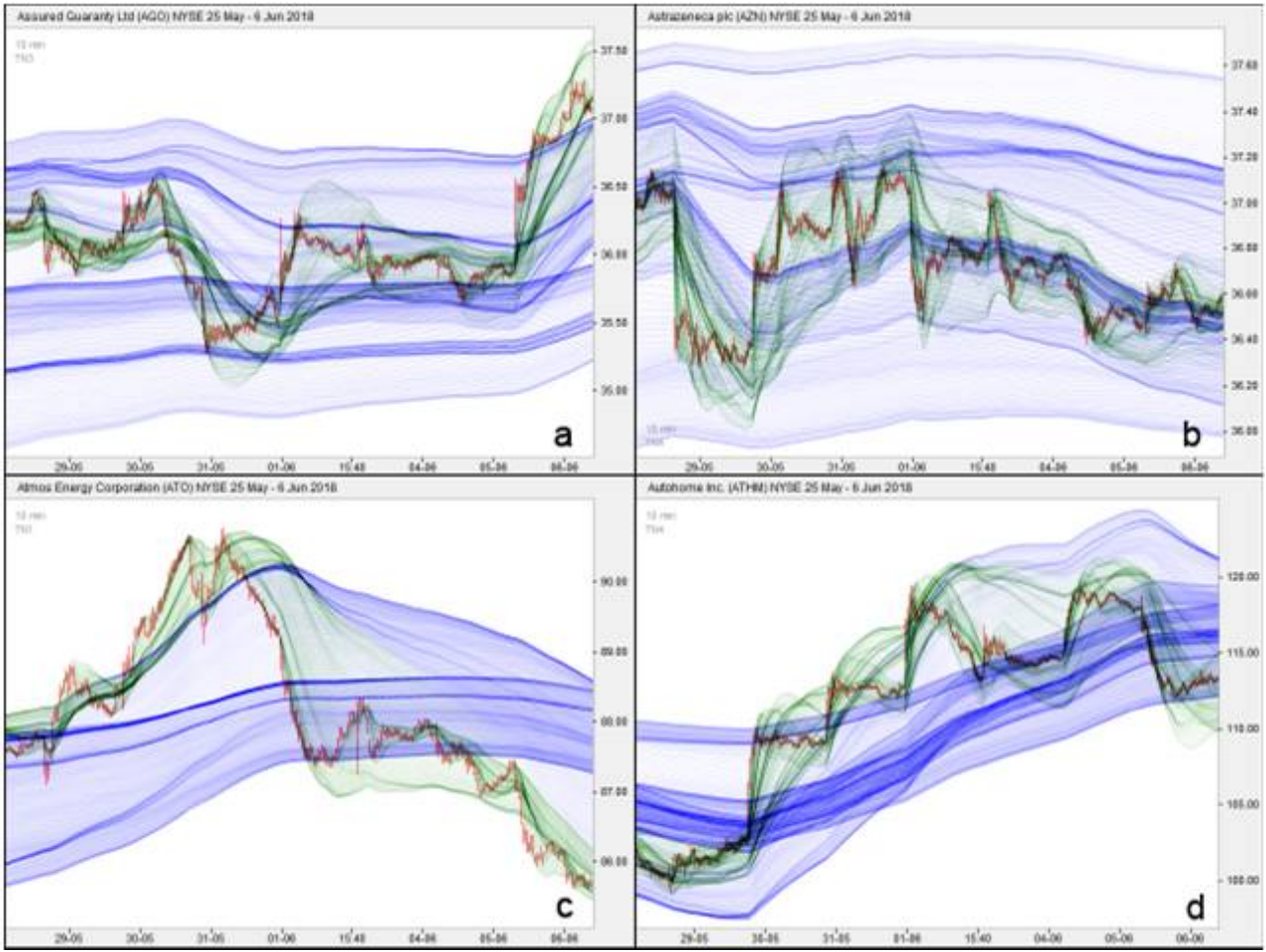

**Fig. Q21. Four "A" NYSE stocks in 10-minute at the same instant, June 6, 2018 16:00.** a) Assured Guaranty, TN3. b) Astrazeneca, TN4. c) Atmos Energy, TN1. d) Autohome, TN4.

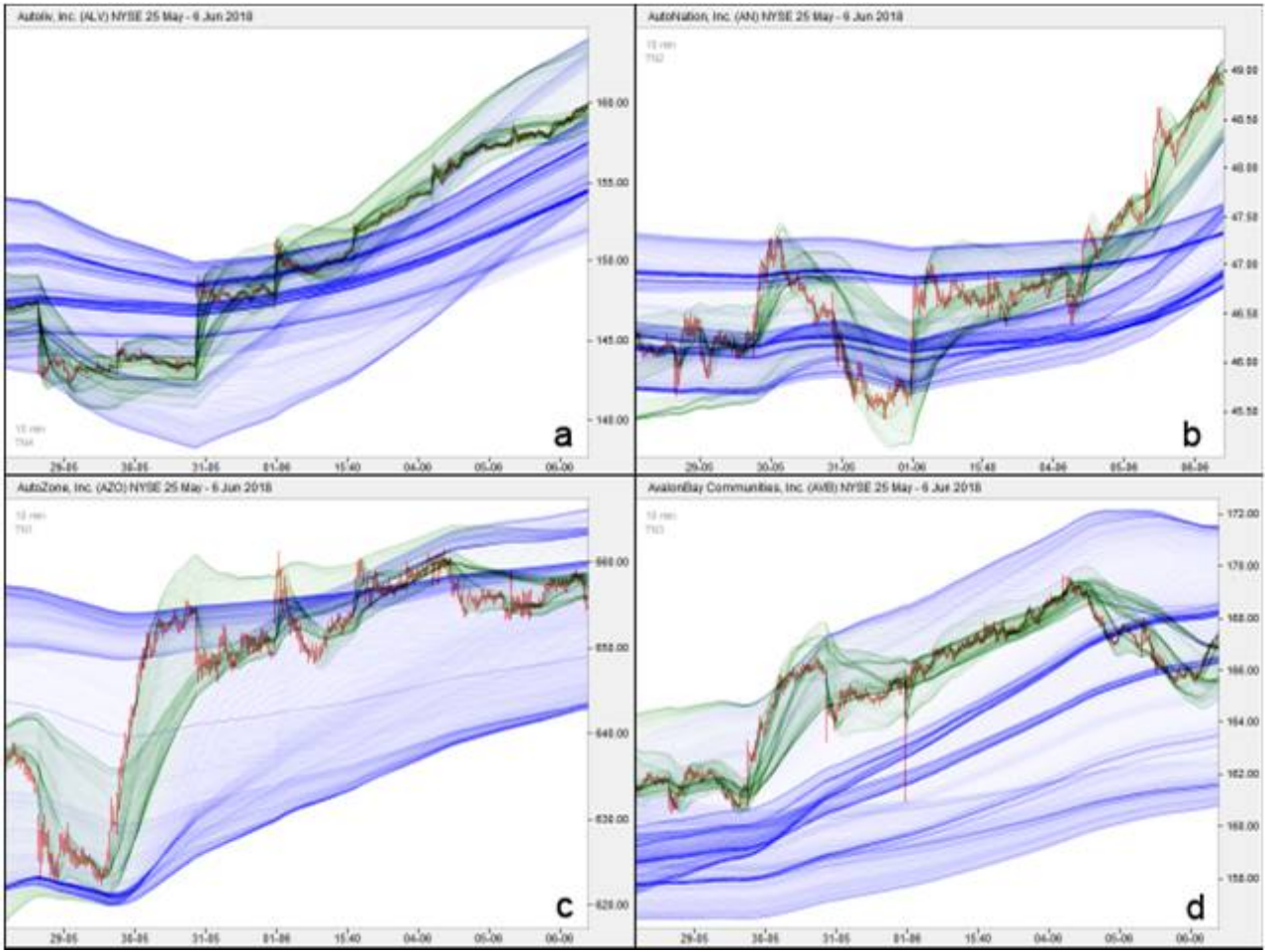

**Fig. Q22. Four "A" NYSE stocks in 10-minute at the same instant, June 6, 2018 16:00.** a) Autoliv, TN4. b) AutoNation, TN2. c) AutoZone, TN1. d) AvalonBay Communities, TN3.

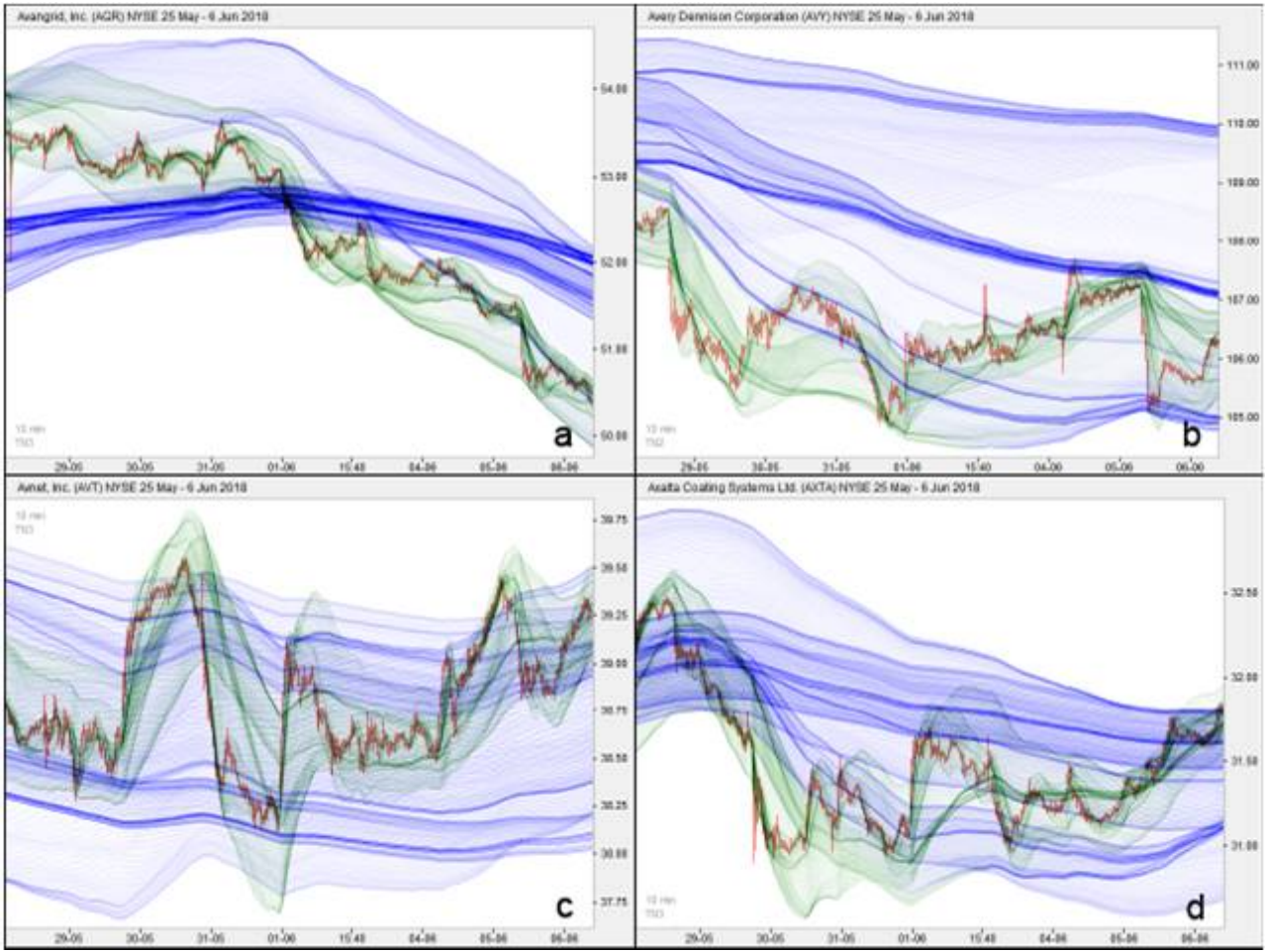

**Fig. Q23. Four "A" NYSE stocks in 10-minute at the same instant, June 6, 2018 16:00.** a) Avangrid, TN4. b) Avery Dennison, TN2. c) Avnet, TN3. d) Axalta Coating Systems, TN3.

## Appendix R: Proof by Totality

In App. Q ("Proof by Simultaneity"), we proposed a first protocol to prove that the characteristic figures drive the price. Using the same logical setup, we will here propose another protocol. We will pick a qualifying chart (probability $\varepsilon$) and show charts of the network under all five sub-types, in four resolutions covering the breadth of granularities of the topological network, and at the same date. Because of $\varepsilon$, none of the 19 other charts are likely to qualify. Yet, all 19 of these charts qualify. Since this comes with a probability equal to $\varepsilon^{20}$, this proves that the price bouncing from one characteristic figure to another cannot be fortuitous, and, therefore, that the characteristic figures drive the price. To make the demonstration even more convincing, we will pick one of the same instruments, at the same date, from App. Q.

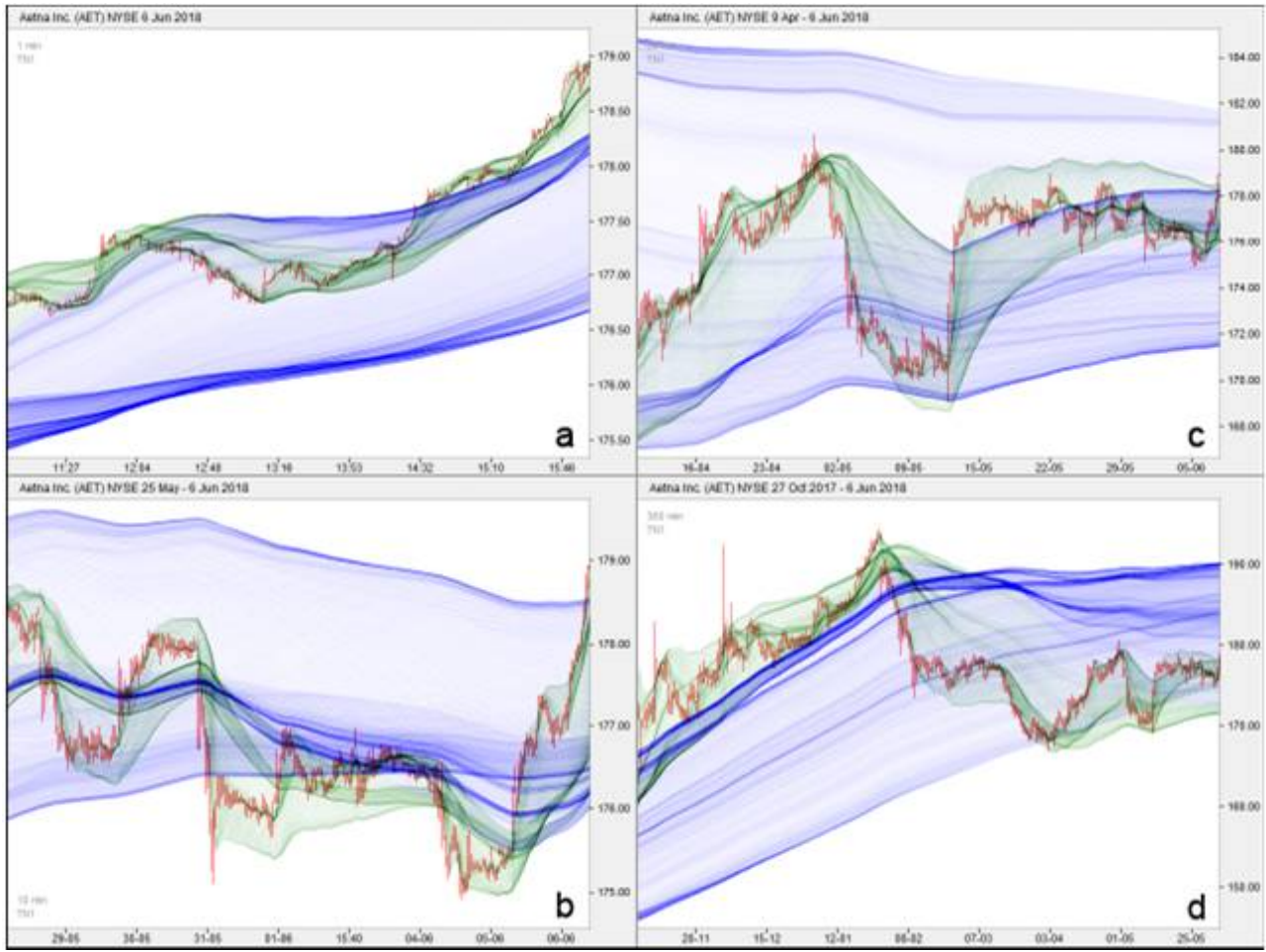

**Fig. R1. AET stock, June 1, 2018, TN1.** a) 1-minute. b) 10-minute. c) 60-minute. d) 360-minute.

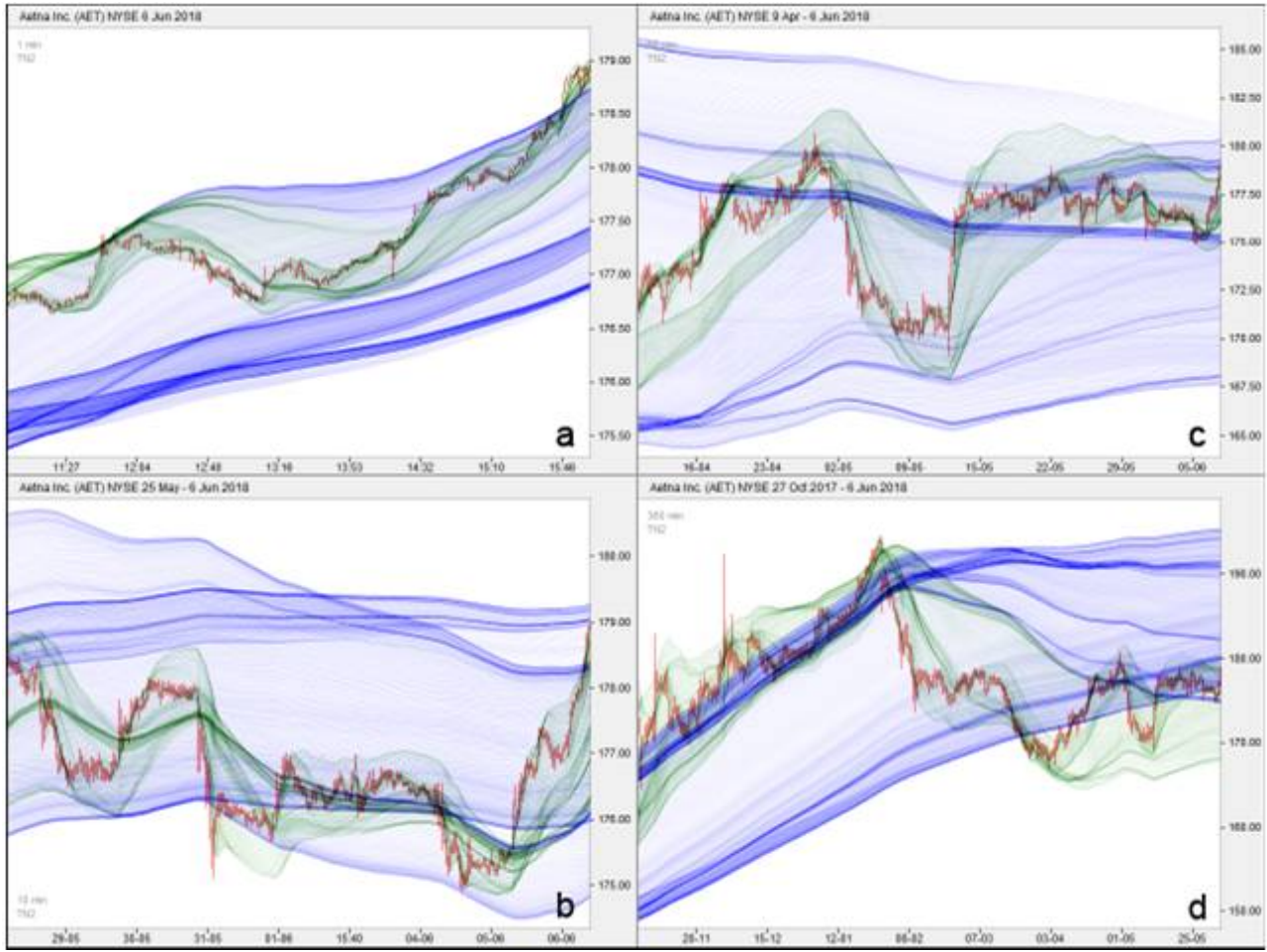

**Fig. R2. AET stock, June 1, 2018, TN2.** a) 1-minute. b) 10-minute. c) 60-minute. d) 360-minute.

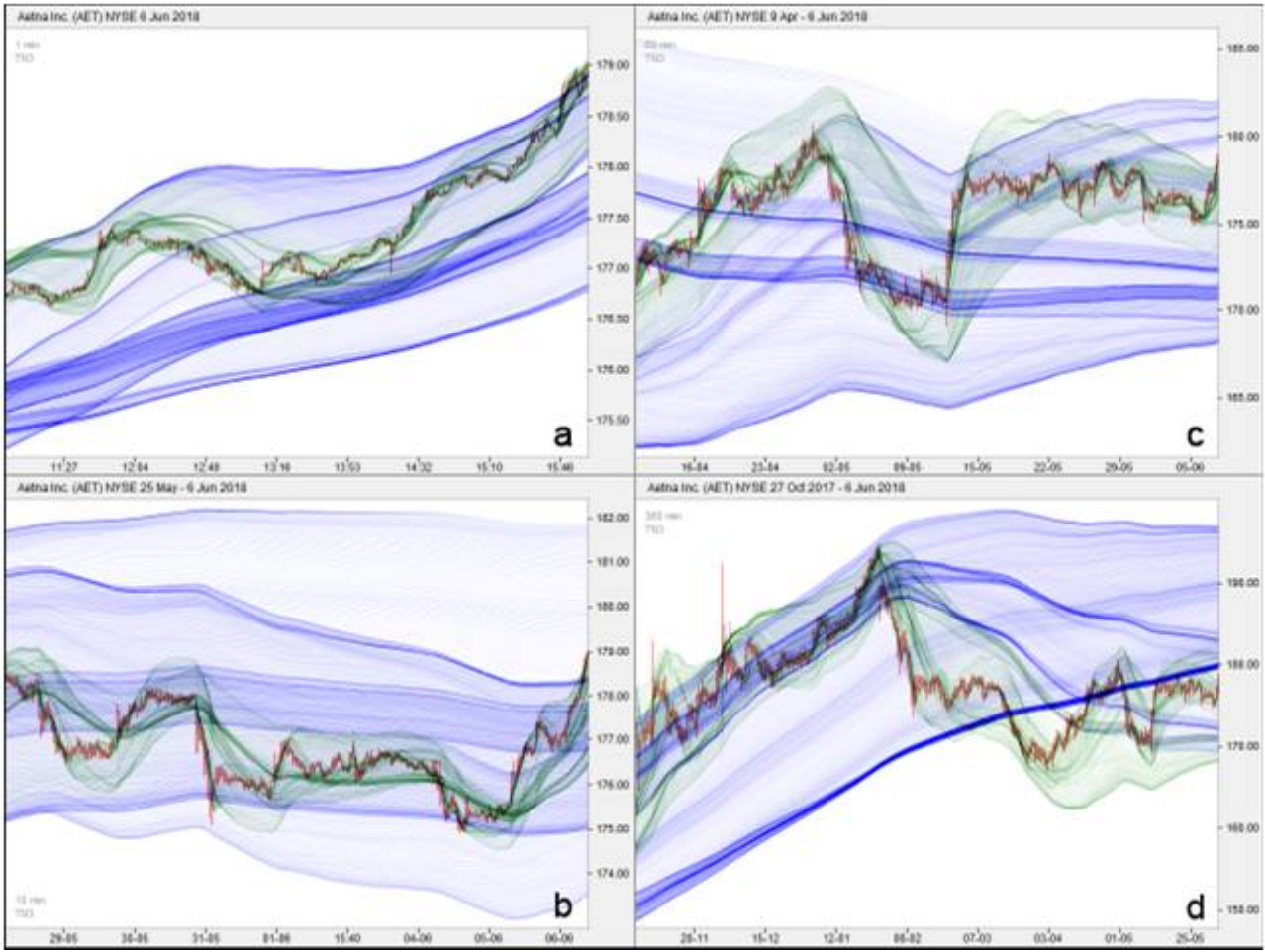

**Fig. R3. AET stock, June 1, 2018, TN3.** a) 1-minute. b) 10-minute. c) 60-minute. d) 360-minute.

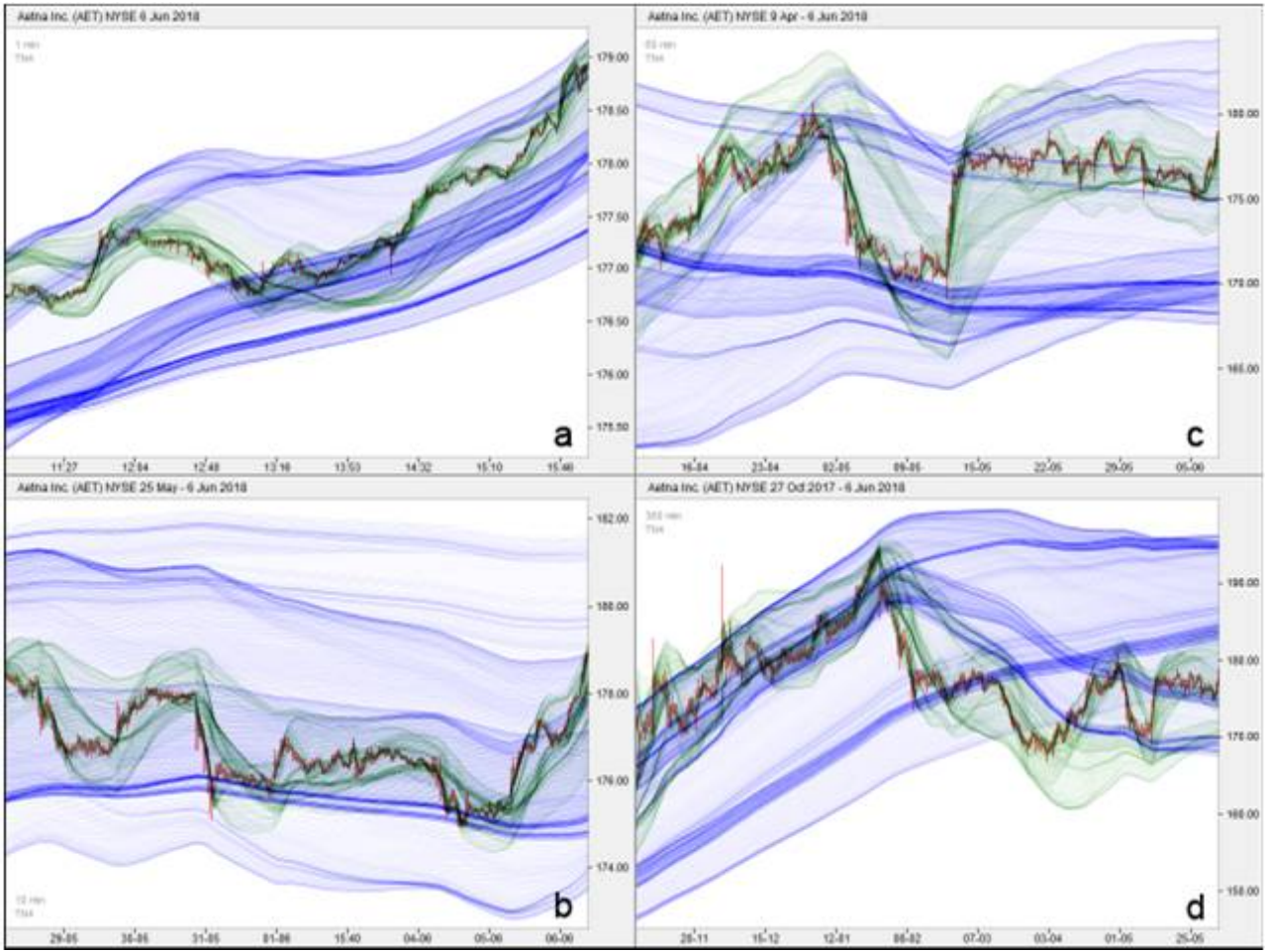

**Fig. R4. AET stock, June 1, 2018, TN4.** a) 1-minute. b) 10-minute. c) 60-minute. d) 360-minute.

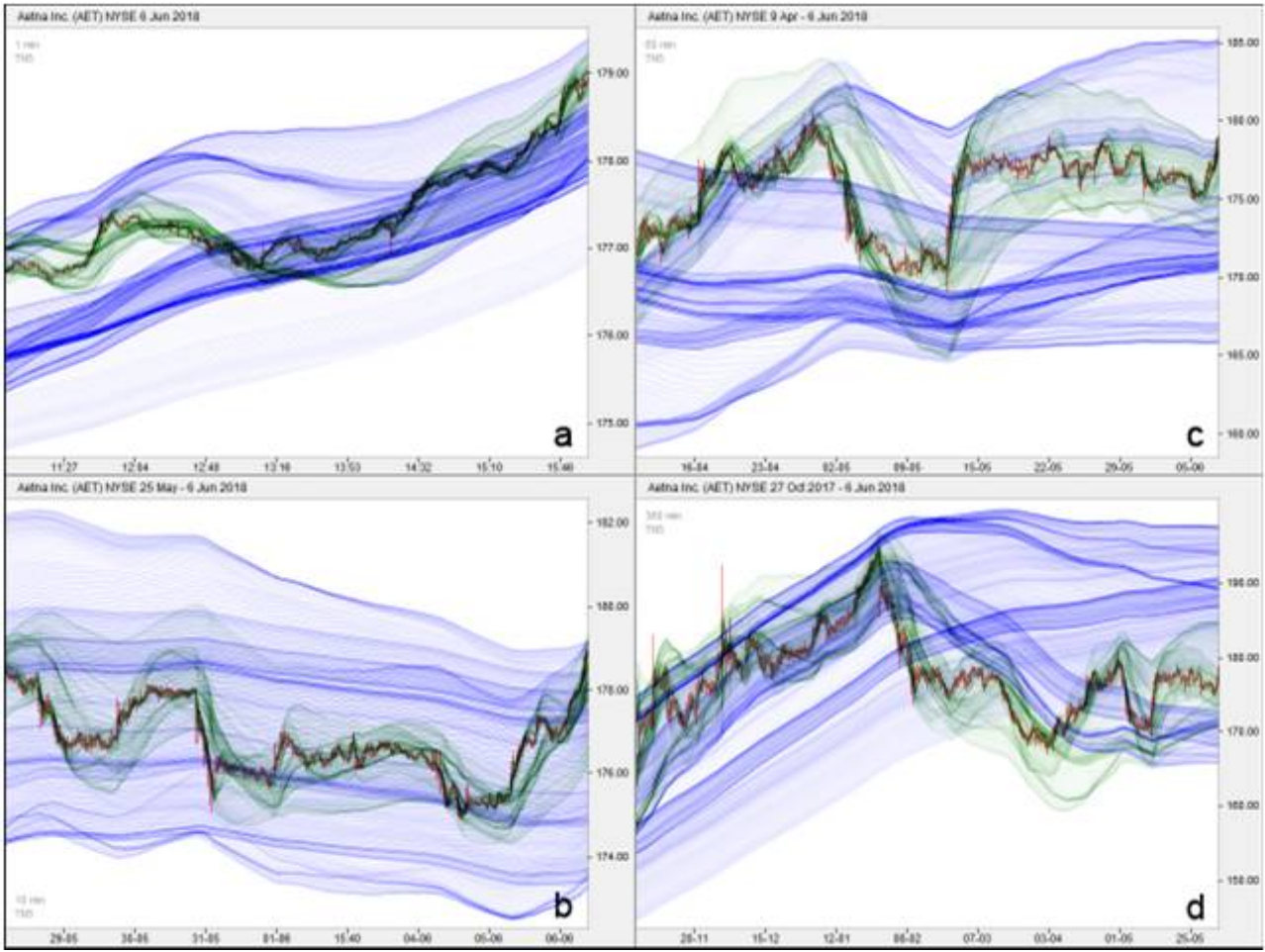

**Fig. R5. AET stock, June 1, 2018, TN5.** a) 1-minute. b) 10-minute. c) 60-minute. d) 360-minute.

## Appendix S: Proof by Consecutiveness

Using the same logical setup as in App. R, we will here propose a third protocol to prove that the characteristic figures drive the price. We will pick a qualifying chart (probability $\varepsilon$) and show the, say, again, 19 adjacent ones (that is, the consecutive charts from the same topological network). Because of $\varepsilon$, it should be unlikely that another one also qualifies. Yet, all the adjacent charts qualify. Since this comes with a probability equal to $\varepsilon^{20}$, this proves that the price bouncing from one characteristic figure to another cannot be fortuitous, and, therefore, that the characteristic figures drive the price. To make the demonstration even more convincing, we will pick one of the charts from App. Q and show the 19 preceding charts.

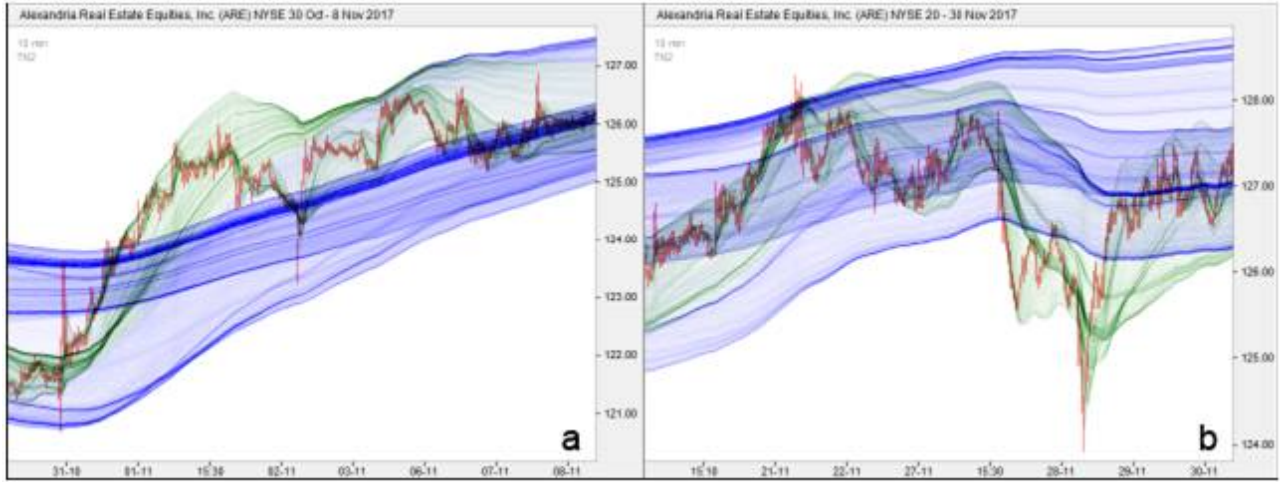

**Fig. S1. ARE stock, 10 minute, TN2.** a) Oct. 30 to Nov. 8, 2017. b) Nov. 8 to Nov. 20, 2017.

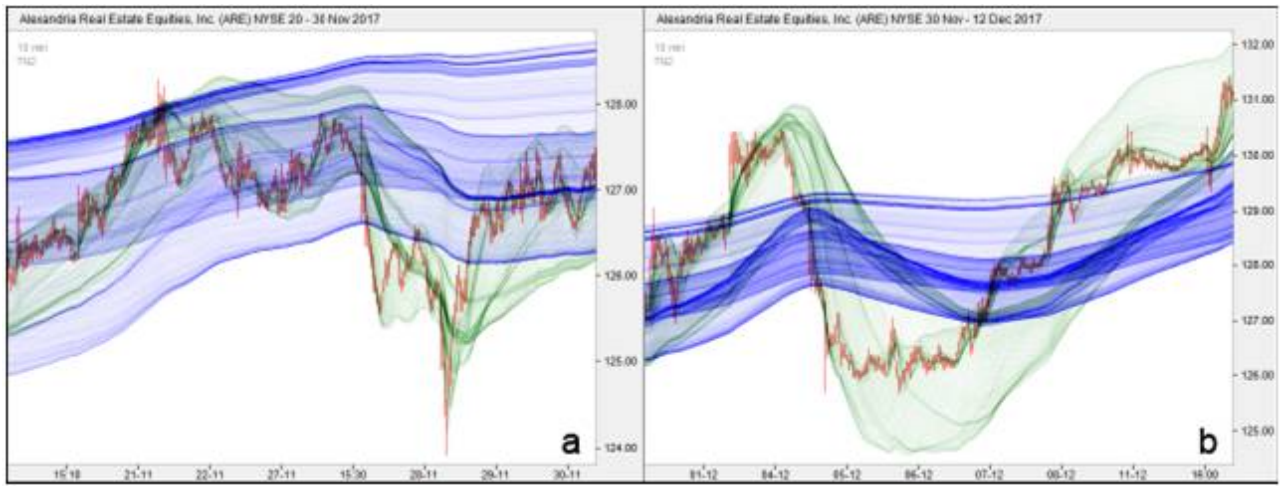

**Fig. S2. ARE stock, 10 minute, TN2.** a) Nov. 20 to Nov. 30, 2017. b) Nov. 30 to Dec. 12, 2017.

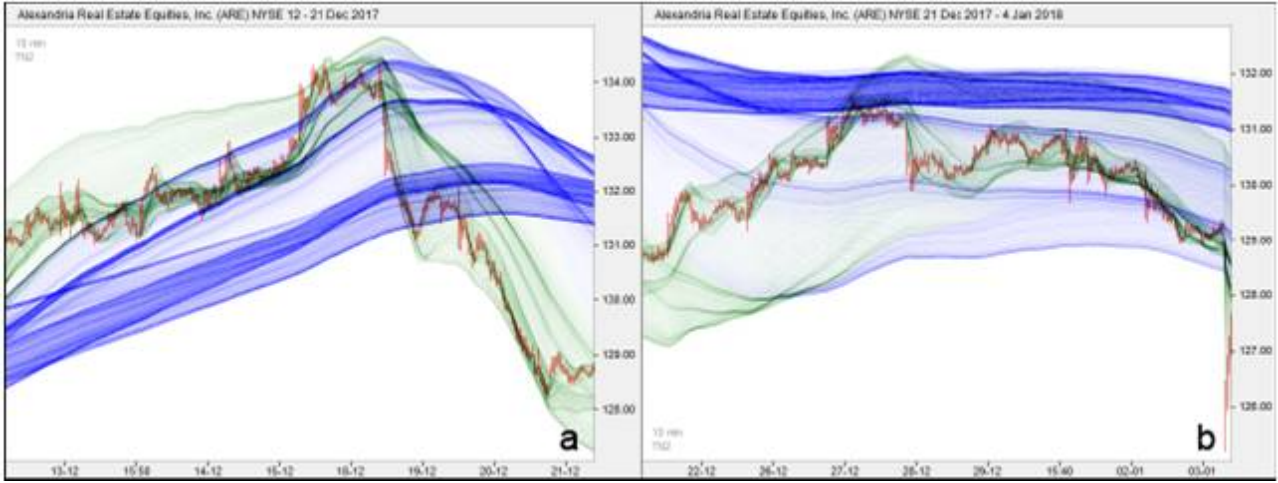

**Fig. S3. ARE stock, 10 minute, TN2.** a) Dec. 12 to Dec. 21, 2017. b) Dec. 21, 2017 to Jan. 4, 2018.

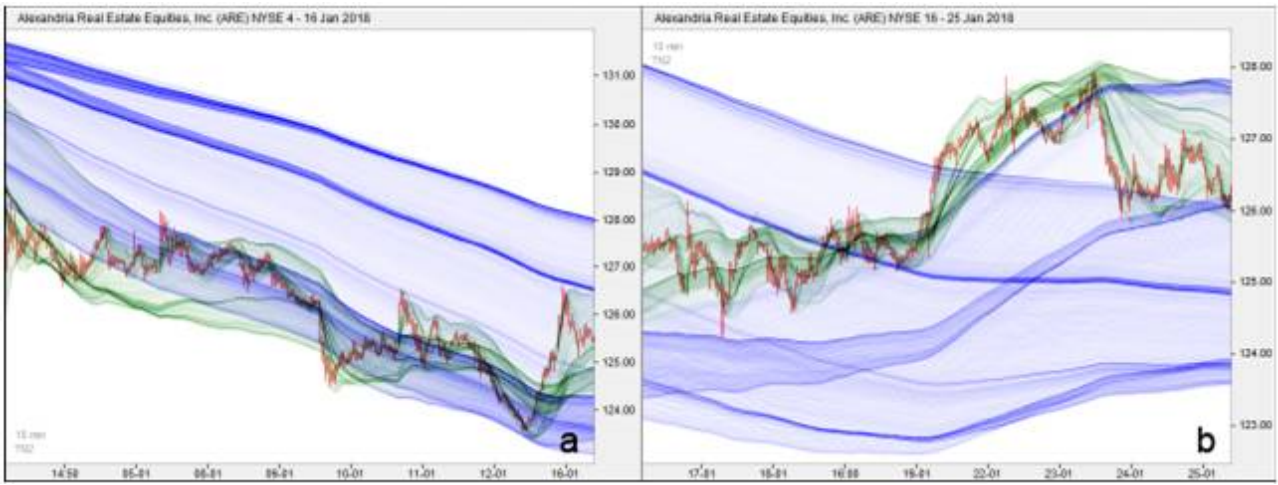

**Fig. S4. ARE stock, 10 minute, TN2.** a) Jan. 4 to Jan. 16, 2018. b) Jan. 16 to Jan. 25, 2018.

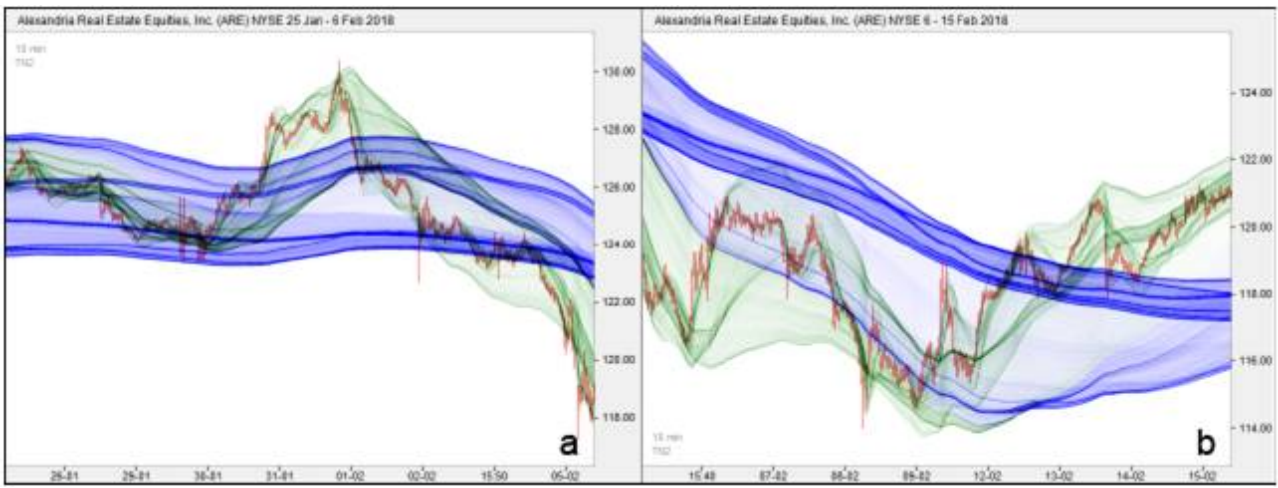

**Fig. S5. ARE stock, 10 minute, TN2.** a) Jan. 25 to Feb. 6, 2018. b) Feb. 6 to Feb. 15, 2018.

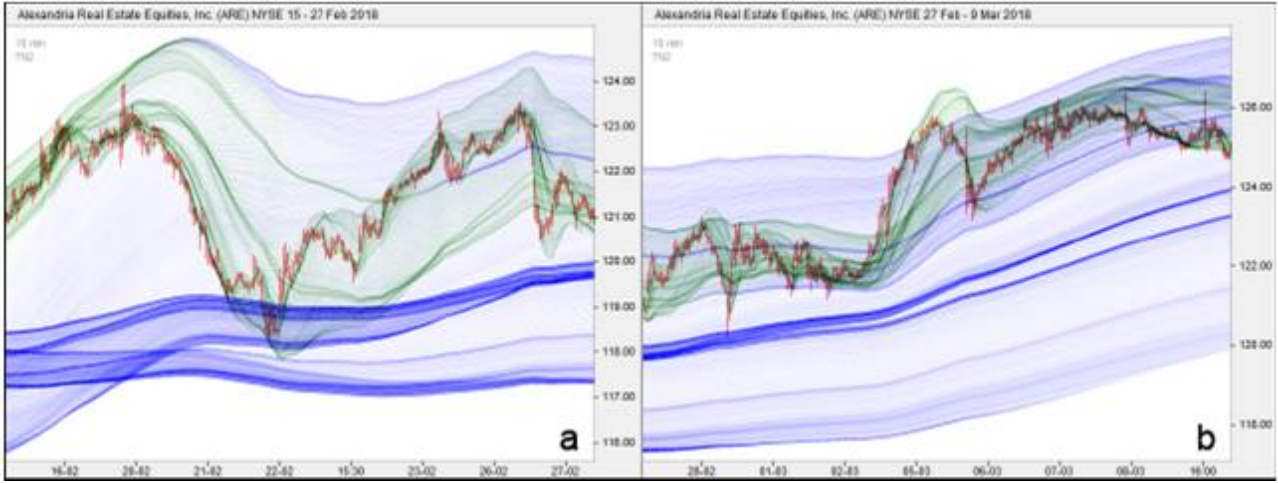

**Fig. S6. ARE stock, 10 minute, TN2.** a) Feb. 15 to Feb. 27, 2018. b) Feb. 27 to Mar. 9, 2018.

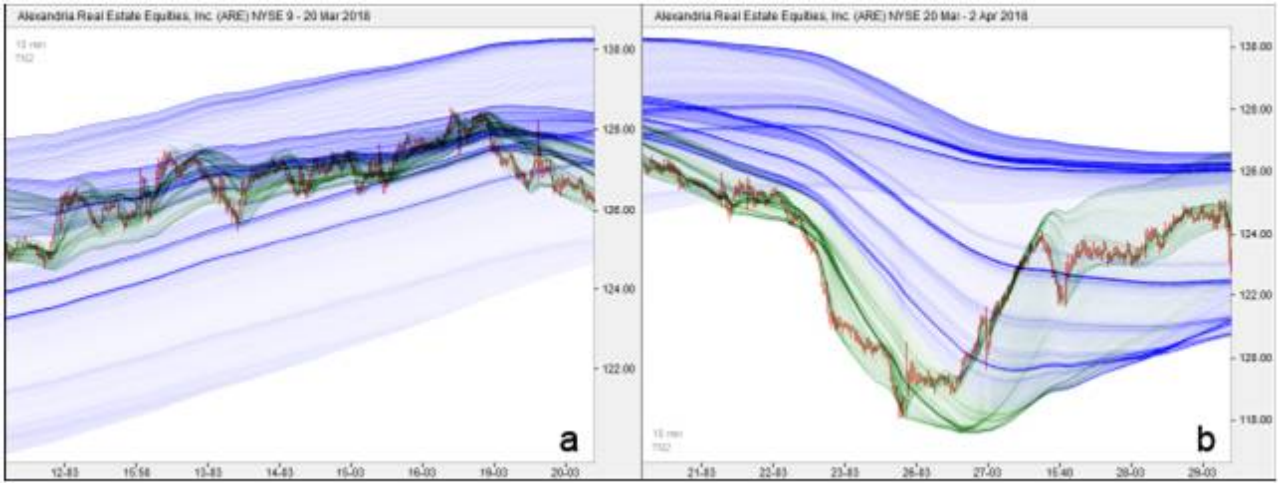

**Fig. S7. ARE stock, 10 minute, TN2.** a) Mar. 9 to Mar. 20, 2018. b) Mar. 20 to Apr. 2, 2018.

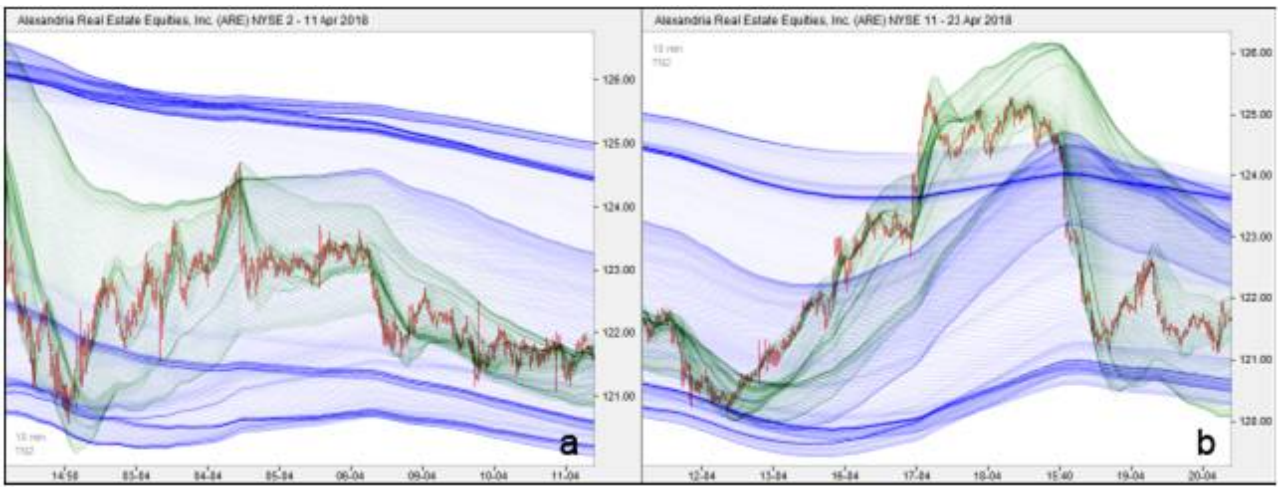

**Fig. S8. ARE stock, 10 minute, TN2.** a) Apr. 2 to Apr. 11, 2018. b) Apr. 11 to Apr. 23, 2018.

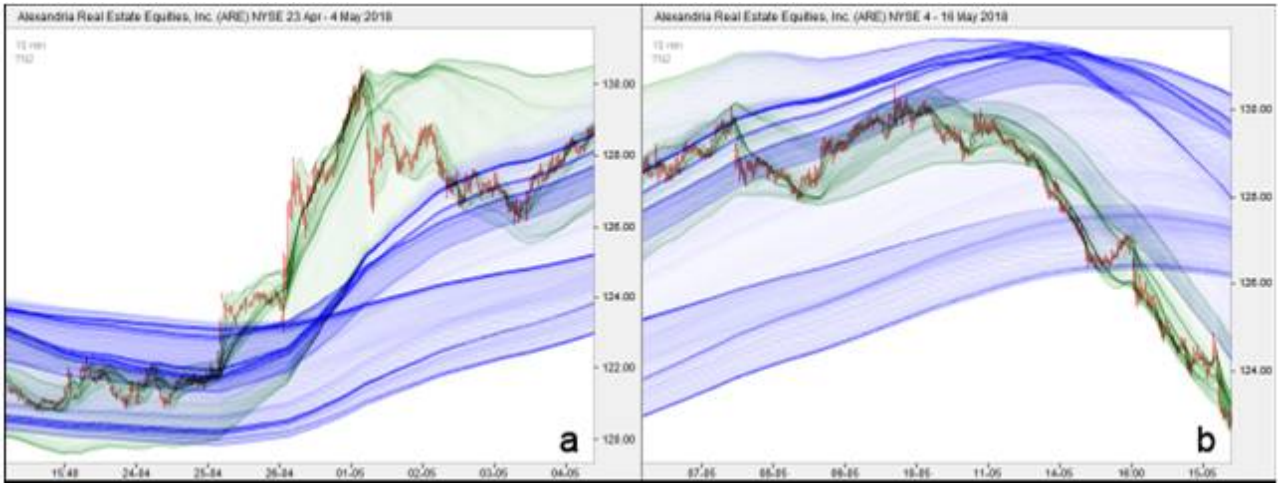
**Fig. S9. ARE stock, 10 minute, TN2.** a) Apr. 23 to May 4, 2018. b) May 4 to May 16, 2018.

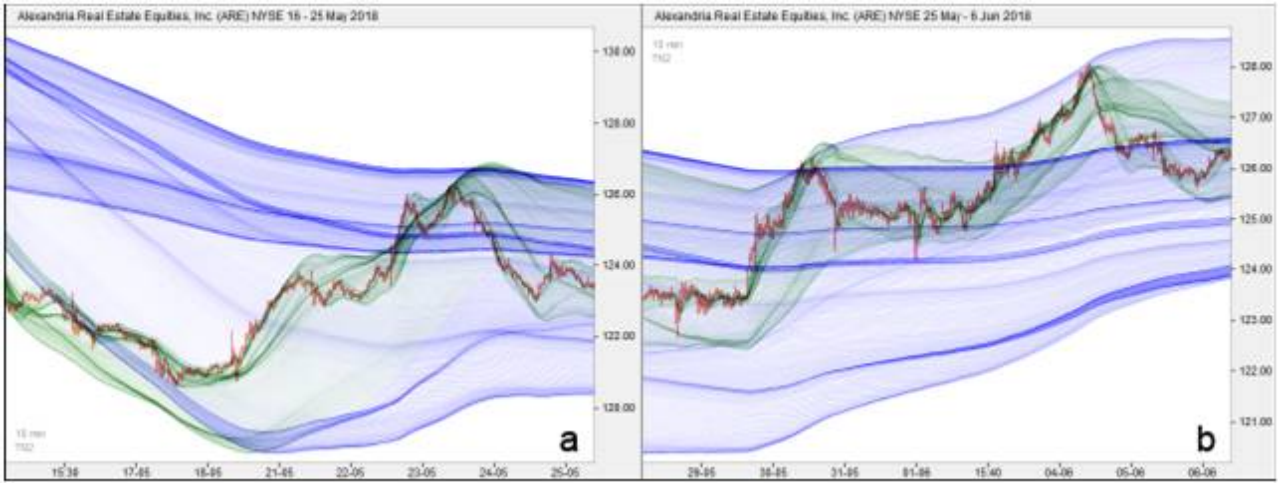
**Fig. S10. ARE stock, 10 minute, TN2.** a) May 16 to May 25, 2018. b) May 25 to June 6, 2018.